\documentclass[modern]{aastex62}
\usepackage[utf8]{inputenc}
\pdfoutput=1

\begin{document}
\title{{\bf Physical Conditions in Shocked Interstellar Gas Interacting with the Supernova Remnant IC~443}\footnote{Based on observations made with the NASA/ESA \emph{Hubble Space Telescope}, obtained from the Mikulski Archive for Space Telescopes (MAST). STScI is operated by the Association of Universities for Research in Astronomy, Inc., under NASA contract NAS5-26555.}}

\correspondingauthor{Adam M.~Ritchey}

\author{Adam M.~Ritchey}
\affiliation{Department of Astronomy, University of Washington, Seattle, WA 98195, USA}
\email{aritchey@astro.washington.edu}

\author{Edward B.~Jenkins}
\affiliation{Princeton University Observatory, Princeton, NJ 08544, USA}
%\email{ebj@astro.princeton.edu}

\author{S.~R.~Federman}
\affiliation{Department of Physics and Astronomy, University of Toledo, Toledo, OH 43606, USA}
%\email{steven.federman@utoledo.edu}

\author{Johnathan S.~Rice}
\affiliation{Department of Physics and Astronomy, University of Toledo, Toledo, OH 43606, USA}
%\email{johnathan.rice@utoledo.edu}

\author{Damiano Caprioli}
\affiliation{Department of Astronomy and Astrophysics, University of Chicago, Chicago, IL 60637, USA}
%\email{caprioli@uchicago.edu}

\author{George Wallerstein}
\affiliation{Department of Astronomy, University of Washington, Seattle, WA 98195, USA}
%\email{wall@astro.washington.edu}

\begin{abstract}
We present the results of a detailed investigation into the physical conditions in interstellar material interacting with the supernova remnant IC~443. Our analysis is based on a comprehensive examination of high-resolution far-ultraviolet spectra obtained with the Space Telescope Imaging Spectrograph onboard the \emph{Hubble Space Telescope} of two stars behind IC~443. One of our targets (HD~43582) probes gas along the entire line of sight through the supernova remnant, while the other (HD~254755) samples material located ahead of the primary supernova shock front. We identify low velocity quiescent gas in both directions and find that the densities and temperatures in these components are typical of diffuse atomic and molecular clouds. Numerous high velocity components are observed in the absorption profiles of neutral and singly-ionized atomic species toward HD~43582. These components exhibit a combination of greatly enhanced thermal pressures and significantly reduced dust-grain depletions. We interpret this material as cooling gas in a recombination zone far downstream from shocks driven into neutral gas clumps. The pressures derived for a group of ionized gas components at high positive velocity toward HD~43582 are lower than those of the other shocked components, pointing to pressure inhomogeneities across the remnant. A strong very high velocity component near $-$620~km~s$^{-1}$ is seen in the absorption profiles of highly-ionized species toward HD~43582. The velocity of this material is consistent with the range of shock velocities implied by observations of soft thermal X-ray emission from IC~443. Moderately high-velocity gas toward HD~254755 may represent shocked material from a separate foreground supernova remnant.
\end{abstract}


\section{INTRODUCTION\label{sec:intro}}
Supernova remnants (SNRs) are widely believed to be the sources responsible for the acceleration of Galactic cosmic rays (GCRs). This still unproven assertion was originally conceived on purely energetic grounds (Ginzburg \& Syrovatskii 1964) because SNRs were among the only known potential sources with the necessary amount of available energy. It was later shown that charged particles could theoretically be accelerated to high energies in SNR shock fronts through the process of diffusive shock acceleration (Bell 1978; Blandford \& Ostriker 1978). When relativistic particles (mainly protons) impinge on interstellar nuclei, the collisions produce neutral pions that quickly decay into $\gamma$-ray photons. Accordingly, the detection of high-energy $\gamma$-ray emission from SNRs has long been considered the most promising means of confirming the SNR paradigm for the origin of GCRs (e.g., Drury et al.~1994). Still, while many GeV and TeV $\gamma$-ray sources have been found to be associated with known SNRs (Acero et al.~2016; H.E.S.S.~Collaboration et al.~2018), it is usually difficult to distinguish between hadronic and leptonic scenarios for producing the $\gamma$-ray emission in individual cases. (Gamma radiation may also be produced in SNRs by bremsstrahlung and inverse Compton emission from relativistic electrons.)

While it has not yet been conclusively demonstrated that the population of Galactic SNRs can fully account for GCR production (Acero et al.~2016; H.E.S.S.~Collaboration et al.~2018), there are several individual remnants, including IC~443, W44, and Tycho's SNR, that are widely regarded as being hadronic cosmic-ray accelerators (Tavani et al.~2010; Abdo et al.~2010a, 2010b; Giuliani et al.~2011; Giordano et al.~2012; Morlino \& Caprioli 2012; Uchiyama et al.~2012; Ackermann et al.~2013). Using four years of data obtained with the Large Area Telescope (LAT) of the \emph{Fermi Gamma-ray Space Telescope}, Ackermann et al.~(2013) detected the characteristic ``pion-decay bump'' in the $\gamma$-ray spectra of IC~443 and W44, providing direct evidence of cosmic-ray acceleration in these sources. More recently, Jogler \& Funk (2016) presented evidence for the pion-decay feature in W51C using five years of \emph{Fermi}-LAT data. Pion-decay $\gamma$-ray emission is enhanced in the vicinity of SNRs interacting with molecular clouds (MCs) due to the increased frequency of collisions between shock-accelerated cosmic rays and interstellar nuclei. The middle-aged SNRs IC~443, W44, and W51C, along with others such as W28, have among the highest GeV luminosities of remnants classified in the first \emph{Fermi}-LAT catalog of SNRs (Acero et al.~2016). These objects provide the best examples of SNR-MC interactions, and have thus been frequently used as test cases for theoretical models and numerical simulations (Chevalier 1999; Uchiyama et al.~2010; Ohira et al.~2011; Tang \& Chevalier 2014, 2015; Lee et al.~2015; Zhang \& Chevalier 2019; Tang 2019).

The reliability of theoretical models of SNR-MC interactions, and their applicability in individual cases, depends on the accuracy of the adopted input parameters, such as the pre-shock and post-shock gas densities (in both the cloud and intercloud regions) and the shock velocities. Frequently, these parameters are derived from observations of atomic and molecular emission lines or from X-ray observations (see, e.g., Tang \& Chevalier 2014, 2015 and references therein). A complementary approach to studying the kinematics and physical conditions of gas in the vicinity of SNRs is to examine interstellar absorption lines seen in the spectra of background stars. Numerous UV absorption-line studies of stars in the Vela SNR (e.g., Jenkins et al.~1976, 1981, 1984; Jenkins \& Wallerstein 1995; Wallerstein et al.~1995; Jenkins et al.~1998; Slavin et al.~2004; Nichols \& Slavin 2004) have demonstrated the power of this technique. However, for SNRs that exhibit strong interactions with MCs, it can be difficult to find background targets that are bright enough for UV and visible absorption-line spectroscopy due to the typically heavy extinction along the lines of sight to potential targets. IC~443 is unique among the GeV-bright SNRs in that there are a number of early-type stars in its vicinity that can serve as background sources for absorption-line spectroscopy.

IC~443 is classified as a mixed-morphology SNR, meaning that it has a shell-like structure at radio wavelengths and a center-filled X-ray morphology (Rho \& Petre 1998). In the northeastern part of the remnant, the supernova shock appears to be encountering mostly atomic gas as evidenced by strong emission from atomic lines characteristic of post-shock recombining material (Fesen \& Kirshner 1980). Throughout the southern portion of the remnant, the shock is interacting with a dense foreground molecular cloud. Several shocked molecular clumps, which delineate the interaction region between the supernova shock and the molecular cloud, have been mapped in CO emission (Denoyer 1979; Huang et al.~1986; Dickman et al.~1992) and have been studied extensively using many different molecular species as tracers (e.g., van Dishoeck et al.~1993; Snell et al.~2005; Reach et al.~2019). The shocked molecular ridge has also been mapped using vibrationally-excited H$_2$ emission (Burton et al.~1988). Shocked H~{\sc i} filaments are also observed and are found to be well correlated with the shocked molecular material in the south and with the bright optical filaments in the northeast (Braun \& Strom 1986a; Lee et al.~2008). The composite Two Micron All Sky Survey (2MASS) image of IC~443 in the $J$, $H$, and $K$ bands from Rho et al.~(2001) clearly illustrates the distinction between shocked atomic gas along the northeastern rim and shocked molecular gas along the southern ridge.

Thermal X-ray emission from IC~443 has generally been described using a two-component ionization equilibrium model with temperatures of $\sim$0.2--0.3~keV for the cold component and $\sim$1.0~keV for the hot component (Petre et al.~1988; Asaoka \& Aschenbach 1994; Troja et al.~2006). If the hot component represents gas in the immediate post-shock region, then the plasma temperature of $\sim$$1.2\times10^7$~K would imply a shock velocity of $\sim$900~km~s$^{-1}$. This velocity is much larger than the value of 65--100~km~s$^{-1}$ derived from shock models applied to spectroscopic observations of the bright optical filaments (Fesen \& Kirshner 1980). A combination of 100~km~s$^{-1}$ J-type shocks and 12--25~km~s$^{-1}$ J or C-type shocks were invoked to account for the molecular emission observed in the shocked molecular clumps (Snell et al.~2005). These facts suggest a scenario wherein the expanding supernova blast wave is encountering a clumpy interstellar medium (ISM). As individual gas clumps are overtaken by the blast wave, secondary shocks are driven into the clumps at velocities much lower than that of the primary shock due to the increase in density. Alternatively, if the primary shock velocity is currently $\sim$100~km~s$^{-1}$, then a model involving heat conduction in the hot interior of the remnant may need to be invoked to account for the X-ray observations (Chevalier 1999).

\begin{figure}[!t]
\centering
\includegraphics[width=0.8\textwidth]{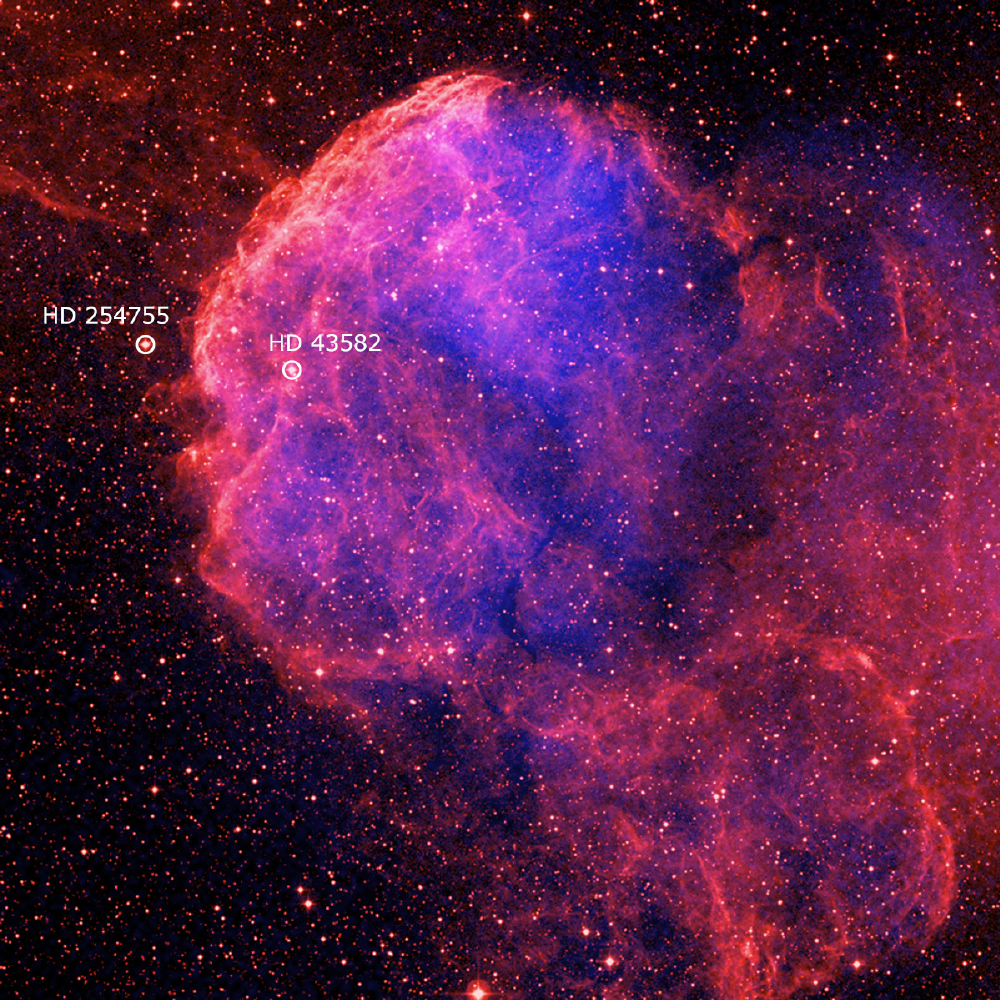}
\caption{Composite image of IC 443. The optical image from the Digitized Sky Survey (red) is shown along with X-ray data (blue) from the \emph{Chandra X-Ray Observatory} (Gaensler et al.~2006) and \emph{ROSAT} (Asaoka \& Aschenbach 1994). The stars targeted for \emph{HST} observations are labeled.\label{fig:image}}
\end{figure}

The first UV absorption-line measurements of gas associated with IC~443 were reported by Gondhalekar \& Phillips (1980), who presented \emph{International Ultraviolet Explorer} (\emph{IUE}) spectra of the star HD~43582, which lies behind one of the bright optical filaments in the northeastern part of the remnant (see Figure~\ref{fig:image}). An analysis of the \emph{IUE} spectrum revealed Fe~{\sc ii} and Mg~{\sc ii} absorption components with (heliocentric) radial velocities as high as +230~km~s$^{-1}$. Welsh \& Sallmen (2003) obtained high-resolution ground-based spectra of three stars probing IC~443 (HD~43582, HD~254577, and HD~254755) using the 2.7 m telescope at McDonald Observatory. Numerous high-velocity Na~{\sc i} and Ca~{\sc ii} absorption components were detected toward HD~43582 and HD~254577 at velocities ranging from $-$97 to +52~km~s$^{-1}$. The spectra obtained by Welsh \& Sallmen (2003) did not reveal any high-velocity absorption toward HD~254755. However, this star lies just outside the optical nebulosity associated with IC~443 (Figure~\ref{fig:image}). More extensive ground-based observations were reported by Hirschauer et al.~(2009), who obtained moderate-resolution spectra of 11 stars near IC~443 with the 3.5 m telescope at Apache Point Observatory (APO). The APO data, acquired using the Astrophysical Research Consortium echelle spectrograph (ARCES), revealed absorption from many different atomic and molecular species over a range of velocities, allowing Hirschauer et al.~(2009) to constrain the physical conditions in the diffuse molecular gas along the various lines of sight. Indriolo et al.~(2010) adopted the Hirschauer et al.~(2009) results in their analysis of cosmic-ray ionization rates inferred from infrared observations of H$_3^+$. The ionization rates for two lines of sight through IC~443 were found to be $\zeta_2\approx2\times10^{-15}\,{\rm s}^{-1}$, or about five times the rate typically found along diffuse molecular cloud sight lines (e.g., Indriolo et al.~2007; Indriolo \& McCall 2012).

Taylor et al.~(2012) obtained very high signal-to-noise (S/N) ratio ground-based spectra of four stars in IC~443 (HD~43582, HD~254477, HD~254577, and HD~254755) using the High Resolution Spectrograph (HRS) of the 9.2 m Hobby-Eberly Telescope (HET) at McDonald Observatory. Their purpose was to study lithium isotope ratios in the diffuse molecular gas in the vicinity of the remnant using the very weak Li~{\sc i}~$\lambda6707$ feature. Taylor et al.~(2012) found evidence of a reduced $^7$Li/$^6$Li ratio in regions more strongly affected by interactions between shock-accelerated particles and molecular gas, a sign of recent Li production by cosmic rays (Ramaty et al.~1997; Lemoine et al.~1998). In the present investigation, we seek to more fully characterize the physical conditions in the shocked interstellar gas interacting with IC~443. To this end, we examine high-resolution far-UV spectra of HD~43582 and HD~254755 obtained using the Space Telescope Imaging Spectrograph (STIS) onboard the \emph{Hubble Space Telescope} (\emph{HST}). The positions of the target stars in relation to the optical and X-ray emission from IC~443 are shown in Figure~\ref{fig:image}. While the line of sight to HD~43582 clearly penetrates the interior region of the SNR, positional and kinematic evidence indicates that the supernova blast wave has not yet reached the bulk of the gas in front of HD~254755 (Welsh \& Sallmen 2003; Hirschauer et al.~2009; Figure~\ref{fig:image}). Thus, by analyzing the absorption features detected along these two closely-spaced sight lines, we are able to evaluate the physical conditions in interstellar material positioned both ahead of and behind the supernova shock front.

A major component of our investigation is an analysis of absorption from collisionally-excited fine-structure levels in C~{\sc i}, O~{\sc i}, and Si~{\sc ii}, which allows us to determine the gas densities, kinetic temperatures, thermal pressures, and ionization fractions in the predominantly neutral gas clumps seen toward HD~43582 and HD~254755. However, our high-resolution \emph{HST}/STIS spectra cover a wide variety of atomic and molecular absorption features, which enable a comprehensive examination of the physical, chemical, and ionization state of the gas detected in the two directions. A re-examination of the HET/HRS data obtained by Taylor et al.~(2012), which cover additional atomic and molecular species, yields a more complete picture of the interaction between the SNR and the surrounding ISM. In Section~\ref{sec:observations}, we describe the \emph{HST}/STIS and HET/HRS observations in more detail and outline the procedures used to process the data. In Section~\ref{sec:profiles}, a profile synthesis routine is used to analyze the absorption profiles of the many different atomic and molecular species that are relevant to our investigation. Various probes of the physical conditions along our target sight lines are described in Section~\ref{sec:conditions}. We comment on the age and distance of IC~443 in Section~\ref{subsec:age_dist}, and discuss the implications of our derivations of physical conditions in Section~\ref{subsec:phys_cond}. Our main conclusions are summarized in Section~\ref{sec:conclusions}. In Appendix~\ref{sec:oi_fine_structure}, we describe in detail our routine that calculates the relative populations of the three O~{\sc i} fine-structure levels for a range of different physical conditions. In Appendix~\ref{sec:temp_var}, we present evidence of time-variable Na~{\sc i} absorption discovered toward HD~43582. The HET/HRS spectra of HD~254477 and HD~254577, which were not observed with \emph{HST}/STIS, are presented in Appendix~\ref{sec:het_profiles}. Detailed profile fitting results are provided in Appendix~\ref{sec:comp_structures}.

\section{OBSERVATIONS AND DATA PROCESSING\label{sec:observations}}

\subsection{\emph{HST}/STIS Observations\label{subsec:hst}}
High-resolution \emph{HST}/STIS spectra of HD~43582 and HD~254755 were acquired under program GO 13709, which was awarded 10 orbits of \emph{HST} observing time in Cycle 22. The data were obtained in 2015 March. Each star was observed employing the E140H grating and a central wavelength setting of 1307~\AA{}, which provides continuous wavelength coverage in the range 1200--1397~\AA{}. The $0.2\times0.09$ arcsec slit yielded a resolving power of $R\approx108,000$ ($\Delta v\approx2.8$~km~s$^{-1}$), while the total exposure times ($\sim$13,000~s per target) resulted in S/N ratios (per pixel) near 1307~\AA{} of $\sim$40 toward HD~43582 and $\sim$15 toward HD~254755. (The latter is the more heavily-reddened of the two stars and is also somewhat fainter, which accounts for the lower S/N achieved.) Additional details concerning the stellar targets and the STIS observations are presented in Table~\ref{tab:observations}.

The raw exposures (five per target) were processed using the CALSTIS pipeline and downloaded via the Mikulski Archive for Space Telescopes (MAST). Upon initial inspection of the pipeline-produced 1D extracted spectra, an issue with the subtraction of scattered light was identified for the first exposure obtained of HD~254755. This was the shortest exposure of the five-orbit visit for HD~254755, and thus had the lowest overall S/N. The scattered light subtraction algorithm within CALSTIS was producing negative flux values for portions of some echelle orders shortward of 1300~\AA{}. A small shift along the echelle dispersion axis applied to the flat-fielded image before the subtraction of scattered light sufficed to resolve the issue. The resulting 1D extracted spectra were then found to be consistent with the four subsequent exposures of this target.

\begin{deluxetable}{lccccccccc}
\tablecolumns{10}
\tabletypesize{\footnotesize}
\tablecaption{Summary of \emph{HST}/STIS Observations\label{tab:observations}}
\tablehead{ \colhead{Star} & \colhead{Sp.~Type} & \colhead{$V$} & \colhead{$E$($B$$-$$V$)} & \colhead{$d$} & \colhead{Dataset} & \colhead{Exp.~Time} & \colhead{Cen.~Wave.} & \colhead{Grating} & \colhead{Slit} \\
\colhead{} & \colhead{} & \colhead{(mag)} & \colhead{(mag)} & \colhead{(kpc)} & \colhead{} & \colhead{(s)} & \colhead{(\AA)} & \colhead{} & \colhead{(arcsec)} }
\startdata
HD~43582 & B0 IIIn & 8.79 & 0.55 & 2.0 & ocki01 & 12807 & 1307 & E140H & $0.2\times0.09$ \\
HD~254755 & O9 Vp & 8.91 & 0.70 & 3.6 & ocki02 & 12807 & 1307 & E140H & $0.2\times0.09$ \\
\enddata
\end{deluxetable}

The individual STIS echelle orders within each observation were merged together and the five separate exposures of each target were co-added to produce a single high-quality STIS spectrum for each star. Throughout this process, individual flux measurements were assigned relative weights according to the inverse squares of their respective uncertainties. These uncertainties were obtained from the (smoothed) error arrays supplied by the CALSTIS routine. (The smoothing was done so as to avoid noise bias wherein positive deviations get less representation than negative ones.) The final flux errors for the composite spectra $\sigma_F$ were calculated from the relation  $\sigma_F=(\sum \sigma^{-2})^{-0.5}$. Before co-adding separate exposures of the same target, the velocity registration of each exposure was compared to that of the others to ensure that the final spectrum would not suffer any degradation in resolution. Finally, small spectral segments surrounding interstellar lines of interest were cut from the composite spectra and normalized via low-order polynomial fits to regions free of interstellar absorption. The many different atomic and molecular absorption features analyzed in this investigation are described in more detail in Section~\ref{sec:profiles}.

\subsection{Auxiliary HET Data\label{subsec:het}}
To supplement our far-UV \emph{HST}/STIS spectra of HD~43582 and HD~254755, we re-examined the very high S/N ratio ground-based spectra that Taylor et al.~(2012) obtained of these stars using the HET/HRS at McDonald Observatory. Since the HET spectra were acquired to study lithium isotope ratios in IC~443, using the very weak Li~{\sc i}~$\lambda6707$ feature, they are characterized by high resolution ($R\approx98,000$) and very high S/N ($\sim$750 per pixel near 6707~\AA{}). Taylor et al.~(2012) describe analyses of the Li~{\sc i}~$\lambda6707$, K~{\sc i}~$\lambda7698$, and CH~$\lambda4300$ features toward HD~43582 and HD~254755 (and also toward HD~254477 and HD~254577). However, the HET observations cover other important interstellar absorption lines, such as Ca~{\sc i}~$\lambda4226$, Ca~{\sc ii}~$\lambda\lambda3933,3968$, Na~{\sc i}~$\lambda\lambda5889,5895$, and CH$^+$~$\lambda4232$. The diffuse interstellar bands (DIBs) at 5780.5 and 5797.1~\AA{} are also covered by the HET data, as is the $A$$-$$X$ (3$-$0) band of C$_2$ near 7719~\AA{}. All of these absorption features are included in our analysis (see Section~\ref{sec:profiles}).

The basic procedures used to reduce the HET observations are described in Taylor et al.~(2012). Those authors employed standard IRAF routines for bias correction, cosmic-ray removal, scattered-light subtraction, flat-fielding, 1D spectral extraction, and wavelength calibration. An additional step employed here is the removal of weak atmospheric absorption lines near the interstellar Na~{\sc i}~$\lambda\lambda5889,5895$ and K~{\sc i}~$\lambda7698$ features. Templates for the atmospheric absorption near the Na~{\sc i} and K~{\sc i} lines were created from HET/HRS observations of the star HD~281159, which, unlike the stars in IC~443, exhibits only very narrow interstellar Na~{\sc i} and K~{\sc i} absorption profiles.\footnote{The HET observations of HD~281159 were obtained between 2012 August and October as part of a separate investigation designed to measure lithium isotope ratios in diffuse molecular clouds (Knauth et al.~2017).} For each of the individual HET exposures of the IC~443 stars (shifted to the geocentric frame of reference), the atmospheric templates were scaled so as to completely remove the telluric features near the interstellar Na~{\sc i} and K~{\sc i} lines and then divided into the science exposure. After correcting for telluric absorption, the individual exposures were Doppler-corrected to the reference frame of the local standard of rest (LSR) and co-added to produce very high S/N ratio summed spectra. (Later these spectra were shifted to the heliocentric frame of reference for comparison with the \emph{HST} data.) All atomic and molecular absorption lines of interest to our investigation were then normalized in the same manner as that described above for the UV lines.

It is important to note that the HET observations of HD~43582 and HD~254755 were acquired between 2008 December and 2010 March and so were not obtained contemporaneously with the \emph{HST} spectra. This could present a problem when comparing the HET and \emph{HST} data since sight lines probing supernova remnants have often been found to exhibit temporal variations in their interstellar absorption profiles (Hobbs et al.~1982, 1991; Danks \& Sembach 1995; Cha \& Sembach 2000; Welty et al.~2008; Rao et al.~2016, 2017; Dirks \& Meyer 2016). Prior to this study, there had not been any reports of temporal variability for sight lines in IC~443. In Appendix~\ref{sec:temp_var}, we take advantage of the fact that the HET observations of HD~43582 were obtained (in service mode) over the course of several months to search for temporal changes in the Na~{\sc i} absorption profiles. We focus on the Na~{\sc i}~$\lambda\lambda5889,5895$ lines because these transitions are strong enough to reveal high velocity gas components and are located in a region of the spectrum characterized by high S/N. While the Ca~{\sc ii}~$\lambda\lambda3933,3968$ absorption features also show numerous high velocity components, the poor S/N below 4000~\AA{} for individual HET exposures limits our ability to identify with confidence any temporal changes in the Ca~{\sc ii} profiles.

\section{RESULTS OF PROFILE SYNTHESIS\label{sec:profiles}}
The combination of far-UV \emph{HST}/STIS spectra and ground-based HET data covering much of the visible spectrum provides us with a wealth of information on the physical, chemical, and ionization state of the gas seen in absorption toward our target stars. The \emph{HST} and HET data together cover numerous transitions of neutral and singly-ionized atomic species, which yield information on the gas-phase abundances, depletions, and physical conditions in the various absorption components detected in the two directions. An examination of the fine-structure excitations of C~{\sc i}, O~{\sc i}, and Si~{\sc ii}, in particular, allows us to derive estimates for the gas densities, kinetic temperatures, and electron fractions in the (mostly) neutral atomic components (Sections~\ref{subsec:oxy_sil_ex} and \ref{subsec:car_ex}). Absorption from multiple bands of CO and C$_2$ at UV and visible wavelengths enables an analysis of molecular excitation and diffuse cloud chemistry, which places constraints on the physical conditions in the molecular components (Section~\ref{subsec:mol_ex}). Finally, the detection of absorption from the highly-ionized species Si~{\sc iii}, Si~{\sc iv}, and N~{\sc v} toward HD~43582 provides information on the kinematics and temperature of the post-shock ionized gas interior to the supernova remnant. Gas densities and thermal pressures for these more highly-ionized components are obtained from an analysis of the fine-structure excitations of C~{\sc ii} (Section~\ref{subsec:car2_ex}).

In this section, we discuss our analyses of the many different absorption profiles relevant to our investigation of gas associated with IC~443. To decompose the absorption profiles into individual components, we employ the multi-component Voigt profile fitting routine ISMOD (see, e.g., Sheffer et al.~2008), which determines best-fitting values for the column density $N$, Doppler $b$-value, and radial velocity $v$ of each component included in the fit through a root mean square (rms) minimizing procedure. Most of the atomic and molecular absorption lines were fitted individually. However, in some cases, more complicated simultaneous fits were required to disentangle complex blended absorption features or to derive consistent results from many different lines or bands of the same species. Each of these cases is discussed in more detail in the following sections.

\subsection{Neutral and Singly-Ionized Atomic Species\label{subsec:atomic}}
\begin{figure}[!t]
\centering
\includegraphics[width=0.49\textwidth]{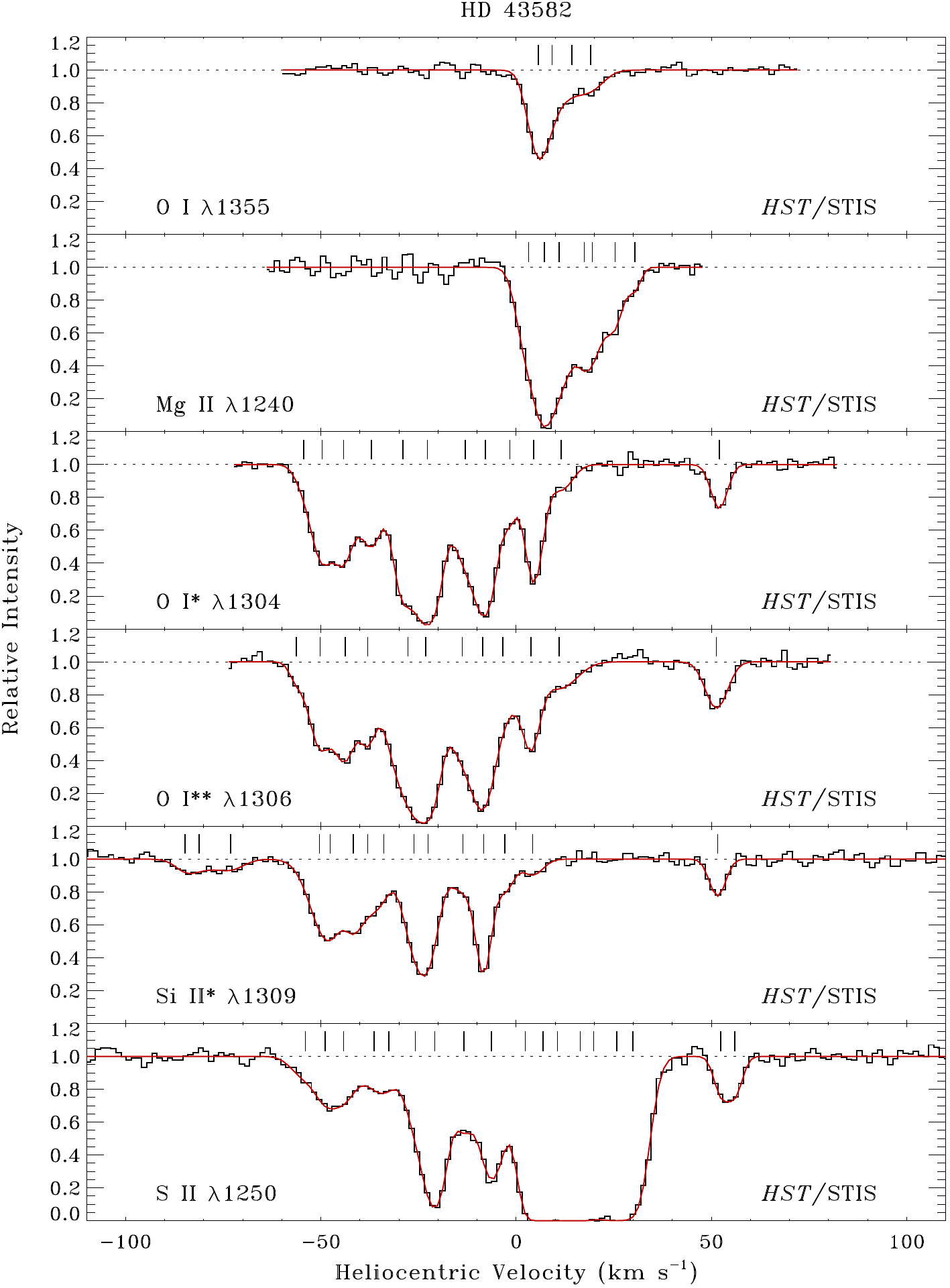}
\includegraphics[width=0.49\textwidth]{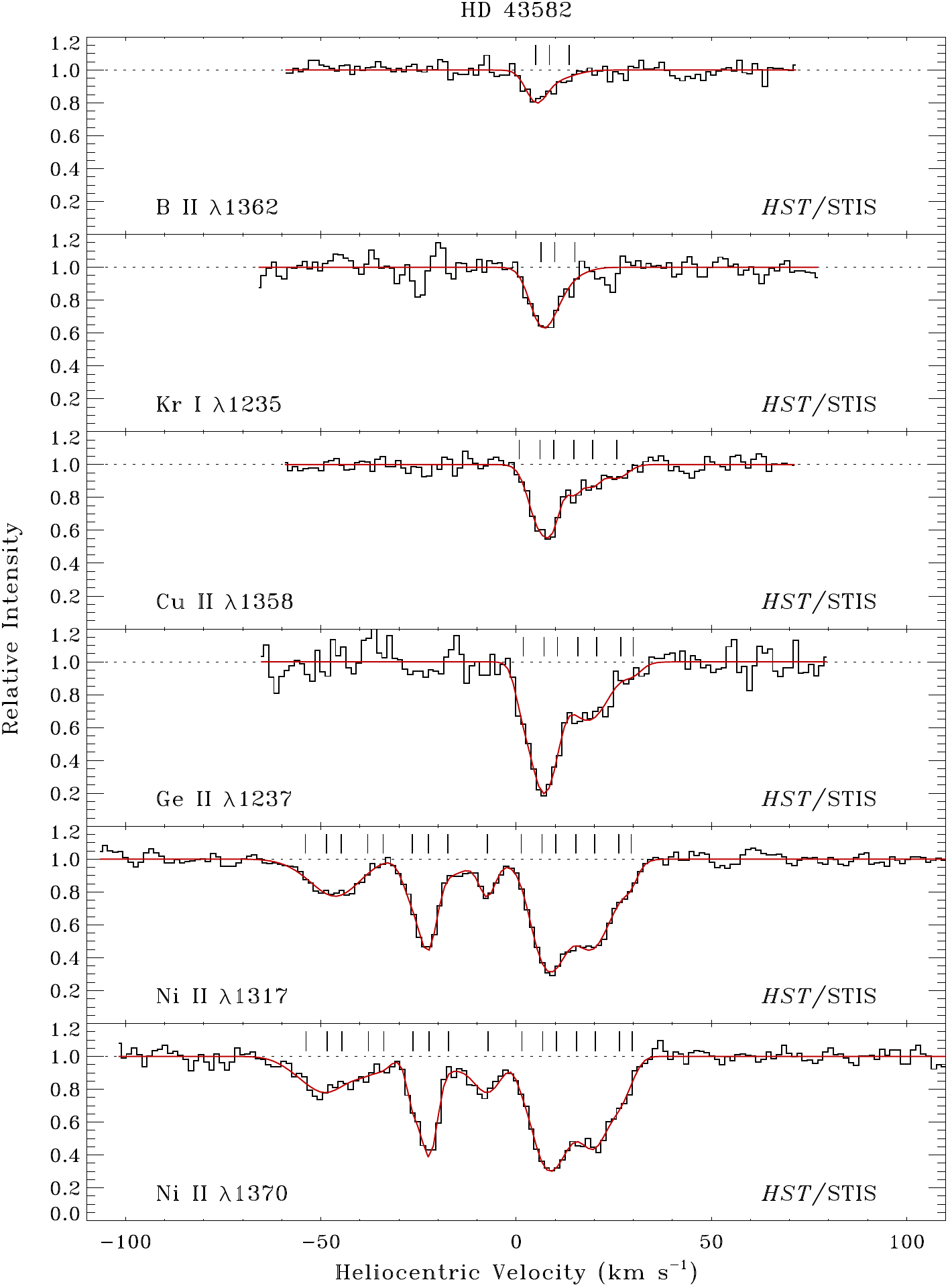}
\caption{Absorption profiles of neutral and singly-ionized atomic species (dominant ions) toward HD~43582 from high-resolution \emph{HST}/STIS spectra. The smooth red curves represent multi-component profile synthesis fits to the observed spectra shown as histograms. Tick marks indicate the positions of the velocity components included in the fits.\label{fig:dominant1}}
\end{figure}

The preferred stage of ionization for atoms within neutral interstellar clouds subject to the average Galactic UV radiation field is generally a function of whether the ionization potential of a given stage is greater than or less than that of neutral hydrogen. Within gas clouds that are interacting with a supernova remnant, enhanced UV radiation from shock fronts may raise the ionization level to stages above the preferred ones (e.g., Jenkins et al.~1998). However, we expect that even in these circumstances the typically dominant atomic species will be useful tracers of gas-phase abundances and physical conditions. Our \emph{HST}/STIS spectra provide access to numerous absorption features from dominant ions, such as O~{\sc i}, Mg~{\sc ii}, Si~{\sc ii}, S~{\sc ii} and Ni~{\sc ii}, as well as from trace neutral species, such as C~{\sc i}, S~{\sc i}, and Cl~{\sc i}, which can help to provide further constraints on physical conditions. The HET data yield information on the trace neutral and singly-ionized species  Na~{\sc i}, K~{\sc i}, Ca~{\sc i}, and Ca~{\sc ii}. The absorption profiles for all of the neutral and singly-ionized atomic species observed toward HD~43582 and HD~254755 (see Figures~\ref{fig:dominant1}--\ref{fig:trace2}) show distinct similarities, but also important differences. We have generally tried to analyze these species concurrently so as to derive consistent component structures among the various absorption profiles.

\subsubsection{Component Structures and Column Densities\label{subsubsec:components}}
\begin{figure}[!t]
\centering
\includegraphics[width=0.49\textwidth]{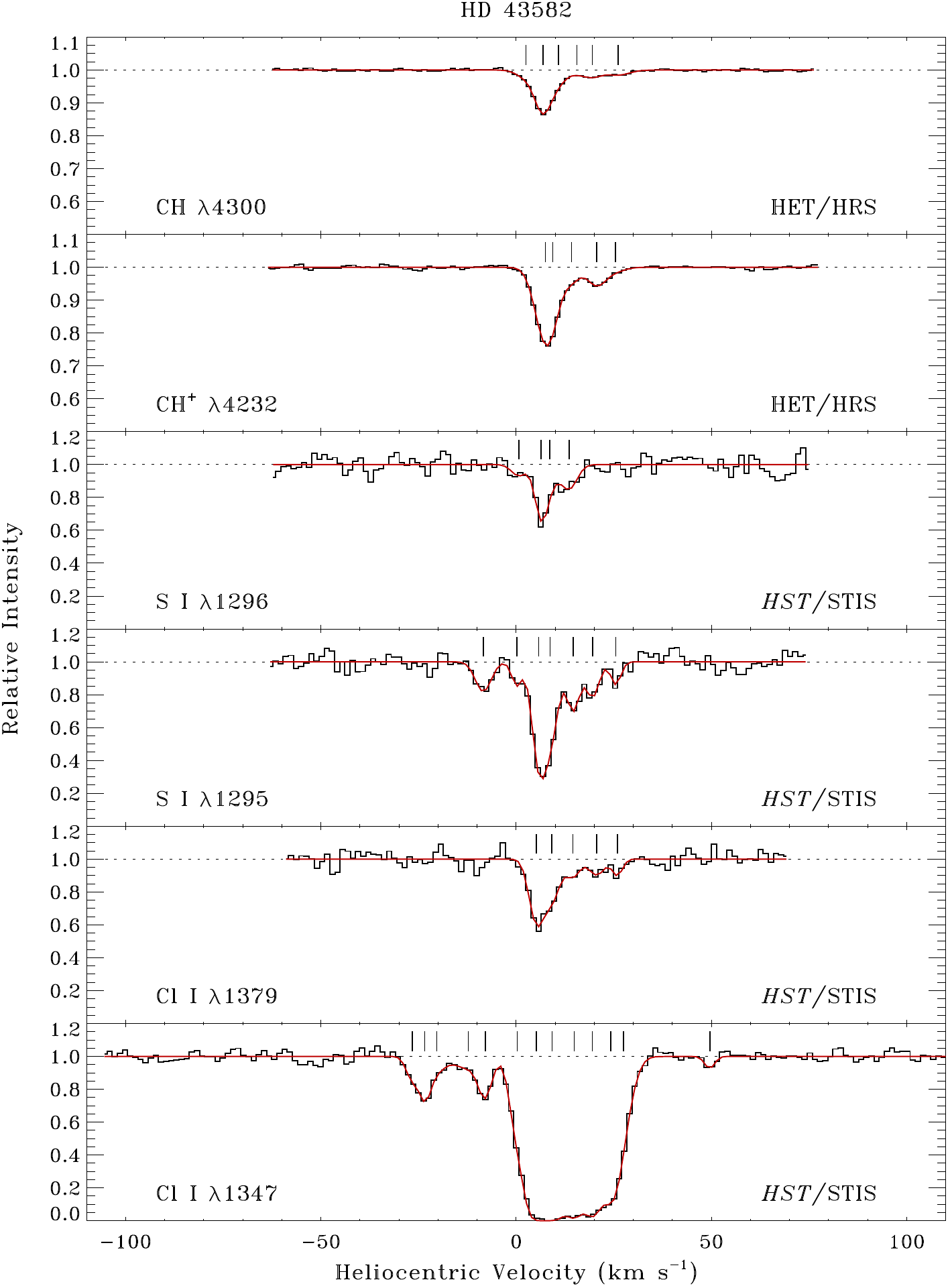}
\includegraphics[width=0.49\textwidth]{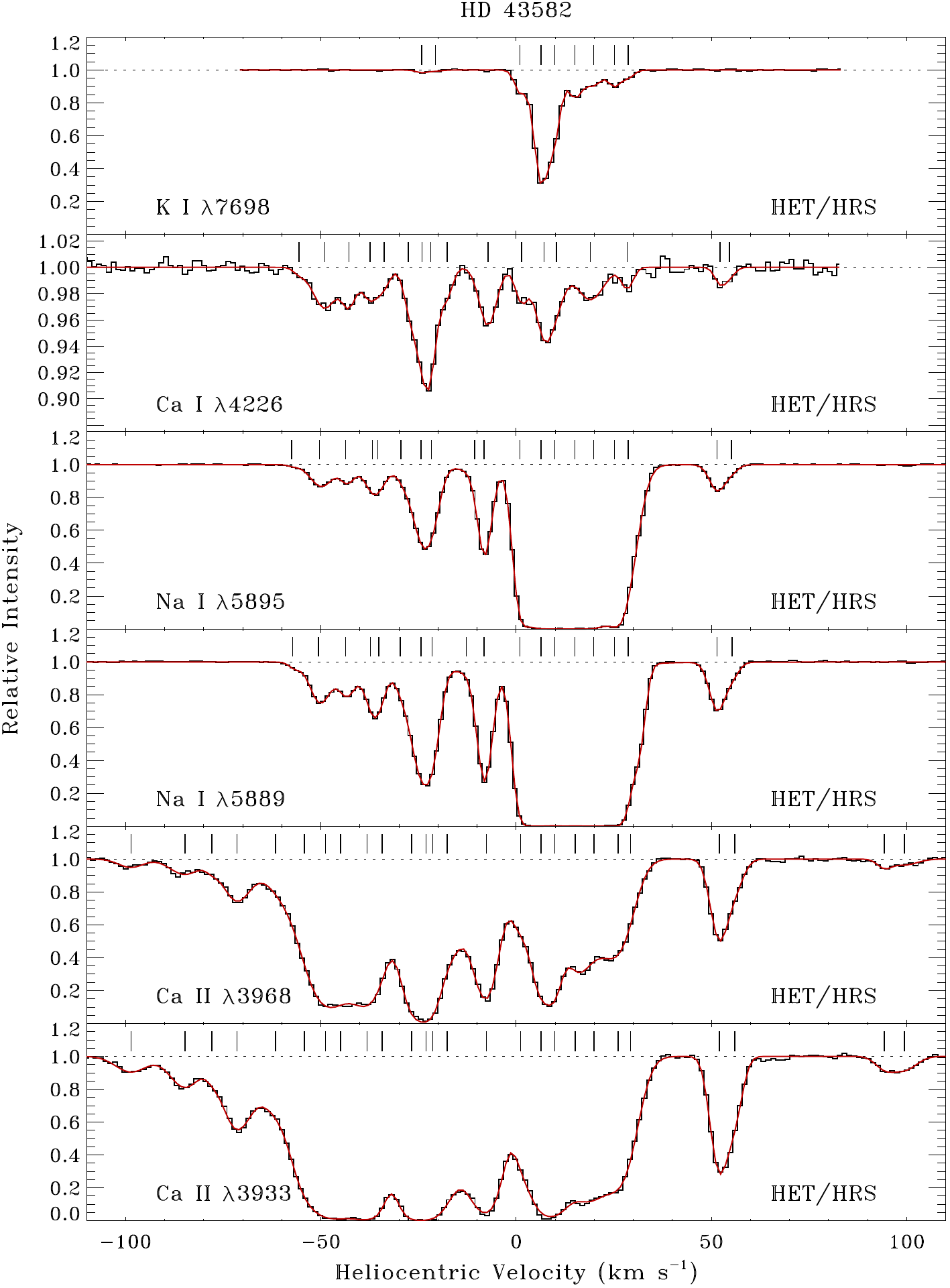}
\caption{Absorption profiles of trace neutral and singly-ionized atomic species, and the molecular species CH and CH$^+$, toward HD~43582 from high-resolution \emph{HST}/STIS spectra and from ground-based data acquired with the HET/HRS. The smooth red curves represent multi-component profile synthesis fits to the observed spectra shown as histograms. Tick marks indicate the positions of the velocity components included in the fits.\label{fig:trace1}}
\end{figure}

\begin{figure}[!t]
\centering
\includegraphics[width=0.49\textwidth]{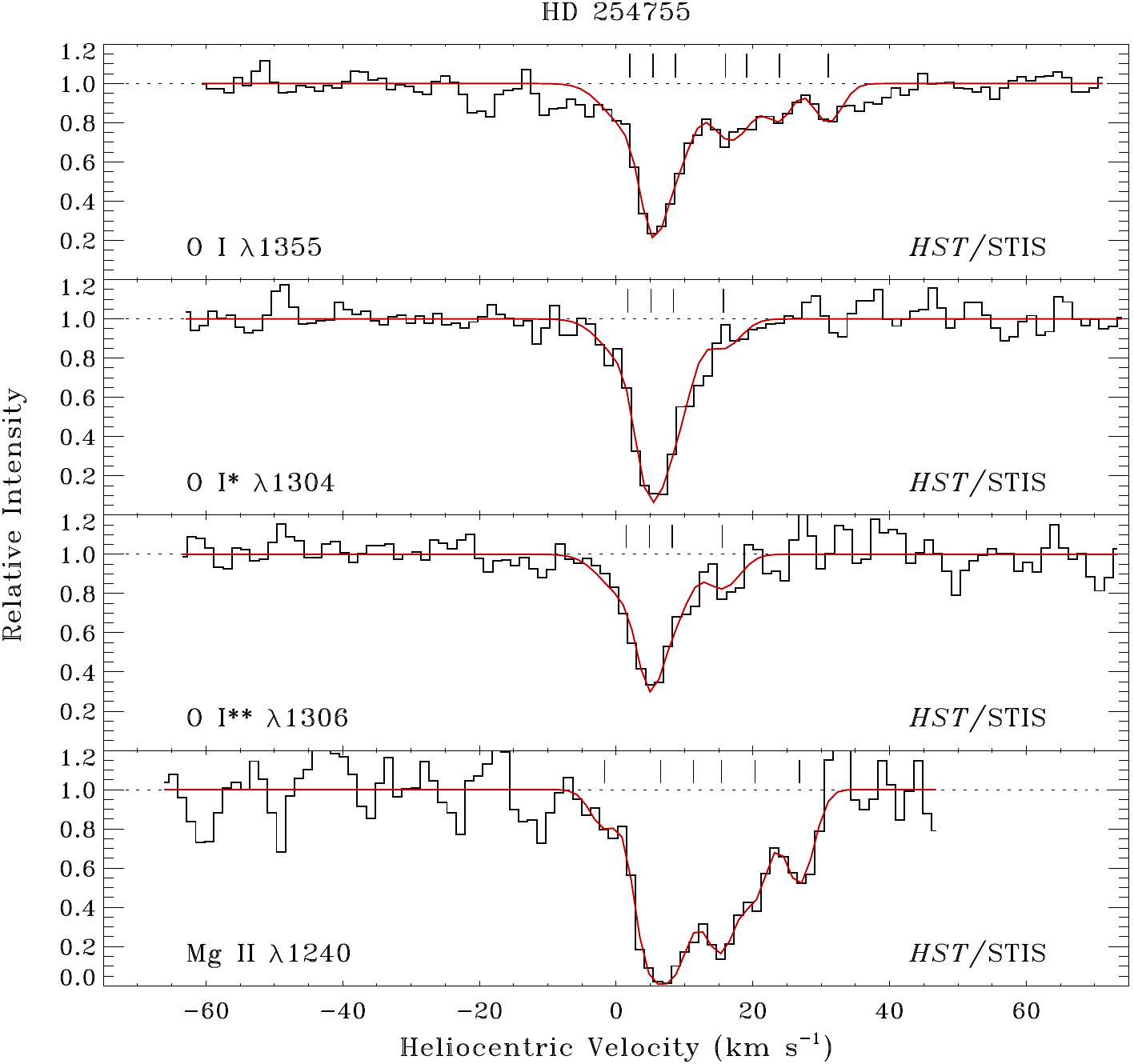}
\includegraphics[width=0.49\textwidth]{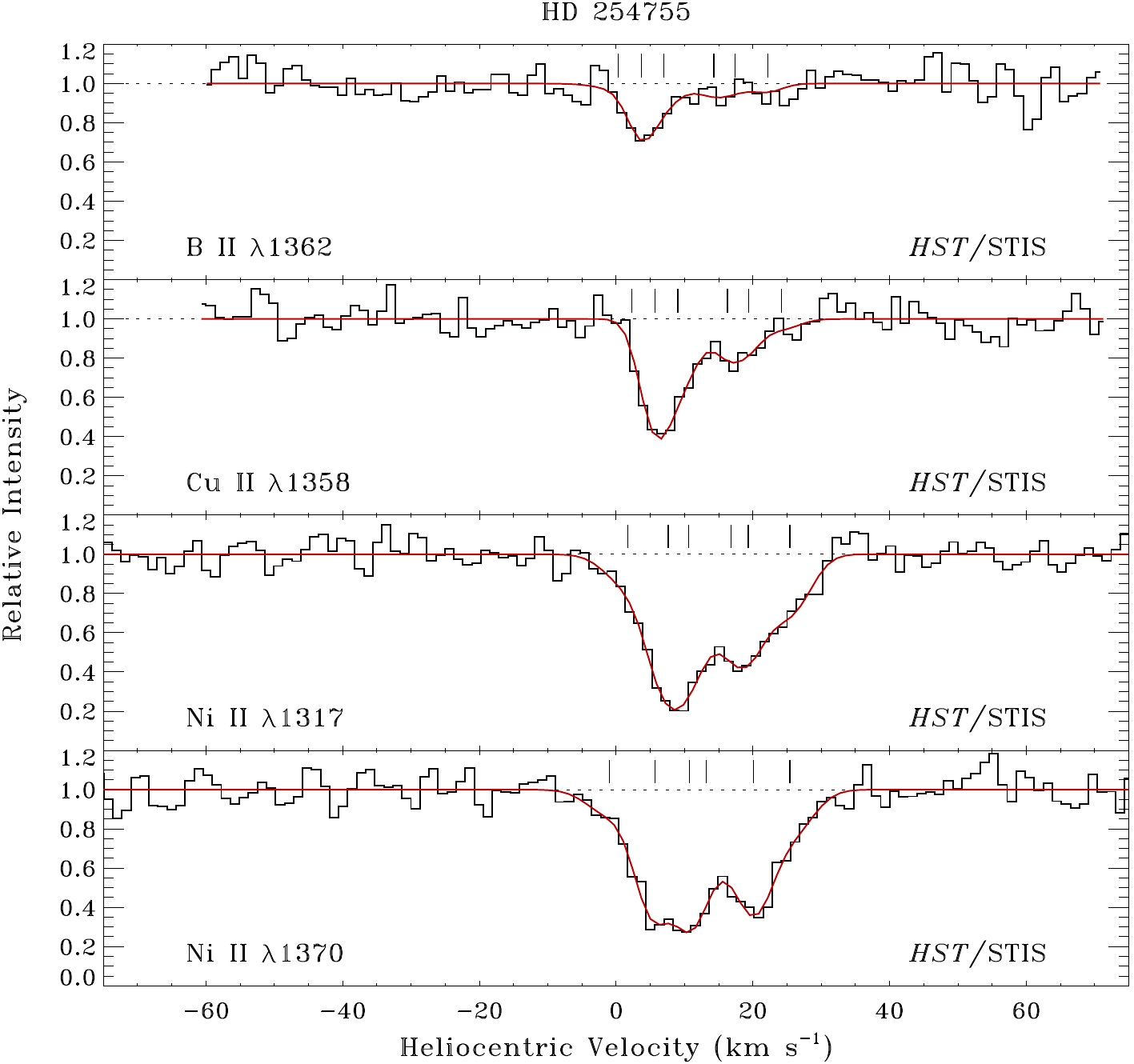}
\caption{Absorption profiles of neutral and singly-ionized atomic species (dominant ions) toward HD~254755 from high-resolution \emph{HST}/STIS spectra. The smooth red curves represent multi-component profile synthesis fits to the observed spectra shown as histograms. Tick marks indicate the positions of the velocity components included in the fits.\label{fig:dominant2}}
\end{figure}

\begin{figure}[!t]
\centering
\includegraphics[width=0.49\textwidth]{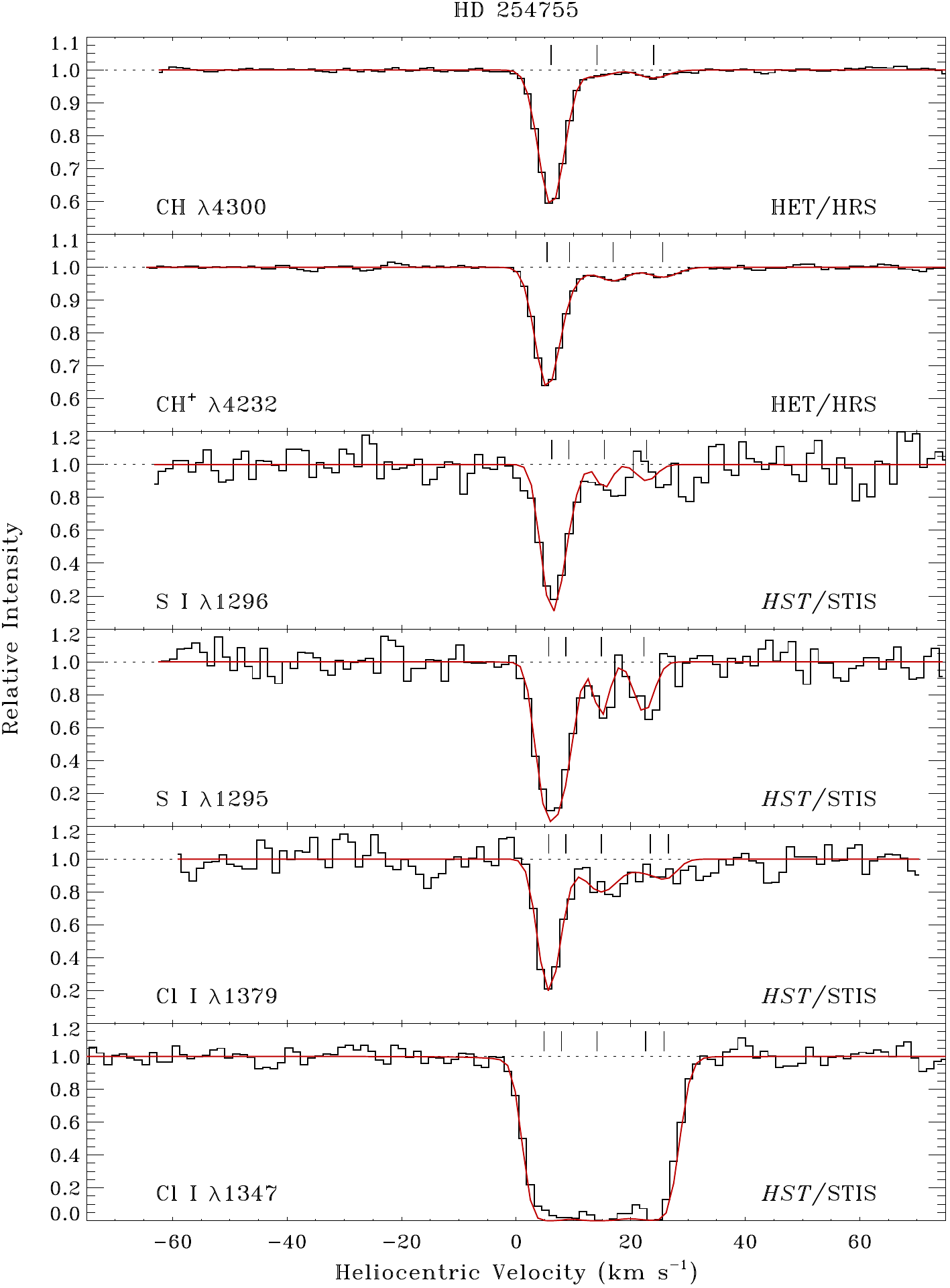}
\includegraphics[width=0.49\textwidth]{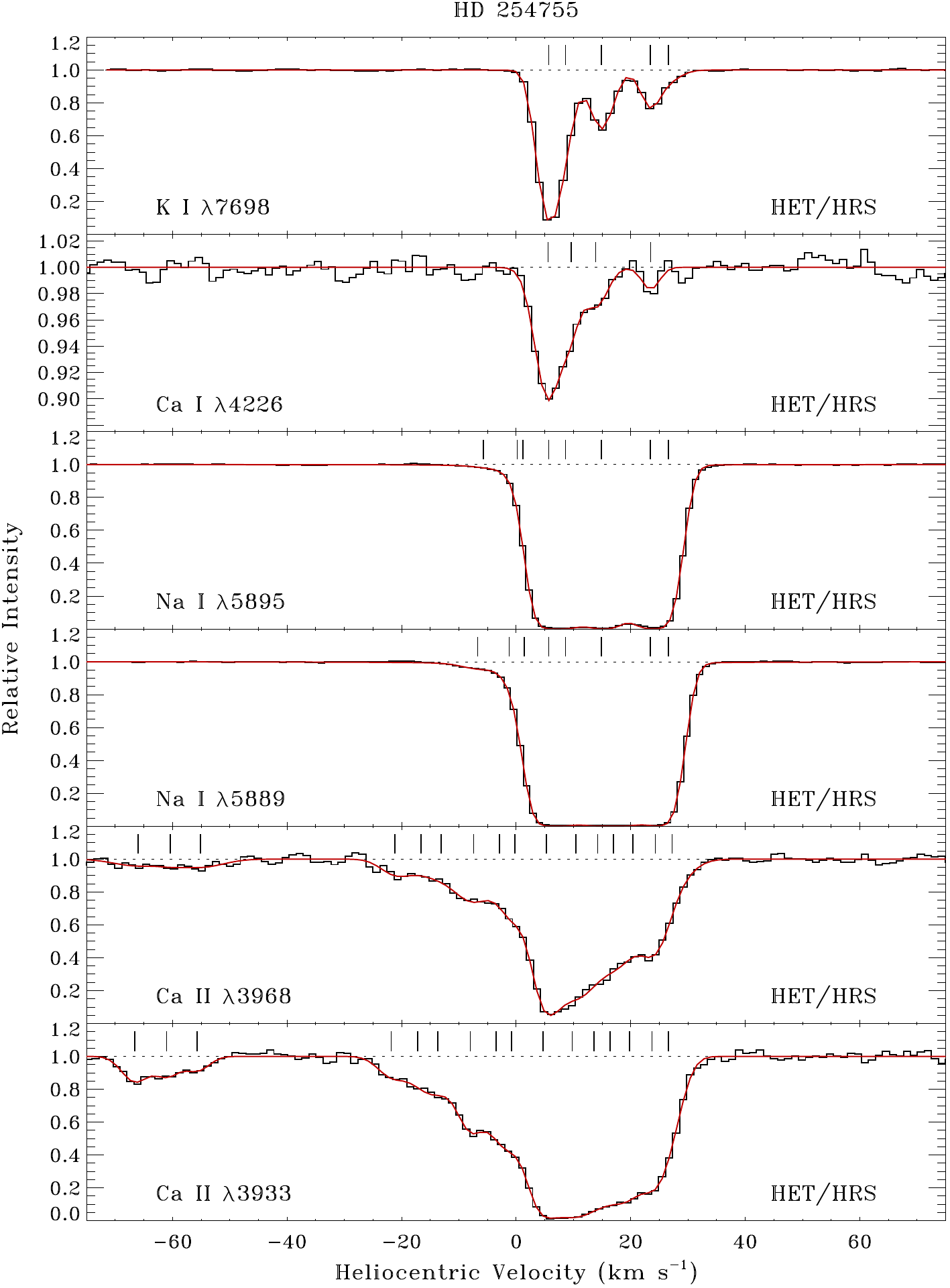}
\caption{Absorption profiles of trace neutral and singly-ionized atomic species, and the molecular species CH and CH$^+$, toward HD~254755 from high-resolution \emph{HST}/STIS spectra and from ground-based data acquired with the HET/HRS. The smooth red curves represent multi-component profile synthesis fits to the observed spectra shown as histograms. Tick marks indicate the positions of the velocity components included in the fits.\label{fig:trace2}}
\end{figure}

\begin{figure}
\centering
\includegraphics[width=0.9\textwidth]{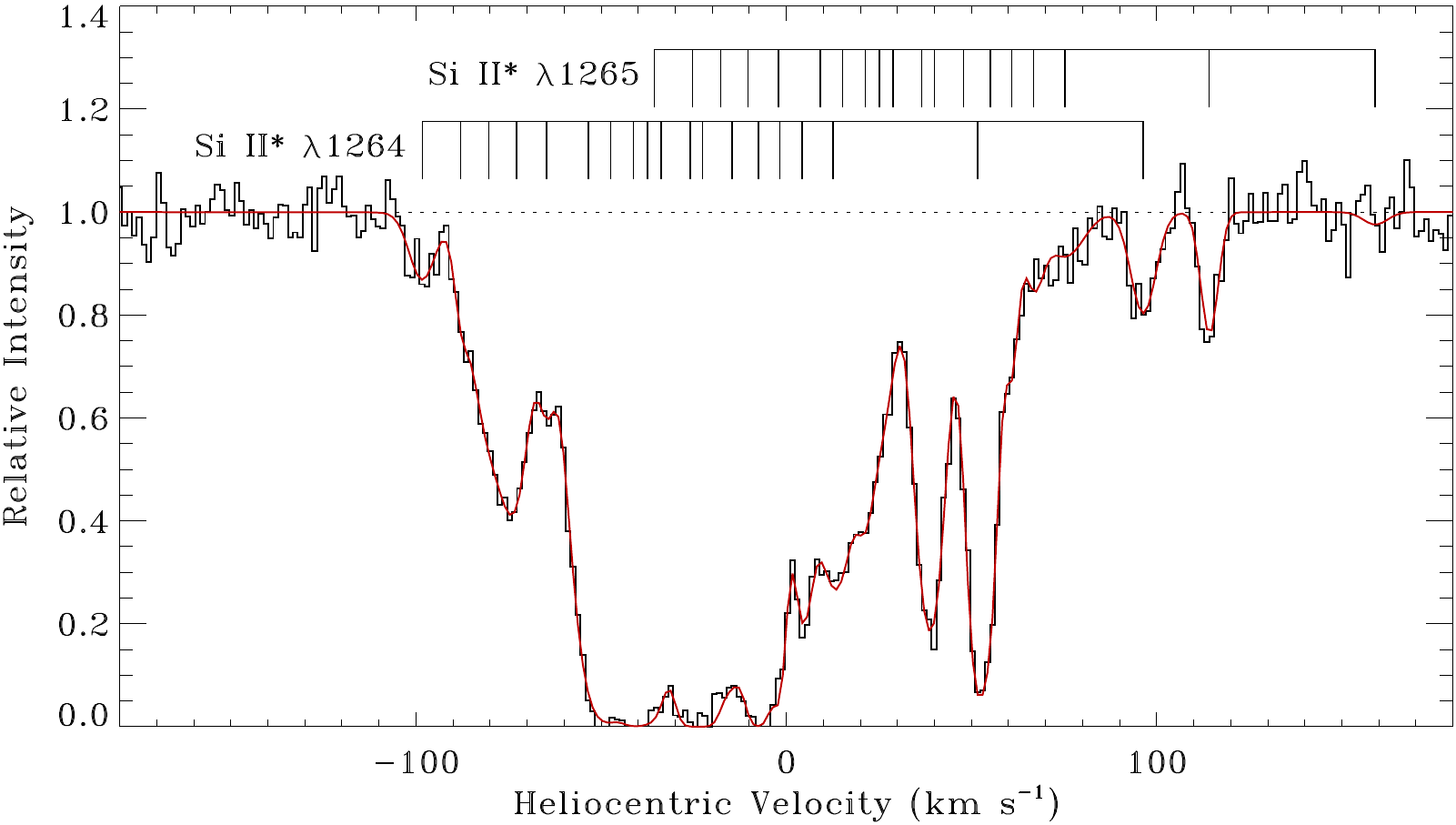}
\caption{Absorption profile of the blended Si~{\sc ii}*~$\lambda\lambda1264,1265$ feature toward HD~43582. The smooth red curve represents a simultaneous multi-component profile synthesis fit to both lines. Two sets of tick marks give the (identical) positions of the velocity components included in the fit. The zero point in velocity corresponds to the wavelength of the Si~{\sc ii}*~$\lambda1264$ transition. The $\lambda1265$ transition is shifted by +62.6~km~s$^{-1}$.\label{fig:si1}}
\end{figure}

\begin{figure}
\centering
\includegraphics[width=0.9\textwidth]{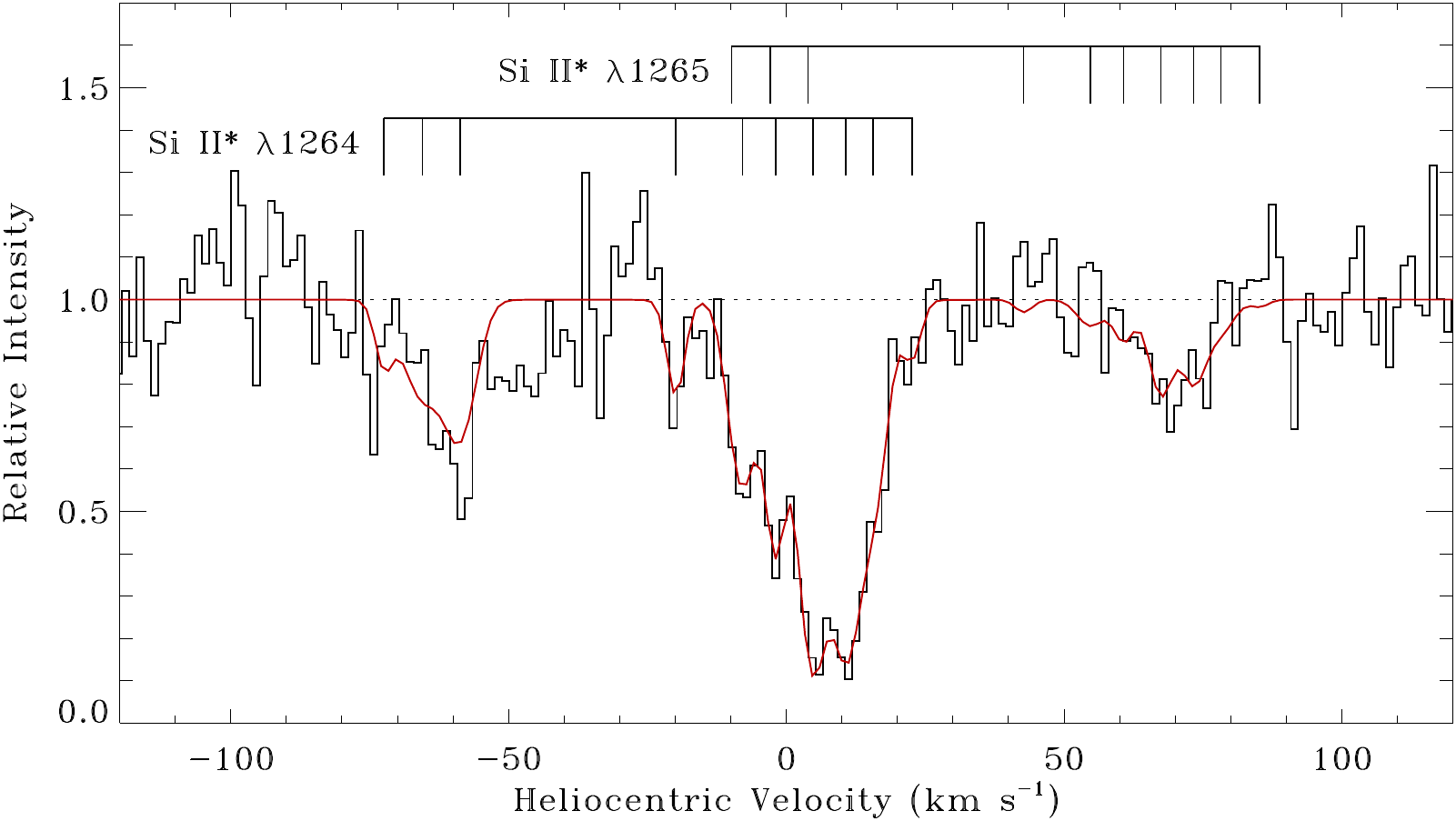}
\caption{Absorption profile of the blended Si~{\sc ii}*~$\lambda\lambda1264,1265$ feature toward HD~254755. The smooth red curve represents a simultaneous multi-component profile synthesis fit to both lines. Two sets of tick marks give the (identical) positions of the velocity components included in the fit. The zero point in velocity corresponds to the wavelength of the Si~{\sc ii}*~$\lambda1264$ transition. The $\lambda1265$ transition is shifted by +62.6~km~s$^{-1}$.\label{fig:si2}}
\end{figure}

Perhaps the most remarkable feature of the \emph{HST}/STIS spectrum of HD~43582 is the strong and unusually complex absorption from excited fine-structure levels in O~{\sc i} and Si~{\sc ii}. The first and second excited levels of the ground electronic state in neutral oxygen (denoted O~{\sc i}* and O~{\sc i}**) have absorption lines at 1304.9~\AA{} and 1306.0~\AA{}, respectively. The absorption profiles of these lines toward HD~43582 consist of multiple distinct components with velocities\footnote{All velocities in this paper are reported in the heliocentric frame of reference. To convert from heliocentric to LSR velocities for the IC~443 stars, subtract 12.1~km~s$^{-1}$.} ranging from $-$55 to +52~km~s$^{-1}$ (Figure~\ref{fig:dominant1}). A similar component structure is seen in the Si~{\sc ii}* line at 1309.3~\AA{}, although the stronger Si~{\sc ii}* transitions at 1264.7 and 1265.0~\AA{} show additional components with velocities ranging from $-$98 to +96~km~s$^{-1}$ (Figure~\ref{fig:si1}). Other species that show many of the same absorption components as the O~{\sc i}* and O~{\sc i}** lines toward HD~43582 include S~{\sc ii}, Ni~{\sc ii}, Ca~{\sc i}, and Na~{\sc i} (Figures~\ref{fig:dominant1} and \ref{fig:trace1}), while Ca~{\sc ii} is more similar to Si~{\sc ii}* $\lambda\lambda1264,1265$ in that it shows many additional components at velocities between $-$100 and +100~km~s$^{-1}$. The absorption profiles of neutral and singly-ionized atomic species toward HD~254755 show more moderate velocities, with much of the absorption falling between 0 and +30~km~s$^{-1}$ (Figures~\ref{fig:dominant2} and \ref{fig:trace2}). Still, the O~{\sc i}* and O~{\sc i}** lines are strong toward HD~254755 at the velocity of the main interstellar absorption component near +6~km~s$^{-1}$, and the Ca~{\sc ii} profile shows a high velocity component near $-$63~km~s$^{-1}$, which is also detected in Si~{\sc ii}* (Figure~\ref{fig:si2}).

Our general philosophy in fitting the absorption profiles of the various neutral and singly-ionized atomic species was to include the fewest number of components in the fit so long as the residuals, after subtracting the synthetic profile from the observed one, were indistinguishable from the noise in the continuum. We started by fitting the relatively simple absorption profiles, such as O~{\sc i}~$\lambda1355$ and K~{\sc i}~$\lambda7698$, and used the results as a starting point to fit other progressively more complex profiles, adding additional components as needed. While a precise correspondence in velocity among absorption components detected in different species is not required, we did expect that similar species would exhibit components at similar velocities, keeping in mind that the velocity resolution of the \emph{HST} and HET spectra is $\sim$3~km~s$^{-1}$. From our full complement of profile synthesis fits, we find that the typical scatter in velocity among ``corresponding'' components is $\sim$1~km~s$^{-1}$. The $b$-values of the individual absorption components included in the fits generally fall between 2 and 5~km~s$^{-1}$ for dominant ions (including Ca~{\sc ii}) and between 1 and 3~km~s$^{-1}$ for trace neutral species.

Column density determinations are dependent on the adopted oscillator strengths ($f$-values) of the observed transitions. We have generally adopted $f$-values from the compilations of Morton (2000, 2003). However, for the Ni~{\sc ii}~$\lambda1317$ and $\lambda1370$ lines we use the empirical $f$-values derived by Jenkins \& Tripp (2006), and for the Ge~{\sc ii}~$\lambda1237$ transition we use the recently-determined experimental $f$-value obtained by Heidarian et al.~(2017). Ideally, one would like to have multiple transitions of varying intrinsic strength for each species so that any saturation in the line profiles can be properly accounted for. A good example is the pair of Cl~{\sc i} lines at 1347.2~\AA{} and 1379.5~\AA{}, which have $f$-values that differ by a factor of $\sim$60. The Cl~{\sc i}~$\lambda1379$ line is useful for probing the highest column density portion of the absorption profile, which for the IC~443 stars corresponds to velocities between 0 and +30~km~s$^{-1}$. Meanwhile, the Cl~{\sc i}~$\lambda1347$ line (toward HD~43582, for example; see Figure~\ref{fig:trace1}) probes lower column density material at higher velocity. By examining both lines in concert, we can constrain both the total column density and the detailed line-of-sight component structure. In many cases, however, there is only one transition available for a given species or all of the available transitions are badly saturated over significant portions of the absorption profile. Examples of the latter situation include the S~{\sc ii}~$\lambda\lambda1250,1253,1259$ triplet and the Na~{\sc i}~$\lambda\lambda5889,5895$ doublet. In these two cases specifically, we can still examine the column densities, velocities, and $b$-values of some of the higher velocity components seen toward HD~43582 (Figures~\ref{fig:dominant1} and \ref{fig:trace1}), but the total (line-of-sight) column densities are undetermined.

When two transitions were available for a given species, we generally fit each line independently so that we would have some way of gauging the consistency and reliability of the profile fitting procedure overall. The median difference in the column densities obtained from two lines of the same species is $\sim$15\%, which is comparable to the typical column density uncertainty of $\sim$0.06~dex for individual components. In some cases, the fits to two lines of the same species were not entirely independent. Depending on the complexity of the absorption profiles and/or the degree to which the spectra were affected by noise, we would occasionally need to constrain the fitting parameters in some way (for example, by holding the relative velocities among the absorption components fixed, as was done for the Ca~{\sc ii}~$\lambda\lambda3933,3968$ doublet and the Ni~{\sc ii}~$\lambda1317$ and $\lambda1370$ lines toward HD~43582). Since the stronger of the two Mg~{\sc ii} lines at 1239.9~\AA{} was found to be moderately (if not heavily) saturated toward both stars (and because the S/N ratios are fairly low near the Mg~{\sc ii} features), we used the component structure found for the weaker line of the Mg~{\sc ii} doublet as a fixed profile template when fitting the stronger line. (This helps to prevent the fitting program from obtaining unusually large column densities when the measured intensities in the line approach zero.)

The velocity separation between the Si~{\sc ii}*~$\lambda1264$ and $\lambda1265$ transitions is only $\sim$63~km~s$^{-1}$. Thus, since the spread in velocity exhibited by the Si~{\sc ii}* absorption components toward our target stars ranges from 100 to 200 km~s$^{-1}$, the observed Si~{\sc ii}*~$\lambda\lambda1264,1265$ profiles are a complicated blend of absorption from both transitions (see Figures~\ref{fig:si1} and \ref{fig:si2}). To properly disentangle the blended profiles, we used a modified version of the profile synthesis routine, which allowed us to fit both Si~{\sc ii}* transitions simultaneously. In these fits, the column densities, $b$-values, and radial velocities of the fitted components were required to be identical between the two Si~{\sc ii}* transitions, but otherwise were allowed to vary as usual. In the case of HD~43582, we were aided by our earlier analysis of the Si~{\sc ii}*~$\lambda1309$ profile, which we used to develop an initial solution for fitting the $\lambda1264$ and $\lambda1265$ transitions. Ultimately, additional components near the periphery of the profile not seen in the $\lambda1309$ line needed to be included to produce an acceptable fit. The Si~{\sc ii}*~$\lambda1309$ feature is not detected toward HD~254755, although the blending between the $\lambda1264$ and $\lambda1265$ lines is also not very severe in this case. However, the low S/N ratio near the Si~{\sc ii}*~$\lambda\lambda1264,1265$ profile toward HD~254755 made it difficult to discern the component structure for the absorption complex near $-$63~km~s$^{-1}$ (Figure~\ref{fig:si2}). This complex is more clearly detected in the Si~{\sc ii}~$\lambda1304$ line. Thus, we adopted the component structure found from the Si~{\sc ii}~$\lambda1304$ complex near $-$63~km~s$^{-1}$ to fit the same absorption feature in the Si~{\sc ii}* profile (see Section~\ref{subsubsec:high_vel2}).

\begin{figure}
\centering
\includegraphics[width=0.9\textwidth]{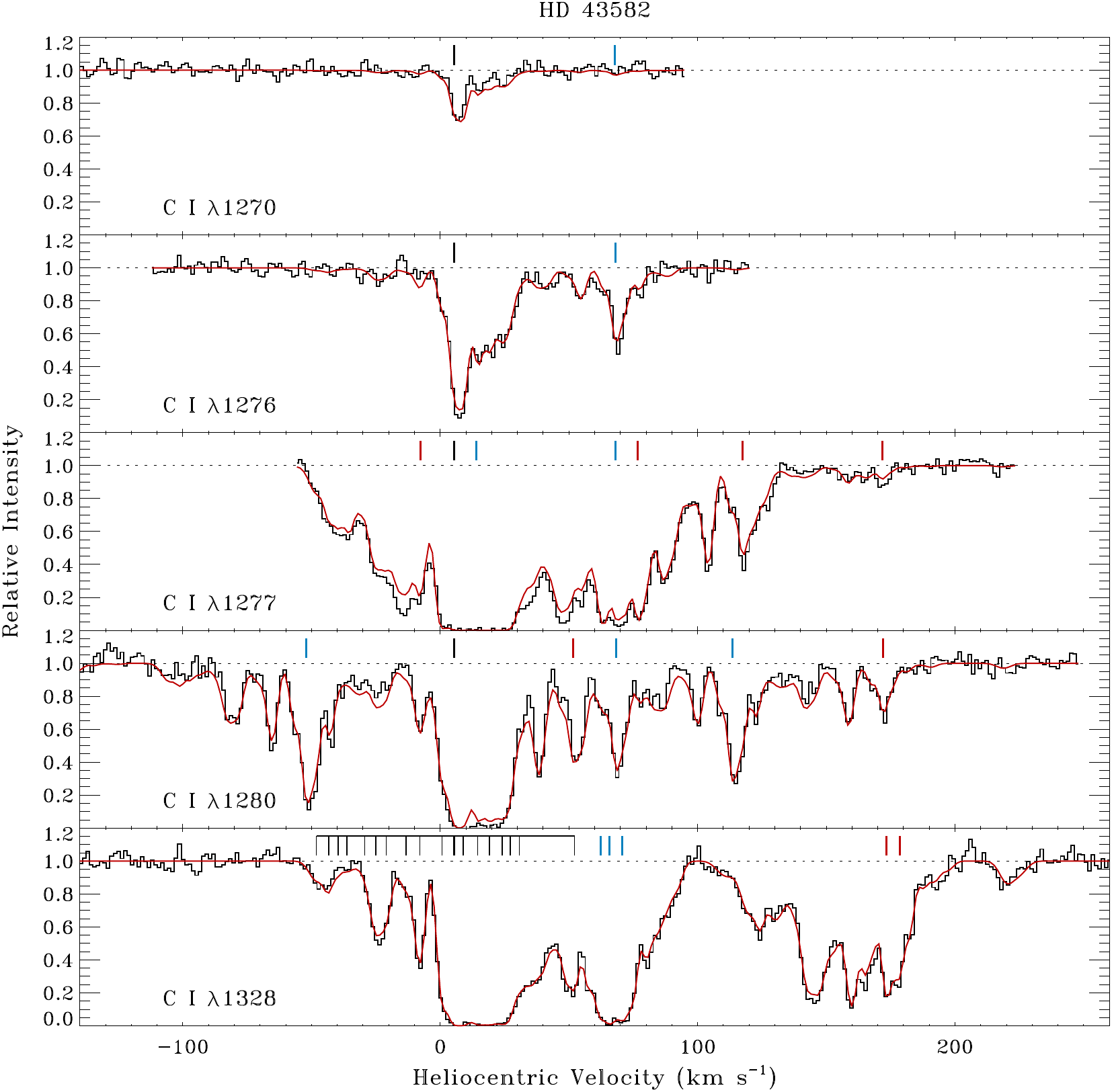}
\caption{Global multi-component profile synthesis fit to the C~{\sc i} multiplets toward HD~43582. As in previous figures, the synthetic profiles are shown as smooth red curves, with histograms representing the observed spectra. Solitary black, blue, and red tick marks indicate the positions of the individual C~{\sc i} transitions arising from the $J=0$, $1$, and $2$ levels, respectively, for the dominant velocity component along the line of sight. The more extensive set of tick marks for the C~{\sc i}~$\lambda1328$ transition shows the complete set of components included in the global fit. The C~{\sc i}~$\lambda1260$ multiplet is excluded from the fit due to overlap with a high velocity component of Si~{\sc ii}~$\lambda1260$.\label{fig:ci1}}
\end{figure}

\begin{figure}
\centering
\includegraphics[width=0.9\textwidth]{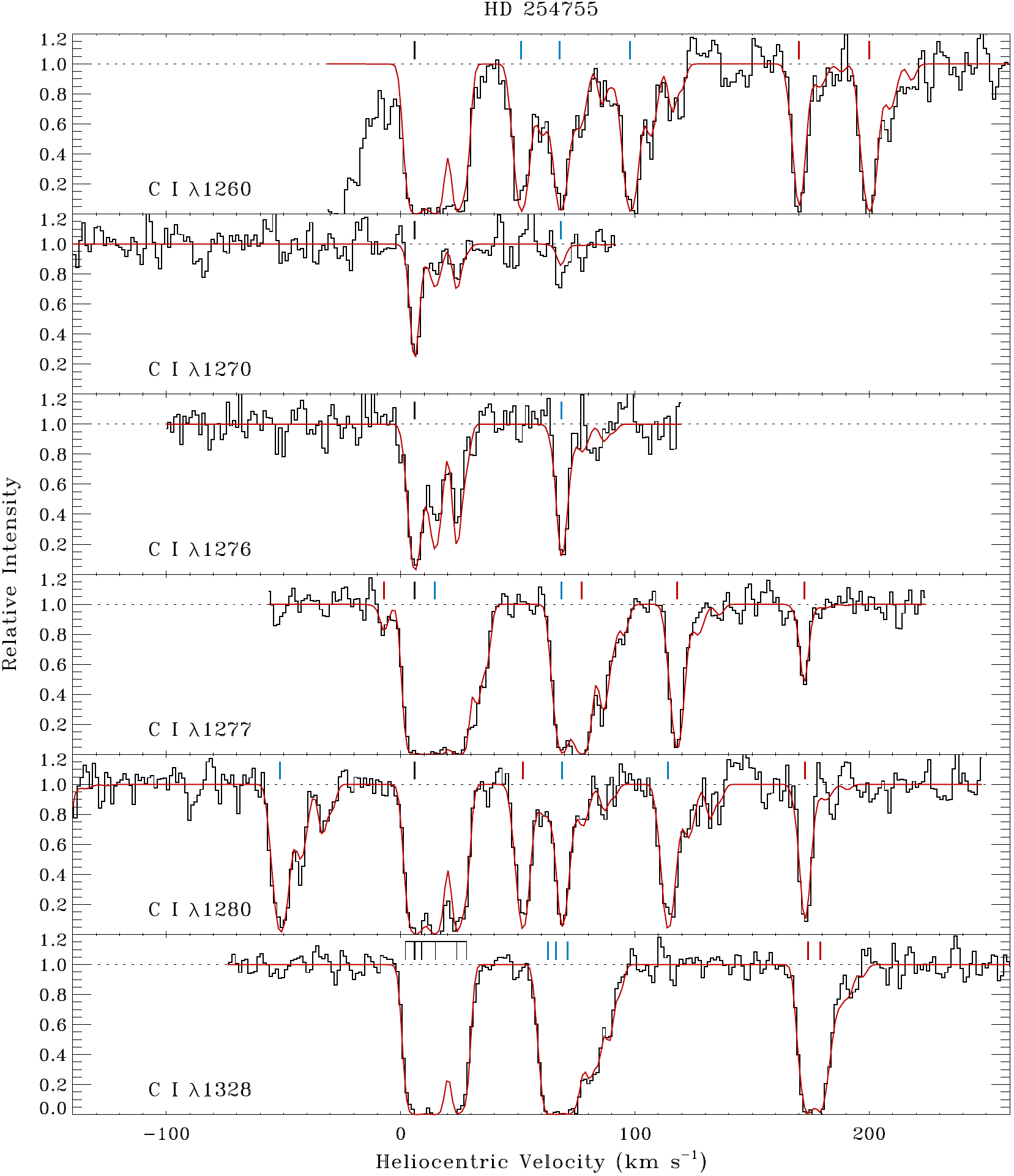}
\caption{Global multi-component profile synthesis fit to the C~{\sc i} multiplets toward HD~254755. As in previous figures, the synthetic profiles are shown as smooth red curves, with histograms representing the observed spectra. Solitary black, blue, and red tick marks indicate the positions of the individual C~{\sc i} transitions arising from the $J=0$, $1$, and $2$ levels, respectively, for the dominant velocity component along the line of sight. The more extensive set of tick marks for the C~{\sc i}~$\lambda1328$ transition shows the complete set of components included in the global fit.\label{fig:ci2}}
\end{figure}

An even more difficult challenge was posed by the many different C~{\sc i} multiplets included within the wavelength coverage of the \emph{HST}/STIS spectra. In Section~\ref{subsec:car_ex}, we describe an analysis of the C~{\sc i} multiplets that employs the same procedure used by Jenkins \& Tripp (2001, 2011) to disentangle the complex overlapping absorptions from the three C~{\sc i} fine-structure levels. That procedure is based on an analysis of the apparent optical depth (AOD) profiles. In this section, however, we analyze the C~{\sc i} multiplets using a profile synthesis approach so as to facilitate a more direct comparison with our fitting results for other neutral and singly-ionized atomic species. For this application, the profile fitting routine was again modified to allow the simultaneous fitting of all of the available C~{\sc i} multiplets. As with the Si~{\sc ii}* program described above, the C~{\sc i} fitting program requires that the $b$-values and radial velocities of the fitted components are identical among the three C~{\sc i} fine-structure levels, but still determines the best-fitting values for these quantities. The fractional column densities of the $J=0$, 1, and 2 levels within each component are not required to be identical, however, and indeed the relative populations among these levels vary quite substantially from one component to the next (see Section~\ref{subsec:car_ex}). The $f$-values adopted for our profile fits to the C~{\sc i} multiplets are the same as those presented in Jenkins \& Tripp (2001, 2011). Initial solutions for the C~{\sc i} component structures were derived from our analyses of the K~{\sc i} and Na~{\sc i} profiles. For HD~43582, the K~{\sc i}~$\lambda7698$ line provided the initial component structure for the high column density portion of the C~{\sc i} absorption profile, while the Na~{\sc i}~$\lambda\lambda5889,5895$ features yielded initial solutions for the higher velocity components (i.e., those that are outside of the saturated cores of the Na~{\sc i} lines; see Figure~\ref{fig:trace1}). For HD~254755, the K~{\sc i}~$\lambda7698$ line alone was sufficient to provide an initial component solution for the C~{\sc i} multiplets. Our global profile synthesis fits to the available C~{\sc i} multiplets toward HD~43582 and HD~254755 are presented in Figures~\ref{fig:ci1} and \ref{fig:ci2}, respectively. (The C~{\sc i}~$\lambda1260$ multiplet is not included in the fit for HD~43582 due to overlap with a high velocity component of Si~{\sc ii}~$\lambda1260$ near +224~km~s$^{-1}$; see Section~\ref{subsubsec:high_vel}.)

Before proceeding with the profile analysis for the O~{\sc i}*~$\lambda1304$ and O~{\sc i}**~$\lambda1306$ lines, we needed to identify and remove the absorption features arising from excited oxygen atoms in Earth's upper atmosphere. The velocity displacements and absorption strengths of telluric lines in an \emph{HST} spectrum vary with the time of the observation and with the zenith angle of the telescope's viewing axis. For our \emph{HST} observations of HD~43582 and HD~254755, any telluric features would be expected to have heliocentric velocities of approximately $-$30~km~s$^{-1}$. The telluric O~{\sc i}* and O~{\sc i}** lines are easily discernible toward HD~254755 since they are well separated from the interstellar absorption and fall almost exactly at the expected velocity. The telluric features toward HD~43582, however, are blended with the complex interstellar absorption profiles of O~{\sc i}* and O~{\sc i}**, although distinct peaks in absorption coincide with the expected positions of the telluric lines. Since the two stars are positioned very close to one another on the sky, and were observed just three days apart (with identical exposure sequences in orbit), the strengths of the telluric lines in the two directions should be very similar. We therefore used our observations of the telluric O~{\sc i}* and O~{\sc i}** lines toward HD~254755 as a template for removing these features from the spectrum of HD~43582. An additional complication arose in the analysis of the O~{\sc i}*~$\lambda1304$ line toward HD~43582. A high velocity absorption component from Si~{\sc ii}~$\lambda1304$ near +96~km~s$^{-1}$ impacts the O~{\sc i}* profile directly. In Section~\ref{subsubsec:high_vel}, we describe in more detail how we modeled this high-velocity component and removed it from the spectrum.

\begin{deluxetable}{lcccccccc}
\tablecolumns{9}
\tablewidth{0pt}
\tabletypesize{\scriptsize}
\tablecaption{Total Line-of-Sight Equivalent Widths and Column Densities\label{tab:column_densities}}
\tablehead{ \colhead{Species} & \colhead{$\lambda$\tablenotemark{a}} & \colhead{log~($f\lambda$)} & \colhead{Ref.} & \multicolumn{2}{c}{HD 43582} & \colhead{} & \multicolumn{2}{c}{HD 254755} \\
\cline{5-6} \cline{8-9} \\
\colhead{} & \colhead{} & \colhead{} & \colhead{} & \colhead{$W_{\lambda}$} & \colhead{log~$N$} & \colhead{} & \colhead{$W_{\lambda}$} & \colhead{log~$N$} \\
\colhead{} & \colhead{(\AA)} & \colhead{} & \colhead{} & \colhead{(m\AA)} & \colhead{} & \colhead{} & \colhead{(m\AA)} & \colhead{} }
\startdata
B~{\sc ii} & 1362.463 & $3.133$ & 1 & \phn$7.9\pm2.3$ & $11.72\pm0.11$ && $12.6\pm4.7$ & $11.93\pm0.14$ \\
C~{\sc i} & \ldots & \ldots & & \ldots & \phn$14.86\pm0.01$\tablenotemark{b} && \ldots & \phn$15.22\pm0.03$\tablenotemark{b} \\
C~{\sc i}* & \ldots & \ldots & & \ldots & \phn$14.48\pm0.01$\tablenotemark{b} && \ldots & \phn$14.78\pm0.02$\tablenotemark{b} \\
C~{\sc i}** & \ldots & \ldots & & \ldots & \phn$14.35\pm0.01$\tablenotemark{b} && \ldots & \phn$14.57\pm0.02$\tablenotemark{b} \\
O~{\sc i} & 1355.598 & $-2.805$\phantom{$-$} & 1 & $26.9\pm1.4$ & $18.26\pm0.03$ && $47.8\pm3.1$ & $18.57\pm0.03$ \\
O~{\sc i}* & 1304.858 & $1.795$ & 1 & $182.1\pm2.1$\phn & $14.71\pm0.03$ && $39.3\pm2.6$ & $14.12\pm0.06$ \\
O~{\sc i}** & 1306.029 & $1.795$ & 1 & $181.6\pm2.5$\phn & $14.70\pm0.02$ && $29.9\pm3.2$ & $13.77\pm0.05$ \\
Mg~{\sc ii} & 1239.925 & $-0.106$\phantom{$-$} & 1 & $95.2\pm2.6$ & $16.51\pm0.04$ && $102.0\pm7.3$\phn & $16.72\pm0.07$ \\
 & 1240.395 & $-0.355$\phantom{$-$} & 1 & $77.3\pm2.5$ & $16.54\pm0.04$ && $82.2\pm6.5$ & $16.72\pm0.06$ \\
 & \ldots & \ldots & & \ldots & \phn$16.53\pm0.03$\tablenotemark{c} && \ldots & \phn$16.72\pm0.05$\tablenotemark{c} \\
Si~{\sc ii}* & 1264.738 & $3.125$ & 1 & $337.2\pm6.4$\phn & $14.05\pm0.03$ && $102.0\pm7.6$\phn & $13.07\pm0.03$ \\
 & 1265.002 & $2.171$ & 1 & $145.8\pm6.4$\phn & $14.05\pm0.02$ && $18.3\pm7.6$ & $13.07\pm0.15$ \\
 & 1309.276 & $2.052$ & 1 & $107.9\pm2.7$\phn & $14.05\pm0.02$ && \ldots & \ldots \\
 & \ldots & \ldots & & \ldots & \phn$14.05\pm0.01$\tablenotemark{c} && \ldots & \phn$13.07\pm0.03$\tablenotemark{c} \\
S~{\sc i} & 1295.653 & $2.052$ & 1 & $36.7\pm1.7$ & $13.61\pm0.03$ && $38.6\pm2.1$ & $14.50\pm0.09$ \\
 & 1296.174 & $1.455$ & 1 & $11.8\pm1.5$ & $13.63\pm0.05$ && $24.6\pm1.6$ & $14.54\pm0.08$ \\
 & \ldots & \ldots & & \ldots & \phn$13.62\pm0.02$\tablenotemark{c} && \ldots & \phn$14.52\pm0.06$\tablenotemark{c} \\
Cl~{\sc i} & 1347.240 & $2.314$ & 1 & $141.9\pm1.4$\phn & $14.68\pm0.05$ && \ldots & \ldots \\
 & 1379.528 & $0.579$ & 1 & $19.4\pm1.4$ & $14.70\pm0.03$ && $29.3\pm2.7$ & $15.05\pm0.06$ \\
 & \ldots & \ldots & & \ldots & \phn$14.69\pm0.03$\tablenotemark{c} && \ldots & \phn$15.05\pm0.06$\tablenotemark{c} \\
Ni~{\sc ii} & 1317.217 & $1.876$ & 2 & $105.5\pm4.2$\phn & $14.21\pm0.02$ && $65.2\pm4.2$ & $14.06\pm0.03$ \\
 & 1370.132 & $1.906$ & 2 & $118.2\pm5.4$\phn & $14.22\pm0.02$ && $72.3\pm5.2$ & $14.05\pm0.03$ \\
 & \ldots & \ldots & & \ldots & \phn$14.21\pm0.01$\tablenotemark{c} && \ldots & \phn$14.05\pm0.02$\tablenotemark{c} \\
Cu~{\sc ii} & 1358.773 & $2.553$ & 1 & $26.5\pm1.9$ & $12.87\pm0.03$ && $30.0\pm3.7$ & $12.96\pm0.05$ \\
Ge~{\sc ii} & 1237.059 & $3.033$ & 3 & $51.0\pm5.0$ & $12.81\pm0.05$ && \ldots & \ldots \\
Kr~{\sc i} & 1235.838 & $2.402$ & 4 & $14.9\pm3.0$ & $12.80\pm0.08$ && \ldots & \ldots \\
CO & \dots & \ldots & & \ldots & \phn$14.06\pm0.06$\tablenotemark{d} && \ldots & \phn$14.86\pm0.02$\tablenotemark{d} \\
\hline
K~{\sc i} & 7698.965 & $3.409$ & 1 & $174.2\pm0.8$\phn & $12.23\pm0.03$ && $223.8\pm0.6$\phn & $12.88\pm0.08$ \\
Ca~{\sc i} & 4226.728 & $3.874$ & 1 & $32.9\pm1.0$ & $11.08\pm0.01$ && $13.1\pm0.7$ & $10.69\pm0.02$ \\
Ca~{\sc ii} & 3933.661 & $3.392$ & 1 & $1266.4\pm5.1$\phn\phn & $13.63\pm0.02$ && $458.3\pm4.7$\phn & $13.13\pm0.03$ \\
 & 3968.467 & $3.092$ & 1 & $976.1\pm3.4$\phn & $13.68\pm0.02$ && $338.8\pm5.9$\phn & $13.18\pm0.03$ \\
 & \ldots & \ldots & & \ldots & \phn$13.66\pm0.01$\tablenotemark{c} && \ldots & \phn$13.15\pm0.02$\tablenotemark{c} \\
C$_2$ & \dots & \ldots & & \ldots & \ldots && \ldots & \phn$13.76\pm0.03$\tablenotemark{d} \\
CH & 4300.313 & $1.338$ & 5 & $17.9\pm0.5$ & $13.35\pm0.01$ && $36.9\pm0.5$ & $13.74\pm0.02$ \\
CH$^+$ & 4232.548 & $1.363$ & 5 & $31.4\pm0.6$ & $13.60\pm0.01$ && $36.1\pm0.7$ & $13.69\pm0.02$ \\
\enddata
\tablenotetext{a}{Vacuum wavelengths are quoted for UV lines. Wavelengths in air are given for visible lines.}
\tablenotetext{b}{Total C~{\sc i}, C~{\sc i}*, and C~{\sc i}** column densities obtained from global profile synthesis fits to the available multiplets.}
\tablenotetext{c}{Final Mg~{\sc ii}, Si~{\sc ii}*, S~{\sc i}, Cl~{\sc i}, Ni~{\sc ii}, and Ca~{\sc ii} column densities obtained by taking the weighted mean of the results from individual lines.}
\tablenotetext{d}{Total CO and C$_2$ column densities obtained from global profile synthesis fits to the available bands.}
\tablerefs{(1) Morton (2003), (2) Jenkins \& Tripp (2006), (3) Heidarian et al.~(2017), (4) Morton (2000), (5) Gredel et al.~(1993).}
\end{deluxetable}

The total (line-of-sight) equivalent widths and column densities resulting from our profile synthesis fits to the atomic and molecular species observed toward HD~43582 and HD~254755 are presented in Table~\ref{tab:column_densities}. Excluded from the table are species such as S~{\sc ii} and Na~{\sc i}, for which the observed transitions are badly saturated over significant portions of the absorption profile. The (1$\sigma$) equivalent width errors listed in Table~\ref{tab:column_densities} account for uncertainties in continuum placement along with the expected statistical variations arising from noise in the spectra. Column density uncertainties for individual components $\sigma_N$ were calculated from the relation $\sigma_N=(\sigma_{N\rm{,obs}}^2+\sigma_{N\rm{,sat}}^2)^{0.5}$, where the observational uncertainties $\sigma_{N\rm{,obs}}$ are proportional to the equivalent width errors (i.e., $\sigma_{N\rm{,obs}}/N=\sigma_{W_{\lambda}}/W_{\lambda}$) and the uncertainties arising from saturation in the line profile $\sigma_{N\rm{,sat}}$ depend on the difference between the fitted column density and the column density one would obtain under the assumption that the line is optically thin. The uncertainties in the total column densities given in Table~\ref{tab:column_densities} were calculated by adding in quadrature the uncertainties in the column densities of the individual components contributing to the totals. For species with two or more transitions listed in Table~\ref{tab:column_densities}, the final column densities were determined by taking the weighted mean of the results from the individual transitions. For C~{\sc i}, C~{\sc i}*, and C~{\sc i}**, the total column densities were obtained from global profile synthesis fits to all of the available C~{\sc i} multiplets. Likewise, the total CO and C$_2$ column densities were obtained from global fits to the available bands (see Section~\ref{subsec:molecular}). Detailed results for individual components derived through profile fitting are presented in Appendix~\ref{sec:comp_structures}.

\subsubsection{Sight Line Depletion Factors\label{subsubsection:depl_factors}}
While our \emph{HST}/STIS observations partially cover the broad H~{\sc i}~Ly$\alpha$ features toward HD~43582 and HD~254755, there are no existing spectroscopic observations of molecular hydrogen in these directions. Thus, the total hydrogen column densities, $N($H$_{\rm tot})=N($H~{\sc i}$)+2N($H$_2)$, cannot be determined directly. As an alternative, we use the methodology described in Section~7 of Jenkins (2009), which allows us to estimate the total hydrogen column densities along our sight lines by examining the relative gas-phase abundances of elements with different depletion behaviors. The methodology involves performing a least-squares linear fit to the variables $y=\log N(X)-\log (X/{\rm H})_{\odot}-B_X+A_Xz_X$ and $x=A_X$, where $N(X)$ is the column density of element $X$ in its preferred ionization stage, $\log (X/{\rm H})_{\odot}$ is the logarithm of the solar abundance of element $X$, and the parameters $A_X$, $B_X$, and $z_X$ describe the depletion trend of element $X$. Within the framework developed by Jenkins (2009), the logarithmic depletion of element $X$, defined as $[X/{\rm H}]=\log [N(X)/N({\rm H}_{\rm tot})]-\log (X/{\rm H})_{\odot}$, varies as a function of the sight line depletion strength factor, denoted $F_*$, according to the relation $[X/{\rm H}]=B_X+A_X(F_*-z_X)$. Thus, a simple linear fit to a plot of $y$ versus $x$ yields the $y$-intercept, which equals $\log N({\rm H}_{\rm tot})$, and the slope, which corresponds to $F_*$.

\begin{figure}
\centering
\includegraphics[width=0.8\textwidth]{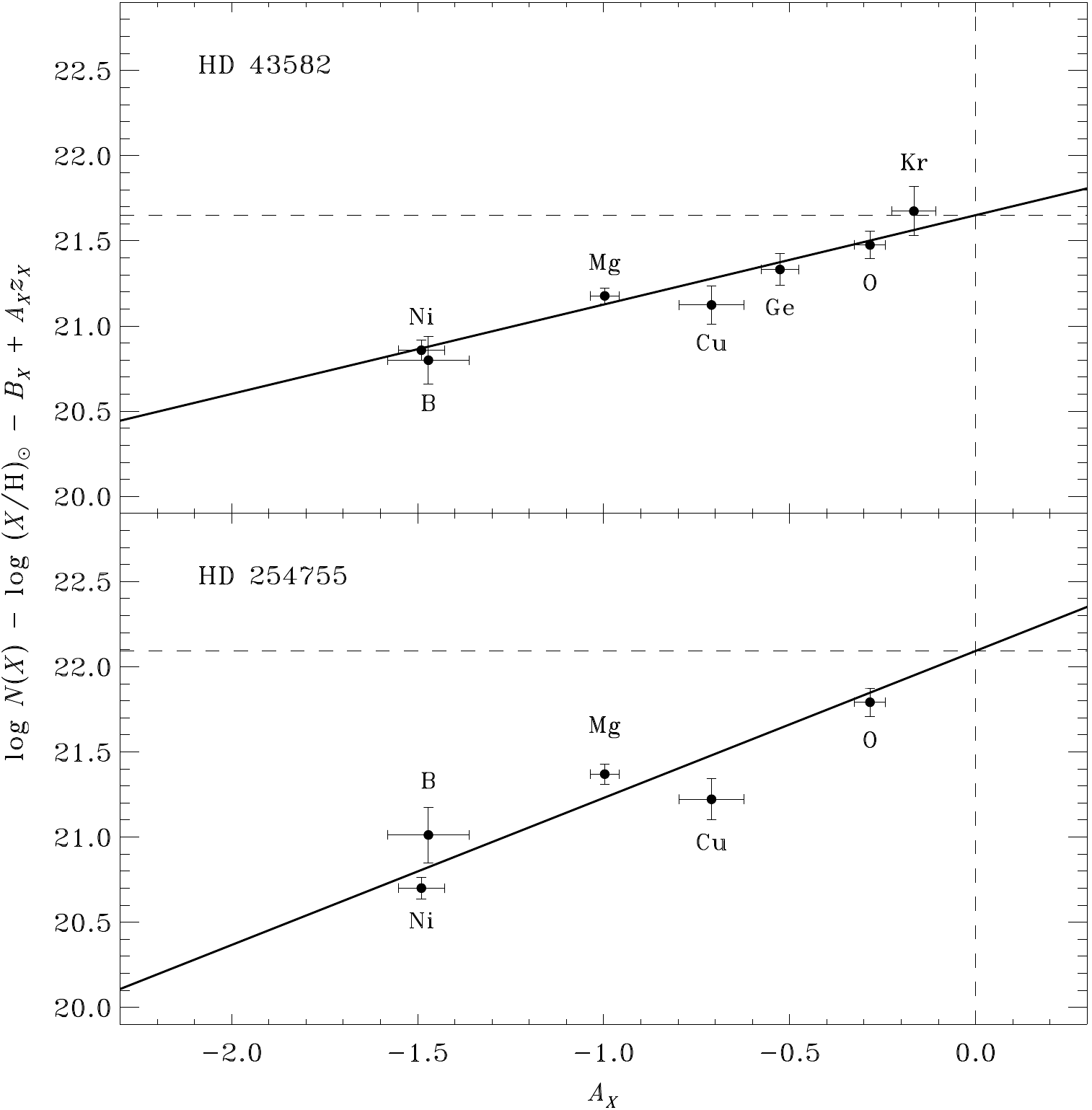}
\caption{Application of the methodology described in Jenkins (2009) to estimate the total hydrogen column density, $N$(H$_{\mathrm{tot}}$), and the depletion strength factor, $F_*$, along the lines of sight to HD~43582 (upper panel) and HD~254755 (lower panel). The $y$-intercept of the best fit line yields $N$(H$_{\mathrm{tot}}$), while the slope yields $F_*$ (see Section 7.1 of Jenkins 2009). The element-specific depletion parameters, $A_X$, $B_X$, and $z_X$ are adopted from Jenkins (2009) for Cu, Mg, and Ni, and from Ritchey et al.~(2018) for Kr, O, Ge, and B.\label{fig:depl_factors}}
\end{figure}

Table~\ref{tab:column_densities} gives total column density measurements for the dominant ions B~{\sc ii}, O~{\sc i}, Mg~{\sc ii}, Ni~{\sc ii}, Cu~{\sc ii}, Ge~{\sc ii}, and Kr~{\sc i}. (The latter two species are not detected toward HD~254755 due to the poor S/N near the Kr~{\sc i}~$\lambda1235$ and Ge~{\sc ii}~$\lambda1237$ transitions.) We use these measurements to estimate the total hydrogen column densities and sight line depletion factors as outlined above. Implicit in this analysis is the assumption that the column densities in the typically dominant ionic stages are good proxies for the total gas-phase elemental abundances. While enhanced radiation from shocks could increase the ionization level to stages not measured here, such a process would mainly affect the abundances in the lower column density material seen at moderate to high velocity (e.g., toward HD~43582). The line-of-sight column densities examined in this section are dominated by low-velocity components, which very well may trace quiescent (unshocked) gas along the lines of sight.

Figure~\ref{fig:depl_factors} presents our application of the Jenkins (2009) methodology to derive synthetic values for $N({\rm H}_{\rm tot})$ and $F_*$ toward HD~43582 and HD~254755. The element-specific depletion parameters $A_X$, $B_X$, and $z_X$ are adopted from Jenkins (2009) for Mg, Ni, and Cu, and from Ritchey et al.~(2018) for B, O, Ge, and Kr. Solar abundances for the elements are obtained from Lodders (2003). The least-squares linear fits, which account for uncertainties in both $y$ and $x$, yield $\log N({\rm H}_{\rm tot})=21.65\pm0.07$ and $F_*=0.52\pm0.07$ for HD~43582\footnote{The Ni~{\sc ii} absorption profile toward HD~43582 shows many components at negative velocities that are not seen in the absorption profiles of other dominant ions such as O~{\sc i}, Mg~{\sc ii}, and Ge~{\sc ii} (Figure~\ref{fig:dominant1}). Since these additional Ni~{\sc ii} components likely result from the destruction of dust grains by shocks, use of the line-of-sight column density of Ni~{\sc ii} in Figure~\ref{fig:depl_factors} could potentially yield misleading values for $F_*$ and $N$(H$_{\rm tot}$). However, when the Ni~{\sc ii} column density is excluded from the analysis for HD~43582, the derived values of $F_*$ and $N$(H$_{\rm tot}$) are not significantly different. We will explore changes in $F_*$ along the line of sight to HD~43582 in Section~\ref{subsec:ni_ca_depl}.} and $\log N({\rm H}_{\rm tot})=22.09\pm0.10$ and $F_*=0.86\pm0.10$ for HD~254755. The dimensionless quantity $F_*$ is defined such that sight lines showing only a very small amount of dust grain depletion have $F_*=0$, while sight lines characterized by strong dust depletions have $F_*=1$ (Jenkins 2009). We therefore find a moderate amount of depletion toward HD~43582 and a fairly strong degree of depletion toward HD~254755. This is consistent with the generally higher molecular column densities observed toward HD~254755 compared to HD~43582 (Section~\ref{subsec:molecular}). The values of $F_*$ derived here will be especially useful in our analysis of O~{\sc i} and Si~{\sc ii} fine-structure excitations (as described in Section~\ref{subsec:oxy_sil_ex}).

\subsubsection{High Velocity Gas toward HD~43582\label{subsubsec:high_vel}}
\begin{figure}
\centering
\includegraphics[width=0.9\textwidth]{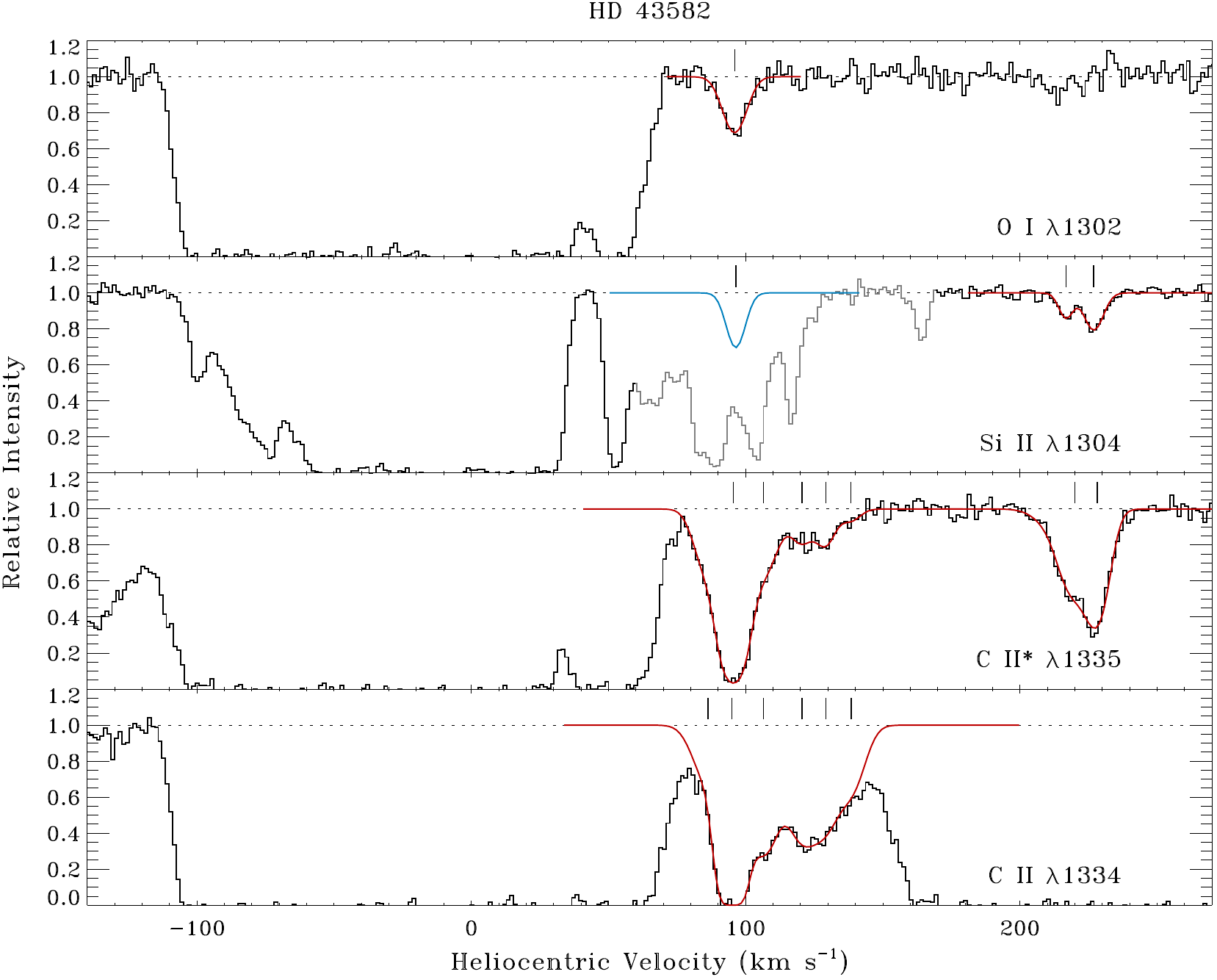}
\caption{Absorption profiles of intrinsically strong lines that probe high-velocity gas toward HD~43582. The smooth red curves represent profile fits to various absorption features where such fits are possible. Tick marks indicate the positions of the velocity components included in the fits. Note the similar component structure of the Si~{\sc ii}~$\lambda1304$ and C~{\sc ii}*~$\lambda1335$ absorption complexes near +224 km s$^{-1}$. This complex is not detected in O~{\sc i}~$\lambda1302$ and cannot be detected in C~{\sc ii}~$\lambda1334$ due to strong saturated absorption from C~{\sc ii}* at the same wavelengths. The other high-velocity absorption complex, which peaks near +96 km s$^{-1}$, is detected in O~{\sc i}, C~{\sc ii}, and C~{\sc ii}*, but cannot be discerned in Si~{\sc ii} due to overlapping absorption from O~{\sc i}*~$\lambda1304$ (gray portion of the histogram). The smooth blue curve represents our best guess as to the strength of this feature in Si~{\sc ii}.\label{fig:high_vel}}
\end{figure}

Most of the analysis described in Sections~\ref{subsubsec:components} and \ref{subsubsection:depl_factors} involved atomic transitions of low to moderate absorption strength since these are the transitions that are not very badly saturated and can therefore be used to examine the total column densities along the lines of sight. Intrinsically strong transitions of abundant elements, on the other hand, can yield information on very low column density material at high positive or negative velocity. We have already seen that the Ca~{\sc ii} and Si~{\sc ii}* profiles toward HD~43582 exhibit numerous distinct components with velocities between $-$100 and +100~km~s$^{-1}$ (Figures~\ref{fig:trace1} and \ref{fig:si1}). Even higher velocities are exhibited by intrinsically strong lines of Si~{\sc ii}, C~{\sc ii}, and C~{\sc ii}*. In Figure~\ref{fig:high_vel}, we show the absorption profiles of the O~{\sc i}~$\lambda1302$ and Si~{\sc ii}~$\lambda1304$ lines along with the partially blended profiles of C~{\sc ii}~$\lambda1334$ and C~{\sc ii}*~$\lambda1335$. Much of the absorption in these transitions is completely saturated, although several distinct high-velocity absorption complexes may be discerned. The complex near +96~km~s$^{-1}$ in the O~{\sc i}, C~{\sc ii}, and C~{\sc ii}* lines is the same as that detected in Ca~{\sc ii} and Si~{\sc ii}*, although the C~{\sc ii} and C~{\sc ii}* features show additional components with velocities as high as 138~km~s$^{-1}$. Another high-velocity absorption complex is seen near +224~km~s$^{-1}$ in Si~{\sc ii} and C~{\sc ii}*. This complex is not detected in O~{\sc i}, however, and cannot be detected in C~{\sc ii} due to strong saturated absorption from C~{\sc ii}* at the same wavelengths.

\begin{deluxetable}{ccccccccccc}
\tablecolumns{11}
\tablewidth{0pt}
\tabletypesize{\scriptsize}
\tablecaption{High Velocity Absorption Components toward HD~43582\label{tab:high_vel}}
\tablehead{ \colhead{$\langle v_{\sun} \rangle$\tablenotemark{a}} & \colhead{$\log N$(C~{\sc ii})} & \colhead{$b$(C~{\sc ii})} & \colhead{$\log N$(C~{\sc ii}*)} & \colhead{$b$(C~{\sc ii}*)} & \colhead{$\log N$(Si~{\sc ii})} & \colhead{$b$(Si~{\sc ii})} & \colhead{$\log N$(Si~{\sc ii}*)} & \colhead{$b$(Si~{\sc ii}*)} & \colhead{$\log N$(O~{\sc i})} & \colhead{$b$(O~{\sc i})} \\
\colhead{(km~s$^{-1}$)} & \colhead{} & \colhead{(km~s$^{-1}$)} & \colhead{} & \colhead{(km~s$^{-1}$)} & \colhead{} & \colhead{(km~s$^{-1}$)} & \colhead{} & \colhead{(km~s$^{-1}$)} & \colhead{} & \colhead{(km~s$^{-1}$)} }
\startdata
\phn+86.3 & $13.02\pm0.06$ & 6.9 & \ldots & \ldots & \ldots & \ldots & \ldots & \ldots & \ldots & \ldots \\
\phn+95.9 & $14.25\pm0.08$ & 4.4 & $13.93\pm0.06$ & 5.6 & [$13.00$] & [4.3] & $11.71\pm0.13$ & 4.3 & $13.37\pm0.06$ & 5.6 \\
     +106.5 & $13.54\pm0.05$ & 6.8 & $13.09\pm0.04$ & 6.5 & \ldots & \ldots & \ldots & \ldots & \ldots & \ldots \\
     +120.6 & $13.43\pm0.05$ & 6.9 & $12.64\pm0.08$ & 5.1 & \ldots & \ldots & \ldots & \ldots & \ldots & \ldots \\
     +129.3 & $13.24\pm0.05$ & 6.1 & $12.64\pm0.07$ & 4.4 & \ldots & \ldots & \ldots & \ldots & \ldots & \ldots \\
     +138.4 & $13.04\pm0.06$ & 6.3 & $12.15\pm0.18$ & 4.4 & \ldots & \ldots & \ldots & \ldots & \ldots & \ldots \\
     +218.4 & \ldots\tablenotemark{b} & \ldots\tablenotemark{b} & $13.25\pm0.04$ & 7.6 & $12.51\pm0.07$ & 3.1 & $\lesssim11.5$\tablenotemark{c} & \ldots & $\lesssim13.0$\tablenotemark{c} & \ldots \\
     +227.6 & \ldots\tablenotemark{b} & \ldots\tablenotemark{b} & $13.33\pm0.04$ & 5.3 & $12.82\pm0.05$ & 4.4 & $\lesssim11.5$\tablenotemark{c} & \ldots & $\lesssim13.0$\tablenotemark{c} & \ldots \\
\enddata
\tablenotetext{a}{Average heliocentric velocity of the components listed.}
\tablenotetext{b}{Absorption at these velocities cannot be detected in the C~{\sc ii}~$\lambda1334$ line due to strong saturated absorption from C~{\sc ii}*~$\lambda1335$ at the same wavelengths.}
\tablenotetext{c}{3$\sigma$ upper limits.}
\end{deluxetable}

The absorption complex near +96~km~s$^{-1}$ cannot be discerned in the Si~{\sc ii}~$\lambda1304$ line due to overlapping absorption from O~{\sc i}*~$\lambda1304$. This presents a problem because there is good reason to suspect that a Si~{\sc ii} absorption component is present at this velocity. In order to properly evaluate the column densities of the O~{\sc i}* components, the contribution to the absorption from Si~{\sc ii} must first be determined. The only other Si~{\sc ii} transition covered by the \emph{HST} observations is the line at 1260.4~\AA{}. However, this feature is affected by blending from the C~{\sc i}~$\lambda1260$ multiplet. Additional information on the high-velocity absorption complexes detected toward HD~43582 may be gleaned from the \emph{IUE} observations first reported by Gondhalekar \& Phillips (1980). The \emph{IUE} spectra of HD~43582 (available from the MAST archive) suffer from low resolution and poor S/N. Nevertheless, these data cover important transitions such as the Mg~{\sc ii} doublet near 2800~\AA{} and the Fe~{\sc ii} line at 2600~\AA{}. Both of the high-velocity absorption complexes identified in our \emph{HST} spectra also appear in the Mg~{\sc ii} and Fe~{\sc ii} lines from \emph{IUE} (although the precise component structures of the high-velocity complexes are very difficult to discern from the \emph{IUE} observations). The Mg~{\sc ii} lines are potentially the most useful for estimating the strength of the Si~{\sc ii}~$\lambda1304$ component at +96~km~s$^{-1}$ since Si and Mg have similar abundances, ionization potentials, and depletion properties. The minimum intensities of the absorption complexes evident near +120 and +220~km~s$^{-1}$ in the Mg~{\sc ii}~$\lambda\lambda2796,2803$ lines are nearly identical, suggesting that the Si~{\sc ii} component at +96~km~s$^{-1}$ may have a minimum intensity similar to that seen for the complex near +224~km~s$^{-1}$.

In Table~\ref{tab:high_vel}, we present the results of profile synthesis fits to the high-velocity absorption complexes detected in the O~{\sc i}, Si~{\sc ii}, C~{\sc ii}, and C~{\sc ii}* lines displayed in Figure~\ref{fig:high_vel}. (Note that our profile fits to the C~{\sc ii}* features account for absorption from both C~{\sc ii}* transitions at 1335.66~\AA{} and 1335.71~\AA{}.) Also included in the table are the results we obtained for the Si~{\sc ii}* component at +96~km~s$^{-1}$ from our simultaneous fit to the Si~{\sc ii}*~$\lambda1264$ and $\lambda1265$ transitions (Figure~\ref{fig:si1}). The Si~{\sc ii}*~$\lambda\lambda1264,1265$ profile toward HD~43582 exhibits only a single, relatively narrow component at +96~km~s$^{-1}$, similar to what is seen in the O~{\sc i}~$\lambda1302$ line. The absorption at this velocity is much stronger in the C~{\sc ii} and C~{\sc ii}* lines. In the following, we will assume that all of the C, O, and Si in the +96~km~s$^{-1}$ component is in the gas-phase and that the relative abundances are equivalent to the solar relative abundances. If we further assume that typical ionization conditions for neutral gas apply, then the O~{\sc i} column density of $\log N($O~{\sc i}$)=13.37$ would imply that  $\log N($Si~{\sc ii}$)\approx12.06$ (accounting for the small amount of Si~{\sc ii}* that is present). Under the same assumption, however, the total C~{\sc ii} column density of $\log N($C~{\sc ii}$_{\rm tot})=\log [N($C~{\sc ii}$)+N($C~{\sc ii}*$)]=14.42$ would imply that  $\log N($Si~{\sc ii}$)\approx13.57$. Evidently, this high-velocity component is not predominantly neutral, but may instead be nearly completely ionized. Considering that O~{\sc i} is a good proxy for H~{\sc i}, while C~{\sc ii} is a tracer of both H~{\sc i} and H~{\sc ii} regions, the measured column densities of O~{\sc i} and C~{\sc ii} for the +96~km~s$^{-1}$ component imply an ionization fraction of  $x(e)=n(e)/n({\rm H}_{\rm tot})\approx0.96$.

The smooth blue curve in the panel for Si~{\sc ii}~$\lambda1304$ in Figure~\ref{fig:high_vel} represents our best guess as to the shape and strength of the Si~{\sc ii} profile for the +96~km~s$^{-1}$ component. To obtain this profile, we adopted the same velocity and $b$-value as determined for the corresponding Si~{\sc ii}* component and assumed a column density of $\log N($Si~{\sc ii}$)=13.00$. This choice of column density produced an absorption profile with a minimum intensity similar to that seen for the absorption complex near +224~km~s$^{-1}$, a constraint which was motivated by the appearance of the Mg~{\sc ii} features in the \emph{IUE} spectrum of HD~43582. Furthermore, our assumed value for $N$(Si~{\sc ii}) yields a Si~{\sc ii}*/Si~{\sc ii} column density ratio of $\sim$$5\times10^{-2}$, which is similar to the ratios we find for the other absorption components observed at moderate to high velocity toward HD~43582 (see Section~\ref{subsec:oxy_sil_ex}). Note that the moderate velocity component near +52~km~s$^{-1}$ in the Si~{\sc ii}~$\lambda1304$ line is partially blended with the blue side of the O~{\sc i}*~$\lambda1304$ profile (Figure~\ref{fig:high_vel}). However, unlike with the Si~{\sc ii} component at +96~km~s$^{-1}$, it was relatively straightforward to model the +52~km~s$^{-1}$ component and remove it from the spectrum so that we could proceed with the analysis of the O~{\sc i}* profile.

\begin{figure}
\centering
\includegraphics[width=0.9\textwidth]{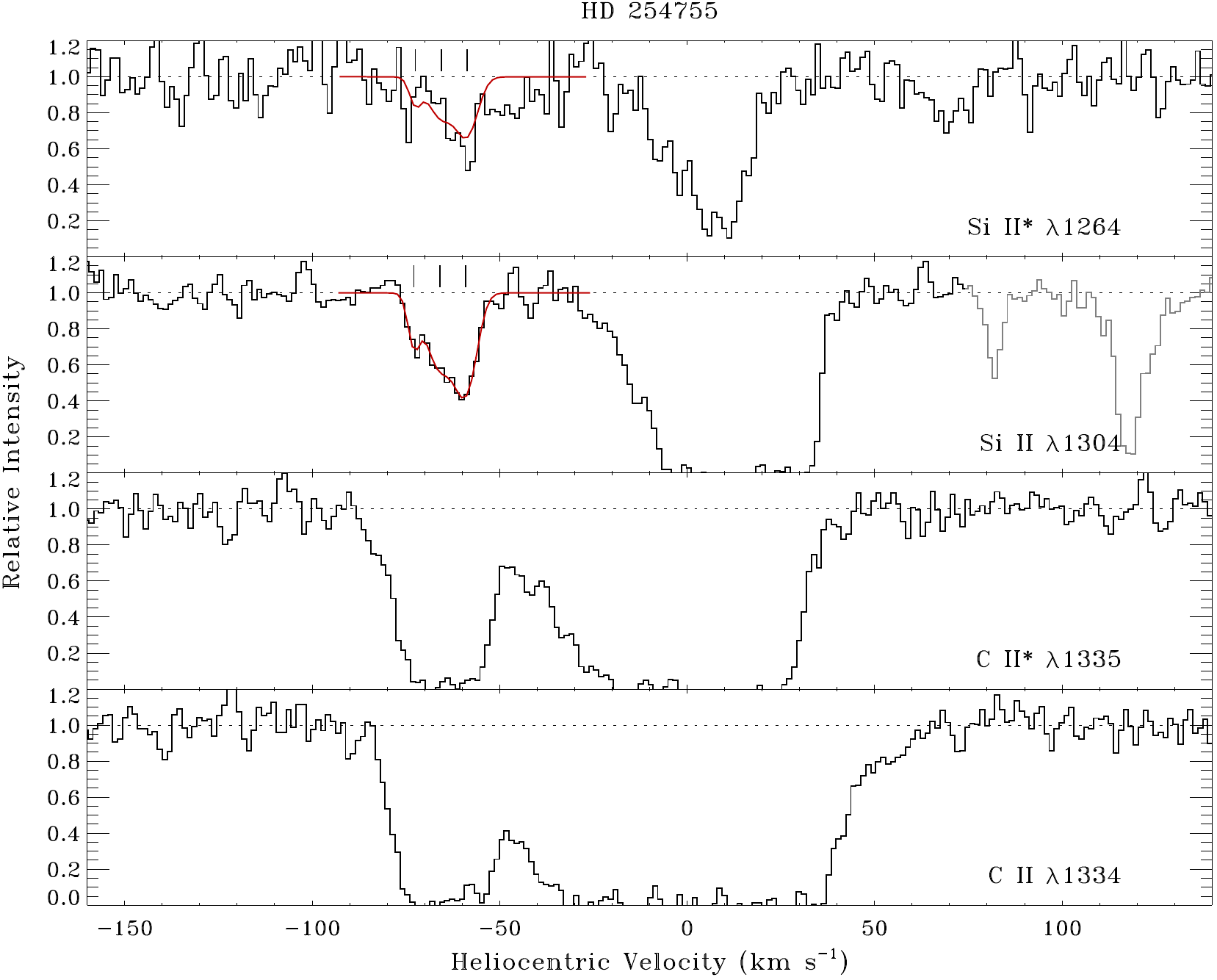}
\caption{Absorption profiles of intrinsically strong lines that probe moderately high-velocity gas toward HD~254755. The smooth red curves in the upper two panels represent profile fits to the absorption complex near $-$63~km~s$^{-1}$. Tick marks indicate the positions of the velocity components included in the fits. (The gray portion of the histogram in the second panel from the top traces telluric and interstellar absorption from O~{\sc i}*~$\lambda1304$.) The component structure obtained from the Si~{\sc ii}~$\lambda1304$ complex near $-$63~km~s$^{-1}$ was used to fit this same absorption feature in the Si~{\sc ii}* profile. The C~{\sc ii}~$\lambda1334$ and C~{\sc ii}*~$\lambda1335$ profiles are too badly saturated to allow any meaningful determinations of column densities.\label{fig:high_vel2}}
\end{figure}

\subsubsection{Moderately High Velocity Gas toward HD~254755\label{subsubsec:high_vel2}}
The absorption complex near $-$63~km~s$^{-1}$ toward HD~254755, which is seen in some singly-ionized species such as Ca~{\sc ii} and Si~{\sc ii}* (Figures~\ref{fig:trace2} and \ref{fig:si2}), merits special attention. While the velocity of this feature is only moderately high compared to the velocities of some of the absorption complexes seen toward HD~43582, it is nonetheless unusual. The line of sight to HD~254755 passes beyond the outer edge of the optical filaments associated with IC~443, which presumably mark the outward expansion of the supernova shock wave as it interacts with atomic gas in the northeastern part of the remnant. Moderately high velocity gas positioned ahead of the shock front would not generally be expected, unless the absorption in question traces a shock precursor. Since IC~443 is in the Galactic anticenter direction ($l=189.1$, $b=+3.0$), differential Galactic rotation would not be capable of producing the moderately high negative velocities seen toward HD~254755. A closer examination of this absorption is therefore warranted.

In Figure~\ref{fig:high_vel2}, we plot the absorption profiles of the Si~{\sc ii}~$\lambda1304$, Si~{\sc ii}*~$\lambda1264$, C~{\sc ii}~$\lambda1334$, and C~{\sc ii}*~$\lambda1335$ lines toward HD~254755, all of which show absorption from the complex near $-$63~km~s$^{-1}$. We do not include the O~{\sc i}~$\lambda1302$ line in this figure because the $-$63~km~s$^{-1}$ component (if present) is severely blended with absorption from P~{\sc ii}~$\lambda1301$ at low velocity. The upper two panels of Figure~\ref{fig:high_vel2} show profile synthesis fits to the $-$63~km~s$^{-1}$ feature in the Si~{\sc ii} and Si~{\sc ii}* lines. The Si~{\sc ii} complex was fitted first and the resulting component structure was used as a fixed template for the corresponding feature in our simulataneous fit of the Si~{\sc ii}*~$\lambda1264$ and $\lambda1265$ lines (Figure~\ref{fig:si2}). These fits yield total column densities of $\log N($Si~{\sc ii}$)=13.58\pm0.04$ and $\log N($Si~{\sc ii}*$)=12.19\pm0.09$ for the absorption complex near $-$63~km~s$^{-1}$. We do not attempt to derive column densities of C~{\sc ii} or C~{\sc ii}* for this feature because the absorption is clearly saturated. However, the Si~{\sc ii}/C~{\sc ii} ratio appears to be consistent with the solar Si/C ratio, although a subsolar ratio is not excluded because the C~{\sc ii} absorption provides only a lower limit to the true C~{\sc ii} abundance. The Si~{\sc ii}*/Si~{\sc ii} column density ratio for the absorption complex near $-$63~km~s$^{-1}$ is $(4.1\pm1.1)\times10^{-2}$, which is comparable to the ratios determined for the high excitation absorption features observed toward HD~43582 (Section~\ref{subsec:oxy_sil_ex}).

\subsection{Molecular Species\label{subsec:molecular}}
Along with the absorption features of neutral and singly-ionized atomic species, Figures~\ref{fig:trace1} and \ref{fig:trace2} show the absorption profiles of the CH~$\lambda4300$ and CH$^+$~$\lambda4232$ lines extracted from our HET spectra of HD~43582 and HD~254755. These molecular species exhibit component structures similar to what we find for the trace neutral species, such as K~{\sc i}, S~{\sc i}, and Cl~{\sc i}, where a single dominant component near +6 or +7~km~s$^{-1}$ contains much of the absorption and weaker components are found at more positive velocities (up to approximately +30~km~s$^{-1}$). The total column densities of both CH and CH$^+$ are larger toward HD~254755, whereas the $N$(CH$^+$)/$N$(CH) ratio is about a factor of two higher toward HD~43582 (see Table~\ref{tab:column_densities}). Hirschauer et al.~(2009), examining moderate-resolution ARCES spectra of stars in IC~443, detected the $R$(0) and $R$(1) lines of CN near 3874~\AA{} toward HD~254755 (but not toward HD~43582). We have re-analyzed the CN data for HD~254755 using a smaller $b$-value compared to that adopted in Hirschauer et al.~(2009), finding a total CN column density of $\log N($CN$)=12.46\pm0.06$. This change was motivated by our analysis of the K~{\sc i}~$\lambda7698$ line (Figures~\ref{fig:trace2}), which resulted in a $b$-value of 1.2~km~s$^{-1}$ for the dominant absorption component toward HD~254755, considerably smaller than the value of 2.2~km~s$^{-1}$ adopted for the CN analysis in Hirschauer et al.~(2009).

The CH column densities listed in Table~\ref{tab:column_densities} can be used to estimate the molecular fractions that characterize the gas toward our target stars by considering the strong empirical correlation known to exist between the column densities of CH and H$_2$ (e.g., Federman 1982; Danks et al.~1984; Welty et al.~2006; Sheffer et al.~2008). We use an updated determination of this correlation (Welty et al.~2014; Ritchey et al.~2015), which indicates that $\log N({\rm H}_2)\approx20.77$ toward HD~43582 and $21.18$ toward HD~254755. The scatter in the relationship between $\log N({\rm CH})$ and $\log N({\rm H}_2)$ is $\sim$0.14 dex (Welty et al.~2006). To estimate the amount of H~{\sc i} toward our target stars, we take advantage of another strong empirical correlation, i.e., between the H~{\sc i} column density and the equivalent width of the $\lambda5780.5$ DIB (e.g., Friedman et al.~2011). From our HET spectra, we find $W_{\lambda}(5780.5)=407.0\pm6.4$~m\AA{} toward HD~43582 and $430.2\pm8.9$~m\AA{} toward HD~254755, which imply respective values of $\log N($H~{\sc i}$)\approx21.45$ and $21.47$. These estimates for $N$(H~{\sc i}) could be too low if the sight lines to HD~43582 and HD~254755 are characterized by stronger than average radiation fields (e.g., Welty et al.~2014). A comparison with the values of $N({\rm H}_{\rm tot})$ derived in Section~\ref{subsubsection:depl_factors} suggests that this may be the case, assuming the $N$(H$_2$) values predicted from $N$(CH) are correct.\footnote{A direct evaluation of the H~{\sc i} Ly$\alpha$ features toward our target stars also seems to indicate that the H~{\sc i} column densities are larger than predicted by the $\lambda5780.5$ DIB. However, these determinations are uncertain because the \emph{HST} spectra cover only a portion of the line. Much of the blue wing of the Ly$\alpha$ profile is missing, making an accurate assessment of the continuum level difficult.} The estimates for $N$(H~{\sc i}) and $N$(H$_2$) derived in this section yield molecular fractions of $f({\rm H}_2)=2N({\rm H}_2)/[N($H~{\sc i}$)+2N({\rm H}_2)]\approx 0.3$ and 0.5 for the sight lines to HD~43582 and HD~254755, respectively. Higher H~{\sc i} column densities would imply somewhat smaller values for the line-of-sight molecular fractions.

\begin{figure}
\centering
\includegraphics[width=0.49\textwidth]{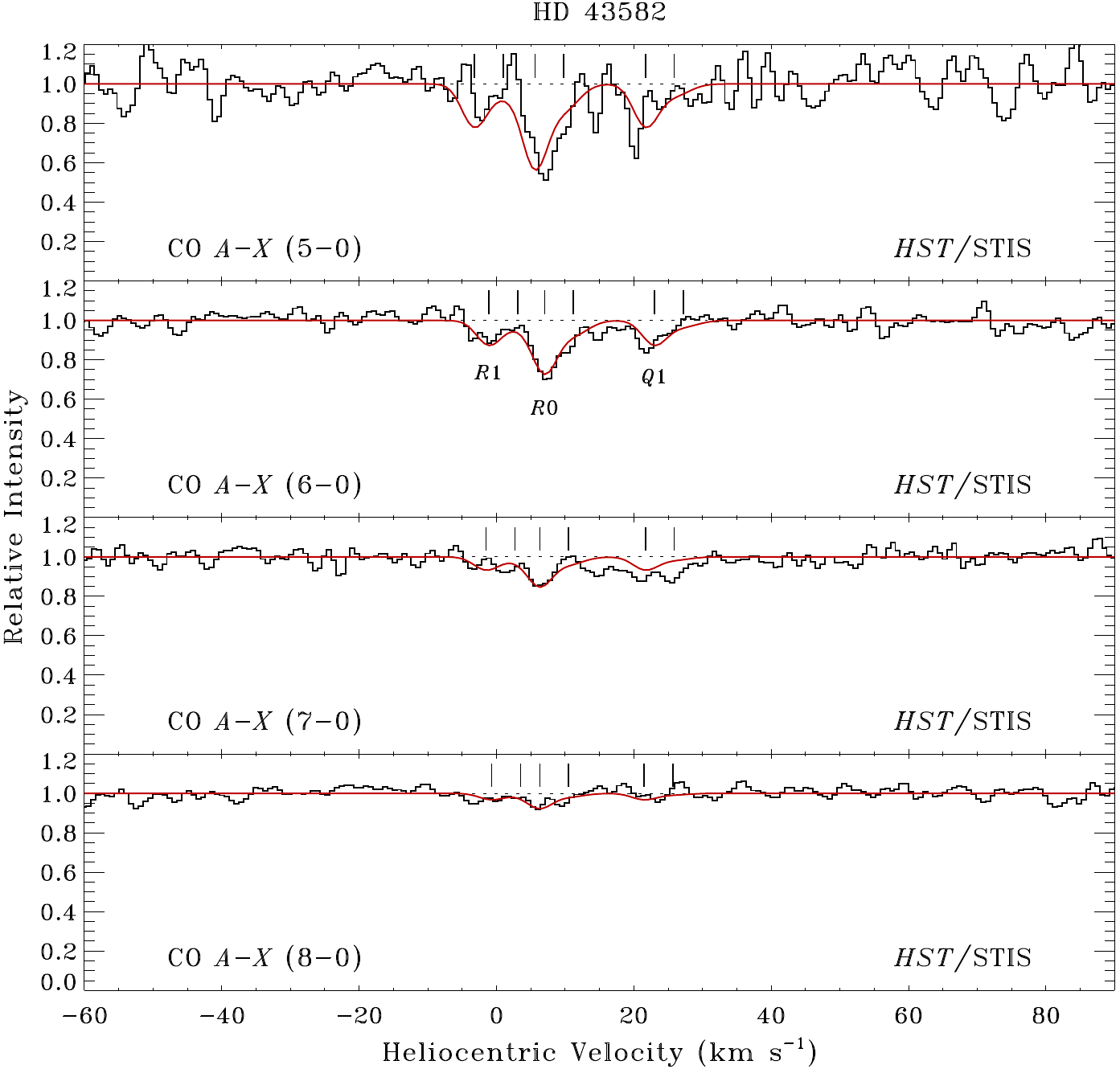}
\includegraphics[width=0.49\textwidth]{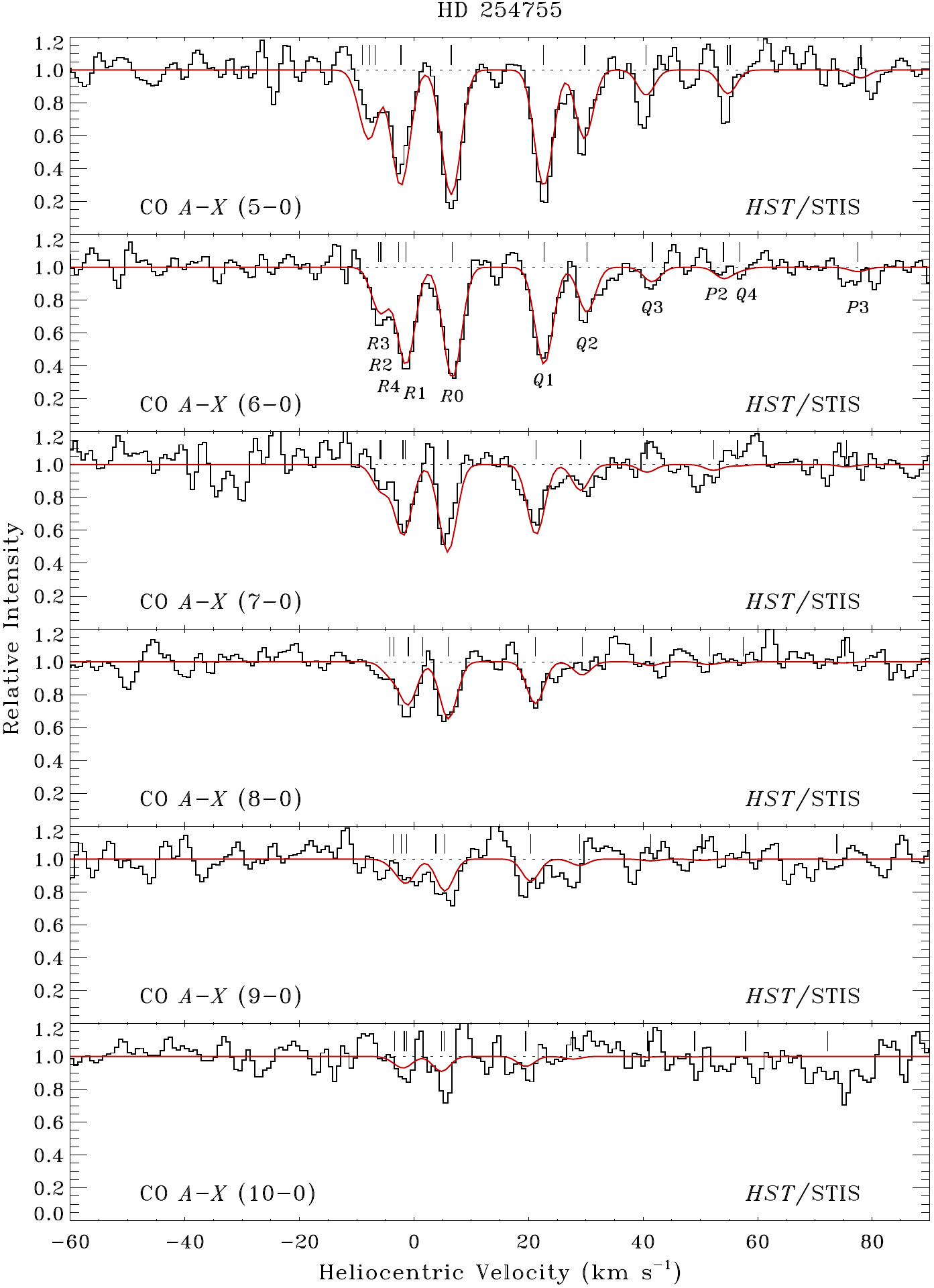}
\caption{Global profile synthesis fits to the observed CO bands toward HD~43582 (left panels) and HD~254755 (right panels) from UV data acquired with \emph{HST}/STIS. As in previous figures, the synthetic profiles are shown as smooth red curves, with histograms representing the observed spectra. For HD 43582, two velocity components are included in the fit for each rotational transition up to $J=1$ (as indicated by tick marks and labels). For HD 254755, a single velocity component is adopted in the fit for each rotational transition up to $J=4$.\label{fig:co_fits}}
\end{figure}

Our \emph{HST}/STIS spectra cover several absorption bands of the CO $A$$-$$X$ system, starting with the (5$-$0) band near 1392~\AA{} (see Figure~\ref{fig:co_fits}). To fit the CO bands consistently, we employed a modified version of the profile fitting routine that allowed us to synthesize all of the detected CO bands simultaneously. (This is essentially the same version of the program used to fit the C~{\sc i} multiplets.) The $f$-values adopted for the CO transitions are the same as those used by Sheffer et al.~(2008). For HD~254755, we included all of the CO $A$$-$$X$ bands from (5$-$0) to (10$-$0) and fit each rotational transition up to $J=4$, adopting a single velocity component with a $b$-value of 1.0~km~s$^{-1}$. For HD~43582, we excluded the (9$-$0) and (10$-$0) bands as there was no detectable absorption from these features and fit the rotational transitions up to $J=1$ only. Two velocity components separated by $\sim$4~km~s$^{-1}$ were adopted in the fit for HD~43582, yielding a component structure similar to that found for the two strongest CH components in this direction. The total column densities resulting from our global fits to the CO bands detected toward HD~43582 and HD~254755 are reported in Table~\ref{tab:column_densities}. As expected, the CO column density is significantly higher toward HD~254755, a result which is consistent with the higher CH column density and the detection of CN and C$_2$ in this direction (see below).

\begin{figure}
\centering
\includegraphics[width=0.9\textwidth]{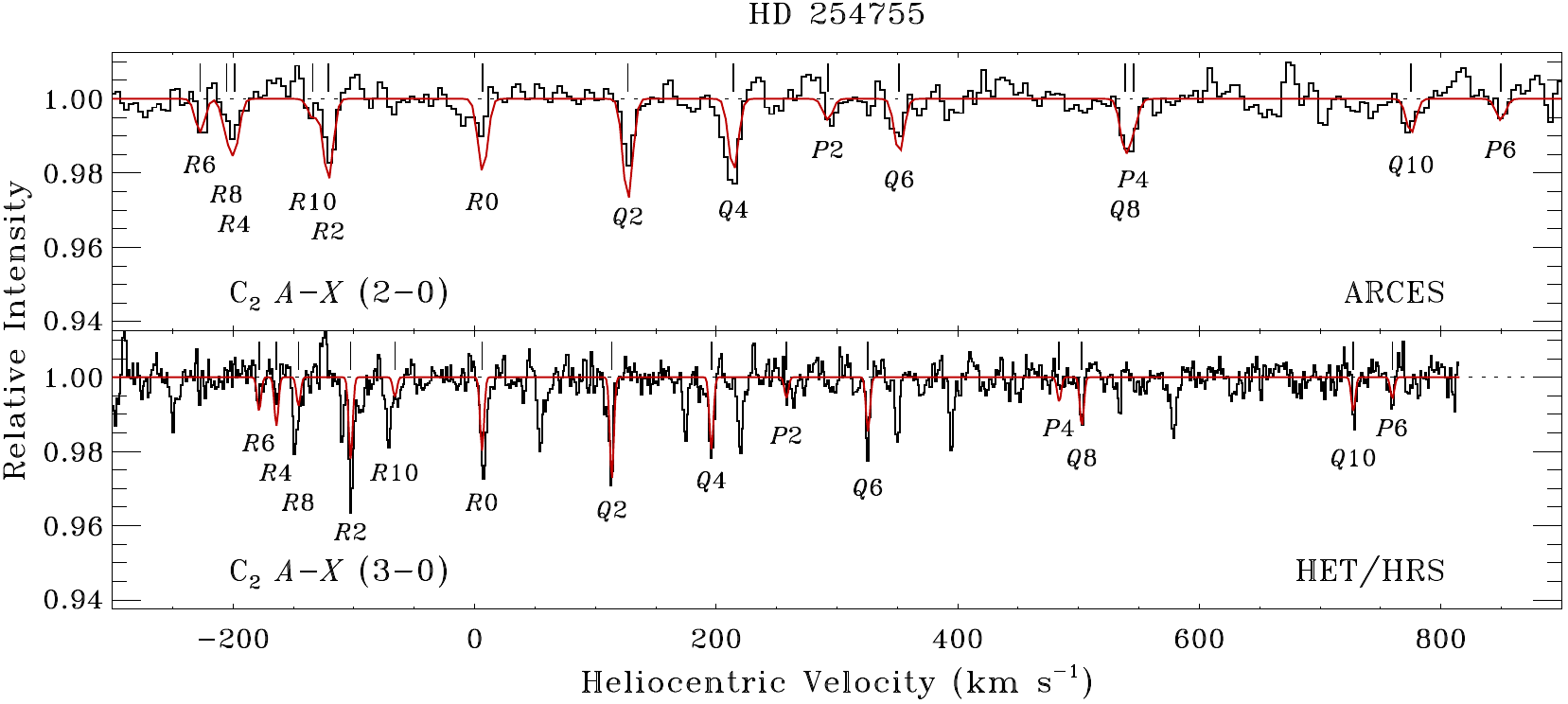}
\caption{Global profile synthesis fit to the $A$$-$$X$ (2$-$0) and (3$-$0) bands of C$_2$ toward HD~254755 from ground-based data acquired with ARCES and with the HET/HRS. As in previous figures, the synthetic profiles are shown as smooth red curves, with histograms representing the observed spectra. A single velocity component is adopted in the fit for each rotational transition up to $J=10$ (as indicated by tick marks and labels). The absorption features in the lower panel that are not included in the model spectrum are telluric features that were not removed when the data were reduced.\label{fig:c2_fits}}
\end{figure}

Hirschauer et al.~(2009) detected the $A$$-$$X$ (2$-$0) band of C$_2$ near 8757~\AA{} toward HD~254755 and derived column densities for the individual rotational levels up to $J=10$. Our HET spectrum of HD~254755 reveals absorption from the C$_2$ $A$$-$$X$ (3$-$0) band near 7719~\AA{} (see Figure~\ref{fig:c2_fits}). In light of the new information provided by the (3$-$0) band, we re-examined the C$_2$ features detected by Hirschauer et al.~(2009) toward HD~254755. We used our profile synthesis routine to perform a simultaneous fit to the (2$-$0) and (3$-$0) bands, including all rotational transitions up to $J=10$. The $f$-values adopted in the C$_2$ fit were obtained from Sonnentrucker et al.~(2007) and Hupe et al.~(2012). As in our fit to the CO bands toward HD~254755, we adopted a single velocity component, and found the $b$-value to be 1.7~km~s$^{-1}$. Note that the velocity resolution of the ARCES data ($\sim$9.5~km~s$^{-1}$) is much lower than that of the HET spectra (as is apparent in Figure~\ref{fig:c2_fits}). Also, there are telluric absorption lines in the vicinity of the (3$-$0) band that were not removed by the data reduction procedure. Special care was required in our modeling of the C$_2$ lines so as to exclude these telluric features. The total C$_2$ column density obtained from our global fit to the (2$-$0) and (3$-$0) bands toward HD~254755 is given in Table~\ref{tab:column_densities}. This more precise result for $N$(C$_2$) represents a 20\% decrease from the value reported in Hirschauer et al.~(2009). In Section~\ref{subsec:mol_ex}, the column densities of the various molecular species discussed here will be used to evaluate the physical conditions in the molecule-bearing gas along our target sight lines through an analysis of CO and C$_2$ excitation and by considering the chemical networks that link the CH, C$_2$, and CN molecules.

\begin{figure}
\centering
\includegraphics[width=0.49\textwidth]{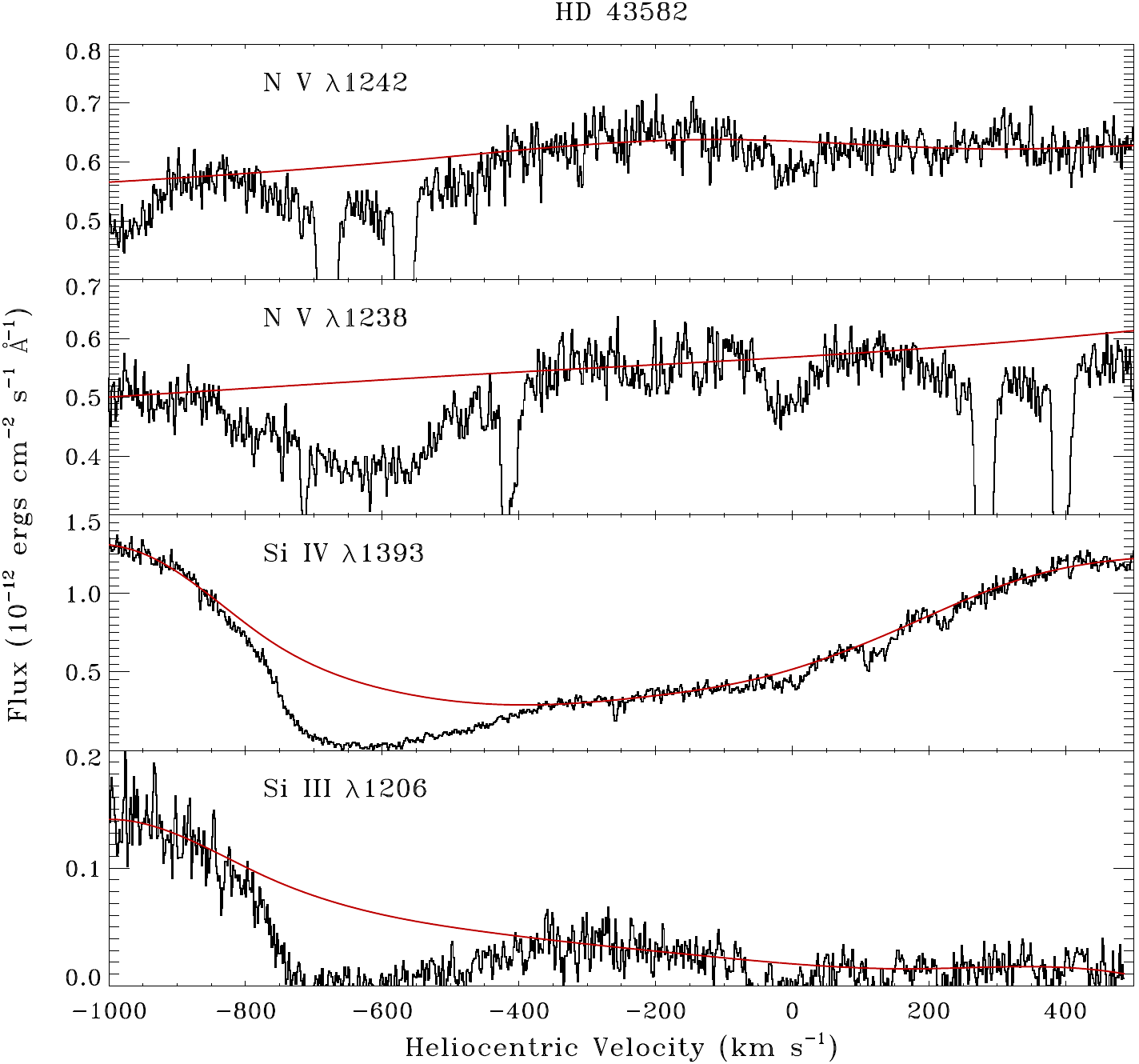}
\includegraphics[width=0.49\textwidth]{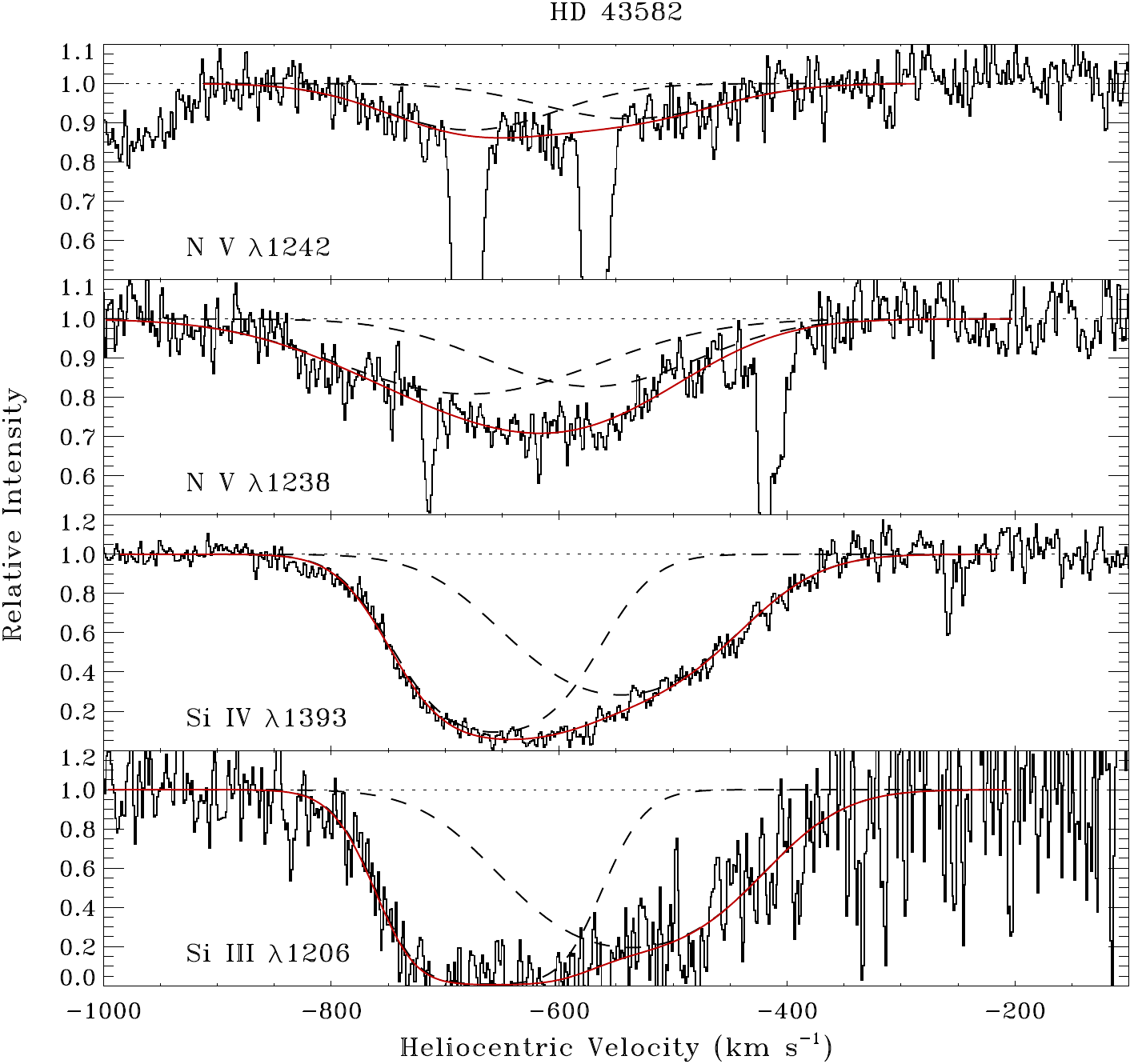}
\caption{Absorption profiles of highly-ionized species toward HD~43582. The unnormalized spectra are shown in the left panels with red lines giving the adopted continuum fits. The right panels present the continuum-normalized spectra for the broad absorption feature near $-$620 km~s$^{-1}$. The smooth red curves in the right panels represent Voigt profile fits to the broad absorption feature with dashed lines showing the contributions from two individual components. (The strong narrow absorption lines in the top two panels are due to Mg~{\sc ii}~$\lambda\lambda1239$, 1240, Ge~{\sc ii}~$\lambda1237$, and Kr~{\sc i}~$\lambda1235$, while the narrow features near $-$250 km~s$^{-1}$ in the third panel from the top arise from the CO $A$$-$$X$~(5$-$0) band near 1392.5~\AA{}.) Note also, in the panels on the left, the weaker absorption from Si~{\sc iii}, Si~{\sc iv}, and N~{\sc v} near 0 km~s$^{-1}$ and the additional Si~{\sc iv} absorption features near +122 km~s$^{-1}$ and +225 km~s$^{-1}$ (see also Figure~\ref{fig:high_ions2}).\label{fig:high_ions}}
\end{figure}

\subsection{Highly-Ionized Species toward HD~43582\label{subsec:high_ions}}
\begin{figure}
\centering
\includegraphics[width=0.7\textwidth]{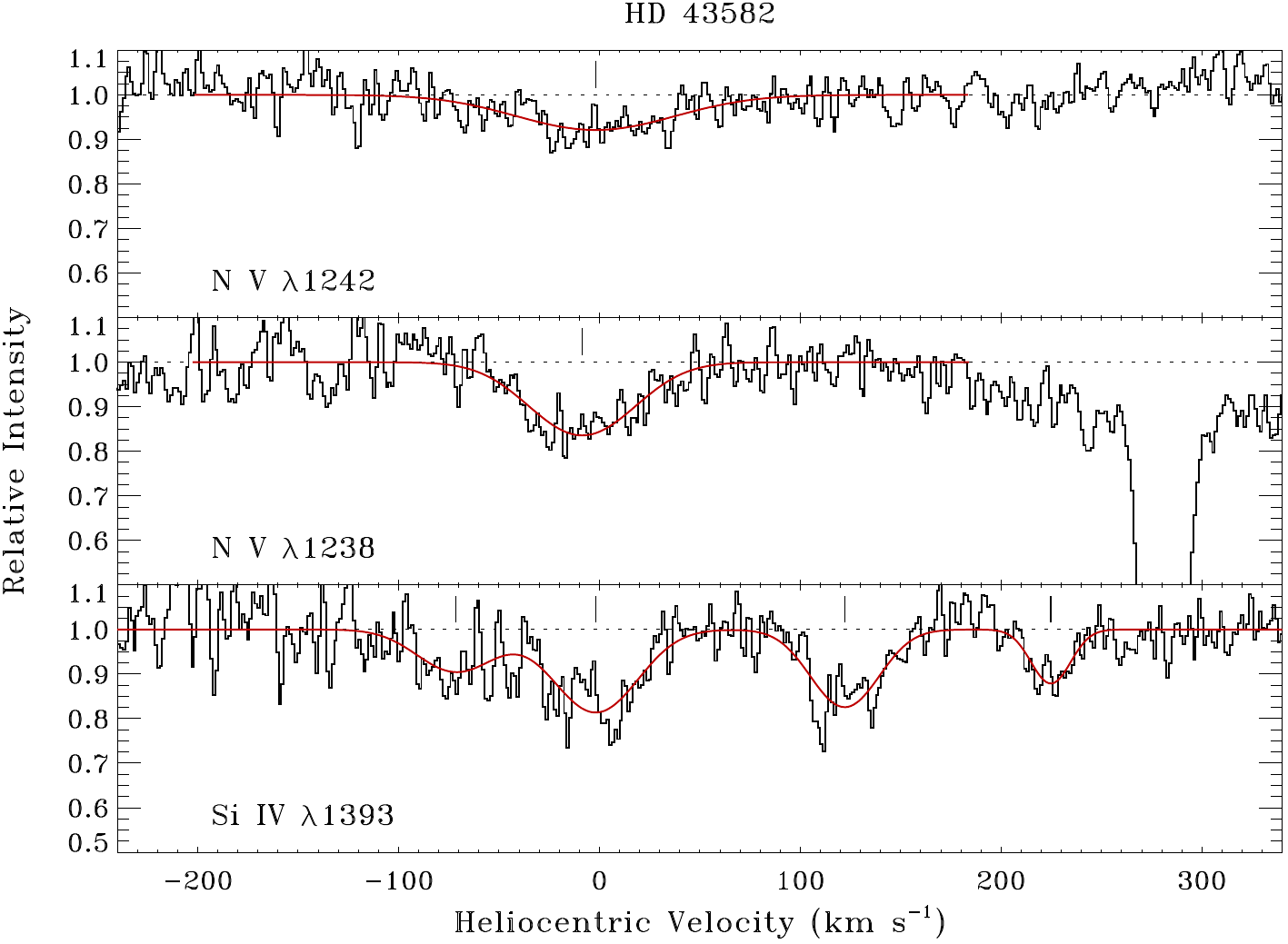}
\caption{Voigt profile fits to the Si~{\sc iv} and N~{\sc v} features observed toward HD~43582 (excluding the very broad absorption feature near $-$620 km~s$^{-1}$). The synthetic profiles are shown as smooth red curves, with histograms representing the observed spectra. Tick marks indicate the positions of the velocity components included in the fits. The strong narrow absorption feature in the second panel is due to Mg~{\sc ii}~$\lambda1239$.\label{fig:high_ions2}}
\end{figure}

Within the coverage of our \emph{HST}/STIS spectra are several transitions from atomic species at stages of ionization much higher that those typically associated with neutral gas. These transitions include the N~{\sc v}~$\lambda\lambda1238,1242$ doublet, the Si~{\sc iv}~$\lambda1393$ line, and the Si~{\sc iii}~$\lambda1206$ feature. (The other line of the Si~{\sc iv} doublet at 1402.8~\AA{} falls just beyond the edge of the detector given our chosen wavelength setting.) Since these highly-ionized species can be useful probes of post-shock gases, we searched for absorption from these transitions in the \emph{HST} spectrum of HD~43582. Absorption near 0~km~s$^{-1}$ is clearly seen in the N~{\sc v} doublet and in the Si~{\sc iv}~$\lambda1393$ line, while two additional Si~{\sc iv} absorption features are detected near +122 and +225~km~s$^{-1}$ (see Figures~\ref{fig:high_ions} and \ref{fig:high_ions2}). (The Si~{\sc iii}~$\lambda1206$ transition probably also exhibits absorption at these velocities. However, the flux from the star at these wavelengths is nearly completely suppressed by the nearby H~{\sc i}~Ly$\alpha$ feature, making it impossible to derive any meaningful measurements from the Si~{\sc iii} line for these components.)

In addition to the relatively narrow absorptions near 0, +122, and +225~km~s$^{-1}$, a very broad absorption feature is detected in the Si~{\sc iii}, Si~{\sc iv}, and N~{\sc v} lines near $-$620~km~s$^{-1}$ (Figure~\ref{fig:high_ions}). When this component was first identified, it was unclear whether the absorption was associated with hot gas in the SNR or was the result of mass loss from the background hot star. The spectral type of HD~43582 is B0 IIIn (Morgan et al.~1955), so the star could potentially show signs of high-velocity mass-loss features. To investigate this possibility, we examined archival STIS spectra of stars with similar spectral types. The available high-resolution STIS spectra of HD~99890, which also has the spectral type B0 IIIn (Garrison et al.~1977), proved to be particularly useful for this comparison. The N~{\sc v}, Si~{\sc iv}, and C~{\sc iv} doublets in the spectrum of HD~99890 have prominent P-Cygni profiles, and also show relatively narrow absorption components near $-$1500~km~s$^{-1}$, which presumably trace small-scale structure in the high-velocity outgoing wind material. A key difference between the spectrum of HD~99890 and that of HD~43582 is that the N~{\sc v} doublet in the latter case shows no detectable emission at the expected location for a P-Cygni profile. Moreover, the broad absorption feature near $-$620~km~s$^{-1}$ in the spectrum of HD~43582 is at a much lower velocity than the narrow wind components detected in the spectrum of HD~99890. Indeed, the high velocities of the narrow absorption components associated with HD~99890, and other stars we examined with archival STIS spectra, are typical of such features in the P-Cygni profiles of early-type stars (e.g., Prinja \& Howarth 1986). If the broad N~{\sc v} component near $-$620~km~s$^{-1}$ traces hot, expanding gas in the SNR rather than outgoing wind material from HD~43582, then the close correspondence in velocity between the N~{\sc v} component and those of Si~{\sc iii} and Si~{\sc iv} argues in favor of a SNR origin for all three features.

\begin{deluxetable}{ccccccc}
\tablecolumns{7}
\tablewidth{0pt}
\tabletypesize{\footnotesize}
\tablecaption{Highly-Ionized Species toward HD~43582\label{tab:high_ions}}
\tablehead{ \colhead{$\langle v_{\sun} \rangle$\tablenotemark{a}} & \colhead{$\log N$(Si~{\sc iii})} & \colhead{$b$(Si~{\sc iii})} & \colhead{$\log N$(Si~{\sc iv})} & \colhead{$b$(Si~{\sc iv})} & \colhead{$\log N$(N~{\sc v})} & \colhead{$b$(N~{\sc v})} \\
\colhead{(km~s$^{-1}$)} & \colhead{} & \colhead{(km~s$^{-1}$)} & \colhead{} & \colhead{(km~s$^{-1}$)} & \colhead{} & \colhead{(km~s$^{-1}$)} }
\startdata
$-$665 & $14.07\pm0.07$ & 72 & $14.25\pm0.06$ & 80 & $14.01\pm0.08$ & 127 \\
$-$546 & $13.80\pm0.11$ & 112 & $14.11\pm0.06$ & 108 & $13.89\pm0.10$ & 116 \\
\phn$-$71 & \ldots\tablenotemark{b} & \ldots\tablenotemark{b} & $12.40\pm0.13$ & 27 & \ldots & \ldots \\
\phn\phn$-$4 & \ldots\tablenotemark{b} & \ldots\tablenotemark{b} & $12.74\pm0.07$ & 28 & $13.37\pm0.10$ & 45 \\
+122 & \ldots\tablenotemark{b} & \ldots\tablenotemark{b} & $12.62\pm0.07$ & 23 & \ldots & \ldots \\
+225 & \ldots\tablenotemark{b} & \ldots\tablenotemark{b} & $12.21\pm0.10$ & 13 & \ldots & \ldots \\
\enddata
\tablenotetext{a}{Average heliocentric velocity of the components listed.}
\tablenotetext{b}{While there seems to be absorption from Si~{\sc iii} at these velocities, the flux from the star is almost completely suppressed by the nearby H~{\sc i}~Ly$\alpha$ line, precluding any meaningful measurement of the absorption.}
\end{deluxetable}

We analyzed the Si~{\sc iii}, Si~{\sc iv}, and N~{\sc v} absorption features detected toward HD~43582 using our Voigt profile fitting routine. We modeled the broad $-$620~km~s$^{-1}$ feature with two velocity components to account for the clear asymmetry in the Si~{\sc iii} and Si~{\sc iv} lines (Figure~\ref{fig:high_ions}). (The N~{\sc v} features at this velocity could have been fit with a single broad component, but two were included for consistency.) We also fit the Si~{\sc iv} and N~{\sc v} components detected at more moderate velocities (Figure~\ref{fig:high_ions2}). The results are presented in Table~\ref{tab:high_ions}. The low velocity Si~{\sc iv} and N~{\sc v} components have an average centroid velocity of $-$4~km~s$^{-1}$, although the absorption extends from $-$60 to +50~km~s$^{-1}$ (in the N~{\sc v} line). A secondary component centered at $-$71~km~s$^{-1}$ is detected in Si~{\sc iv}. These low velocity absorption components from highly-ionized species cover essentially the same range in velocity exhibited by the neutral and singly-ionized species discussed in Section~\ref{subsubsec:components}. The difference is that the absorption from the high ions is broad and smooth (as far as we can tell given the limitations of the data), while the low ionization species exhibit numerous distinct components. The Si~{\sc iv} components at +122 and +225~km~s$^{-1}$ appear to be related to the high velocity absorption components detected in O~{\sc i}, Si~{\sc ii}, Si~{\sc ii}*, C~{\sc ii}, and C~{\sc ii}* (Section~\ref{subsubsec:high_vel}). However, while the low ionization species, including O~{\sc i} and C~{\sc ii}, feature a prominent component at +96~km~s$^{-1}$, the Si~{\sc iv} absorption near +122~km~s$^{-1}$ corresponds more directly to the weaker components seen only in C~{\sc ii} and C~{\sc ii}* between +100 and +140~km~s$^{-1}$ (Table~\ref{tab:high_vel} and Figure~\ref{fig:high_vel}).

The Doppler broadening parameters derived through profile synthesis can be used to constrain the temperatures of the gases responsible for the absorption. Since the $b$-value is related to the kinetic temperature $T$ by the equation $b^2=(2kT)/m+2v_t^2$, where $k$ is Boltzmann's constant, $m$ is the atomic mass, and $v_t$ is the turbulent velocity, the measured $b$-value gives an upper limit to $T$ when $v_t$ is set equal to zero. For the two velocity components used to fit the broad absorption feature near $-$620~km~s$^{-1}$, the measured $b$-values of the Si~{\sc iii}, Si~{\sc iv}, and N~{\sc v} components indicate that $T\lesssim1.4$--$2.1\times10^7\,$K. The smaller $b$-values of the components observed at more moderate velocities (between $-$100 and +240~km~s$^{-1}$) suggest that for these components $T\lesssim0.3$--$1.7\times10^6\,$K. While in principle the $b$-values determined for two different atomic species observed within the same velocity component may be used to derive estimates for the temperature and turbulent velocity simultaneously, such an analysis presumes that the two species probe the same physical conditions. This may not be true of the highly-ionized species observed toward HD~43582. Indeed, for two of the three components in Table~\ref{tab:high_ions} with measurements for two or more ions, the $b$-values reported in the table are inconsistent with a single solution for $T$ and $v_t$ for all species. Rather, the N~{\sc v} components seem to be broader than what would be expected based on the Si~{\sc iii} and Si~{\sc iv} components, implying that the N~{\sc v} absorption probes hotter and/or more turbulent gas than does Si~{\sc iv} or Si~{\sc iii}.

\begin{figure}
\centering
\includegraphics[width=0.7\textwidth]{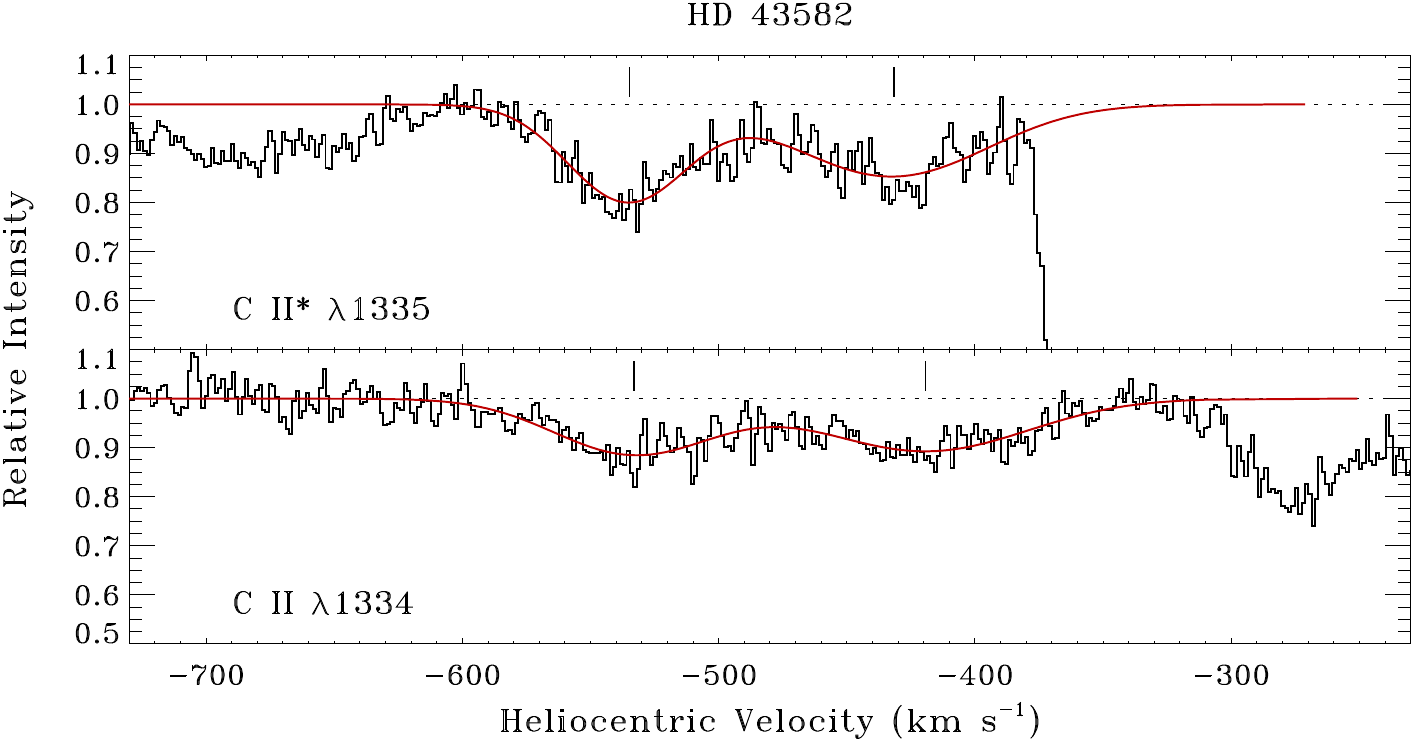}
\caption{Voigt profile fits to the high negative velocity C~{\sc ii} and C~{\sc ii}* features observed toward HD~43582. The synthetic profiles are shown as smooth red curves, with histograms representing the observed spectra. Tick marks indicate the positions of the velocity components included in the fits. Since the velocity separation between the C~{\sc ii} and C~{\sc ii}* transitions is only 264~km~s$^{-1}$, the two sets of components overlap with one another within the velocity range plotted. The strong absorption feature in the upper panel marks the blue edge of the saturated portion of the C~{\sc ii}~$\lambda1334$ line (see Figure~\ref{fig:high_vel}).\label{fig:vhv_c2}}
\end{figure}

Gnat \& Sternberg (2007) present theoretical models for solar composition gas cooling radiatively at constant pressure. Based on these models, we find that the relative abundances of Si~{\sc iii}, Si~{\sc iv}, and N~{\sc v} for the high negative velocity absorption feature observed toward HD~43582 near $-$620~km~s$^{-1}$ (taking the sums of the components at $-$665 and $-$546~km~s$^{-1}$; Table~\ref{tab:high_ions}) match the predictions for radiatively cooling gas at a temperature of $\log T\approx4.9$. At this temperature, the models indicate that absorption from C~{\sc ii} should also be present. We therefore reexamined the \emph{HST}/STIS spectrum of HD~43582 to search for this absorption. Indeed, we find two broad and shallow C~{\sc ii} (and C~{\sc ii}*) components centered at $-$534 and $-$425~km~s$^{-1}$ (Figure~\ref{fig:vhv_c2}). While the velocities of these features are somewhat lower than those of the corresponding features in Si~{\sc iii}, Si~{\sc iv}, and N~{\sc v}, the C~{\sc ii} absorption still falls within the bounds of the very broad absorption exhibited by the more highly-ionized species.\footnote{The high negative velocity C~{\sc ii} and C~{\sc ii}* features cannot be misidentified stellar (photospheric) lines because the line widths are much too narrow. The full widths at half maximum (FWHMs) are in the range 50--90~km~s$^{-1}$ for the C~{\sc ii} and C~{\sc ii}* features, while for the stellar absorption lines (which we can clearly identify in our \emph{HST} and ground-based spectra of HD~43582) the FWHMs are 200 to 400~km~s$^{-1}$. Moreover, the near perfect alignment in velocity space between the two sets of components (Figure~\ref{fig:vhv_c2}) makes us confident of their identification as due to C~{\sc ii} and C~{\sc ii}*.}

Voigt profile fits to the high negative velocity C~{\sc ii} and C~{\sc ii}* features discovered toward HD~43582 yield column densities of $\log N($C~{\sc ii}$)=13.31\pm0.10$ and $\log N($C~{\sc ii}*$)=13.43\pm0.06$ for the component at $-$534~km~s$^{-1}$ and $\log N($C~{\sc ii}$)=13.39\pm0.10$ and $\log N($C~{\sc ii}*$)=13.50\pm0.08$ for the component at $-$425~km~s$^{-1}$. The fitted $b$-values for the two velocity components are $b($C~{\sc ii}$)=43$ and 55~km~s$^{-1}$ and $b($C~{\sc ii}*$)=31$ and 51~km~s$^{-1}$. The total C~{\sc ii} column density of $\log N($C~{\sc ii}$_{\rm tot})=14.02\pm0.04$ for the two velocity components together is very close to the model prediction of $\log N($C~{\sc ii}$_{\rm tot})\approx14.07$ for radiatively cooling gas at $T\approx8.7\times10^4\,$K (based on the observed column densities of Si~{\sc iii}, Si~{\sc iv}, and N~{\sc v} for the component at $-$546~km~s$^{-1}$; Table~\ref{tab:high_ions}). The C~{\sc ii} absorption therefore supports the idea that the very high negative velocity absorption toward HD~43582 probes material in a post-shock cooling layer associated with the SNR. If this material is indeed at a temperature of $\sim$$9\times10^4\,$K, then the measured $b$-values suggest that the line widths are dominated by turbulent rather than thermal broadening.

\section{DERIVATIONS OF PHYSICAL CONDITIONS\label{sec:conditions}}
The primary aim of our investigation into interstellar gas associated with IC~443 is to derive estimates for the physical conditions in the material, which may help to better constrain models of the interaction taking place between the SNR and the surrounding medium. In Section~\ref{sec:profiles}, we presented the results of our profile fitting analysis for many different atomic and molecular species observed toward HD~43582 and HD~254755. In this section, we will use those results, and other more specialized analyses, to derive estimates for various physical quantities, including gas densities, kinetic temperatures, thermal pressures, and ionization fractions. We begin with an analysis of O~{\sc i} and Si~{\sc ii} fine-structure excitations, which will yield an extensive set of physical conditions for the many distinct absorption components seen in low ionization species toward our target stars.

\subsection{O~{\sc i} and Si~{\sc ii} Excitations\label{subsec:oxy_sil_ex}}
The relative populations of the three O~{\sc i} fine-structure levels are sensitive probes of gas density and kinetic temperature (e.g., Keenan \& Berrington 1988). The upper two fine-structure levels of neutral oxygen are populated primarily by collisions with atomic hydrogen and other particle constituents (see Appendix~\ref{sec:oi_fine_structure}). However, the excitation energies of the upper two levels are considerably larger for O~{\sc i} than for C~{\sc i}, another neutral atomic species with three fine-structure levels within its ground $^3$P term. (The upper O~{\sc i} levels have excitation energies of 158.27 and 226.98~cm$^{-1}$, while the upper levels in C~{\sc i} have energies of 16.42 and 43.41~cm$^{-1}$.) Thus, while C~{\sc i} fine-structure lines are readily observed in UV spectra (e.g., Jenkins \& Tripp 2001, 2011), detectable absorption from the O~{\sc i} fine-structure levels is much less common. The O~{\sc i}*~$\lambda1304$ and O~{\sc i}**~$\lambda1306$ lines are typically strong only in cases involving high density and/or high temperature. An advantage to working with the O~{\sc i} fine-structure lines rather than those of C~{\sc i}, however, is that the O~{\sc i} lines probe the dominant stage of ionization of oxygen in neutral gas, while the C~{\sc i} lines arise from a trace neutral species. Furthermore, O~{\sc i} is closely associated with H~{\sc i} due to a strong charge exchange reaction (Stancil et al.~1999). The physical conditions obtained from an analysis of O~{\sc i} excitations should therefore be representative of the conditions in the bulk of the H~{\sc i} gas.

The ground 3s$^2$3p $^2$P term of Si~{\sc ii} is split into two fine-structure levels with an energy separation between the ground level ($^2$P$_{1/2}$) and the excited level ($^2$P$_{3/2}$) of 287.24~cm$^{-1}$. As with the O~{\sc i} (and C~{\sc i}) fine-structure levels, the upper level of Si~{\sc ii} is collisionally-excited. However, collisions with electrons are more important for Si~{\sc ii} excitation than are collisions with neutral hydrogen (and other particle constituents), except in cases of very low fractional ionization (e.g., Keenan et al.~1985). Absorption from Si~{\sc ii}* along typical interstellar sight lines most likely probes a mixture of neutral and ionized gas regions. However, the striking similarity in the absorption profiles of the O~{\sc i}*~$\lambda1304$, O~{\sc i}**~$\lambda1306$, and Si~{\sc ii}*~$\lambda1309$ lines toward HD~43582 (Figure~\ref{fig:dominant1}) suggests that all three of these species trace the same gas structures along this particular line of sight. Our method of extracting physical conditions for the (predominantly) neutral gas components seen toward HD~43582 and HD~254755 will rely on measurements of the relative populations of the O~{\sc i} and Si~{\sc ii} fine-structure levels from our profile fitting results and also from an analysis of the apparent column density profiles of the relevant species.

\subsubsection{Basic Analysis Scheme\label{subsubsec:oxy_sil_analysis}}
Knowledge of the populations of all three O~{\sc i} fine-structure levels yields a unique solution for the gas density and kinetic temperature (in most cases). The reason for this is that there is a (nearly) one-to-one mapping between the density and temperature and the ratios $n($O~{\sc i}*$)/n($O~{\sc i}**$)$ and $[n($O~{\sc i}*$)+n($O~{\sc i}**$)]/n($O~{\sc i}$_{\rm tot})$, where $n($O~{\sc i}$_{\rm tot})=n($O~{\sc i}$)+n($O~{\sc i}*$)+n($O~{\sc i}**$)$. The ratio of the populations in the two excited levels of O~{\sc i} with respect to each other is sensitive to the temperature of the gas, while the ratio of the sum of the excited level populations with respect to the total amount of O~{\sc i} present yields the pressure (and hence the density). However, when we first examined the $N$(O~{\sc i}*)/$N$(O~{\sc i}**) ratios for the components observed toward HD~43582, we found that some of the ratios were below the minimum value of $\sim$0.9 that would be expected if the gas were completely neutral. We subsequently discovered that an enhanced level of ionization in the gas could provide a natural explanation for the low O~{\sc i}*/O~{\sc i}** ratios. Thus, in order to obtain acceptable solutions for the densities and temperatures of the various gas components, using the O~{\sc i} excitation ratios described above, we require knowledge of the ionization level in the gas, which we extract from the corresponding $N$(Si~{\sc ii}*)/$N$(Si~{\sc ii}) ratios.

In Appendix~\ref{sec:oi_fine_structure}, we provide a detailed description of our calculations of the relative populations of the three O~{\sc i} fine-structure levels for various physical conditions. A somewhat simplified approach is adopted in the case of Si~{\sc ii} excitations, which we describe here. The rate coefficient for de-excitations by electron impact from the $J=3/2$ to $J=1/2$ level of the Si~{\sc ii} ground term is given by
\begin{equation}
k_{2,1}(e,T)=\frac{8.63\times10^{-6}}{g_2T^{0.5}}\Omega_{1,2}(T)\;{\rm cm}^3\,{\rm s}^{-1}
\end{equation}
(e.g., Aggarwal \& Keenan 2014), where $\Omega_{1,2}(T)$ is the effective collision strength as a function of temperature. The corresponding rate coefficient for excitations by electron impact is obtained through the principle of detailed balance, $k_{1,2}(e,T)=(g_2/g_1)k_{2,1}(e,T)\exp(-E_{1,2}/kT)$, where the energy difference is equivalent to $E_{1,2}/k=413.27$~K. The statistical weights of the ground and excited levels are $g_1=2$ and $g_2=4$. The effective collision strength at a given temperature is interpolated from the data provided by Aggarwal \& Keenan (2014), who calculated values of $\Omega_{1,2}$ in the range $3.7<\log T<5.5$. For excitations by collisions with atomic hydrogen, we use the expression given by Barinovs et al.~(2005), which is
\begin{equation}
k_{1,2}({\rm H}^0,T)=(1.75\times10^{-11})(43.5+1.78\sqrt{T}+0.005T)\exp(-E_{1,2}/kT)\;{\rm cm}^3\,{\rm s}^{-1}~.
\end{equation}
We also consider excitations by collisions with free protons using the rate coefficients $k_{1,2}({\rm H}^+,T)$ tabulated by Bely \& Faucher (1970), although these become important only for $T\gtrsim10^4$~K. The de-excitation rate coefficients for H$^0$ and H$^+$ impacts, $k_{2,1}({\rm H}^0,T)$ and $k_{2,1}({\rm H}^+,T)$, are then obtained through the principle of detailed balance. We ignore direct excitation of the upper Si~{\sc ii} fine-structure level by radiation, which would require a strong nearby infrared source, as well as indirect excitation by fluorescence, because the allowed Si~{\sc ii} transitions are optically thick. Applying the condition for equilibrium between collisional excitations to the upper $J=3/2$ level and collisional and spontaneous de-excitations to the lower $J=1/2$ level results in an equation for the Si~{\sc ii}*/Si~{\sc ii} population ratio
\begin{equation}
\frac{n({\rm Si~\textsc{ii}\text{*}})}{n({\rm Si~\textsc{ii}})}=\frac{n(e)k_{1,2}(e,T)+n({\rm H}^+)k_{1,2}({\rm H}^+,T)+n({\rm H}^0)k_{1,2}({\rm H}^0,T)}{n(e)k_{2,1}(e,T)+n({\rm H}^+)k_{2,1}({\rm H}^+,T)+n({\rm H}^0)k_{2,1}({\rm H}^0,T)+A_{2,1}}~,
\end{equation}
where the spontaneous decay rate is $A_{2,1}=2.17\times10^{-4}\,{\rm s}^{-1}$ (Nussbaumer 1977). If the temperature and neutral hydrogen density are known quantities, then Equation (3) can be used to determine the electron density from the measured Si~{\sc ii}*/Si~{\sc ii} ratio.

Our analysis scheme for the O~{\sc i} and Si~{\sc ii} excitations is as follows. We use the calculations described in Appendix~\ref{sec:oi_fine_structure} to determine the unique combination of temperature $T$ and total hydrogen density, $n({\rm H}_{\rm tot})=n({\rm H}^0)+2n({\rm H}_2)+n({\rm H}^+)$, consistent with the measured values of $N($O~{\sc i}*$)/N($O~{\sc i}**$)$ and $[N($O~{\sc i}*$)+N($O~{\sc i}**$)]/N($O$_{\rm tot})$, where $N($O$_{\rm tot})=N($O~{\sc i}$_{\rm tot})+N($O~{\sc ii}$)$. (In Section~\ref{subsubsec:oxy_sil_profiles}, we describe our method for determining the total column density of oxygen in both neutral and ionized forms.) For these initial calculations of $T$ and $n({\rm H}_{\rm tot})$, we adopt an arbitrary value for the electron fraction $x(e)=n(e)/n({\rm H}_{\rm tot})$. The proton density is then set to $n({\rm H}^+)=n(e)-2\times10^{-4}n({\rm H}_{\rm tot})$, where the second term accounts for the fact that a small number of free electrons are created by the photoionization of heavy elements, such as carbon, rather than hydrogen. The preliminary values we calculate for $T$ and $n({\rm H}_{\rm tot})$ are then used in conjunction with the measured value of $N$(Si~{\sc ii}*)/$N$(Si~{\sc ii}) and Equation (3) to determine $n(e)$. Since the newly calculated value of $n(e)$ will likely differ from the one adopted in the initial O~{\sc i} excitation calculations, we repeat the O~{\sc i} calculations using the newly derived value of $n(e)$. These calculations yield improved values for $T$ and $n({\rm H}_{\rm tot})$, which are again used along with Equation (3) to derive an updated value for $n(e)$. This procedure is repeated until the adopted and derived values of the electron fraction converge. (The threshold for convergence is set to a level of $\sim$0.1\%.)

\subsubsection{Apparent Column Density Profiles\label{subsubsec:oxy_sil_profiles}}
In Section~\ref{subsubsec:components}, we analyzed the absorption profiles of the O~{\sc i}, O~{\sc i}*, O~{\sc i}**, and Si~{\sc ii}* lines (among others) by means of a profile fitting routine, which assumes a Voigt profile for each fitted component. We will use the column densities derived in that section, along with the analysis scheme outlined above, to obtain estimates for the physical conditions in those components. First, we employ a more specialized analysis procedure that will allow us to examine the physical conditions as a function of velocity along our lines of sight, without making any assumptions regarding the number of absorption components or the shapes of the underlying profiles. Our method is analogous to the one used by Jenkins \& Tripp (2001, 2011) to examine C~{\sc i} excitations in the diffuse ISM. We start by constructing apparent column density profiles for the relevant species using the apparent optical depth (AOD) method (Savage \& Sembach 1991). Normalized absorption profiles are converted into profiles of apparent optical depth $\tau_a$ as a function of velocity via the relation $\tau_a(v)=\ln[I_0(v)/I(v)]$, where $I(v)$ is the observed intensity in the line and $I_0(v)=1$ is the intensity in the (normalized) continuum. Errors in apparent optical depth are given by $\sigma_{\tau_a}(v)=\sigma_{I}(v)/I(v)$, where $\sigma_{I}(v)$ represents the error in the normalized intensity as a function of velocity. Wherever the apparent optical depth is not too large (i.e., where $\tau_a\lesssim2.5$), we calculate the apparent column density per unit velocity from the relation 
\begin{equation}
N_a(v)=3.768\times10^{14}\frac{\tau_a(v)}{f\lambda}\;{\rm cm}^{-2}({\rm km~s}^{-1})^{-1}~,
\end{equation}
where $f$ is the transition oscillator strength and $\lambda$ is expressed in \AA{}. Errors in apparent column density are calculated from $\sigma_{N_a}(v)=[\sigma_{\tau_a}(v)/\tau_a(v)]N_a(v)$. All of the derived apparent column density profiles (and their corresponding error arrays) are resampled to a common velocity grid with a spacing of 0.5~km~s$^{-1}$. We then retain only those velocity channels where $N_a\geq3\sigma_{N_a}$. (For HD~254755, this threshold is lowered to $2\sigma_{N_a}$ as a result of the lower S/N achieved.)

\begin{figure}
\centering
\includegraphics[width=0.9\textwidth]{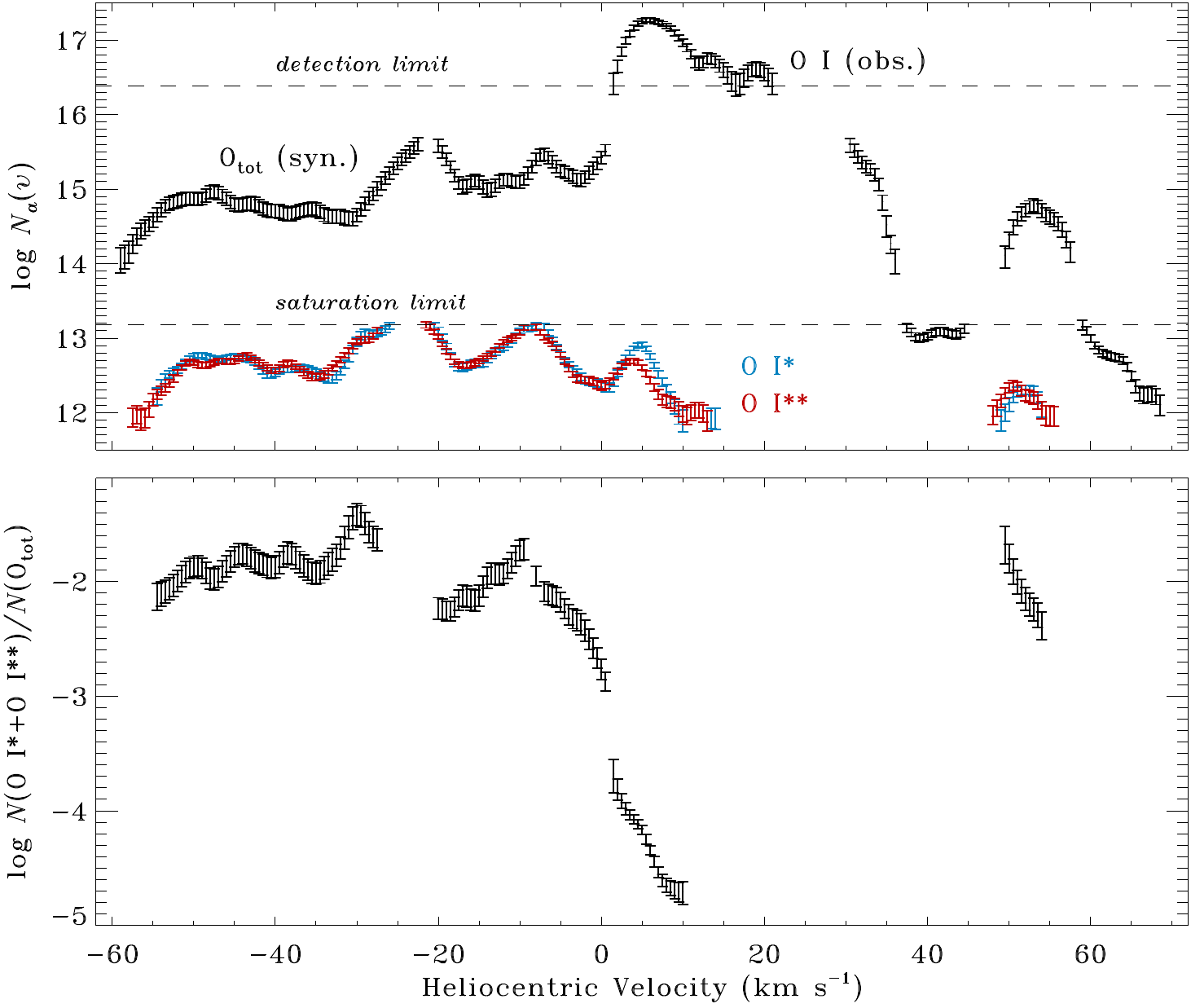}
\caption{Upper panel: Apparent column density profiles of O~{\sc i} (or O$_{\mathrm{tot}}$), O~{\sc i}*, and O~{\sc i}** toward HD~43582. The observed O~{\sc i}~$\lambda1355$ line is used to define the O~{\sc i} profile at the highest column densities since the O~{\sc i}~$\lambda1302$ line is useful for tracing only the low column density portions of the profile. (The detection limit for the O~{\sc i}~$\lambda1355$ line and the saturation limit for the O~{\sc i}~$\lambda1302$ feature are indicated.) At intermediate column densities, a synthetic profile (for O$_{\mathrm{tot}}$) is calculated from observations of the S~{\sc ii}~$\lambda1250$ and $\lambda1259$ lines. The O~{\sc i}* and O~{\sc i}** profiles are obtained from the observed O~{\sc i}*~$\lambda1304$ and O~{\sc i}**~$\lambda1306$ lines. Lower panel: The corresponding apparent column density ratio [$N$(O~{\sc i}*)~+~$N$(O~{\sc i}**)]/$N$(O$_{\mathrm{tot}}$) plotted as a function of velocity.\label{fig:oxy_profiles}}
\end{figure}

In Figure~\ref{fig:oxy_profiles}, we plot the apparent column density profiles of O~{\sc i}, O~{\sc i}*, and O~{\sc i}** for the line of sight to HD~43582. As can be seen in the figure, the weak O~{\sc i}~$\lambda1355$ line traces only the highest column density portion of the O~{\sc i} absorption profile. The 3$\sigma$ detection limit for this feature corresponds to $\log N_a($O~{\sc i}$)\approx16.4$. On the other hand, the strong O~{\sc i}~$\lambda1302$ line reaches an apparent optical depth of $\sim$2.5 at $\log N_a($O~{\sc i}$)\approx13.2$ (as indicated by the dashed line labeled ``saturation limit'' in the figure). Thus, we have no direct measure of the apparent O~{\sc i} column density per unit velocity for values between these two limits. Unfortunately, most of the velocity channels where we measure apparent column densities for O~{\sc i}* and O~{\sc i}** (via the observed lines at 1304.9~\AA{} and 1306.0~\AA{}, respectively) are precisely those where we are unable to measure O~{\sc i} directly. (Note that the ``saturation limit'' in Figure~\ref{fig:oxy_profiles} also applies to the O~{\sc i}*~$\lambda1304$ and O~{\sc i}**~$\lambda1306$ lines since these transitions have the same $f$-value as O~{\sc i}~$\lambda1302$.) To solve this problem, we calculate a synthetic profile for the total oxygen column density from observations of the S~{\sc ii}~$\lambda1250$ and $\lambda1259$ lines. A direct comparison of the apparent S~{\sc ii} column densities derived from the three lines of the S~{\sc ii} triplet indicates that for all of the velocity channels where $\tau_a\lesssim2.5$ the S~{\sc ii} lines are unsaturated (i.e., their apparent column densities agree with one another within their mutual uncertainties). A further comparison between the apparent S~{\sc ii} and Ni~{\sc ii} column densities indicates that for $v_{\sun}<+1\,{\rm km~s}^{-1}$ all of the S (and O) is in the gas phase (see Section~\ref{subsec:ni_ca_depl}). We therefore calculate synthetic values for $N_a($O$_{\rm tot})$ from $N_a($S~{\sc ii}$)$ assuming solar relative abundances.\footnote{Since the gas we are observing in the O~{\sc i}* and O~{\sc i}** lines toward HD~43582 may be partially ionized, the S~{\sc ii} column density is a proxy for the total oxygen column density $N($O$_{\rm tot})=N($O~{\sc i}$_{\rm tot})+N($O~{\sc ii}$)$. While enhanced UV radiation (e.g., from shocks) may ionize some of the neutral O (and H) atoms, the S$^+$ ions, which have an ionization potential of 23.3~eV, will be much less affected. For velocity channels where O~{\sc i} is directly observed via the O~{\sc i}~$\lambda1355$ line, we assume that the gas is almost entirely neutral such that $N($O$_{\rm tot})\approx N($O~{\sc i}$)$.} By using a combination of the weakest line and the strongest line of S~{\sc ii}, we can calculate $N_a($O$_{\rm tot})$ for nearly all of the velocity channels where O~{\sc i}* and O~{\sc i}** are directly measured (see Figure~\ref{fig:oxy_profiles}).

\begin{figure}
\centering
\includegraphics[width=0.9\textwidth]{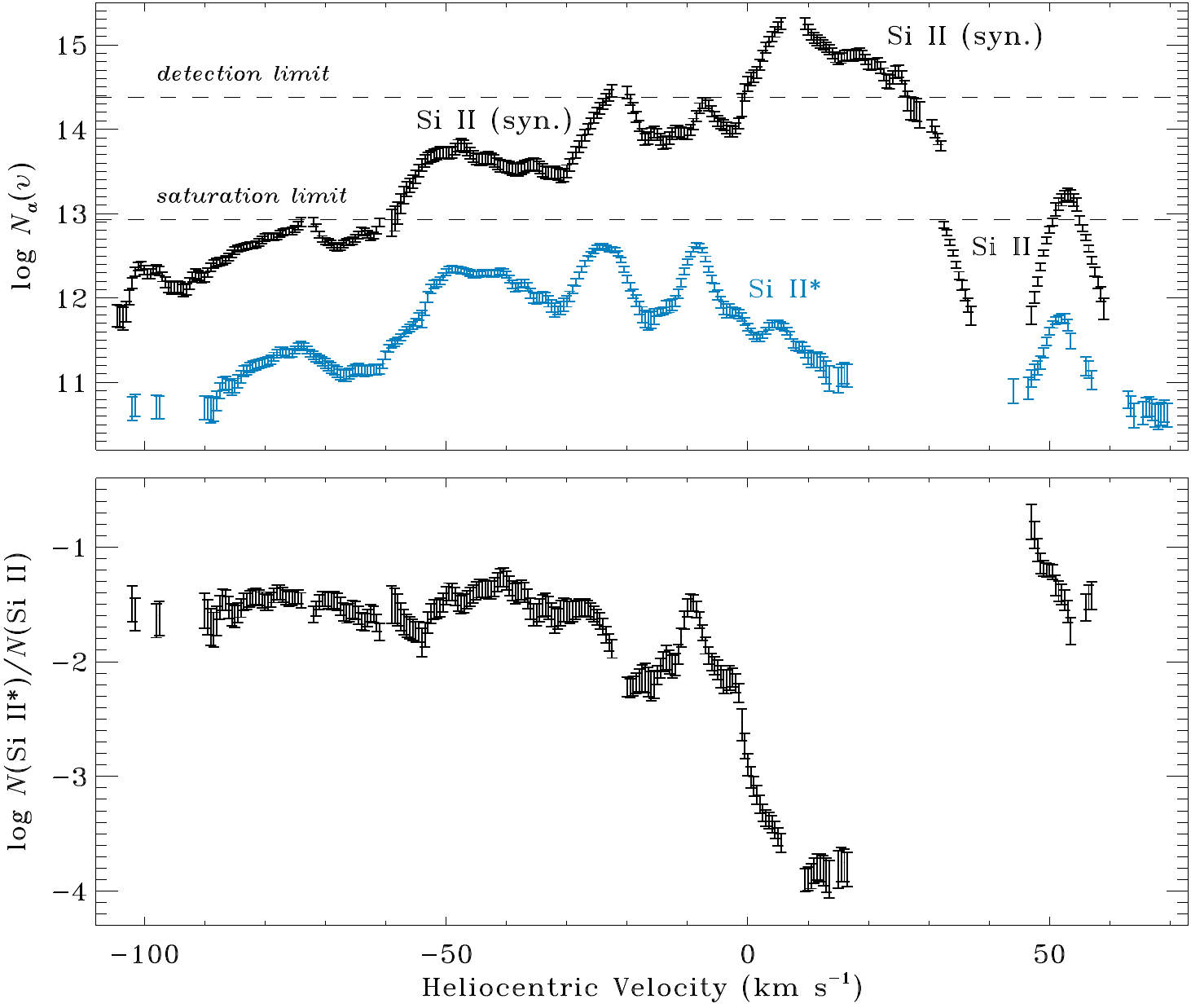}
\caption{Upper panel: Apparent column density profiles of Si~{\sc ii} and Si~{\sc ii}* toward HD~43582. The Mg~{\sc ii}~$\lambda1240$ line is used to calculate a synthetic profile for Si~{\sc ii} at the highest column densities since the Si~{\sc ii}~$\lambda1304$ line is useful for tracing only the low column density portions of the profile. (The detection limit for the Mg~{\sc ii}~$\lambda1240$ line and the saturation limit for the Si~{\sc ii}~$\lambda1304$ feature are indicated.) At intermediate column densities, a synthetic Si~{\sc ii} profile is calculated from observations of the S~{\sc ii}~$\lambda1250$ and $\lambda1259$ lines. The Si~{\sc ii}* profile is obtained from a combination of the Si~{\sc ii}*~$\lambda1309$ line at high column densities and the Si~{\sc ii}*~$\lambda1264$ line at low column densities. Lower panel: The corresponding apparent column density ratio $N$(Si~{\sc ii}*)/$N$(Si~{\sc ii}) plotted as a function of velocity.\label{fig:sil_profiles}}
\end{figure}

Figure~\ref{fig:sil_profiles} presents the derived apparent column density profiles of Si~{\sc ii} and Si~{\sc ii}* for the line of sight to HD~43582. Since there are no Si~{\sc ii} transitions within our wavelength coverage that are weak enough to trace the high column density portion of the Si~{\sc ii} absorption profile, we use the Mg~{\sc ii}~$\lambda1240$ line as a proxy for Si~{\sc ii}, assuming solar relative abundances and adopting a depletion strength factor of $F_*=0.53$ (see Section~\ref{subsubsection:depl_factors}). This applies only to those velocity channels where the Mg~{\sc ii} line shows detectable absorption (i.e., within the velocity interval $-1.0\leq v_{\sun}\leq +28.5\,{\rm km~s}^{-1}$). For the low column density portions of the Si~{\sc ii} absorption profile (i.e., for apparent column densities below the ``saturation limit'' indicated in Figure~\ref{fig:sil_profiles}), we use the Si~{\sc ii}~$\lambda1304$ line directly. However, as in the case of O~{\sc i} discussed above, there is a considerable range in $N_a$(Si~{\sc ii}) (i.e., $12.9<\log N_a <14.4$) where the Mg~{\sc ii} line is too weak to detect and the strong Si~{\sc ii} transition is badly saturated. Thus, we again use the S~{\sc ii}~$\lambda1250$ and $\lambda1259$ lines to calculate a synthetic profile for $N_a$(Si~{\sc ii}) in this intermediate column density regime. For velocity channels with $-59.0\leq v_{\sun}\leq -1.5\,{\rm km~s}^{-1}$, we assume that all of the Si is in the gas phase. This choice produces a synthetic Si~{\sc ii} profile that connects almost seemlessly to the observed Si~{\sc ii} profile at low column density and the profile inferred from Mg~{\sc ii} at high column density. For the velocity component near +52~km~s$^{-1}$, which is not detected in Ni~{\sc ii} (Figure~\ref{fig:dominant1} and Section~\ref{subsec:ni_ca_depl}) and thus may be characterized by a somewhat higher degree of dust-grain depletion compared to the components at negative velocity, we adopt a depletion strength factor of $F_*\approx0.2$, which is the value at which S becomes undepleted (Jenkins 2009). For the Si~{\sc ii}* profile, we use a combination of the Si~{\sc ii}*~$\lambda1309$ line at high column densities and the Si~{\sc ii}*~$\lambda1264$ line at low column densities. Because the $\lambda1264$ line is partially blended with $\lambda1265$ (Figure~\ref{fig:si1}), we first need to remove the $\lambda1265$ contribution from the optical depth profile, which we do by creating a synthetic profile of the $\lambda1265$ line from the unblended $\lambda1309$ feature.

For the line of sight to HD~254755, the O~{\sc i}*~$\lambda1304$ and O~{\sc i}**~$\lambda1306$ lines are detected only within a fairly narrow range in velocity (i.e., $0.0\leq v_{\sun}\leq +12.0\,{\rm km~s}^{-1}$). For these velocity channels, the O~{\sc i}~$\lambda1355$ line provides a direct determination of the column density of O~{\sc i} in the lowest level of excitation. The Si~{\sc ii}*~$\lambda1264$ line toward HD~254755 exhibits detectable absorption over the interval $-10.5\leq v_{\sun}\leq +18.0\,{\rm km~s}^{-1}$. Since the Si~{\sc ii}~$\lambda1304$ line is badly saturated over much of this velocity range, we again use the Mg~{\sc ii}~$\lambda1240$ line as a proxy for Si~{\sc ii}, adopting a depletion strength factor of $F_*=0.86$ (as indicated by the analysis described in Section~\ref{subsubsection:depl_factors}). As discussed in Section~\ref{subsubsec:high_vel2}, absorption from Si~{\sc ii}*~$\lambda1264$ is also detectable at moderately high velocity toward HD~254755 (i.e., within the velocity interval $-64.0\leq v_{\sun}\leq -56.5\,{\rm km~s}^{-1}$; see Figures~\ref{fig:si2} and \ref{fig:high_vel2}). The Si~{\sc ii}~$\lambda1304$ line is weak enough to be directly measurable in these velocity channels and indicates a rather high degree of Si~{\sc ii} excitation. Values of the $N_a$(Si~{\sc ii}*)/$N_a$(Si~{\sc ii}) ratio for these moderately high velocity channels range from $\sim$$4\times10^{-2}$ to $\sim$$9\times10^{-2}$, a range which is comparable to (if not somewhat higher than) the range exhibited by the high excitation, negative velocity gas seen toward HD~43582. Unfortunately, since there is no detectable absorption from the O~{\sc i}* or O~{\sc i}** lines at these moderately high negative velocities toward HD~254755, the analysis described in this section does not apply. We will return to a discussion of the moderately high velocity gas seen toward HD~254755 in Section~\ref{subsubsec:precursor}.

\subsubsection{Derived Physical Conditions\label{subsubsec:oxy_sil_results}}
For all of the velocity channels toward HD~43582 and HD~254755 where we have valid measurements of the apparent column densities of O~{\sc i}, O~{\sc i}*, O~{\sc i}**, Si~{\sc ii}, and Si~{\sc ii}* (or their proxies), we apply the analysis scheme described in Section~\ref{subsubsec:oxy_sil_analysis} to obtain values for $n({\rm H}_{\rm tot})$, $T$, and $n(e)$. Since the derived quantities are particularly sensitive to the measured O~{\sc i}*/O~{\sc i}** ratios, we retain final results only for those velocity channels where the measured value of $N_a($O~{\sc i}*$)/N_a($O~{\sc i}**$)$ is greater than (or equal to) three times the associated error. (Errors in the O~{\sc i} and Si~{\sc ii} excitation ratios are derived through standard error propagation based on the errors in the apparent column densities.) The results for individual velocity channels meeting these requirements are given in Table~\ref{tab:oxy_sil_results}. In addition to the derived values for the total hydrogen density, electron density, and kinetic temperature, the table lists the corresponding thermal pressure for each velocity channel. In Section~\ref{subsubsec:components}, we described our procedure for removing telluric absorption from the O~{\sc i}* and O~{\sc i}** profiles toward HD~43582. The affected velocity channels are those in the interval $-33.0\leq v_{\sun}\leq -27.5\,{\rm km~s}^{-1}$. As can be seen in Figure~\ref{fig:oxy_profiles}, these velocities exhibit the highest overall level of O~{\sc i} excitation in the line of sight, implying extreme physical conditions (with $T>10^5\,{\rm K}$). While we believe that our removal of telluric absorption was as effective as possible given the circumstances, we cannot be completely confident that the derived physical conditions for these velocity channels are accurate (since there may be some residual blending with telluric absorption). Thus, we omit these velocities from Table~\ref{tab:oxy_sil_results}.

\startlongtable
\begin{deluxetable}{lcccccccc}
\tablecolumns{9}
\tablewidth{0pt}
\tabletypesize{\scriptsize}
\tablecaption{Observed and Calculated Quantities Obtained for Individual Velocity Channels\label{tab:oxy_sil_results}}
\tablehead{ \colhead{Star} & \colhead{$v_{\sun}$} & \colhead{$N$(O~{\sc i}*)/$N$(O~{\sc i}**)} & \colhead{log~[(O~{\sc i}*+O~{\sc i}**)/O$_{\mathrm{tot}}$]} & \colhead{log~(Si~{\sc ii}*/Si~{\sc ii})} & \colhead{log~$n$(H$_{\mathrm{tot}}$)} & \colhead{log~$T$} & \colhead{log~($p/k$)} & \colhead{log~$n$($e$)} \\
\colhead{} & \colhead{(km s$^{-1}$)} & \colhead{} & \colhead{} & \colhead{} & \colhead{} & \colhead{} & \colhead{} & \colhead{} }
\startdata
HD~43582   & $-$54.5   & $0.97\pm0.29$   & $-$$2.12\pm0.10$   & $-$$1.73\pm0.08$   & 2.85   & 3.39   & 6.24   &  0.83 \\
   & $-$54.0   & $1.17\pm0.27$   & $-$$2.10\pm0.09$   & $-$$1.82\pm0.11$   & 3.44   & 2.71   & 6.15   &  0.57 \\
   & $-$53.5   & $1.32\pm0.25$   & $-$$2.08\pm0.09$   & $-$$1.75\pm0.09$   & 3.70   & 2.53   & 6.23   &  0.78 \\
   & $-$53.0   & $1.33\pm0.22$   & $-$$2.05\pm0.09$   & $-$$1.65\pm0.08$   & 3.73   & 2.52   & 6.25   &  0.96 \\
   & $-$52.5   & $1.23\pm0.19$   & $-$$2.02\pm0.08$   & $-$$1.57\pm0.08$   & 3.61   & 2.63   & 6.24   &  1.00 \\
   & $-$52.0   & $1.11\pm0.15$   & $-$$1.99\pm0.08$   & $-$$1.54\pm0.08$   & 3.41   & 2.84   & 6.25   &  0.98 \\
   & $-$51.5   & $1.04\pm0.13$   & $-$$1.95\pm0.08$   & $-$$1.53\pm0.08$   & 3.29   & 3.06   & 6.34   &  0.95 \\
   & $-$51.0   & $1.04\pm0.13$   & $-$$1.91\pm0.08$   & $-$$1.52\pm0.07$   & 3.30   & 3.09   & 6.39   &  0.95 \\
   & $-$50.5   & $1.10\pm0.13$   & $-$$1.88\pm0.08$   & $-$$1.48\pm0.07$   & 3.49   & 2.89   & 6.37   &  1.00 \\
   & $-$50.0   & $1.21\pm0.14$   & $-$$1.86\pm0.08$   & $-$$1.42\pm0.07$   & 3.71   & 2.69   & 6.40   &  1.13 \\
   & $-$49.5   & $1.34\pm0.16$   & $-$$1.86\pm0.08$   & $-$$1.38\pm0.07$   & 3.90   & 2.55   & 6.45   &  1.26 \\
   & $-$49.0   & $1.38\pm0.16$   & $-$$1.88\pm0.08$   & $-$$1.39\pm0.07$   & 3.94   & 2.51   & 6.45   &  1.30 \\
   & $-$48.5   & $1.32\pm0.15$   & $-$$1.92\pm0.08$   & $-$$1.42\pm0.07$   & 3.84   & 2.55   & 6.39   &  1.23 \\
   & $-$48.0   & $1.22\pm0.14$   & $-$$1.96\pm0.08$   & $-$$1.47\pm0.07$   & 3.64   & 2.65   & 6.30   &  1.11 \\
   & $-$47.5   & $1.12\pm0.13$   & $-$$1.97\pm0.08$   & $-$$1.49\pm0.07$   & 3.45   & 2.81   & 6.27   &  1.04 \\
   & $-$47.0   & $1.06\pm0.12$   & $-$$1.95\pm0.08$   & $-$$1.48\pm0.07$   & 3.34   & 2.97   & 6.31   &  1.03 \\
   & $-$46.5   & $1.06\pm0.12$   & $-$$1.90\pm0.08$   & $-$$1.45\pm0.07$   & 3.38   & 3.00   & 6.37   &  1.05 \\
   & $-$46.0   & $1.12\pm0.13$   & $-$$1.86\pm0.08$   & $-$$1.43\pm0.07$   & 3.54   & 2.85   & 6.39   &  1.07 \\
   & $-$45.5   & $1.18\pm0.14$   & $-$$1.82\pm0.08$   & $-$$1.40\pm0.07$   & 3.69   & 2.75   & 6.44   &  1.10 \\
   & $-$45.0   & $1.15\pm0.13$   & $-$$1.78\pm0.08$   & $-$$1.38\pm0.07$   & 3.66   & 2.81   & 6.48   &  1.11 \\
   & $-$44.5   & $1.04\pm0.12$   & $-$$1.75\pm0.08$   & $-$$1.36\pm0.07$   & 3.41   & 3.18   & 6.58   &  1.14 \\
   & $-$44.0   & $0.94\pm0.11$   & $-$$1.75\pm0.08$   & $-$$1.35\pm0.07$   & 2.76   & 3.91   & 6.66   &  1.50 \\
   & $-$43.5   & $0.91\pm0.10$   & $-$$1.77\pm0.08$   & $-$$1.36\pm0.07$   & 2.53   & 4.10   & 6.63   &  1.60 \\
   & $-$43.0   & $0.93\pm0.11$   & $-$$1.79\pm0.08$   & $-$$1.37\pm0.07$   & 2.71   & 3.90   & 6.61   &  1.49 \\
   & $-$42.5   & $0.95\pm0.11$   & $-$$1.81\pm0.08$   & $-$$1.37\pm0.07$   & 2.86   & 3.74   & 6.59   &  1.41 \\
   & $-$42.0   & $0.92\pm0.12$   & $-$$1.83\pm0.08$   & $-$$1.34\pm0.07$   & 2.60   & 3.95   & 6.55   &  1.57 \\
   & $-$41.5   & $0.85\pm0.11$   & $-$$1.84\pm0.08$   & $-$$1.30\pm0.07$   & 2.24   & 4.32   & 6.56   &  1.79 \\
   & $-$41.0   & $0.82\pm0.12$   & $-$$1.86\pm0.08$   & $-$$1.26\pm0.07$   & 2.18   & 4.47   & 6.65   &  1.88 \\
   & $-$40.5   & $0.87\pm0.12$   & $-$$1.87\pm0.08$   & $-$$1.25\pm0.07$   & 2.30   & 4.17   & 6.46   &  1.78 \\
   & $-$40.0   & $0.93\pm0.13$   & $-$$1.86\pm0.08$   & $-$$1.29\pm0.07$   & 2.74   & 3.76   & 6.50   &  1.55 \\
   & $-$39.5   & $0.93\pm0.13$   & $-$$1.82\pm0.08$   & $-$$1.33\pm0.08$   & 2.73   & 3.83   & 6.56   &  1.52 \\
   & $-$39.0   & $0.87\pm0.12$   & $-$$1.77\pm0.08$   & $-$$1.37\pm0.08$   & 2.29   & 4.38   & 6.66   &  1.71 \\
   & $-$38.5   & $0.83\pm0.11$   & $-$$1.74\pm0.08$   & $-$$1.40\pm0.08$   & 2.18   & 4.63   & 6.81   &  1.76 \\
   & $-$38.0   & $0.86\pm0.11$   & $-$$1.75\pm0.08$   & $-$$1.39\pm0.08$   & 2.25   & 4.48   & 6.73   &  1.72 \\
   & $-$37.5   & $0.95\pm0.12$   & $-$$1.78\pm0.08$   & $-$$1.37\pm0.08$   & 2.90   & 3.73   & 6.63   &  1.39 \\
   & $-$37.0   & $1.07\pm0.14$   & $-$$1.82\pm0.08$   & $-$$1.39\pm0.08$   & 3.46   & 3.00   & 6.46   &  1.11 \\
   & $-$36.5   & $1.16\pm0.15$   & $-$$1.86\pm0.08$   & $-$$1.44\pm0.08$   & 3.62   & 2.78   & 6.39   &  1.05 \\
   & $-$36.0   & $1.20\pm0.17$   & $-$$1.89\pm0.08$   & $-$$1.53\pm0.08$   & 3.66   & 2.71   & 6.37   &  0.94 \\
   & $-$35.5   & $1.19\pm0.17$   & $-$$1.92\pm0.08$   & $-$$1.58\pm0.08$   & 3.64   & 2.70   & 6.34   &  0.86 \\
   & $-$35.0   & $1.14\pm0.17$   & $-$$1.92\pm0.08$   & $-$$1.58\pm0.09$   & 3.54   & 2.79   & 6.32   &  0.85 \\
   & $-$34.5   & $1.04\pm0.16$   & $-$$1.91\pm0.09$   & $-$$1.54\pm0.09$   & 3.30   & 3.09   & 6.39   &  0.91 \\
   & $-$34.0   & $0.91\pm0.14$   & $-$$1.87\pm0.09$   & $-$$1.50\pm0.09$   & 2.51   & 4.02   & 6.53   &  1.40 \\
   & $-$33.5   & $0.80\pm0.13$   & $-$$1.84\pm0.09$   & $-$$1.49\pm0.09$   & 2.07   & 4.66   & 6.73   &  1.68 \\
   & $-$20.0   & $1.23\pm0.16$   & $-$$2.23\pm0.08$   & $-$$2.21\pm0.08$   & 3.43   & 2.61   & 6.04   & $-$0.27\phantom{$-$} \\
   & $-$19.5   & $1.23\pm0.15$   & $-$$2.23\pm0.08$   & $-$$2.22\pm0.07$   & 3.43   & 2.61   & 6.03   & $-$0.27\phantom{$-$} \\
   & $-$19.0   & $1.16\pm0.14$   & $-$$2.25\pm0.08$   & $-$$2.20\pm0.08$   & 3.28   & 2.71   & 6.00   & $-$0.41\phantom{$-$} \\
   & $-$18.5   & $1.04\pm0.13$   & $-$$2.25\pm0.08$   & $-$$2.18\pm0.08$   & 3.02   & 3.00   & 6.02   & $-$0.42\phantom{$-$} \\
   & $-$18.0   & $0.96\pm0.13$   & $-$$2.21\pm0.08$   & $-$$2.15\pm0.09$   & 2.69   & 3.50   & 6.19   & $-$0.01\phantom{$-$} \\
   & $-$17.5   & $0.93\pm0.13$   & $-$$2.16\pm0.08$   & $-$$2.13\pm0.10$   & 2.39   & 3.92   & 6.31   &  0.41 \\
   & $-$17.0   & $0.94\pm0.14$   & $-$$2.11\pm0.08$   & $-$$2.12\pm0.11$   & 2.51   & 3.85   & 6.35   &  0.23 \\
   & $-$16.5   & $0.98\pm0.14$   & $-$$2.12\pm0.08$   & $-$$2.16\pm0.11$   & 2.89   & 3.37   & 6.26   & $-$0.81\phantom{$-$} \\
   & $-$16.0   & $1.01\pm0.15$   & $-$$2.15\pm0.08$   & $-$$2.20\pm0.11$   & 3.02   & 3.14   & 6.17   & $-$0.67\phantom{$-$} \\
   & $-$15.5   & $0.99\pm0.14$   & $-$$2.16\pm0.08$   & $-$$2.18\pm0.10$   & 2.90   & 3.31   & 6.21   & $-$0.80\phantom{$-$} \\
   & $-$15.0   & $0.93\pm0.13$   & $-$$2.11\pm0.08$   & $-$$2.11\pm0.07$   & 2.41   & 3.95   & 6.36   &  0.41 \\
   & $-$14.5   & $0.88\pm0.12$   & $-$$2.04\pm0.08$   & $-$$2.05\pm0.07$   & 1.98   & 4.48   & 6.45   &  0.89 \\
   & $-$14.0   & $0.91\pm0.11$   & $-$$1.96\pm0.08$   & $-$$1.99\pm0.08$   & 2.15   & 4.38   & 6.53   &  0.87 \\
   & $-$13.5   & $0.94\pm0.11$   & $-$$1.92\pm0.08$   & $-$$1.95\pm0.09$   & 2.63   & 3.93   & 6.56   &  0.46 \\
   & $-$13.0   & $0.95\pm0.11$   & $-$$1.92\pm0.08$   & $-$$1.99\pm0.09$   & 2.64   & 3.93   & 6.56   &  0.25 \\
   & $-$12.5   & $0.90\pm0.10$   & $-$$1.93\pm0.08$   & $-$$2.03\pm0.09$   & 2.04   & 4.53   & 6.57   &  0.86 \\
   & $-$12.0   & $0.85\pm0.09$   & $-$$1.92\pm0.08$   & $-$$2.01\pm0.08$   & 1.84   & 4.78   & 6.62   &  1.03 \\
   & $-$11.5   & $0.86\pm0.09$   & $-$$1.88\pm0.08$   & $-$$1.93\pm0.08$   & 1.93   & 4.72   & 6.65   &  1.12 \\
   & $-$11.0   & $0.92\pm0.10$   & $-$$1.83\pm0.08$   & $-$$1.78\pm0.07$   & 2.40   & 4.26   & 6.65   &  1.07 \\
   & $-$10.5   & $1.01\pm0.11$   & $-$$1.77\pm0.08$   & $-$$1.63\pm0.07$   & 3.26   & 3.36   & 6.62   &  0.56 \\
   & $-$10.0   & $1.04\pm0.12$   & $-$$1.72\pm0.08$   & $-$$1.52\pm0.07$   & 3.42   & 3.22   & 6.63   &  0.73 \\
   &  \phn$-$9.5   & $0.96\pm0.12$   & $-$$1.71\pm0.08$   & $-$$1.48\pm0.07$   & 2.83   & 3.92   & 6.75   &  1.29 \\
   &  \phn$-$8.0   & $1.03\pm0.14$   & $-$$1.95\pm0.08$   & $-$$1.63\pm0.06$   & 3.26   & 3.10   & 6.36   &  0.77 \\
   &  \phn$-$7.0   & $1.38\pm0.17$   & $-$$2.08\pm0.08$   & $-$$1.86\pm0.07$   & 3.77   & 2.48   & 6.25   &  0.57 \\
   &  \phn$-$6.5   & $1.28\pm0.15$   & $-$$2.10\pm0.08$   & $-$$1.94\pm0.07$   & 3.62   & 2.56   & 6.18   &  0.28 \\
   &  \phn$-$6.0   & $1.13\pm0.12$   & $-$$2.12\pm0.08$   & $-$$2.00\pm0.07$   & 3.35   & 2.77   & 6.12   &  0.03 \\
   &  \phn$-$5.5   & $1.04\pm0.11$   & $-$$2.13\pm0.08$   & $-$$2.03\pm0.07$   & 3.11   & 3.04   & 6.15   & $-$0.08\phantom{$-$} \\
   &  \phn$-$5.0   & $1.01\pm0.11$   & $-$$2.17\pm0.08$   & $-$$2.06\pm0.08$   & 3.01   & 3.13   & 6.14   & $-$0.07\phantom{$-$} \\
   &  \phn$-$4.5   & $1.06\pm0.12$   & $-$$2.23\pm0.08$   & $-$$2.12\pm0.08$   & 3.08   & 2.95   & 6.02   & $-$0.13\phantom{$-$} \\
   &  \phn$-$4.0   & $1.15\pm0.14$   & $-$$2.28\pm0.08$   & $-$$2.16\pm0.09$   & 3.24   & 2.72   & 5.96   & $-$0.14\phantom{$-$} \\
   &  \phn$-$3.5   & $1.20\pm0.16$   & $-$$2.31\pm0.08$   & $-$$2.16\pm0.09$   & 3.31   & 2.63   & 5.94   &  0.00 \\
   &  \phn$-$3.0   & $1.19\pm0.18$   & $-$$2.33\pm0.08$   & $-$$2.13\pm0.09$   & 3.28   & 2.64   & 5.92   &  0.16 \\
   &  \phn$-$2.5   & $1.09\pm0.17$   & $-$$2.37\pm0.08$   & $-$$2.13\pm0.07$   & 3.04   & 2.83   & 5.87   &  0.17 \\
   &  \phn$-$2.0   & $0.99\pm0.16$   & $-$$2.42\pm0.08$   & $-$$2.18\pm0.07$   & 2.72   & 3.18   & 5.89   &  0.16 \\
   &  \phn$-$1.5   & $0.96\pm0.16$   & $-$$2.49\pm0.08$   & $-$$2.26\pm0.07$   & 2.49   & 3.38   & 5.87   &  0.15 \\
   &  \phn$-$1.0   & $1.00\pm0.17$   & $-$$2.57\pm0.08$   & $-$$2.53\pm0.12$   & 2.60   & 3.12   & 5.73   & $-$0.94\phantom{$-$} \\
   &  \phn$-$0.5   & $1.06\pm0.19$   & $-$$2.66\pm0.08$   & $-$$2.72\pm0.10$   & 2.69   & 2.89   & 5.58   & $-$1.01\phantom{$-$} \\
   &   \phn+0.0   & $1.04\pm0.19$   & $-$$2.76\pm0.08$   & $-$$2.89\pm0.09$   & 2.54   & 2.95   & 5.49   & $-$1.15\phantom{$-$} \\
   &   \phn+0.5   & $0.93\pm0.17$   & $-$$2.87\pm0.08$   & $-$$3.00\pm0.08$   & 1.87   & 3.71   & 5.58   & $-$1.82\phantom{$-$} \\
   &   \phn+1.5   & $0.84\pm0.13$   & $-$$3.68\pm0.12$   & $-$$3.15\pm0.07$   & 0.63   & 4.11   & 4.74   & $-$0.17\phantom{$-$} \\
   &   \phn+2.0   & $0.93\pm0.13$   & $-$$3.81\pm0.08$   & $-$$3.24\pm0.07$   & 1.23   & 3.38   & 4.60   & $-$0.60\phantom{$-$} \\
   &   \phn+2.5   & $1.06\pm0.13$   & $-$$3.91\pm0.06$   & $-$$3.31\pm0.07$   & 1.55   & 2.83   & 4.38   & $-$0.74\phantom{$-$} \\
   &   \phn+3.0   & $1.17\pm0.13$   & $-$$3.98\pm0.05$   & $-$$3.36\pm0.07$   & 1.72   & 2.61   & 4.33   & $-$0.71\phantom{$-$} \\
   &   \phn+3.5   & $1.29\pm0.14$   & $-$$4.03\pm0.04$   & $-$$3.38\pm0.07$   & 1.88   & 2.46   & 4.34   & $-$0.61\phantom{$-$} \\
   &   \phn+4.0   & $1.46\pm0.15$   & $-$$4.07\pm0.04$   & $-$$3.41\pm0.07$   & 2.08   & 2.32   & 4.40   & $-$0.46\phantom{$-$} \\
   &   \phn+4.5   & $1.72\pm0.18$   & $-$$4.11\pm0.04$   & $-$$3.46\pm0.07$   & 2.33   & 2.18   & 4.51   & $-$0.23\phantom{$-$} \\
   &   \phn+5.0   & $2.00\pm0.22$   & $-$$4.17\pm0.04$   & $-$$3.52\pm0.07$   & 2.51   & 2.08   & 4.60   & $-$0.03\phantom{$-$} \\
   &   \phn+5.5   & $2.16\pm0.25$   & $-$$4.25\pm0.04$   & $-$$3.58\pm0.07$   & 2.55   & 2.04   & 4.59   &  0.06 \\
   &  +50.5   & $0.70\pm0.15$   & $-$$1.90\pm0.10$   & $-$$1.21\pm0.06$   & 2.19   & 4.92   & 7.12   &  2.09 \\
   &  +51.0   & $0.85\pm0.18$   & $-$$1.99\pm0.09$   & $-$$1.33\pm0.08$   & 2.15   & 4.19   & 6.34   &  1.72 \\
   &  +51.5   & $1.00\pm0.22$   & $-$$2.07\pm0.09$   & $-$$1.38\pm0.08$   & 3.09   & 3.12   & 6.21   &  1.25 \\
   &  +52.0   & $1.16\pm0.26$   & $-$$2.14\pm0.09$   & $-$$1.41\pm0.08$   & 3.40   & 2.69   & 6.09   &  1.26 \\
   &  +52.5   & $1.22\pm0.28$   & $-$$2.19\pm0.09$   & $-$$1.45\pm0.08$   & 3.46   & 2.60   & 6.06   &  1.25 \\
   &  +53.0   & $1.23\pm0.29$   & $-$$2.23\pm0.09$   & $-$$1.55\pm0.09$   & 3.45   & 2.58   & 6.03   &  1.15 \\
   &  +53.5   & $1.15\pm0.31$   & $-$$2.28\pm0.09$   & $-$$1.72\pm0.10$   & 3.25   & 2.70   & 5.95   &  0.88 \\
HD~254755   &   \phn+2.5   & $1.62\pm0.46$   & $-$$4.03\pm0.10$   & $-$$3.23\pm0.18$   & 2.38   & 2.20   & 4.58   & $-$0.02\phantom{$-$} \\
   &   \phn+3.0   & $1.77\pm0.46$   & $-$$4.12\pm0.08$   & $-$$3.46\pm0.17$   & 2.43   & 2.15   & 4.57   & $-$0.14\phantom{$-$} \\
   &   \phn+3.5   & $1.83\pm0.45$   & $-$$4.17\pm0.07$   & $-$$3.57\pm0.18$   & 2.42   & 2.13   & 4.55   & $-$0.21\phantom{$-$} \\
   &   \phn+4.0   & $1.79\pm0.43$   & $-$$4.20\pm0.07$   & $-$$3.55\pm0.18$   & 2.36   & 2.14   & 4.50   & $-$0.21\phantom{$-$} \\
   &   \phn+4.5   & $1.73\pm0.42$   & $-$$4.21\pm0.07$   & $-$$3.38\pm0.18$   & 2.31   & 2.15   & 4.47   & $-$0.05\phantom{$-$} \\
   &   \phn+9.0   & $1.88\pm0.63$   & $-$$4.38\pm0.08$   & $-$$3.52\pm0.16$   & 2.28   & 2.10   & 4.37   & $-$0.04\phantom{$-$} \\
\enddata
\end{deluxetable}

\begin{figure}
\centering
\includegraphics[width=1.0\textwidth]{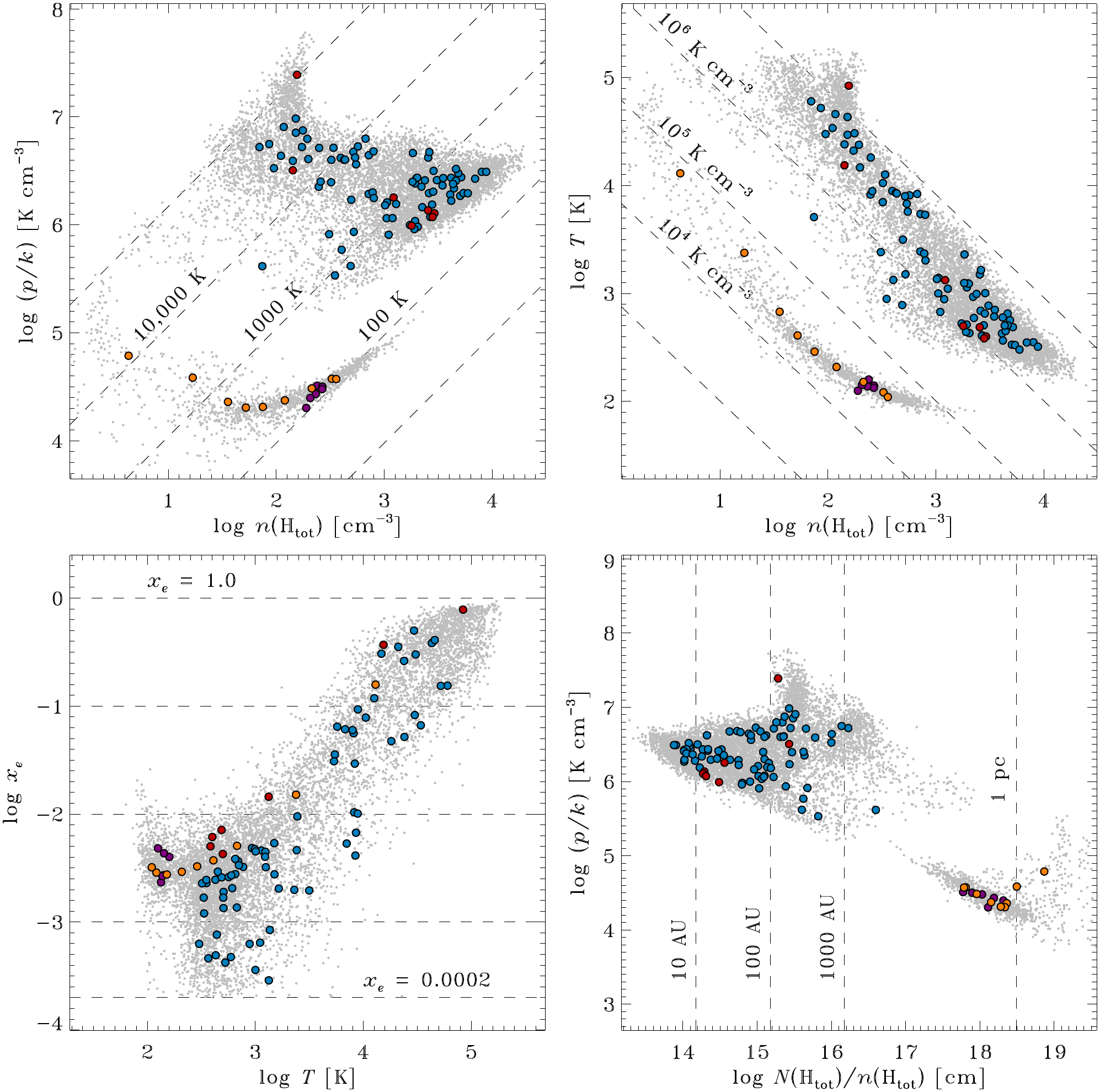}
\caption{Pairwise plots of the physical conditions derived for individual velocity bins toward HD~43582 and HD~254755. Upper left: Thermal pressure versus total hydrogen density, with lines of constant temperature indicated by diagonal dashed lines. Upper right: Kinetic temperature versus total hydrogen density, with lines of constant pressure indicated by diagonal dashed lines. Lower left: Fractional ionization versus kinetic temperature. Lower right: Thermal pressure versus cloud ``size'', defined as $N$(H$_{\mathrm{tot}}$)/$n$(H$_{\mathrm{tot}}$), with size scales of 1~pc, 1000~AU, and 100~AU indicated by vertical dashed lines. The different colors of the plotted points represent different velocity ranges for the two directions. For HD~43582, blue, orange, and red points denote velocities (in km~s$^{-1}$) in the range: $v_{\sun}<+1$, $+1<v_{\sun}<+6$, and $+50<v_{\sun}<+54$, respectively. For HD~254755, purple points represent velocities in the range $+2<v_{\sun}<+9$. The underlying distribution of light gray points gives the outcomes obtained when the input excitation ratios are drawn randomly from normal distributions with standard deviations equal to the measured 1$\sigma$ uncertainties.\label{fig:phys_cond}}
\end{figure}

In Figure~\ref{fig:phys_cond}, we present pairwise plots of the physical conditions derived for individual velocity channels toward HD~43582 and HD~254755. The data points are color coded according to the following scheme. For HD~43582, blue, orange, and red points denote velocities in the range $v_{\sun}< +1\,{\rm km~s}^{-1}$, $+1<v_{\sun}<+6\,{\rm km~s}^{-1}$, and $+50<v_{\sun}<+54\,{\rm km~s}^{-1}$, respectively. For HD~254755, purple points represent velocities in the interval $+2<v_{\sun}<+9\,{\rm km~s}^{-1}$. In this way, we can distinguish the results obtained for gas at low velocity (orange and purple points) from the results derived for velocity channels at high negative and positive velocities (blue and red points, respectively). For many of the quantities plotted in Figure~\ref{fig:phys_cond}, there is a clear segregation in the outcomes between the low velocity gas and the gas seen at higher velocities toward HD~43582. In order to evaluate the uncertainties in the derived physical quantities listed in Table~\ref{tab:oxy_sil_results} and plotted in Figure~\ref{fig:phys_cond}, we employed a Monte Carlo technique based on the errors in the input excitation ratios. (The reason is that the outcomes for density and temperature are strongly correlated with one another, making an analytical evaluation of the uncertainties cumbersome.) For each velocity channel, we drew 100 random combinations of the input excitation ratios from normal distributions with means equal to the measured values and standard deviations equal to the 1$\sigma$ uncertainties. We then proceeded with the analysis in the same way as we did for the measured values. The resulting distribution of outcomes is shown as the set of small light gray points in each of the four panels of Figure~\ref{fig:phys_cond}. These distributions provide a good indication of the range of possible physical conditions that are consistent with our observations.

The upper two panels of Figure~\ref{fig:phys_cond} show the derived thermal pressures and kinetic temperatures plotted against the total hydrogen densities. The low velocity gas toward HD~43582 and HD~254755 is characterized by a pressure of $\log\,(p/k)\approx4.4\pm0.2$, densities in the range $1.8\lesssim\log n({\rm H}_{\rm tot})\lesssim2.8$, and temperatures in the range $2.0\lesssim\log T\lesssim2.3$. In contrast, the mean thermal pressure in the higher velocity material toward HD~43582 is $\log\,(p/k)\approx6.3\pm0.3$, significantly higher than that which characterizes the low velocity gas. The mean and 1$\sigma$ standard deviations quoted in this section were obtained from Gaussian fits to the distribution functions of the various physical quantities (separating the low velocity gas from the higher velocity material toward HD~43582). The pressure distributions are more symmetric than those of density or temperature, and are well characterized by lognormal distributions (at least in the central portions of the distributions). The temperature and density distribution functions have significant tails toward high temperature and low density. These skewed distributions closely resemble the asymmetric uncertainties associated with the temperatures and densities derived for individual velocity channels. As the value of the O~{\sc i}*/O~{\sc i}** ratio drops below $\sim$0.9, the temperature derived from the O~{\sc i} excitation analysis increases rapidly while the density decreases. Thus, for measured values of the O~{\sc i}*/O~{\sc i}** ratio that are close to 0.9, the observational uncertainties, which are typically $\sim$15\%, will produce highly asymmetric errors in temperature and density. This seems to be the cause of the significant tail in the distribution of points that represents the low velocity gas toward HD~43582 and HD~254755 (Figure~\ref{fig:phys_cond}), and it may also (at least partially) explain the considerable spread in the outcomes for density and temperature that characterizes the high pressure gas toward HD~43582. The bulk of the high pressure material has a mean density of $\log n({\rm H}_{\rm tot})\approx3.3\pm0.4$ and a mean temperature of $\log T\approx2.7\pm0.3$. However, the spread in outcomes has a secondary peak near $\log n({\rm H}_{\rm tot})\approx2.2$ and $\log T\approx4.6$.

In the lower left panel of Figure~\ref{fig:phys_cond}, we plot the ionization fractions obtained for individual velocity channels against the corresponding kinetic temperatures. A clear trend emerges wherein the ionization fraction increases with increasing kinetic temperature. The bulk of the material seen in the two directions, including both the low velocity unshocked gas and the higher velocity material toward HD~43582, has an ionization fraction $x(e)\lesssim0.01$. However, most of the velocity channels with temperatures in excess of $10^4$~K have $x(e)\gtrsim0.1$. This trend is most likely a consequence of collisional ionization occurring in the recently-shocked, high-temperature, compressed gas, although enhanced UV radiation from shocks or an elevated flux of cosmic rays in the shocked material could also contribute to the increase in ionization. In the lower right panel of Figure~\ref{fig:phys_cond}, we show the thermal pressures plotted against a measure of the size (or thickness) of the gas clumps, defined as $N({\rm H}_{\rm tot})/n({\rm H}_{\rm tot})$. Values of $N({\rm H}_{\rm tot})$ are calculated for individual velocity channels from the corresponding values of $N_a$(O~{\sc i}) (for the low velocity gas) or $N_a$(S~{\sc ii}) (for the higher velocity material toward HD~43582), assuming solar relative abundances and adopting the same depletion factors as in Section~\ref{subsubsec:oxy_sil_profiles}. Note that the sizes defined in this way refer to 0.5~km~s$^{-1}$ intervals, which was an arbitrary choice. The main purpose of this panel is to show the change in the relative cloud thicknesses between the low velocity material toward HD~43582 and HD~254755 and the high pressure gas toward HD~43582. Estimates of the absolute sizes of the various cloud components may be obtained from the profile fitting results discussed in the next section.

\subsubsection{Comparison with the Results of Profile Fitting\label{subsubsec:oxy_sil_comparison}}
The analysis described above, which is based on an examination of the apparent column densities per unit velocity of O~{\sc i} and Si~{\sc ii} in their ground and excited fine-structure levels, provides a detailed picture of the physical conditions in the gas toward HD~43582 and HD~254755 and how those conditions change as a function of velocity along the lines of sight. However, there are some drawbacks to this analysis. One issue is that certain velocity ranges are excluded from consideration because the relevant absorption lines have too high an optical depth. (The cutoff we imposed in apparent optical depth is $\tau_a\lesssim2.5$.) For instance, the absorption component near $-$24~km~s$^{-1}$ toward HD~43582 is not fully represented in the AOD analysis because the O~{\sc i}* and O~{\sc i}** lines are too strong near the peak. (Note the gap in coverage near this velocity in Figure~\ref{fig:oxy_profiles}.) Likewise, the highest column density portions of the absorption profiles near +7~km~s$^{-1}$ are excluded from the analysis because the Mg~{\sc ii} lines (toward both stars) are too strong at these velocities. Another more general issue is that the apparent column densities in any given velocity channel may arise from multiple physically distinct cloud components that happen to overlap in velocity. This would be especially likely near the boundaries between two adjacent velocity components, where the wings of the absorption profiles overlap due to the combined effects of thermal and turbulent broadening within each distinct component.

We can attempt to address these issues by examining the results of our profile fitting analysis (Section~\ref{subsubsec:components}). Through profile synthesis, we can derive reliable column densities even for absorption components that are moderately saturated, such as the $-$24~km~s$^{-1}$ component in O~{\sc i}* and O~{\sc i}** toward HD~43582 and the +7~km~s$^{-1}$ in Mg~{\sc ii} toward both stars. Moreover, profile fitting offers a natural means of disentangling complex absorption profiles so that we can isolate the amount of material in each distinct component even when the individual components overlap with one another. (Granted, this picture of individual distinct interstellar clouds may be an oversimplification of the true situation. In reality, components that appear to be distinct may instead be arbitrary depictions of density fluctuations or velocity crowding within a turbulent but otherwise coherent medium.) In switching to a profile synthesis approach, we lose much of the detail afforded by the AOD analysis in terms of how the physical conditions change as a function of velocity, and the accuracy of the profile fitting results will depend to some degree on whether the underlying models of the absorption profiles are correct. Still, a comparison between the results of the AOD analysis and those based on profile synthesis may prove insightful.

In Table~\ref{tab:comp_column_densities}, we list the column densities of the species relevant to our analysis of physical conditions for distinct velocity components identified in the absorption profiles observed toward HD~43582 and HD~254755. While the profile fits themselves included many additional subcomponents (so as to properly model subtle asymmetries in the line profiles; see Figures~\ref{fig:dominant1}--\ref{fig:si2} and Appendix~\ref{sec:comp_structures}), we add together in Table~\ref{tab:comp_column_densities} the column densities of any closely-spaced subcomponents that constitute a single distinct feature. In this way, we seek to minimize the impact of small discrepancies in how the absorption profiles of different species are subdivided into individual components. We use the Na~{\sc i} and/or K~{\sc i} profiles as guides for identifying these distinct velocity components. (The Na~{\sc i} and K~{\sc i} absorption features are intrinsically narrow and the HET data are characterized by very high S/N.) We identify six distinct absorption components in the Na~{\sc i}~$\lambda\lambda5889,5895$ profiles toward HD~43582, excluding the components that constitute the saturated cores of the lines (see Figure~\ref{fig:trace1}). The K~{\sc i}~$\lambda7698$ feature toward HD~43582 indicates that the saturated portions of the Na~{\sc i} profiles are composed of three additional distinct components. Likewise, the K~{\sc i} feature toward HD~254755 indicates the presence of three distinct absorption components. The velocities given for these components in Table~\ref{tab:comp_column_densities} are the column-density weighted mean velocities of the individual subcomponents that contribute to the sums (averaged over all the species showing absorption from a given component).

%\begin{longrotatetable}
\begin{deluxetable}{lccccccccc}
\tablecolumns{10}
\tablewidth{0pt}
\tabletypesize{\scriptsize}
\tablecaption{Column Densities for Distinct Velocity Components\label{tab:comp_column_densities}}
\tablehead{ \colhead{Star} & \colhead{$\langle v_{\sun} \rangle$\tablenotemark{a}} & \colhead{log~$N$(O~{\sc i})} & \colhead{log~$N$(O~{\sc i}*)} & \colhead{log~$N$(O~{\sc i}**)} & \colhead{log~$N$(Mg~{\sc ii})} & \colhead{log~$N$(Si~{\sc ii}*)} & \colhead{log~$N$(S~{\sc ii})} & \colhead{log~$N$(O$_{\mathrm{tot}}$)\tablenotemark{b}} & \colhead{log~$N$(Si~{\sc ii})\tablenotemark{c}} \\
\colhead{} & \colhead{(km~s$^{-1}$)} & \colhead{} & \colhead{} & \colhead{} & \colhead{} & \colhead{} & \colhead{} & \colhead{} & \colhead{} }
\startdata
HD~43582 & $-$50.3 & \ldots & $13.57\pm0.04$ & $13.51\pm0.04$ & \ldots & $13.27\pm0.02$ & $14.18\pm0.09$ & $15.68\pm0.11$ & $14.53\pm0.10$ \\
 & $-$43.4 & \ldots & $13.48\pm0.04$ & $13.54\pm0.05$ & \ldots & $12.80\pm0.03$ & $14.07\pm0.09$ & $15.57\pm0.11$ & $14.42\pm0.10$ \\
 & $-$36.5 & \ldots & $13.49\pm0.04$ & $13.27\pm0.04$ & \ldots & $13.05\pm0.03$ & $14.16\pm0.11$ & $15.66\pm0.12$ & $14.51\pm0.12$ \\
 & $-$23.8 & \ldots & $14.33\pm0.05$ & $14.36\pm0.05$ & \ldots & $13.54\pm0.02$ & $15.02\pm0.07$ & $16.52\pm0.09$ & $15.37\pm0.08$ \\
 & \phn$-$8.7 & \ldots & $14.14\pm0.04$ & $14.13\pm0.04$ & \ldots & $13.45\pm0.02$ & $14.85\pm0.06$ & $16.35\pm0.08$ & $15.20\pm0.07$ \\
 & \phn+6.9 & $18.15\pm0.03$ & $13.69\pm0.04$ & $13.54\pm0.04$ & $16.39\pm0.04$ & $12.71\pm0.05$ & [sat.] & $18.15\pm0.03$ & $16.35\pm0.05$ \\
 & +17.7 & $17.57\pm0.07$ & \dots & \ldots & $15.84\pm0.03$ & \ldots & [sat.] & $17.57\pm0.07$ & $15.80\pm0.04$ \\
 & +26.2 & \ldots & \ldots & \ldots & $15.31\pm0.05$ & \ldots & [sat.] & \dots & $15.28\pm0.05$ \\
 & +52.3 & \ldots & $12.99\pm0.05$ & $13.13\pm0.05$ & \ldots & $12.62\pm0.04$ & $14.12\pm0.07$ & $15.59\pm0.10$ & $14.04\pm0.08$ \\
HD~254755 & \phn+6.1 & $18.42\pm0.04$ & $14.10\pm0.06$ & $13.72\pm0.05$ & $16.58\pm0.06$ & $12.81\pm0.04$ & [sat.] & $18.42\pm0.04$ & $16.50\pm0.07$ \\
 & +16.1 & $17.77\pm0.08$ & $12.78\pm0.18$ & $12.86\pm0.15$ & $16.00\pm0.05$ & $12.07\pm0.08$ & [sat.] & $17.77\pm0.08$ & $15.92\pm0.06$ \\
 & +24.4 & $17.34\pm0.11$ & \ldots & \ldots & $15.48\pm0.08$ & \ldots & [sat.] & $17.34\pm0.11$ & $15.40\pm0.08$ \\
\enddata
\tablenotetext{a}{Column-density weighted mean heliocentric velocity averaged over all species showing absorption from that component.}
\tablenotetext{b}{The total oxygen column density, $N$(O$_{\mathrm{tot}}$)~$\equiv$~$N$(O~{\sc i})~+~$N$(O~{\sc i}*)~+~$N$(O~{\sc i}**)~+~$N$(O~{\sc ii}), is inferred from the S~{\sc ii} column density, or obtained directly from the observed column density of O~{\sc i}.}
\tablenotetext{c}{The column density of Si~{\sc ii} in the ground ($^2P_{1/2}$) level is inferred from the S~{\sc ii} column density, or from the column density of Mg~{\sc ii} when available.}
\end{deluxetable}
%\end{longrotatetable}

The column densities listed in Table~\ref{tab:comp_column_densities} for O~{\sc i}, O~{\sc i}*, O~{\sc i}**, Mg~{\sc ii}, Si~{\sc ii}*, and S~{\sc ii} are those obtained directly through profile synthesis. (Note that we do not list S~{\sc ii} column densities for components that are badly saturated.) The values given for $N$(O$_{\rm tot}$) and $N$(Si~{\sc ii}) are computed in the same way as described in Section~\ref{subsubsec:oxy_sil_profiles}. That is, the total oxygen column density is obtained directly from the observed column density of O~{\sc i}, and the Si~{\sc ii} column density is inferred from that of Mg~{\sc ii}, for all of the components where the O~{\sc i}~$\lambda1355$ and Mg~{\sc ii}~$\lambda\lambda1239,1240$ lines are directly observed (i.e., for components with velocities between 0 and +30~km~s$^{-1}$). For the higher velocity components toward HD~43582, the S~{\sc ii} column density is used to compute synthetic values for $N$(O$_{\rm tot}$) and $N$(Si~{\sc ii}). In each of these calculations, we assume solar relative abundances and adopt the same depletion factors as in Section~\ref{subsubsec:oxy_sil_profiles}.

%\begin{longrotatetable}
\begin{deluxetable}{lccccccccc}
\tablecolumns{10}
\tablewidth{0pt}
\tabletypesize{\scriptsize}
\tablecaption{Excitation Ratios and Derived Physical Conditions for Distinct Velocity Components\label{tab:comp_results}}
\tablehead{ \colhead{Star} & \colhead{$\langle v_{\sun} \rangle$} & \colhead{$N$(O~{\sc i}*)/$N$(O~{\sc i}**)} & \colhead{log~[(O~{\sc i}*+O~{\sc i}**)/O$_{\mathrm{tot}}$]} & \colhead{log~(Si~{\sc ii}*/Si~{\sc ii})} & \colhead{log~$N$(H$_{\rm tot}$)\tablenotemark{a}} & \colhead{$n$(H$_{\mathrm{tot}}$)} & \colhead{$T$} & \colhead{$n$($e$)} & \colhead{log~($p/k$)} \\
\colhead{} & \colhead{(km~s$^{-1}$)} & \colhead{} & \colhead{} & \colhead{} & \colhead{} & \colhead{(cm$^{-3}$)} & \colhead{(K)} & \colhead{(cm$^{-3}$)} & \colhead{} }
\startdata
HD~43582 & $-$50.3 & $1.15\pm0.15$ & $-$$1.84\pm0.11$ & $-$$1.26\pm0.10$ & $18.92\pm0.10$ & \phn4200 & \phn\phn610 & 22.\phn\phn & $6.45\pm0.16$ \\
 & $-$43.4 & $0.86\pm0.13$ & $-$$1.75\pm0.12$ & $-$$1.62\pm0.11$ & $18.81\pm0.10$ & \phn\phn140 & 41000\tablenotemark{b} & 30.\phn\phn & $6.88\pm0.28$ \\
 & $-$36.5 & $1.65\pm0.23$ & $-$$1.97\pm0.13$ & $-$$1.46\pm0.12$ & $18.90\pm0.11$ & 14000 & \phn\phn210 & 30.\phn\phn & $6.53\pm0.14$ \\
 & $-$23.8 & $0.95\pm0.16$ & $-$$1.87\pm0.10$ & $-$$1.83\pm0.08$ & $19.76\pm0.08$ & \phn\phn480 & \phn8300 & 5.3 & $6.65\pm0.24$ \\
 & \phn$-$8.7 & $1.03\pm0.14$ & $-$$1.91\pm0.09$ & $-$$1.75\pm0.07$ & $19.59\pm0.07$ & \phn1900 & \phn1400 & 2.4 & $6.45\pm0.12$ \\
 & \phn+6.9 & $1.41\pm0.20$ & $-$$4.23\pm0.04$ & $-$$3.65\pm0.06$ & $21.52\pm0.06$ & \phn\phn\phn72 & \phn\phn230 & \phn0.17 & $4.19\pm0.12$ \\
 & +52.3 & $0.71\pm0.12$ & $-$$2.22\pm0.10$ & $-$$1.42\pm0.09$ & $18.86\pm0.08$ & \phn\phn\phn83 & 54000\tablenotemark{b} & 65.\phn\phn & $6.92\pm0.45$ \\
HD~254755 & \phn+6.1 & $2.41\pm0.45$ & $-$$4.17\pm0.06$ & $-$$3.69\pm0.08$ & $21.89\pm0.06$ & \phn\phn680 & \phn\phn\phn95 & 1.4 & $4.74\pm0.17$ \\
\enddata
\tablenotetext{a}{The total hydrogen column density associated with each component is derived from the O~{\sc i} column density (when available) or the S~{\sc ii} column density, assuming solar relative abundances and adopting the same depletion factors as in Section~\ref{subsubsec:oxy_sil_profiles}.}
\tablenotetext{b}{The temperatures listed here for the $-$43 and +52~km~s$^{-1}$ components toward HD~43582, which were derived from the observed O~{\sc i} and Si~{\sc ii} excitation ratios, appear to be too large based on other constraints available from the data. For example, the $b$-values of the O~{\sc i}* and O~{\sc i}** features at $-$43 and +52~km~s$^{-1}$ imply maximum temperatures of $\sim$$9600\,$K and $\sim$$11000\,$K, respectively.}
\end{deluxetable}
%\end{longrotatetable}

In Table~\ref{tab:comp_results}, we give the ratios $N($O~{\sc i}*$)/N($O~{\sc i}**$)$, $[N($O~{\sc i}*$)+N($O~{\sc i}**$)]/N($O$_{\rm tot})$, and $N($Si~{\sc ii}*$)/N($Si~{\sc ii}$)$ for each of the components from Table~\ref{tab:comp_column_densities} with reliable determinations of these quantities. (The components near +17 and +25~km~s$^{-1}$ toward both stars are excluded from Table~\ref{tab:comp_results} either because the excited lines of O~{\sc i} and Si~{\sc ii} are not detected or because the error in the O~{\sc i}*/O~{\sc i}** ratio is too large to yield meaningful results.) Using the same analysis scheme as outlined in Section~\ref{subsubsec:oxy_sil_analysis}, we derive values for $n({\rm H}_{\rm tot})$, $T$, and $n(e)$ for each component and present the results in Table~\ref{tab:comp_results}. We also list in Table~\ref{tab:comp_results} the total hydrogen column density associated with each component (as determined from the O~{\sc i} or S~{\sc ii} column density and the depletion factors as described above). The uncertainties in the thermal pressures listed in Table~\ref{tab:comp_results} are derived from Gaussian fits to the pressure distribution functions using the same Monte Carlo technique as described in Section~\ref{subsubsec:oxy_sil_results}.

In two cases, the temperatures derived through the O~{\sc i} and Si~{\sc ii} excitation analysis appear to be too large based on other constraints that the observations provide. For the $-$43 and +52~km~s$^{-1}$ components toward HD~43582, the excitation analysis yields $T\approx4.1\times10^4\,$K and $5.4\times10^4\,$K, respectively. However, for collisionally-ionized gas, cooling radiatively at constant pressure, the fractional abundance of neutral oxygen drops below $\sim$0.003 for $T\gtrsim3\times10^4\,$K (Gnat \& Sternberg 2007). Thus, if the temperatures of the $-$43 and +52~km~s$^{-1}$ components were equal to the values indicated by the excitation analysis, the O~{\sc i} column densities of the components would be so low that they should be directly measurable via the strong line at 1302.2~\AA{}. Since the O~{\sc i}~$\lambda1302$ line is completely opaque at these velocities (see Figure~\ref{fig:high_vel}), we conclude that extreme levels of oxygen ionization for these particular components are not allowed by the observations.

Additional constraints on the temperatures of the $-$43 and +52~km~s$^{-1}$ components are provided by the measured line widths. Our profile fits to the O~{\sc i}* and O~{\sc i}** features that correspond to the $-$43~km~s$^{-1}$ component yield respective $b$-values of 2.8 and 3.2~km~s$^{-1}$. Assuming only thermal broadening contributes to these line widths, the maximum temperature implied would be $\sim$$9.6\times10^3\,$K (much less than the value obtained from the excitation analysis). The $-$43~km~s$^{-1}$ component is blended with adjacent components at $-$50 and $-$37~km~s$^{-1}$ (see Figure~\ref{fig:dominant1}), making the $b$-values obtained for these components somewhat uncertain. The +52~km~s$^{-1}$ component, however, is completely isolated, yet still we find that the line widths are too small for the temperature of the gas to be as high as that indicated by the excitation analysis. The $b$-values of the O~{\sc i}* and O~{\sc i}** features that correspond to the +52~km~s$^{-1}$ component are 2.3 and 3.4~km~s$^{-1}$, respectively, indicating a maximum temperature of $\sim$$1.1\times10^4\,$K.

The issue may simply be that the populations of the excited O~{\sc i} levels for these problematic components are not known as precisely as we would like. Both components have $N($O~{\sc i}*$)/N($O~{\sc i}**$)\lesssim0.9$, meaning that a small change in the value of this ratio has a significant impact on the derived temperature. For example, if we adjust the O~{\sc i}*/O~{\sc i}** ratios upward by approximately 1$\sigma$ (to values of 0.95 and 0.84 for the $-$43 and +52~km~s$^{-1}$ components, respectively), we find temperatures in agreement with the values implied by the assumption of thermal broadening of the line profiles. (If the temperatures were reduced in this manner, the densities derived from the excitation analysis would increase accordingly. For example, if the temperature of the $-$43~km~s$^{-1}$ component were reduced to $\sim$$9600\,$K, the density would increase to $\sim$$540\,{\rm cm}^{-3}$ such that the corresponding thermal pressure would decrease by only $\sim$0.1~dex.) For all of the other components in Table~\ref{tab:comp_results}, the temperatures obtained from the excitation analysis are less than the maximum temperatures implied by the widths of the components, indicating that a combination of thermal and turbulent motions contributes to line broadening.

The physical conditions obtained from the column densities derived through profile synthesis (Table~\ref{tab:comp_results}) are broadly consistent with the results of the AOD analysis (Table~\ref{tab:oxy_sil_results} and Figure~\ref{fig:phys_cond}). The main low velocity absorption components toward HD~43582 and HD~254755 (at +7 and +6~km~s$^{-1}$, respectively) exhibit densities and temperatures typical of diffuse atomic and molecular clouds (e.g., Snow \& McCall 2006). The higher velocity components toward HD~43582 exhibit a range of higher densities and/or temperatures, the main distinction being that these components are characterized by a significantly higher thermal pressure compared to the main low velocity components, as already indicated by the AOD analysis. In Section~\ref{subsec:phys_cond}, we discuss the physical conditions derived here and in Section~\ref{subsubsec:oxy_sil_results} in more detail in the context of the interaction between the supernova shock and the interstellar gas in the vicinity of IC~443. First, however, we examine what other information regarding the physical conditions in the gas interacting with IC~443 may be gleaned from our \emph{HST} and HET spectra of HD~43582 and HD~254755.

\subsection{C~{\sc i} Excitations\label{subsec:car_ex}}
The relative populations of the collisionally-excited fine-structure levels in C~{\sc i} are even more sensitive probes of local physical conditions than those in O~{\sc i} and Si~{\sc ii}. Due to the lower excitation energies (and low spontaneous decay rates), the upper two C~{\sc i} fine-structure levels may be significantly populated even in gas at relatively low density and temperature. However, the ease with which the upper C~{\sc i} levels are excited becomes a problem for highly-compressed gas, such as that seen toward HD~43582. When the thermal pressure is high enough, the relative populations of the C~{\sc i} fine-structure levels approach values determined by the level degeneracies. In such instances, the C~{\sc i} levels are no longer very useful for precise determinations of local physical conditions. In this section, we examine the C~{\sc i} level populations in the gas toward HD~43582 and HD~254755, using both the results of our profile synthesis fits (Section~\ref{subsubsec:components}) and the outcomes from a more specialized procedure similar to the AOD analysis described above.

\begin{figure}
\centering
\includegraphics[width=0.9\textwidth]{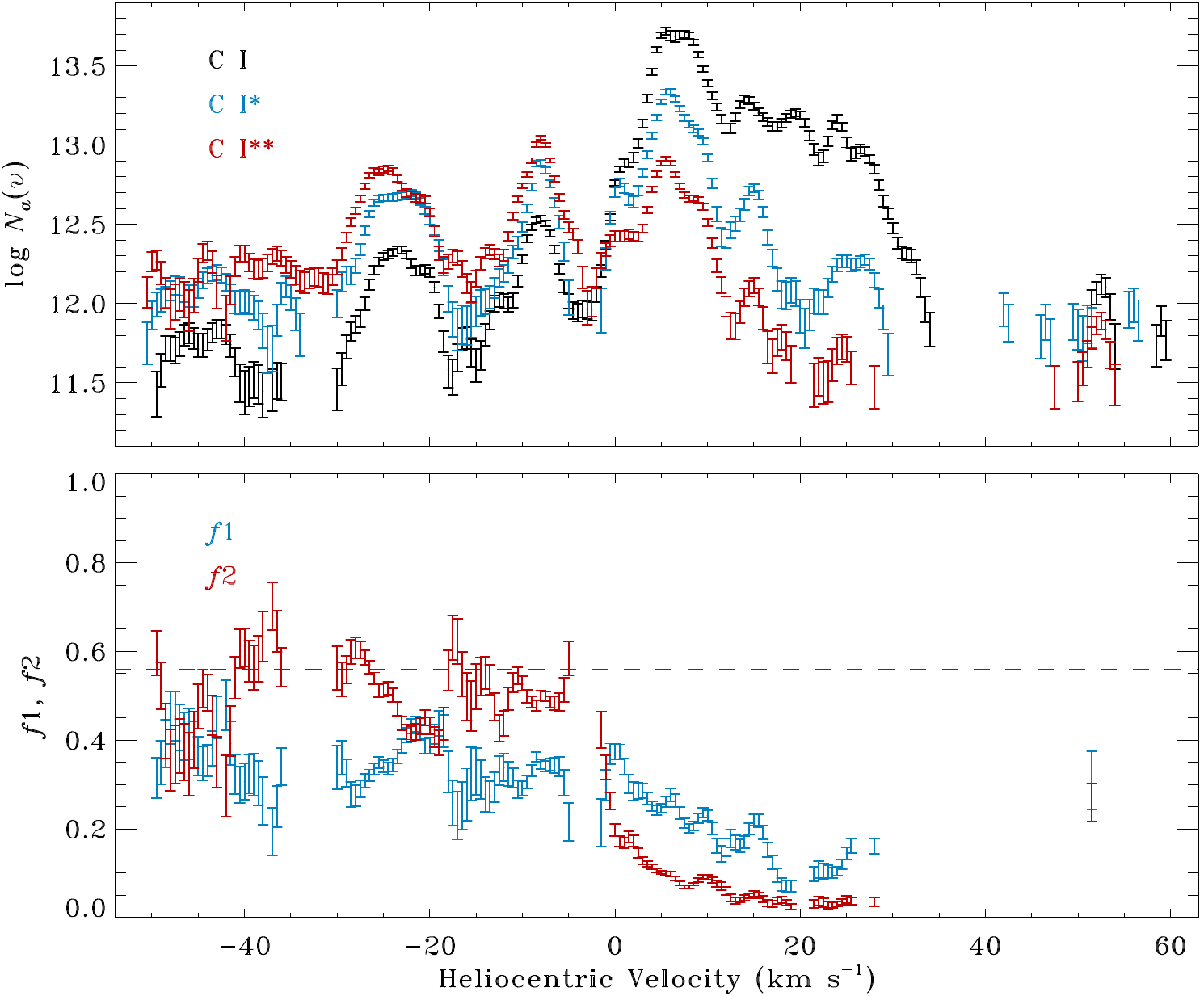}
\caption{Upper panel: Derived apparent column density profiles of C~{\sc i}, C~{\sc i}*, and C~{\sc i}** toward HD~43582. Lower panel: The corresponding relative C~{\sc i} fine-structure populations plotted as a function of velocity. The blue and red dashed lines indicate the respective values of $f1$ and $f2$ that result when the C~{\sc i} levels are populated according to their relative statistical weights.\label{fig:car_profiles}}
\end{figure}

Jenkins \& Tripp (2001, 2011) devised a method for disentangling the complex absorptions created by the overlapping C~{\sc i} lines in \emph{HST} spectra. Normalized C~{\sc i} absorption profiles are converted into profiles of apparent optical depth as a function of velocity (Savage \& Sembach 1991). Then, a system of linear equations (as specified in Jenkins \& Tripp 2001) is solved to reveal the apparent column density profile of each of the three C~{\sc i} fine-structure levels independent of the other two. All of the C~{\sc i} multiplets available from the \emph{HST} observations are used in the calculation except for the C~{\sc i}~$\lambda1260$ multiplet in the case of HD~43582 since this feature is affected by overlapping absorption from Si~{\sc ii}~$\lambda1260$ at high velocity (see Section~\ref{subsubsec:high_vel}). The C~{\sc i} $f$-values adopted for these calculations are the same as those used by Jenkins \& Tripp (2001, 2011). The resulting apparent column density profiles of C~{\sc i}, C~{\sc i}*, and C~{\sc i}** toward HD~43582 are shown in the upper panel of Figure~\ref{fig:car_profiles}. In the lower panel of the figure, we show how the quantities $f1\equiv N($C~{\sc i}*$)/N($C~{\sc i}$_{\rm tot})$ and $f2\equiv N($C~{\sc i}**$)/N($C~{\sc i}$_{\rm tot})$, where $N($C~{\sc i}$_{\rm tot})=N($C~{\sc i}$)+N($C~{\sc i}*$)+N($C~{\sc i}**$)$, change as a function of velocity along the line of sight. The dashed blue and red lines indicate the respective values of $f1$ and $f2$ that result when the C~{\sc i} levels are populated according to their relative statistical weights. Note that this condition is found for many of the velocity channels with $v_{\sun}\leq -5\,{\rm km~s}^{-1}$.

Jenkins \& Tripp (2011) describe a test that they implemented to determine if the solutions for the apparent column densities of the C~{\sc i} levels were being distorted by unresolved saturated portions of the absorption profiles (see Appendix A.2.4 of Jenkins \& Tripp 2011). The test amounts to varying the influence of the stronger C~{\sc i} lines in obtaining solutions for the apparent column densities. If the outcomes change significantly when the stronger lines are given less weight, then one can surmise that unresolved saturated absorption structures are distorting the column density solutions. When we apply this test to the outcomes for the C~{\sc i} levels toward HD~254755, we find that for the main low velocity absorption feature near +6~km~s$^{-1}$ the column density solutions are indeed being distorted by unresolved saturation and that this affects all three of the C~{\sc i} levels. Reliable solutions for the apparent C~{\sc i} column densities toward HD~254755 can only be obtained for velocity channels with $v_{\sun}> +12\,{\rm km~s}^{-1}$. For the main low velocity component toward HD~254755, we must instead rely on the C~{\sc i} column densities derived through profile synthesis.

\begin{deluxetable}{lccccccc}
\tablecolumns{8}
\tablewidth{0pt}
\tabletypesize{\scriptsize}
\tablecaption{Relative C~{\sc i} Fine-Structure Populations for Distinct Velocity Components\label{tab:car_components}}
\tablehead{ \colhead{Star} & \colhead{$\langle v_{\sun} \rangle$} & \colhead{log~$N$(C~{\sc i})} & \colhead{log~$N$(C~{\sc i}*)} & \colhead{log~$N$(C~{\sc i}**)} & \colhead{log~$N$(C~{\sc i}$_{\rm tot}$)} & \colhead{$f1$} & \colhead{$f2$} \\
\colhead{} & \colhead{(km~s$^{-1}$)} & \colhead{} & \colhead{} & \colhead{} & \colhead{} & \colhead{} & \colhead{} }
\startdata
HD~43582  &    $-$48.1 & $12.43\pm0.05$ & $12.91\pm0.03$ & $13.03\pm0.02$ & $13.33\pm0.02$ & $0.377\pm0.029$ & $0.496\pm0.034$ \\
          &    $-$43.3 & $12.44\pm0.04$ & $12.76\pm0.03$ & $12.84\pm0.02$ & $13.19\pm0.02$ & $0.370\pm0.028$ & $0.451\pm0.032$ \\
          &    $-$37.3 & $12.24\pm0.07$ & $12.81\pm0.03$ & $12.96\pm0.03$ & $13.24\pm0.02$ & $0.374\pm0.035$ & $0.525\pm0.040$ \\
          &    $-$24.0 & $13.23\pm0.02$ & $13.64\pm0.01$ & $13.79\pm0.01$ & $14.09\pm0.01$ & $0.355\pm0.011$ & $0.506\pm0.016$ \\
          & \phn$-$9.0 & $13.28\pm0.02$ & $13.61\pm0.01$ & $13.80\pm0.01$ & $14.09\pm0.01$ & $0.331\pm0.013$ & $0.512\pm0.021$ \\
          &   \phn+6.0 & $14.61\pm0.02$ & $14.19\pm0.01$ & $13.77\pm0.01$ & $14.79\pm0.01$ & $0.250\pm0.010$ & $0.095\pm0.004$ \\
          &      +15.9 & $14.23\pm0.02$ & $13.42\pm0.01$ & $12.78\pm0.03$ & $14.31\pm0.02$ & $0.129\pm0.007$ & $0.030\pm0.003$ \\
          &      +25.0 & $14.02\pm0.02$ & $13.11\pm0.02$ & $12.44\pm0.06$ & $14.08\pm0.02$ & $0.107\pm0.007$ & $0.023\pm0.003$ \\
          &      +52.2 & $12.62\pm0.04$ & $12.45\pm0.07$ & $12.66\pm0.04$ & $13.06\pm0.03$ & $0.243\pm0.044$ & $0.396\pm0.050$ \\
HD~254755 &   \phn+6.0 & $15.05\pm0.04$ & $14.72\pm0.02$ & $14.55\pm0.02$ & $15.30\pm0.02$ & $0.264\pm0.019$ & $0.179\pm0.013$ \\
          &      +14.8 & $14.45\pm0.04$ & $13.64\pm0.02$ & $13.17\pm0.03$ & $14.53\pm0.03$ & $0.129\pm0.011$ & $0.043\pm0.005$ \\
          &      +24.6 & $14.40\pm0.03$ & $13.42\pm0.02$ & $12.54\pm0.08$ & $14.45\pm0.03$ & $0.093\pm0.008$ & $0.012\pm0.003$ \\
\enddata
\end{deluxetable}

\begin{figure}
\centering
\includegraphics[width=0.9\textwidth]{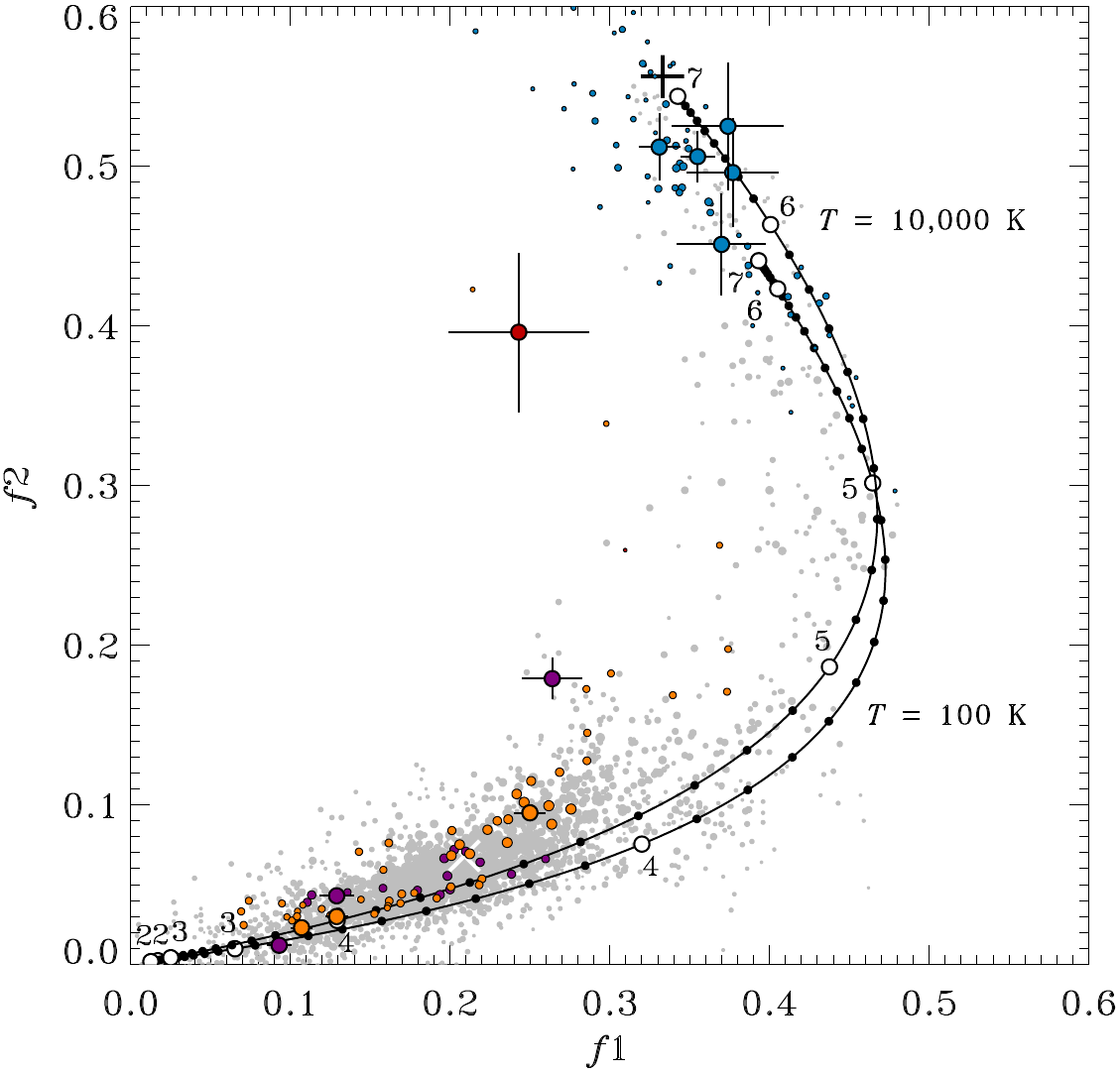}
\caption{Measurements of the relative C~{\sc i} fine-structure populations for velocity components toward HD~43582 and HD~254755. The outcomes for $f1$ and $f2$ derived from profile synthesis fits to the C~{\sc i} multiplets are shown as large colored circles with error bars, while the measurements of $f1$ and $f2$ obtained from the apparent column density profiles of C~{\sc i}, C~{\sc i}*, and C~{\sc i}** are represented by smaller colored circles. (The colors themselves have the same meaning as in Figure~\ref{fig:phys_cond}.) The underlying distribution of small light gray points represents the measurements of $f1$ and $f2$ obtained by Jenkins \& Tripp (2011) for sight lines probing the local Galactic ISM. The sizes of the AOD points and those from Jenkins \& Tripp (2011) are proportional to their corresponding values of log~$N$(C~{\sc i}$_{\rm tot}$). The white ``$\times$'' located at $f1=0.209$, $f2=0.068$ represents the weighted average of all of the measurements obtained by Jenkins \& Tripp (2011). The curves represent theoretical expectations for $f1$ and $f2$ for two different temperatures (as indicated) and for thermal pressures in the range log ($p/k$)~=~2.0--7.0. The points plotted along the curves indicate values of log ($p/k$) separated by 0.1 dex, with integer values represented by large white circles. The ``$+$'' sign at $f1=0.333$, $f2=0.556$ marks the location where the C~{\sc i} levels are populated in proportion to the level degeneracies.\label{fig:car_f1f2}}
\end{figure}

In Table~\ref{tab:car_components}, we present the C~{\sc i}, C~{\sc i}*, and C~{\sc i}** column densities obtained from our global fits to the C~{\sc i} multiplets (Section~\ref{subsubsec:components} and Appendix~\ref{sec:comp_structures}) for the same set of distinct absorption components as identified in Table~\ref{tab:comp_column_densities}. As in Table~\ref{tab:comp_column_densities}, the C~{\sc i} column densities listed in Table~\ref{tab:car_components} are obtained by taking the sum of the column densities of any closely-spaced components that together constitute a single distinct feature. Also presented in the table are the values of $f1$ and $f2$ that correspond to the column densities obtained for each distinct component. In Figure~\ref{fig:car_f1f2}, the measured values of $f1$ and $f2$ toward HD~43582 and HD~254755 (as determined both from the analysis of apparent column densities and from the column densities derived through profile synthesis) are plotted along with curves that give the theoretical expectations for these quantities for gas in which collisional excitations are balanced by a combination of collisional de-excitations and spontaneous radiative decays of the excited levels. The two curves shown are for $T=100\,$K and $T=10,000\,$K and for thermal pressures in the range $2.0\geq\log\,(p/k)\geq7.0$. Also shown in the figure are the measurements of $f1$ and $f2$ obtained by Jenkins \& Tripp (2011) for individual velocity channels along a collection of $\sim$90 sight lines probing the local Galactic ISM.

It is difficult to derive specific values for the thermal pressure in the gas toward HD~43582 and HD~254755 using only the relative C~{\sc i} fine-structure populations because the temperature of the C~{\sc i}-bearing material is not well constrained. Although we have estimates for temperature from the excitation analysis involving O~{\sc i} and Si~{\sc ii}, the C~{\sc i}-bearing material is likely confined to regions that are colder and denser compared to the (presumably) larger volumes where the bulk of the dominant ions reside. Nevertheless, it is clear from Figure~\ref{fig:car_f1f2} that the physical conditions implied by the observed C~{\sc i} excitations are broadly consistent with the results obtained from the analysis of the O~{\sc i} and Si~{\sc ii} excitations. The range of $f1$ and $f2$ values that we find for the low velocity gas toward HD~43582 (orange points in Figure~\ref{fig:car_f1f2}) is similar to the distribution of values exhibited by the low pressure gas observed by Jenkins \& Tripp (2011; light gray points). Note that the C~{\sc i} level populations for the components near +15 and +25~km~s$^{-1}$ toward both of our targets indicate rather low pressures (i.e., $3.1\lesssim\log\,(p/k)\lesssim3.4$), which explains why we were unable to reliably detect the excited O~{\sc i} and Si~{\sc ii} absorption features associated with these components. The main low velocity component toward HD~254755 (purple point near $f1=0.26$ and $f2=0.18$) is somewhat unusual in that it lies considerably above the distribution of points that represents the general ISM.

The negative velocity absorption components toward HD~43582 (blue points in Figure~\ref{fig:car_f1f2}) have $f1$ and $f2$ values that are clustered near the equilibrium curves but at very high pressure. Again, without knowing the temperature of this material in the region where the C~{\sc i} resides, it is difficult to derive any specific values for $\log\,(p/k)$. (For temperatures ranging from 300~K to 10,000~K, the derived thermal pressures would lie in the range $6.0\lesssim\log\,(p/k)\lesssim7.0$.) Regardless, it is clear that the C~{\sc i} excitations for these negative velocity components are consistent in a general sense with the high pressures derived from the analysis of the O~{\sc i} and Si~{\sc ii} excitations. The absorption component at +52~km~s$^{-1}$ toward HD~43582 (red point near $f1=0.24$ and $f2=0.40$) is found to lie significantly far from the equilibrium curves. Jenkins \& Tripp (2011) interpreted outcomes for $f1$ and $f2$ like those we find for the +52~km~s$^{-1}$ component toward HD~43582 and the main low velocity component toward HD~254755 as representing a superposition of gases with the same radial velocity but with different physical conditions. Evidently, the low pressure gas that constitutes the bulk of the material associated with the main low velocity component toward HD~254755 is accompanied by a small but not insignificant amount of gas at high pressure. Likewise, the +52~km~s$^{-1}$ component toward HD~43582 seems to be characterized by a mixture of gases at both very high and very low pressure.

\subsection{Nickel and Calcium Depletion Levels\label{subsec:ni_ca_depl}}
\begin{figure}
\centering
\includegraphics[width=0.49\textwidth]{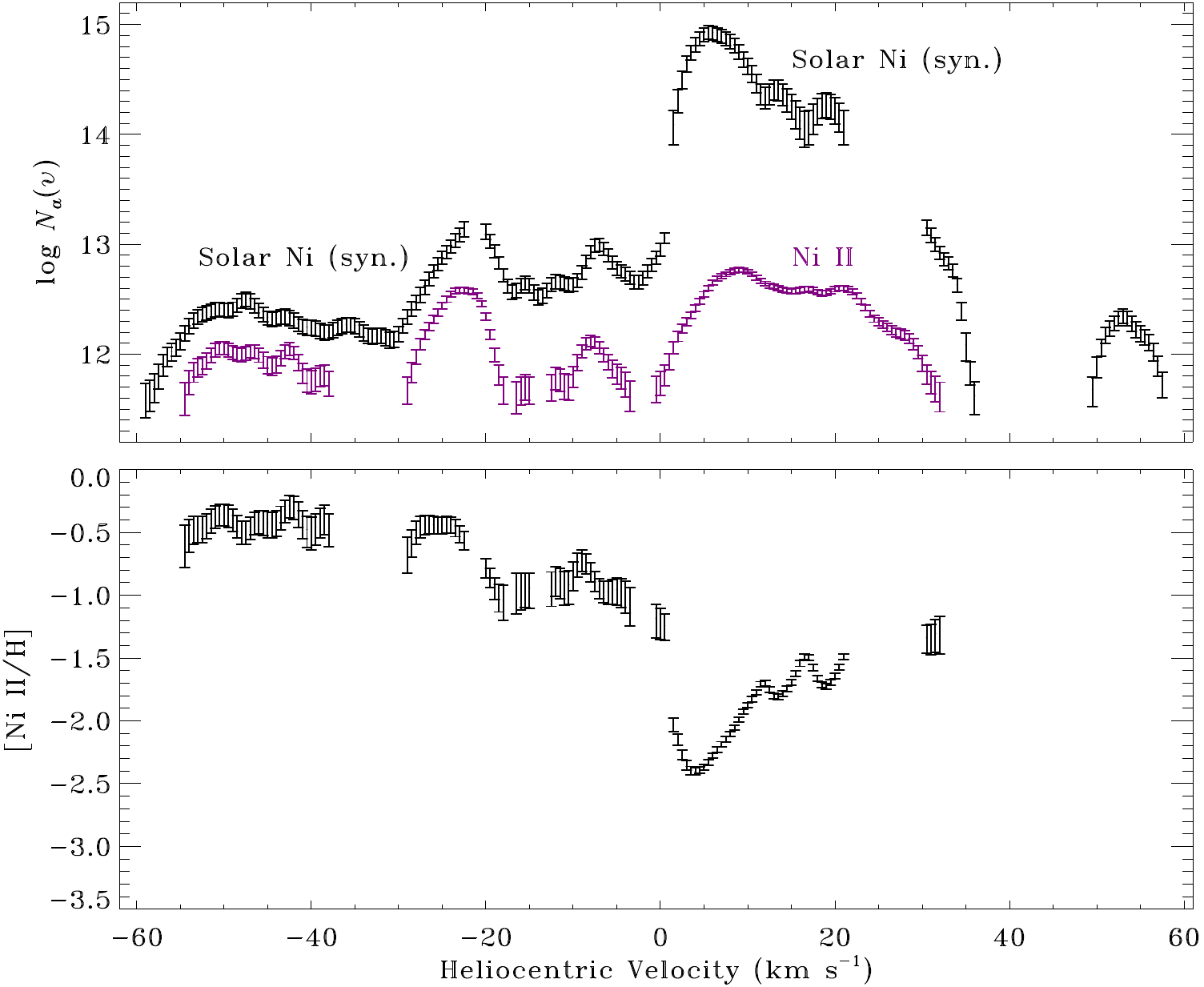}
\includegraphics[width=0.49\textwidth]{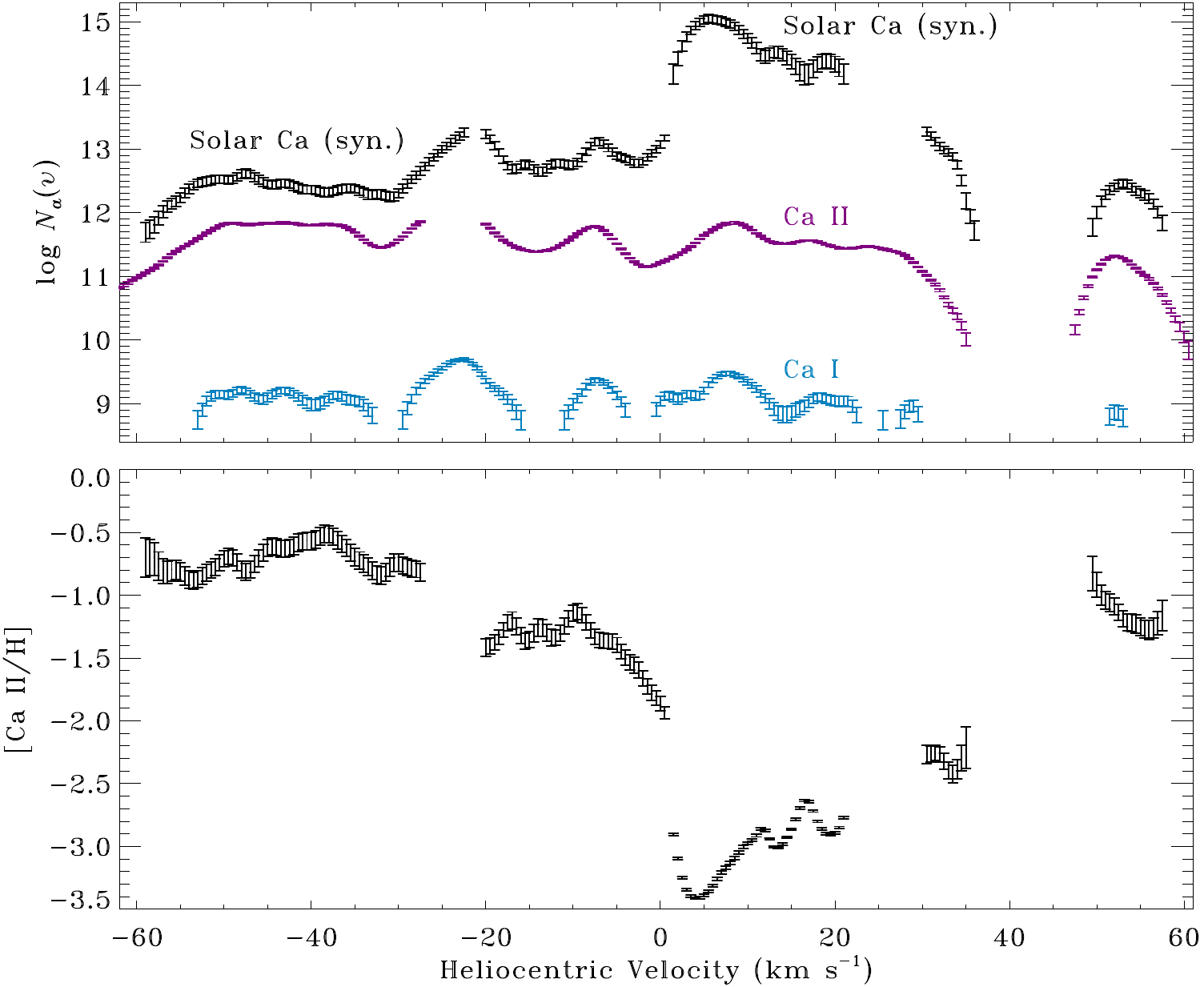}
\caption{Upper panels: Apparent column density profiles used to examine the gas-phase depletions of Ni (left) and Ca (right) toward HD~43582. The Ni~{\sc ii} profile is constructed by taking the weighted mean of the apparent column densities traced by the Ni~{\sc ii}~$\lambda1317$ and $\lambda1370$ lines, whereas the Ca~{\sc i} and Ca~{\sc ii} profiles are obtained directly from the observed Ca~{\sc i}~$\lambda4226$ and Ca~{\sc ii}~$\lambda3968$ lines. Also shown in each panel is a synthetic profile that represents an estimate for the total amount of Ni or Ca (in the gas and dust phases combined) at a given velocity, assuming the total abundances are solar. These synthetic profiles are based on observations of the O~{\sc i}~$\lambda1355$ line at the highest column densities and the S~{\sc ii}~$\lambda1250$ and $\lambda1259$ lines at lower column densities. Lower panels: The corresponding Ni~{\sc ii} and Ca~{\sc ii} depletion factors, [$X$/H]~=~log~($X$/H)~$-$~log~($X$/H)$_{\sun}$, plotted as a function of velocity.\label{fig:nic_cal_depl}}
\end{figure}

In Section~\ref{subsubsection:depl_factors}, we used measurements of the total column densities of various dominant ions to derive estimates for the total hydrogen column densities and depletion factors along the lines of sight to our target stars. The $F_*$ values derived using the methodology of Section~\ref{subsubsection:depl_factors} mainly pertain to the low velocity gas toward HD~43582 and HD~254755 since most of the absorption lines that yielded measurements of total column density are relatively weak and are only detectable in the high column density material at low velocity. There is evidence, however, from the Ni~{\sc ii} absorption profiles toward HD~43582 (Figure~\ref{fig:dominant1}) that the gas at higher velocities along this line of sight exhibits a significantly reduced amount of dust grain depletion. A direct comparison between the apparent column densities of Ni~{\sc ii} and S~{\sc ii} for velocity channels with $v_{\sun}<+1\,{\rm km~s}^{-1}$ toward HD~43582 yields values of the ratio $\log\,[N_a($Ni~{\sc ii}$)/N_a($S~{\sc ii}$)]$ ranging from $-2.0$ to $-1.3$. Considering the depletion trends for Ni and S, as revealed by the analysis of Jenkins (2009), a ratio of $\log\,[N_a($Ni~{\sc ii}$)/N_a($S~{\sc ii}$)]\approx-2.0$ would imply $F_*\approx-0.1$. Since $[{\rm S}/{\rm H}]=0$ at $F_*=0.19$ (Jenkins 2009), we can safely assume that all of the S in these velocity channels is undepleted and we can use $N_a($S~{\sc ii}$)$ as a proxy for $N({\rm H}_{\rm tot})$. For velocity channels with $+1.5\leq v_{\sun} \leq+21.0\,{\rm km~s}^{-1}$, where the S~{\sc ii} lines are heavily saturated, we use $N_a($O~{\sc i}$)$ as a proxy for $N({\rm H}_{\rm tot})$, adopting $F_*=0.53$ (as in Section~\ref{subsubsec:oxy_sil_profiles}). By comparing the observed apparent column densities of Ni~{\sc ii} to our proxies for $N({\rm H}_{\rm tot})$, we obtain the Ni~{\sc ii} depletion factor [Ni~{\sc ii}/H] as a function of velocity, as shown in the lefthand panels of Figure~\ref{fig:nic_cal_depl}.

As the lower left panel of Figure~\ref{fig:nic_cal_depl} makes clear, the depletion of Ni varies quite significantly with velocity from a maximum of $[$Ni~{\sc ii}$/{\rm H}]=-2.4$ at  $v_{\sun}=+4\,{\rm km~s}^{-1}$ to a minimum of $[$Ni~{\sc ii}$/{\rm H}]\approx-0.4$ for velocities in the range $-55< v_{\sun}<-22\,{\rm km~s}^{-1}$. A second plateau at $[$Ni~{\sc ii}$/{\rm H}]\approx-0.9$ is seen for velocities in the interval $-21< v_{\sun}<+1\,{\rm km~s}^{-1}$. Since the velocity channels found to have reduced depletions correspond precisely with those where we find enhanced pressures (Section~\ref{subsubsec:oxy_sil_results}), we interpret this as clear evidence of dust grain destruction by grain-grain collisions or shock sputtering. The dust destruction appears to be most efficient in the gas at the highest (negative) velocity. For this material, $\sim$40\% of the Ni resides in the gas phase, compared to $\sim$0.4\% for the low velocity gas where the depletion reaches its maximum. For material in that intermediate velocity range between $-21$ and $+1\,{\rm km~s}^{-1}$, we find that $\sim$13\% of the Ni is in the gas phase. These calculations presume that most of the gas-phase Ni can be accounted for by considering just the singly-ionized form. If enhanced radiation from shocks has converted some of the Ni~{\sc ii} into Ni~{\sc iii}, then the true gas-phase abundances (and implied dust grain destruction efficiencies) would be higher than stated here. (Note that, in the presence of enhanced radiation, Ni would have a stronger tendency to be doubly-ionized than S, which serves as our proxy for the total amount of hydrogen present in the material with $v_{\sun}<+1\,{\rm km~s}^{-1}$.)

Another refractory element that shows prominent absorption at negative velocities toward HD~43582 is Ca, in the form of both Ca~{\sc i} and Ca~{\sc ii} (Figure~\ref{fig:trace1}). However, since the ionization potential of Ca~{\sc ii} (11.9~eV) is less than that of H~{\sc i}, we must consider how much Ca there might be in the form of Ca~{\sc iii} before we can examine any gas-phase Ca depletions. While a proper treatment of the ionization balance of Ca should consider the equilibrium of a three-ion system (e.g., Weingartner \& Draine 2001), a simplified approach is adopted here. If the ionization rate of Ca~{\sc i} is balanced by the rate of recombinations between Ca~{\sc ii} and free electrons, then we have
\begin{equation}
n({\rm Ca~\textsc{i}})[(I/I_0)\Gamma_0({\rm Ca~\textsc{i}})+C({\rm Ca~\textsc{i}},T)n(e)]=n({\rm Ca~\textsc{ii}})\alpha_e({\rm Ca~\textsc{ii}},T)n(e)~,
\end{equation}
where $\Gamma_0({\rm Ca~\textsc{i}})=3.7\times10^{-10}\,{\rm s}^{-1}$ (Welty et al.~2003) is the photoionization rate of Ca~{\sc i} for the average interstellar radiation field, $I/I_0$ is a factor specifying the enhancement in the radiation field density over that of the average field, $C({\rm Ca~\textsc{i}},T)$ is the temperature-dependent rate coefficient for the collisional ionization of Ca~{\sc i} (Shull \& Van Steenberg 1982), and $\alpha_e({\rm Ca~\textsc{ii}},T)$ is the radiative plus dielectronic recombination coefficient for Ca~{\sc ii} as a function of temperature (Shull \& Van Steenberg 1982). A second equation may be obtained by considering the processes that contribute to and deduct from the amount of Ca~{\sc iii} present. In this case, we have
\begin{eqnarray}
n({\rm Ca~\textsc{ii}})[(I/I_0)\Gamma_0({\rm Ca~\textsc{ii}})+C({\rm Ca~\textsc{ii}},T)n(e)]=n({\rm Ca~\textsc{iii}})[\alpha_e({\rm Ca~\textsc{iii}},T)n(e)+\nonumber \\
\alpha_g({\rm Ca~\textsc{iii}},n(e),I,T)n({\rm H}_{\rm tot})]~,
\end{eqnarray}
where the photoionization rate of Ca~{\sc ii} for the average field is equal to $\Gamma_0({\rm Ca~\textsc{ii}})=1.4\times10^{-12}\,{\rm s}^{-1}$ (Welty et al.~2003) and $\alpha_g({\rm Ca~\textsc{iii}},n(e),I,T)$ represents the ``grain-assisted'' recombination rate coefficient for Ca~{\sc iii} normalized to the total hydrogen density (Weingartner \& Draine 2001). Note that an analogous term involving the grain-assisted recombination of Ca~{\sc ii} is not included in Equation (5) because any Ca$^+$ ions neutralized in this way will most likely stick to the grains following charge exchange, meaning that this process will not affect the amount of Ca~{\sc i} present in the gas.\footnote{Using the terminology and notation of Weingartner \& Draine (2001), this is equivalent to assuming that the sticking probability for Ca is equal to $s=1$.} The temperature-dependent rate coefficients for the collisional ionization of Ca~{\sc ii} and for the radiative and dielectronic recombination of Ca~{\sc iii} with free electrons are (again) obtained from Shull \& Van Steenberg (1982).

For velocity channels where both Ca~{\sc i} and Ca~{\sc ii} are detected (see Figure~\ref{fig:nic_cal_depl}), we can use Equation (5) along with the values of $T$, $n(e)$, and $n({\rm H}_{\rm tot})$ that were derived from the analysis of O~{\sc i} and Si~{\sc ii} excitations (Table~\ref{tab:oxy_sil_results}) to determine corresponding values for the radiation field enhancement factor $I/I_0$. These determinations can then be used in conjunction with Equation (6) to obtain estimates for the amount of Ca~{\sc iii} that accompanies the Ca~{\sc i} and Ca~{\sc ii}. Applying this technique to the line of sight toward HD~43582, we find that for all of the velocity channels with $\log T\lesssim3.7$ (which is the majority of cases), Ca~{\sc ii} remains the dominant Ca species, with Ca~{\sc iii} contributing no more than $\sim$15\% to the total gas-phase abundance (and Ca~{\sc i} contributing less than 1\%). This is consistent with the assertion by Welty et al.~(2003) that Ca~{\sc ii} should be the dominant form of Ca wherever Ca~{\sc i} is detected. Thus, if we ignore the contributions from Ca~{\sc i} and Ca~{\sc iii}, and examine only the Ca~{\sc ii} depletion factors [Ca~{\sc ii}/H], as is done in the lower righthand panel of Figure~\ref{fig:nic_cal_depl}, we will underestimate the gas-phase Ca abundance by no more than $\sim$0.07~dex, which is comparable to the typical uncertainty in [Ca~{\sc ii}/H] of $\sim$0.06 dex.

However, an interesting problem arises for velocity channels with $\log T\gtrsim3.7$. Use of Equations (5) and (6) indicates that for these channels the Ca~{\sc iii} abundance increases steadily with temperature and quickly becomes larger than the assumed value for the total Ca abundance in the gas and dust phases combined (which is based on the observed abundance of S~{\sc ii}). The reason this occurs is that for $\log T\gtrsim3.7$ dielectronic recombination becomes  the dominant recombination mechanism leading to Ca~{\sc i} and the recombination rate increases rapidly with $T$. This leads to extreme values for the radiation field enhancement factor $I/I_0$, which subsequently yield high values for $N$(Ca~{\sc iii}). This could be another indication that the high temperatures derived for some velocity channels are not very realistic, and that lower temperatures, that would still fall within the error arrays associated with the O~{\sc i} excitation analysis, are more likely. Another consideration is that the three Ca species probably occupy somewhat different volumes along the line of sight, with Ca~{\sc ii} and Ca~{\sc iii} distributed more broadly than Ca~{\sc i} (e.g., Welty et al.~2003; Pan et al.~2005). (Moreover, the regions in which the different Ca species reside are likely characterized by somewhat different physical conditions, whereas we have implicitly assumed that the densities and temperatures derived from the O~{\sc i} and Si~{\sc ii} excitations apply equally well to both the Ca~{\sc i} and Ca~{\sc ii}-bearing gas.) Incidentally, for velocity channels with $\log T\lesssim3.7$, the derived values of $I/I_0$ range from $\sim$0.03 to $\sim$60, with the highest values associated with the highest positive and negative velocities, suggesting that there may indeed be some enhanced radiation from shocks at these velocities.

It is clear from Figure~\ref{fig:nic_cal_depl} that the changes in the Ca~{\sc ii} depletions toward HD~43582 closely mirror the changes seen in the depletions of Ni~{\sc ii}, although Ca~{\sc ii} is generally more severely depleted than Ni~{\sc ii} at all velocities. The Ca~{\sc ii} depletion factors vary from a maximum of $[$Ca~{\sc ii}$/{\rm H}]=-3.4$ at  $v_{\sun}=+4\,{\rm km~s}^{-1}$ to a minimum of $[$Ca~{\sc ii}$/{\rm H}]\approx-0.8$ for velocities in the range $-60< v_{\sun}<-27\,{\rm km~s}^{-1}$. As with Ni~{\sc ii}, the Ca~{\sc ii} depletions exhibit a secondary plateau for velocities between $-21$ and $+1\,{\rm km~s}^{-1}$ where $[$Ca~{\sc ii}$/{\rm H}]\approx-1.4$. This level of depletion is similar to that seen for velocity channels in the range $+49< v_{\sun}<+58\,{\rm km~s}^{-1}$ where $[$Ca~{\sc ii}$/{\rm H}]\approx-1.1$. The stronger depletions for Ca~{\sc ii} compared to Ni~{\sc ii}, even in the highly-compressed gas toward HD~43582, are consistent with the notion that the dust-phase Ca primarily resides within the cores of grains, which may be more resistant to destruction by shocks than are the outer layers.

\subsection{Molecular Excitation and Chemical Analysis\label{subsec:mol_ex}}
The molecular column densities derived in Section~\ref{subsec:molecular} can be used to evaluate the physical conditions in the molecule-bearing gas along our target sight lines. We list in Table~\ref{tab:mol_results} the column densities that pertain to the dominant molecular absorption components observed toward HD~43582 and HD~254755. We add to this list results for the sight lines to HD~254477 and HD~254577, both of which were also observed by Taylor et al.~(2012) using the HET/HRS. (These two additional sight lines probe material associated with the southern molecular ridge in IC~443. See Figure~1 of Taylor et al.~2012 for the positions of these stars on the sky.) From those HET data, we have derived updated values for the column densities of CH and CH$^+$ toward HD~254477 and HD~254577 through profile fits to the CH~$\lambda4300$ and CH$^+$~$\lambda4232$ lines. We have also redetermined the C$_2$ column density toward HD~254577 by simultaneously fitting the C$_2$ $A$$-$$X$ (2$-$0) and (3$-$0) bands (as was done for HD~254755; see Section~\ref{subsec:molecular}). (Since only one C$_2$ rotational level is observed toward HD~254477, we have simply reduced the C$_2$ column density from Hirschauer et al.~(2009) by a factor of 1.4 to reflect a change in the adopted $f$-value.) We have also revised the CN column densities toward HD~254477 and HD~254577 from Hirschauer et al.~(2009) by adopting somewhat smaller $b$-values. Note that no CO results exist for HD~254477 or HD~254577 because these stars were not observed with \emph{HST}. (Fits to the absorption profiles of the atomic and molecular transitions covered by the HET observations of HD~254477 and HD~254577 are presented in Appendix~\ref{sec:het_profiles}.)

The analyses of C$_2$ excitation and CH, C$_2$, and CN chemistry described here are essentially updates to similar analyses performed for these sight lines by Hirschauer et al.~(2009). Beyond the improved column density determinations that have resulted from our use of the HET data obtained by Taylor et al.~(2012), which have much higher spectral resolution compared to the ARCES data examined by Hirschauer et al.~(2009), the present analysis benefits from several additional refinements. The $f$-value for the C$_2$ $A$$-$$X$ (2$-$0) band is now well determined (e.g., Kokkin et al.~2007; Schmidt \& Bacskay 2007) and has a value $\sim$40\% higher than that used before. This change affects both the column density and the density inferred from C$_2$ excitation. Quantum calculations of the excitation cross sections for C$_2$ (e.g., Najar et al.~2008, 2009) yield values about twice as large as used before. As noted in Hupe et al.~(2012), these two changes roughly offset one another, but we use the more precise values in the present analysis. New information on rate coefficients for C$_2$ and CN chemistry is also available. The main change in these values compared to those adopted in previous efforts (e.g., Federman et al.~1994; Sheffer et al.~2008; Hirschauer et al.~2009) is that many of the coefficients now have measured temperature dependencies and those dependencies are reproduced by improved theoretical calculations (see McElroy et al.~2013). The CO excitation analysis performed here is based on the work of Goldsmith (2013), but makes use of a finer grid of kinetic temperatures (P.~Goldsmith 2016, private communication).

\begin{deluxetable}{lcccccccccc}
\tablecolumns{11}
\tablewidth{0pt}
\tabletypesize{\scriptsize}
\tablecaption{Molecular Column Densities and Derived Physical Conditions\label{tab:mol_results}}
\tablehead{ \colhead{Star} & \colhead{$\langle v_{\sun} \rangle$} & \colhead{$N$(CH$^+$)} & \colhead{$N$(CH)} & \colhead{$N$(CN)} & \colhead{$N$(C$_2$)} & \colhead{$N$(CO)} & \colhead{$T$(C$_2$)\tablenotemark{a}} & \colhead{$n_{\mathrm{H}}$(C$_2$)\tablenotemark{a}} & \colhead{$n_{\mathrm{H}}$(Chem)\tablenotemark{b}} & \colhead{$n_{\mathrm{H}}$(CO)\tablenotemark{c}} \\
\colhead{} & \colhead{(km~s$^{-1}$)} & \colhead{(10$^{13}$~cm$^{-2}$)} & \colhead{(10$^{13}$~cm$^{-2}$)} & \colhead{(10$^{13}$~cm$^{-2}$)} & \colhead{(10$^{13}$~cm$^{-2}$)} & \colhead{(10$^{13}$~cm$^{-2}$)} & \colhead{(K)} & \colhead{(cm$^{-3}$)} & \colhead{(cm$^{-3}$)} & \colhead{(cm$^{-3}$)} }
\startdata
HD~43582 & +7.4 & 3.0 & 1.8 & \ldots\tablenotemark{d} & \ldots\tablenotemark{d} & 12. & \ldots & \ldots & \ldots & 375 \\
HD~254477 & +8.9 & 1.2 & 4.7 & 0.56 & 4.5 & \ldots\tablenotemark{e} & 50\tablenotemark{f} & \ldots & 250 & \ldots \\
HD~254577 & +5.4 & 2.7 & 3.3 & 0.95 & 8.0 & \ldots\tablenotemark{e} & 50 & 300 & 300 & \ldots \\
HD~254755 & +6.0 & 4.2 & 5.0 & 0.29 & 5.8 & 72. & 30 & 150 & 150 & 1200 \\
\enddata
\tablenotetext{a}{Kinetic temperature and total hydrogen density from an analysis of C$_2$ excitation. Values of $n_{\rm H}({\rm C}_2)$ reported here assume $I_{\rm IR}\approx1$ and $n({\rm H})\approx n({\rm H}_2)$.}
\tablenotetext{b}{Total hydrogen density from an analysis of diffuse cloud chemistry involving CH, C$_2$, and CN.}
\tablenotetext{c}{Total hydrogen density from an analysis of CO excitation, assuming $n({\rm H})\approx n({\rm H}_2)$.}
\tablenotetext{d}{No CN or C$_2$ absorption is detected toward HD~43582.}
\tablenotetext{e}{No UV data are available on CO toward HD~254477 or HD~254577.}
\tablenotetext{f}{The value of $T$(C$_2$) cannot be confirmed since only one C$_2$ rotational level is observed. A value of 50~K is assumed.}
\end{deluxetable}

Comparisons between the measured values of the relative C$_2$ rotational populations and the values predicted by models (e.g., van Dishoeck \& Black 1982; van Dishoeck 1984) yield estimates for the kinetic temperature in the C$_2$-bearing material $T$(C$_2$) and the quantity $n\sigma/I_{\rm IR}$, where $n=n({\rm H})+n({\rm H}_2)$ is the density of collision partners, $\sigma$ is the cross section for collisional de-excitation of C$_2$, and $I_{\rm IR}$ is the radiation field enhancement factor in the near infrared. Assuming $I_{\rm IR}\approx1$ and $n({\rm H})\approx n({\rm H}_2)$, we find total hydrogen densities $n_{\rm H}=n({\rm H})+2n({\rm H}_2)$ of $\sim$300~cm$^{-3}$ toward HD~254577 and $\sim$150~cm$^{-3}$ toward HD~254755 and kinetic temperatures of 50~K and 30~K, respectively. The densities derived from C$_2$ excitation for these sight lines are comparable to those we obtain from an analysis of the chemical networks that link the CH, CN, and C$_2$ molecules (see Table~\ref{tab:mol_results}). For the chemical analysis, we follow the methodology of Federman et al.~(1994) in deriving gas densities from a comparison between the observed and predicted column densities of CN and C$_2$ (with updated rate coefficients as noted above). However, as in Hirschauer et al.~(2009), we were unable to reproduce the relatively high C$_2$ column densities observed toward the stars in IC~443 within the framework of our simple chemical model. We therefore restricted the analysis to reproducing just the CN column densities.

The relative CO rotational populations derived through profile fitting yield estimates for $n({\rm H}_2)$ (see Goldsmith 2013), from which we calculate values for $n_{\rm H}$ by (again) assuming that $n({\rm H})\approx n({\rm H}_2)$. The gas density obtained from our analysis of CO excitation toward HD~254755 is significantly higher than those based on C$_2$ excitation and CN chemistry (Table~\ref{tab:mol_results}). However, when we consider the physical conditions derived from the analysis of O~{\sc i} and Si~{\sc ii} excitation (Table~\ref{tab:comp_results}), the density obtained from CO for this sight line seems more realistic. The O~{\sc i} and Si~{\sc ii} analysis gives $n_{\rm H}\approx680\,{\rm cm}^{-3}$ and $T\approx95\,{\rm K}$ for the dominant absorption component toward HD~254755. We would generally expect dominant ions like O~{\sc i} and Si~{\sc ii} to be more broadly distributed along the line of sight than molecular species like CO, C$_2$, and CN, and the molecules to be confined to colder and denser regions within the cloud. The density derived from CO excitation toward HD~254755 ($n_{\rm H}\approx1200\,{\rm cm}^{-3}$) is more in line with these expectations. The gas density implied by the relative C$_2$ rotational populations would be larger if the value of $I_{\rm IR}$ were significantly greater than 1. As an example, our analysis of the Ca~{\sc i}/Ca~{\sc ii} ratio for the dominant absorption component toward HD~254755 yields $I/I_0\approx6$ (following the same procedure as outlined in Section~\ref{subsec:ni_ca_depl}). If a similar scale factor applies to the radiation field strength at wavelengths near the C$_2$ $A$$-$$X$ bands, then our derived value of $n_{\rm H}({\rm C}_2)$ would be in much better agreement with $n_{\rm H}({\rm CO})$.

One issue that may be affecting the chemical analysis is that some fraction of the observed C$_2$ and CH molecules could be associated with CH$^+$ chemistry rather than with CN. The CH$^+$ column densities toward the stars probing IC~443 are relatively large (Table~\ref{tab:mol_results}; Hirschauer et al.~2009), suggesting that nonthermal processes, such as the dissipation of turbulence (Godard et al.~2014) or the propagation of Alfv\'{e}n waves (Federman et al.~1996) or magnetohydrodynamic shocks (Pineau des For\^{e}ts et al.~1986; Draine \& Katz 1986), are active in these regions. Recall that the analysis of C~{\sc i} excitation for the main low velocity component toward HD~254755 suggested the presence of a small amount of gas at high pressure (Section~\ref{subsec:car_ex}), which may be an indication that a weak shock is present in this material. Whatever the mechanism is that produces the CH$^+$, the presence of a significant amount of CH$^+$ in the gas has consequences for the chemistry. Reactions between CH$^+$ and H$_2$, leading to CH$_2^+$ and CH$_3^+$, followed by the dissociative recombination of CH$_3^+$, will produce CH. Further gas-phase reactions will then produce C$_2$ and CO. As much as half of the observed CH, C$_2$, and CO toward the stars in Table~\ref{tab:mol_results} showing CN absorption may be associated with CH$^+$ chemistry with the remaining amounts associated with CN. This could help to resolve some of the discrepancies in the density estimates noted above. However, there are a number of uncertainties involved in trying to disentangle the various chemical pathways, making it difficult to obtain any specific results. Moreover, our simple chemical model does not account for any time-dependent effects, which may ultimately be important.

Indriolo et al.~(2010) adopted the gas densities computed by Hirschauer et al.~(2009) in their analysis of cosmic ray ionization rates in IC~443 from absorption line measurements of H$_3^+$. In particular, Indriolo et al.~(2010) assume that $n_{\rm H}=325\,{\rm cm}^{-3}$ toward HD~254577 and find an ionization rate of molecular hydrogen equal to $\zeta_2=2.6\times10^{-15}\,{\rm s}^{-1}$, the highest in their survey of sight lines in IC~443. While the density adopted by Indriolo et al.~(2010) for HD~254577 is comparable to what we find from our analyses of C$_2$ excitation and CN chemistry, those values may underestimate the true density of the molecular material (as discussed above). If the molecular gas density toward HD~254577 were closer to what we find from CO excitation toward HD~254755, then the inferred cosmic ray ionization rate would be even higher than derived by Indriolo et al.~(2010) since $\zeta_2$ scales linearly with $n_{\rm H}$ (for an assumed value of the ionization fraction). To gain further clarity on these issues, it would be useful to obtain CO observations for additional sight lines in IC~443.

\subsection{C~{\sc ii} Excitations\label{subsec:car2_ex}}
For all of the high and very high velocity absorption components observed toward HD~43582 (i.e., those with $v_{\sun}\gtrsim+100\,{\rm km~s}^{-1}$ or $v_{\sun}\lesssim-100\,{\rm km~s}^{-1}$), the techniques described in the preceding sections that we have used to obtain physical conditions do not apply because the relevant species (e.g., O~{\sc i}* and O~{\sc i}**) are not detected. However, most of these components exhibit significant absorption from both C~{\sc ii} and C~{\sc ii}*. As with Si~{\sc ii}, the upper fine-structure level of C~{\sc ii} is populated primarily by collisions with electrons. As such, the C~{\sc ii}*/C~{\sc ii} column density ratio is a useful probe of the electron density particularly in predominantly ionized gas (e.g., Keenan et al.~1986).

By considering the balance between collisional excitations to the upper fine-structure level of C~{\sc ii} ($^2$P$_{3/2}$) and collisional and spontaneous de-excitations to the lower level ($^2$P$_{1/2}$), we can derive an equation analogous to Equation (3) for the C~{\sc ii}*/C~{\sc ii} population ratio:
\begin{equation}
\frac{n({\rm C~\textsc{ii}\text{*}})}{n({\rm C~\textsc{ii}})}=\frac{n(e)k_{1,2}(e,T)+n({\rm H}^+)k_{1,2}({\rm H}^+,T)+n({\rm H}^0)k_{1,2}({\rm H}^0,T)}{n(e)k_{2,1}(e,T)+n({\rm H}^+)k_{2,1}({\rm H}^+,T)+n({\rm H}^0)k_{2,1}({\rm H}^0,T)+A_{2,1}}~.
\end{equation}
For these calculations, the rate coefficients for excitation by electron impact $k_{1,2}(e,T)$ are derived from the expression given by Goldsmith et al.~(2012). For excitations by collisions with atomic hydrogen, we use rate coefficients $k_{1,2}({\rm H}^0,T)$ derived from the expression provided by Barinovs et al.~(2005). The rate coefficients for excitations by collisions with free protons $k_{1,2}({\rm H}^+,T)$ are interpolated from the results tabulated by Foster et al.~(1996). The corresponding de-excitation rate coefficients are obtained through the principle of detailed balance. The spontaneous decay rate from the $J=3/2$ level to the $J=1/2$ level of C~{\sc ii} is $A_{2,1}=2.29\times10^{-6}\,{\rm s}^{-1}$ (Nussbaumer \& Storey 1981).

The temperatures of the high and very high velocity absorption components may be obtained by comparing the relative abundances of the different ions observed to the nonequilibrium models of Gnat \& Sternberg (2007) for solar composition gas cooling radiatively at constant pressure. We use the nonequilibrium (isobaric) models because for $T\lesssim5\times10^6\,$K the gas cools more rapidly than it can recombine (Gnat \& Sternberg 2007). As an example, the $N$(Si~{\sc iv})/$N$(Si~{\sc iii}) ratio for the broad absorption component at $-$546~km~s$^{-1}$ (Table~\ref{tab:high_ions}) is consistent with a temperature of $\sim$$8.7\times10^4\,$K. At this temperature, the model predictions for the column densities of N~{\sc v} and C~{\sc ii} match the observed column densities within $\sim$0.05~dex (see Section~\ref{subsec:high_ions}). For the high positive velocity components toward HD~43582, different ion ratios are used to derive temperatures. For the +96~km~s$^{-1}$ component (Table~\ref{tab:high_vel}), the $N$(C~{\sc ii}$_{\rm tot}$)/$N$(O~{\sc i}) ratio indicates that the temperature is $\sim$$1.7\times10^4\,$K. (Note that in this case the measured line width of the O~{\sc i}~$\lambda1302$ feature is consistent with a temperature as high as $\sim$$3.0\times10^4\,$K.) Since the group of C~{\sc ii} and C~{\sc ii}* components between +100 and +140~km~s$^{-1}$ (Table~\ref{tab:high_vel}) span a similar range in velocity as the Si~{\sc iv} component at +122~km~s$^{-1}$ (Table~\ref{tab:high_ions}), we use the $N$(C~{\sc ii}$_{\rm tot}$)/$N$(Si~{\sc iv}) ratio for this group of components to derive a temperature of $\sim$$5.0\times10^4\,$K. Finally, the $N$(Si~{\sc ii})/$N$(Si~{\sc iv}) ratio for the sum of the two Si~{\sc ii} components at +218 and +228~km~s$^{-1}$ and the Si~{\sc iv} component at +225~km~s$^{-1}$ yields a temperature of $\sim$$1.9\times10^4\,$K. Each of these calculations for temperature also yields predictions for the relative abundances of H$^0$ and H$^+$ (along with He$^0$, He$^+$, He$^{++}$, etc). In all cases, we find that the gas is nearly completely ionized with hydrogen ionization fractions in the range $x({\rm H}^+)\approx0.97$--$1.00$. The total hydrogen column density $N($H$_{\rm tot})=N($H$^0)+N($H$^+)\approx N($H$^+)$ can also be obtained from the observed column density of a given ion and the fractional abundance of that ion as predicted by the model.

\begin{deluxetable}{lccccccccc}
\tablecolumns{10}
\tablewidth{0pt}
\tabletypesize{\scriptsize}
\tablecaption{Physical Conditions Derived for High Velocity Absorption Components\label{tab:high_vel_results}}
\tablehead{ \colhead{Star} & \colhead{$\langle v_{\sun} \rangle$} & \colhead{Ratio\tablenotemark{a}} & \colhead{Value\tablenotemark{a}} & \colhead{$N$(C~{\sc ii}*)/$N$(C~{\sc ii})} & \colhead{log~$N$(H$_{\rm tot}$)} & \colhead{$n$(H$_{\mathrm{tot}}$)} & \colhead{$T$} & \colhead{$n$($e$)} & \colhead{log~($p/k$)} \\
\colhead{} & \colhead{(km~s$^{-1}$)} & \colhead{} & \colhead{} & \colhead{} & \colhead{} & \colhead{(cm$^{-3}$)} & \colhead{(K)} & \colhead{(cm$^{-3}$)} & \colhead{} }
\startdata
HD~43582 & $-$665 & log~(Si~{\sc iv}/Si~{\sc iii}) & $0.18\pm0.09$  & \ldots\tablenotemark{b}  & $19.87\pm0.02$ & \dots\tablenotemark{b} & 82000  &\ldots\tablenotemark{b}  & \ldots\tablenotemark{b} \\
 & $-$546 & log~(Si~{\sc iv}/Si~{\sc iii}) & $0.31\pm0.12$ & $1.32\pm0.30$ & $19.76\pm0.05$ & 140.\phn & 87000 & 160.\phn & $7.44\pm0.26$ \\
 & \phn+96 & log~(C~{\sc ii}$_{\rm tot}$/O~{\sc i}) & $1.05\pm0.08$ & $0.48\pm0.13$ & $18.21\pm0.01$ & 16. & 17000 & 17. & $5.77\pm0.13$ \\
 & +122 & log~(C~{\sc ii}$_{\rm tot}$/Si~{\sc iv}) & $1.43\pm0.08$ & $0.25\pm0.03$ & $18.52\pm0.05$ & \phn\phn8.8 & 50000 & 10. & $5.99\pm0.05$ \\
 & +225 & log~(Si~{\sc ii}/Si~{\sc iv}) & $0.78\pm0.11$ & $0.19\pm0.02$ & $18.20\pm0.11$ & \phn\phn5.3 & 19000 & \phn\phn5.8 & $5.34\pm0.05$ \\
\enddata
\tablenotetext{a}{Logarithmic column density ratio (and value of the ratio) used to obtain the temperature and total hydrogen column density listed for each component. Values of $T$ and $N$(H$_{\rm tot}$) are derived by comparing the observed column density ratio to model predictions for solar composition gas cooling radiatively at constant pressure (Gnat \& Sternberg 2007).}
\tablenotetext{b}{Neither C~{\sc ii} nor C~{\sc ii}* are detected at this velocity so we are unable to evaluate the density and pressure of the material.}
\end{deluxetable}

For all of the high and very high velocity absorption components where both C~{\sc ii} and C~{\sc ii}* are detected, we can use the temperatures and ionization fractions obtained above to derive the electron densities (and hence the total hydrogen densities) through use of Equation (7). For the +225~km~s$^{-1}$ component, however, it is not possible to directly measure the column density of C~{\sc ii} because this high velocity absorption is blended with saturated absorption from C~{\sc ii}* at low velocity (see Section~\ref{subsubsec:high_vel}). Thus, we use the model prediction for the total C~{\sc ii} column density at $T\approx1.9\times10^4\,$K, which is $\log N($C~{\sc ii}$_{\rm tot})\approx14.39$, and the observed column density of C~{\sc ii}*, which is $\log N($C~{\sc ii}*$)=13.60$ for the sum of the two components at +218 and +228~km~s$^{-1}$ (Table~\ref{tab:high_vel}), to obtain a C~{\sc ii} column density of $\log N($C~{\sc ii}$)\approx14.32$. The densities, temperatures, and thermal pressures derived from these calculations are summarized in Table~\ref{tab:high_vel_results}. Compared to the pressures obtained for the components seen at more moderate velocities toward HD~43582 (Table~\ref{tab:comp_results}), the pressures derived for the high positive velocity components are $\sim$1.0~dex lower. For the very high negative velocity absorption feature, however, the pressure is higher by $\sim$0.5~dex.

\section{DISCUSSION\label{sec:discussion}}

\subsection{Comments on the Age and Distance of IC~443\label{subsec:age_dist}}
The heliocentric distance to IC~443 is usually quoted as 1.5~kpc since the progenitor of the SNR is presumed to have been a member of the Gem OB1 stellar association (e.g., Humphreys 1978). In the recent literature, this value for the distance is often attributed to Welsh \& Sallmen (2003), who examined Na~{\sc i} and Ca~{\sc ii} absorption profiles toward stars in IC~443 and concluded that the remnant must be at a distance similar to that of the stars showing complex high-velocity absorption (i.e., HD~43582 and HD~254577). While the association of IC~443 with Gem OB1 is probably correct, the distance to the stellar association is in need of revision. In Table~\ref{tab:gaia_dist}, we give the distances to stars in Gem OB1 as determined from the parallaxes measured by the \emph{Gaia} satellite (DR2; Anders et al.~2019). The list of stars for inclusion in Table~\ref{tab:gaia_dist} comes from a combination of Humphreys (1978) and Hirschauer et al.~(2009), from which we have included only those stars with \emph{Gaia} parallaxes known to 4$\sigma$ precision or better. Unfortunately, this selection criterion excludes our \emph{HST} target HD~254755. The \emph{Gaia} DR2 distance for this star is 3.6~kpc (Anders et al.~2019), yet the 1$\sigma$ confidence interval ranges from 2.2~kpc to 5.7~kpc. For the members of Gem OB1 with well-determined distances (Table~\ref{tab:gaia_dist}), the average distance as determined by \emph{Gaia} is 2.1~kpc. The means of the lower and upper bounds on the distances for these stars indicate a probable range of 1.8~kpc to 2.3~kpc for the distance to the Gem OB1 association. If the progenitor of IC~443 was indeed a member of Gem OB1, then this range probably represents the best estimate for the distance to the remnant. Note also that HD~43582 and HD~254577, both of which are clearly positioned behind the remnant because they each show complex high-velocity interstellar absorption (Welsh \& Sallmen 2003; Hirschauer et al.~2009; this work), have \emph{Gaia} DR2 distances between 1.8~kpc and 2.4~kpc. This range is also consistent with a recent kinematic determination of the distance to IC~443 of 1.9~kpc (Ambrocio-Cruz et al.~2017) and with an estimate of 1.8~kpc based on dust extinction (Zhao et al.~2020).

\begin{deluxetable}{lcccccc}
\tablecolumns{7}
\tabletypesize{\footnotesize}
\tablecaption{\emph{Gaia} Distances to Stars in Gem OB1 \label{tab:gaia_dist}}
\tablehead{\colhead{Star} & \multicolumn{2}{c}{Gal. Coord.} & \colhead{} & \multicolumn{3}{c}{Dist.\tablenotemark{a}} \\
\cline{2-3} \cline{5-7} \\
\colhead{} & \colhead{$l$} & \colhead{$b$} & \colhead{} & \colhead{Lower} & \colhead{Best} & \colhead{Upper} }
\startdata
HD~42088 & 190.04 & 0.48 && 1.5 & 1.7 & 1.9 \\
HD~42379 & 189.28 & 1.34 && 2.5 & 3.1 & 3.8 \\
HD~42400 & 189.87 & 1.04 && 1.6 & 1.8 & 1.9 \\
HD~43078 & 189.08 & 2.50 && 1.9 & 2.1 & 2.4 \\
HD~43384 & 187.99 & 3.53 && 2.1 & 2.3 & 2.6 \\
HD~43582 & 189.06 & 3.23 && 1.8 & 2.0 & 2.4 \\
HD~43703 & 188.82 & 3.52 && 1.3 & 1.4 & 1.5 \\
HD~43753 & 188.87 & 3.59 && 1.8 & 2.0 & 2.3 \\
HD~43818 & 188.49 & 3.87 && 2.1 & 2.4 & 2.7 \\
HD~254042 & 187.59 & 3.45 && 2.0 & 2.3 & 2.6 \\
HD~254346 & 189.36 & 2.79 && 2.1 & 2.4 & 2.8 \\
HD~254577 & 189.27 & 3.09 && 1.9 & 2.1 & 2.4 \\
HD~254700 & 188.84 & 3.45 && 1.8 & 1.9 & 2.1 \\
HD~255055 & 188.69 & 3.87 && 1.7 & 1.8 & 2.0 \\
HD~256035 & 189.42 & 4.33 && 1.7 & 1.8 & 2.0 \\
ALS~8828 & 188.77 & 2.91 && 1.7 & 1.9 & 2.2 \\
\hline
mean value & & && 1.8 & 2.1 & 2.3 \\
\enddata
\tablenotetext{a}{Distance (in kpc) estimated using the parallax measured by the \emph{Gaia} satellite (DR2; Anders et al.~2019). The central column gives the distance estimate, while the other two columns give the lower and upper bounds on the confidence interval of the estimated distance. The bottom row gives the mean values for Gem OB1.}
\end{deluxetable}

Estimates of the age of IC~443 have varied significantly, with most estimates in the range 3,000 to 30,000~yr (Petre et al.~1988; Chevalier 1999; Lee et al.~2008; Troja et al.~2008). Petre et al.~(1988), examining X-ray emission from IC~443, determined the age to be $\sim$3,000~yr by equating the temperature derived from the X-ray data (i.e., $\sim$$1.2\times10^7\,$K for the hot component in their two-component ionization equilibrium model) with the post-shock temperature. Petre et al.~(1988) based their analysis on the model of Gulliford (1974), who assumed that the IC~443 blast wave was expanding into a medium with a density gradient that extended from the southwest to the northeast. Chevalier (1999) presented a model for IC~443 wherein the remnant is assumed to be in the radiative phase and the velocity of the radiative shell is taken to be the velocity determined from observations of the bright optical filaments in the northeast (i.e., $\sim$$100\,{\rm km~s}^{-1}$; Fesen \& Kirshner 1980). The radiative shell model of Chevalier (1999) implies an age of $\sim$30,000~yr for IC~443 and assumes that the supernova shock has been propagating in the interclump medium of a molecular cloud with a (uniform) pre-shock density of $\sim$15~cm$^{-3}$. Lee et al.~(2008) considered a similar model to that of Chevalier (1999) and (likewise) adopted the velocity determined from the optical filaments as the primary shock velocity, finding an age of $\sim$20,000~yr for IC~443. Lee et al.~(2008) argue that the X-ray temperature does not represent the current shock velocity since it is biased toward the hot interior of the remnant. Their argument invokes thermal conduction in the hot interior to explain the high X-ray temperature.

Troja et al.~(2008), examining hard X-ray thermal emission from IC~443, found evidence for a ring-like structure of hot metal-enriched plasma, which they interpret as stellar ejecta re-heated by the supernova reverse shock. The ring is positioned just north of a pulsar wind nebula (PWN) discovered in the southern portion of IC~443 (G189.22+2.90; Olbert et al.~2001; Gaensler et al.~2006). From the relative sizes of the shells representing the forward shock and the reverse shock, Troja et al.~(2008) deduce a dynamic age of $\sim$4,000~yr for IC~443. Their model assumes the SNR is expanding in a uniform ambient medium with a pre-shock density of $\sim$0.25~cm$^{-3}$. (A slightly different model with a non-uniform density distribution gave a similar result.) A physical association between IC~443 and the PWN found within its borders has been difficult to establish (e.g., Leahy 2004; Gaensler et al.~2006; Swartz et al.~2015). Primarily, this is due to the significant offset between the location of the pulsar and the geometric center of the remnant (and the fact that the symmetry axis of the PWN is oriented $\sim$50$\degr$ away from the direction expected if the pulsar originated at the remnant's center). However, recent observations by Greco et al.~(2018) provide intriguing evidence for a direct connection between IC~443 and the PWN. Greco et al.~(2018) discovered a jet-like structure in X-ray emission extending away from IC~443 to the northwest. The authors associate this jet with metal-rich ejecta expelled by the SN progenitor. Combining an estimate of the proper motion of the pulsar (Gaensler et al.~2006) with an assumed age of 4,000~yr for the remnant (Troja et al.~2008), Greco et al.~(2018) calculate what the position of the pulsar would have been at the time of the SN explosion, finding that this position is nearly perfectly aligned with the jet of ejecta discovered in X-rays.

An obvious implication of the observations of Greco et al.~(2018) is that the SN explosion that led to the formation of IC~443 did not take place near the center of the shell seen at radio and optical wavelengths, as is usually assumed. If this interpretation is correct, then the misalignment could plausibly be attributed to the interaction of the evolving SNR with its surroundings. Troja et al.~(2006) considered a scenario wherein the remnant has primarily evolved in a low density environment, possibly a wind-blown cavity created by the massive star progenitor (see also Braun \& Strom 1986b), and only recently has begun to interact with a moderately dense cloud of neutral (atomic) hydrogen in the northeast. In this model, the low density cavity is surrounded by a torus-like molecular cloud (see Figure~9 of Troja et al.~2006), which may have impeded the expansion of the remnant toward the southeast. Another implication of the apparent alignment between the pulsar and the X-ray emitting jet is the young inferred age for IC~443, which agrees nicely with the age derived from the X-ray temperature and morphology (Petre et al.~1988; Troja et al.~2008). The main cause of the discrepancy between the younger ages derived from the X-ray data and the much larger values obtained by others (e.g., Chevalier 1999; Lee et al.~2008) is the difference in the values adopted to represent the primary shock velocity $v_s$. Both Chevalier (1999) and Lee et al.~(2008) (among others) assume that $v_s\approx100\,{\rm km~s}^{-1}$ as determined from shock models applied to observations of the bright optical filaments (Fesen \& Kirshner 1980). However, if the X-ray temperature of $\sim$$1.2\times10^7\,$K (Petre et al.~1988) is indicative of the temperature of the gas in the immediate post-shock region $T_s$, then the inferred shock velocity would be $v_s=(16kT_s/3\mu)^{1/2}\approx940\,{\rm km~s}^{-1}$ (where $\mu=0.604m_{\rm H}$ is the mean mass per particle assuming the pre-shock gas is fully ionized). If the hot X-ray component represents relic radiation from a time when the remnant was younger and the velocity of the shock was higher, as has been suggested for other mixed-morphology SNRs (e.g., Rho \& Borkowski 2002), the temperature of the cold X-ray component, which is $\sim$$2.2\times10^6\,$K according to Petre et al.~(1988), might still imply that $v_s\approx400\,{\rm km~s}^{-1}$.

Direct kinematic evidence for velocities as high as those implied by the X-ray observations has so far been lacking for IC~443. Thus, it is not surprising that many previous investigators adopted the velocity of the optical filaments as the primary shock velocity and sought other means to explain the high X-ray temperatures. However, with our observations of N~{\sc v}, Si~{\sc iv}, Si~{\sc iii}, and C~{\sc ii} absorption spanning a range in velocity from $-$400 to $-$800~km~s$^{-1}$ toward HD~43582 (Section~\ref{subsec:high_ions}), we may finally have direct evidence of gas motions in IC~443 that are comparable to the velocity of the blast wave (as inferred from the X-ray data). The very high velocity, blueshifted gas toward HD~43582 appears as an isolated, very broad shell-like component in the N~{\sc v}, Si~{\sc iv}, and Si~{\sc iii} lines (see Figure~\ref{fig:high_ions}). The absorption breaks up into two narrower and somewhat lower velocity components in C~{\sc ii} (Figure~\ref{fig:vhv_c2}). While much of the thermal X-ray emission from IC~443 has been described as centrally-peaked (which is one of the defining characteristics of mixed-morphology SNRs; Rho \& Petre 1998), Troja et al.~(2006) detected a partial shell-like structure in soft X-ray emission, which is very well correlated with the bright radio and optical filaments in the northeastern part of the remnant. Interestingly, one of the contours that defines this partial shell structure directly encircles the star HD~43582 (see Figure~2 of Troja et al.~2006). Thus, we may have discovered the absorption counterpart to the soft X-ray shell described by Troja et al.~(2006).

The soft X-ray shell was interpreted by Troja et al.~(2006) as tracing the interaction between the SN shock and a moderately dense H~{\sc i} cloud to the northeast. Troja et al.~(2006) found that the temperature of the soft X-ray component varies from $\sim$$3.4\times10^6\,$K to $\sim$$7.7\times10^6\,$K depending on which region is being sampled. The variation is such that regions that are farther from the presumed explosion center have lower temperatures and higher electron densities. (The densities vary from $\sim$1~cm$^{-3}$ to $\sim$2.5~cm$^{-3}$.) The implication is that as the blast wave encountered the interstellar cloud it experienced a density gradient, which caused the shock wave to decelerate as the density increased. Note that the X-ray temperatures described above correspond to shock velocities in the range $v_s\approx500$--$750\,{\rm km~s}^{-1}$, which is very nearly the range we see for the high negative velocity gas toward HD~43582. In this interpretation, the interaction of the blast wave with the outer diffuse parts of the H~{\sc i} cloud gives rise to the soft X-ray emission. At some distance behind the shock front, the shock-heated plasma, accelerated by the blast wave, will have cooled to the temperature that characterizes the material containing the N~{\sc v}, Si~{\sc iv}, Si~{\sc iii}, and C~{\sc ii} ions (i.e., $\sim$$8.2$--$8.7\times10^4\,$K; Table~\ref{tab:high_vel_results}), allowing us to see these species in absorption. As the shock front encounters the denser parts of the cloud, the impact drives secondary shocks through the denser cloud material. These secondary shocks move at even lower velocities due to the further increase in density, and are ultimately responsible for the bright optical filaments that are prominent in the northeast. This shocked cloud material also gives rise to the absorption we see from the many low-ionization species observed at moderately high velocity toward HD~43582. In the next section, we explore the physical conditions in both the shocked and quiescent components observed toward HD~43582 and HD~254755 in more detail.

\subsection{Physical Conditions in the Gas toward HD~43582 and HD~254755\label{subsec:phys_cond}}
In Section~\ref{subsec:oxy_sil_ex}, we obtained a detailed set of physical conditions for the gas toward HD~43582 and HD~254755 from an analysis of O~{\sc i} and Si~{\sc ii} excitations as a function of velocity. We supplemented this information with an examination of C~{\sc i} excitations along the lines of sight (Section~\ref{subsec:car_ex}), and also studied changes in the gas-phase depletions of Ni and Ca toward HD~43582 (Section~\ref{subsec:ni_ca_depl}). An analysis of CO and C$_2$ rotational excitations and diffuse cloud chemistry (Section~\ref{subsec:mol_ex}) provided estimates for the physical conditions in the molecular material toward our targets stars, while an examination of C~{\sc ii} excitations (Section~\ref{subsec:car2_ex}) yielded physical conditions for the ionized gas components seen at high positive and negative velocity toward HD~43582. Here, we discuss these results and their implications in the context of the interaction between IC~443 and the interstellar gas in its vicinity.

\subsubsection{Low Velocity Quiescent Gas\label{subsubsec:quiescent}}
The main low velocity absorption components toward HD~43582 and HD~254755 near +6 or +7~km~s$^{-1}$ have densities and temperatures characteristic of diffuse atomic and molecular clouds (e.g., Snow \& McCall 2006). Our analysis of the O~{\sc i} fine-structure lines indicates that $n({\rm H}_{\rm tot})\approx70$--$700\,{\rm cm}^{-3}$ and $T\approx100$--$200\,$K for these components (Table~\ref{tab:comp_results}). The relative CO rotational populations indicate higher densities for the associated molecular material (Table~\ref{tab:mol_results}), suggesting a stratified cloud structure. The thermal pressures of $\log\,(p/k)\approx4.2$--$4.7$ (as indicated by the O~{\sc i} analysis) are higher than the typical pressure that characterizes sight lines probing the local Galactic ISM (i.e., $\log\,(p/k)\approx3.6$; Jenkins \& Tripp 2011) but are consistent with the pressures that define the upper end of the thermal equilibrium curve for the cold neutral medium (CNM; e.g., Wolfire et al.~2003). The relatively large gas-phase depletions of Ni and Ca in these components are typical of the depletions seen in cold diffuse molecular clouds (e.g., Crinklaw et al.~1994; Cartledge et al.~2006; Jenkins 2009). The sizes of the clouds are in the range $N({\rm H}_{\rm tot})/n({\rm H}_{\rm tot})\approx4$--$15\,$pc, (based on the values of $N({\rm H}_{\rm tot})$ listed in Table~\ref{tab:comp_results}).

The other low velocity components seen toward HD~43582 and HD~254755 near +15 and +25~km~s$^{-1}$ are characterized by significantly lower pressures and somewhat less severe Ni and Ca depletions compared to the main low velocity components along these sight lines (see Figures~\ref{fig:car_f1f2} and \ref{fig:nic_cal_depl}). These other low velocity components may therefore be probing diffuse line-of-sight material unassociated with IC~443.\footnote{With pathlengths of 2.0 and 3.6~kpc for the sight lines to HD~43582 and HD~254755, respectively, and at a Galactic longitude of $189.1\degr$, differential Galactic rotation would be expected to produce gas components with heliocentric velocities between +12 and +22~km~s$^{-1}$.} As for the main low velocity components, the available evidence suggests that this material may be positioned close to IC~443, perhaps in the immediate foreground of the remnant. Lee et al.~(2012) mapped the locations of a number of small molecular clouds, bright in $^{12}$CO emission, that they identified as the ambient clouds with which IC~443 is currently interacting. These small clouds, with velocities in the range $v_{\sun}\approx+4$ to $+9\,{\rm km~s}^{-1}$, may or may not be associated with the more extensive distribution of molecular gas that covers much of the foreground of IC~443, extending from the northwest to the southeast (Lee et al.~2012; Su et al.~2014). Lee et al.~(2012) interpreted the small bright clouds as the dense cores of parental molecular clouds that survived destruction during the pre-supernova evolution of the progenitor star of IC~443. These cores are now being impacted by the SNR and give rise to the shocked molecular clumps located throughout the southern portion of the remnant. It is likely that the main low velocity absorption components toward HD~43582 and HD~254755 probe the lower density diffuse envelopes of these small molecular clouds since the velocities of the absorption components are nearly identical to those of the clouds and the sight lines to our targets pass close to several of these structures. The sight line to HD~43582 in particular passes very close to two of the small clouds (labelled SC 10 and SC 11 by Lee et al.~2012), and these two clouds are themselves coincident with a shocked molecular clump (labelled H by Huang et al.~1986).

Taylor et al.~(2012) found that the $^7$Li/$^6$Li ratio in the main low velocity component toward HD~43582 is a factor of $\sim$2 lower than in the corresponding component toward HD~254755. The low $^7$Li/$^6$Li ratio, which approaches the value of $\sim$1.5 expected from cosmic ray spallation and fusion reactions (Ramaty et al.~1997; Lemoine et al.~1998), was interpreted as evidence of recent Li production by cosmic rays accelerated by the SNR shock. Since the main low velocity components toward HD~43582 and HD~254755 have thermal pressures typical of the CNM (indicating that this is unshocked, quiescent gas), the low $^7$Li/$^6$Li ratio toward HD~43582 suggests a scenario wherein shock-accelerated cosmic rays are streaming out ahead of the shock and impacting unshocked gas in the vicinity of the remnant (e.g., Gabici et al.~2009; Caprioli et al.~2009). Molecular clouds illuminated by cosmic rays will be bright in $\gamma$-ray emission produced by neutral pion decays. However, while IC~443 is among the brightest SNRs at GeV energies (e.g., Tavani et al.~2010; Abdo et al.~2010b; Acero et al.~2016), the high GeV luminosity may instead arise from shocked clouds overtaken by the SNR blast wave (e.g., Uchiyama et al.~2010). Unfortunately, we are unable to determine the Li isotope ratios in the many shocked components toward HD~43582 because the associated column densities are too low to produce detectable absorption from the very weak Li~{\sc i}~$\lambda6707$ feature (see Taylor et al.~2012).

Boron is another product of cosmic-ray spallation. Thus, since our \emph{HST} observations cover the B~{\sc ii} resonance line at 1362.5~\AA{}, we tried to determine whether the low $^7$Li/$^6$Li ratio in the main low velocity component toward HD~43582 is accompanied by an enhanced B abundance. Since B is mainly synthesized by the spallation of interstellar C and O nuclei under the bombardment of cosmic-ray protons (e.g., Ramaty et al.~1997), the B/O ratio should be elevated in a region of cosmic-ray overdensity. For the main low velocity components toward HD~43582 and HD~254755, we find respective B/O ratios of $\log\,({\rm B}/{\rm O})=-6.48\pm0.10$ and $-6.62\pm0.14$. Thus, while the nominal ratios indicate an enhancement of 0.14~dex toward HD~43582, the two ratios are consistent with one another at the 1$\sigma$ level. Moreover, a slight increase in the B/O ratio toward HD~43582 could easily be explained by a small reduction in the amount of dust grain depletion for the gas in that direction (Ritchey et al.~2011). Indeed, the measured B~{\sc ii} column densities for the sight lines to HD~43582 and HD~254755 are fully consistent with the relative gas-phase abundances of the other dominant ions used to deduce the depletion factors for these directions (see Figure~\ref{fig:depl_factors}). It may be that measurements of the column densities of dominant ions such as B~{\sc ii} for the main low velocity components are contaminated by unrelated foreground material making it difficult to discern the signature of cosmic-ray spallation. The trace neutral species Li~{\sc i} is a better probe of the molecular material that may be interacting more directly with IC~443.

\subsubsection{High Velocity Shocked Material toward HD~43582\label{subsubsec:shocked}}
All of the moderately high velocity absorption components toward HD~43582 (i.e., those with velocities outside of the range $0\lesssim v_{\sun}\lesssim+30\,{\rm km~s}^{-1}$ that are detected in the O~{\sc i}*, O~{\sc i}**, and Si~{\sc ii}* profiles) exhibit thermal pressures that are significantly elevated compared to the main low velocity component at +7~km~s$^{-1}$. Our analysis of the O~{\sc i} and Si~{\sc ii} excitations exhibited by these moderately high velocity components yields thermal pressures in the range $\log\,(p/k)\approx6.5$--$6.9$ (Table~\ref{tab:comp_results}). Such high values for the thermal pressures are corroborated by our measurements of the relative C~{\sc i} fine-structure populations, which approach the regime where the C~{\sc i} levels are populated according to the level degeneracies (Figure~\ref{fig:car_f1f2} and Table~\ref{tab:car_components}). These high pressure, moderately high velocity components also exhibit relatively low levels of dust grain depletion, indicative of dust destruction by grain-grain collisions or shock sputtering (Figure~\ref{fig:nic_cal_depl}). The characteristic sizes of these cloud components range from $N({\rm H}_{\rm tot})/n({\rm H}_{\rm tot})\approx0.0002\,$pc to $0.04\,$pc ($40\,$AU to $8000\,$AU).

While the thermal pressures exhibited by the moderately high velocity absorption components toward HD~43582 are relatively consistent with one another, the combinations of density and temperature that yield those pressures vary significantly from one component to the next. As an example, we find $n({\rm H}_{\rm tot})\approx14$,$000\,{\rm cm}^{-3}$ and $T\approx210\,$K for the component at $-$37~km~s$^{-1}$, while $n({\rm H}_{\rm tot})\approx480\,{\rm cm}^{-3}$ and $T\approx8300\,$K for the adjacent component at $-$24~km~s$^{-1}$ (Table~\ref{tab:comp_results}). Fesen \& Kirshner (1980) found a similar range of thermal pressures from observations of the bright optical filaments distributed throughout the northeastern and southwestern regions of IC~443. From various emission line ratios, Fesen \& Kirshner (1980) obtained densities in the range $n(e)\approx100$--$480\,{\rm cm}^{-3}$ and temperatures ranging from $T_e\approx8000\,$K to $15$,$000\,$K. The associated thermal pressures of the optical filaments are in the range $\log\,(p/k)\approx6.4$--$7.0$. The relative consistency in the thermal pressures exhibited by the moderately high velocity clouds toward HD~43582 (Table~\ref{tab:comp_results} and Figure~\ref{fig:phys_cond}) suggests that this material has been cooling approximately isobarically after being heated by a shock. The differences in density and temperature between different velocity components could therefore be indicative of differences in how far along the cooling and recombination has proceeded. Since the gas traced by optical emission exhibits a similar range of thermal pressures but a higher degree of ionization, this material represents the more recently shocked gas that has yet to recombine.

The destruction of dust grains via shock sputtering is expected to occur in an extended zone downstream from the shock such that the depletions in the immediate post-shock gas should be the same as in the pre-shock medium (Draine \& McKee 1993). Since all of the high pressure components seen at moderately high velocity toward HD~43582 exhibit low levels of dust depletion (Figure~\ref{fig:nic_cal_depl}), we take this as further evidence that the absorption arises from gas in a cooling region far downstream from the shock. With this assumption, we can use the observed densities, temperatures, and velocities of the shocked components to derive more fundamental quantities pertaining to the interaction between the SN shock and the surrounding ISM. When an interstellar cloud is overtaken by the blast wave of a SNR, a shock is driven into the cloud at a velocity $v_c\approx v_b(\rho_0/\rho_c)^{1/2}$ (e.g., McKee \& Cowie 1975), where $v_b$ is the velocity of the blast wave and $\rho_0$ and $\rho_c$ are the pre-shock densities in the intercloud medium and the interstellar cloud, respectively. For a purely atomic cloud (i.e., $\gamma=5/3$), the velocity of the gas immediately behind the cloud shock will be $\sim$$\frac{3}{4}v_c$, while the velocity of the material further downstream in the cooling region will be nearly equal to $v_c$. Following Fesen \& Kirshner (1980), the velocity of the gas in the cooling region will approach $v_c$ according to the relation
\begin{equation}
v_{\rm obs}=v_c\Bigg(1-\frac{1}{4}\frac{T_{\rm obs}}{T_c}\Bigg)
\end{equation}
where $T_{\rm obs}$ is the observed temperature of the cooling region and $T_c$ is the temperature of the gas in the region immediately behind the shock propagating through the cloud.

If we combine Equation (8) with the relation $v_c=(16kT_c/3\mu)^{1/2}$, we can obtain values for both $v_c$ and $T_c$ from the observed velocities and temperatures of the high pressure components seen toward HD~43582. We will start by considering the quantities listed for the shocked components in Table~\ref{tab:comp_results} (excluding the components at $-$43 and +52~km~s$^{-1}$ since the temperatures, and corresponding densities, of these components are regarded as uncertain; see Section~\ref{subsubsec:oxy_sil_comparison}). The observed velocities of the shocked clouds from Table~\ref{tab:comp_results}, relative to the systemic velocity of the ambient gas, which we take to be $v_{\rm sys}=+7\,{\rm km~s}^{-1}$, range from $v_{\rm obs}=|v-v_{\rm sys}|=16\,{\rm km~s}^{-1}$ to $57\,{\rm km~s}^{-1}$. (These are simply the components of the cloud velocities projected onto the line of sight. If the shocks have significant components in the transverse direction, then the actual cloud velocities would of course be higher.) The observed temperatures of the shocked clouds range from $\sim$210~K to $\sim$8300~K. Applying the equations noted above, we find that the shock velocities are essentially the same as the observed velocities (i.e., $v_{\rm obs}\approx v_c$), and the post-shock temperatures range from $\sim$7700~K for the 16~km~s$^{-1}$ shock to $\sim$94,000~K for the 57~km~s$^{-1}$ shock. (For these calculations, we have assumed that the pre-shock cloud medium is completely neutral, i.e., that $\mu=1.260m_{\rm H}$.)

If the shock-heated gas has been cooling approximately isobarically, we can use the observed densities and temperatures of the shocked clouds along with the post-shock temperatures derived above to estimate the post-shock densities, which we find to be in the range 13--210~cm$^{-3}$. Finally, if we assume a shock compression factor of $\sim$4, we can estimate the pre-shock densities of the clouds overtaken by the blast wave. The pre-shock cloud densities obtained in this way are in the range 3--53~cm$^{-3}$, where higher densities are associated with the slower shocks. While this type of analysis is only approximate (e.g., the shock velocities could be higher if there are significant transverse components to the propagation directions; conversely, the post-shock temperatures could be lower depending on the degree of pre-ionization in the pre-shock medium), the results are nevertheless consistent with estimates made by others. Fesen \& Kirshner (1980), for example, obtained pre-shock densities in the range 10--20~cm$^{-3}$ based on a similar analysis applied to their observations of the bright optical filaments.

If a dynamical pressure equilibrium exists between the hot intercloud medium and the shocked clouds, as in the model of McKee \& Cowie (1975) for example, then the pre-shock cloud densities may be used to estimate the pre-shock density in the intercloud medium through the relation $n_0\approx n_c(v_c/v_b)^2$. Using the various combinations of pre-shock cloud density and shock velocity described above, and assuming that the blast wave velocity is in the range $v_b\approx500$--$750\,{\rm km~s}^{-1}$ (as implied by the temperatures found for the soft X-ray shell component; Troja et al.~2006), we find that $n_0\approx0.02$--$0.06\,{\rm cm}^{-3}$. These estimate for $n_0$ are somewhat lower than the value of $\sim$0.25~cm$^{-3}$ obtained by Troja et al.~(2008). Moreover, there are indications that the moderately high velocity clouds observed toward HD~43582 are not in pressure equilibrium with the other shocked components seen along the line of sight. Recall that the pressure we obtained for the very high negative velocity absorption component toward HD~43582 (i.e., $\log\,(p/k)\approx7.4$; Table~\ref{tab:high_vel_results}) is elevated compared to the range of pressures exhibited by the components at more moderate velocities. If we repeat the analysis described above for the component at $-$546~km~s$^{-1}$ (using the quantities listed for this component in Table~\ref{tab:high_vel_results}), we find a post-shock temperature of $4.2\times10^6\,$K for a shock velocity of 560~km~s$^{-1}$, a post-shock density of 2.9~cm$^{-3}$, and a pre-shock density of 0.71~cm$^{-3}$. These results are in very good agreement with the values obtained from the X-ray analysis (Troja et al.~2006, 2008).

The absorption components at high positive velocity toward HD~43582 exhibit thermal pressures that are somewhat lower than those of the moderately high velocity clouds. Our analysis of the C~{\sc ii} excitations exhibited by the components near +96, +122, and +225~km~s$^{-1}$ yields thermal pressures in the range $\log\,(p/k)\approx5.3$--$6.0$ (Table~\ref{tab:high_vel_results}). While the observed velocities of these clouds range from $v_{\rm obs}=|v-v_{\rm sys}|=89\,{\rm km~s}^{-1}$ to $218\,{\rm km~s}^{-1}$, there is evidence that the shocks producing these motions have significant transverse components. The total hydrogen column densities associated with the high positive velocity clouds are in the range $\log N({\rm H}_{\rm tot})\approx18.2$--18.5 (Table~\ref{tab:high_vel_results}). However, according to the models of Sutherland \& Dopita (2017), the cooling column reaches similar values only for shocks with velocities of $v_s\approx240$--$290\,{\rm km~s}^{-1}$, implying line-of-sight projection factors in the range $\cos\theta=v_{\rm obs}/v_s\approx0.4$--0.9. Adopting these shock velocities, we find post-shock temperatures of $0.8$--$1.2\times10^6\,$K, post-shock densities ranging from 0.12 to 0.37~cm$^{-3}$, and pre-shock densities in the range 0.03--0.09~cm$^{-3}$.

In general, the results discussed in this section tend to confirm a scenario for IC~443 wherein the SN shock is interacting with a highly inhomogenous ISM. In particular, the shocked clouds on the far side of the remnant seem to have had much lower pre-shock densities compared to the complex of clouds located on the near side. This is in agreement with the general appearance of the interstellar absorption profiles toward HD~43582, where the absorption from intrinsically strong lines such as O~{\sc i}~$\lambda1302$ and C~{\sc ii}~$\lambda1334$ is completely saturated out to approximately $-$100~km~s$^{-1}$, yet is relatively weak at high positive velocities (Figure~\ref{fig:high_vel}). Lee et al.~(2008) found shocked H~{\sc i} emission at velocities ranging from $-$100 to +50~km~s$^{-1}$ associated with both the southern molecular ridge in IC~443 and the northeastern filaments (see also Braun \& Strom 1986a). This range in velocity is very similar to the range exhibited by the low-ionization species observed toward HD~43582. While the H~{\sc i} emission near the shocked molecular ridge likely results from the dissociation of H$_2$ molecules following the passage of a shock into molecular gas (e.g., Burton et al.~1988), Lee et al.~(2008) attribute the H~{\sc i} emission coincident with the northeastern filaments to recombined H~{\sc i} behind a shock driven into atomic gas. Meaburn et al.~(1990) obtained H$\alpha$ profiles for a number of slit positions covering the main filamentary structures in IC~443. They found that the ionized gas exhibits velocities in the range $-$200 to +250~km~s$^{-1}$. Even higher velocities were reported by Lozinskaya (1979), who detected amorphous H$\alpha$ emission with velocities as high as $-$350~km~s$^{-1}$. Taking geometrical projection effects into account, Lozinskaya (1979) suggested that the outward expansion velocity of this fast moving ionized material is in the range 400--500~km~s$^{-1}$. Absorption from highly-ionized species toward HD~43582 confirms the presence of ionized gas at both high positive and high negative velocities.

The density inhomogeneities and the wide range of ionized and neutral gas motions detected throughout the IC~443 region are illustrative of the challenges faced in attempting to model the evolution of the SNR. In particular, the radiative shell model of Chevalier (1999), wherein the remnant is assumed to be in the radiative phase and the velocity of the radiative shell is $\sim$100~km~s$^{-1}$, does not account for the much higher velocities exhibited by the ionized gas components seen throughout the remnant (Lozinskaya 1979; Meaburn et al.~1990; this work). The differences in the thermal pressures found for the various shocked components detected toward HD~43582 could result from energy losses if only certain portions of the SNR have entered the radiative phase. Thermal evaporation from the clouds embedded in the hot post-shock material could also lead to differences in pressure among the shocked components (e.g., McKee \& Ostriker 1977). The moderately high velocity clouds may have been shocked more recently since the energy losses are less severe for these components compared to the components at high positive velocity. This is consistent with the scenario put forth by Troja et al.~(2006) wherein the remnant has evolved primarily in a low density environment and has only recently begun to interact with a foreground H~{\sc i} cloud to the northeast.

\subsubsection{Moderately High Velocity Gas toward HD~254755\label{subsubsec:precursor}}
One puzzling aspect of our observations is the presence of a significant amount of gas at moderately high negative velocity toward HD~254755. As discussed in Section~\ref{subsubsec:high_vel2}, a prominent absorption complex is seen near $-$63~km~s$^{-1}$ in the C~{\sc ii}, C~{\sc ii}*, Si~{\sc ii}, Si~{\sc ii}*, and Ca~{\sc ii} profiles toward HD~254755 (Figures~\ref{fig:trace2}, \ref{fig:si2}, and \ref{fig:high_vel2}). While we do not have specific estimates of the physical conditions in this material, we do know that the Si~{\sc ii}*/Si~{\sc ii} ratio is similar to the ratios measured for the shocked components at moderately high velocity toward HD~43582. The sight line to HD~254755 passes $\sim$$2.5\arcmin$ beyond the outer edge of IC~443, as indicated by the extent of the bright optical filaments in the northeast. (An angular distance of $2.5\arcmin$ corresponds to an on-sky distance of 1.1~pc at 1.5~kpc or 1.5~pc at 2.1~kpc.) If the northeast filaments mark the present location of the SN shock front, then the moderately high velocity material we observe ahead of the shock could potentially be associated with a shock precursor. Precursors can be induced by efficient cosmic-ray acceleration (e.g., Malkov \& Drury 2001) or by a ``neutral return flux'' when the shock is propagating into a partially-neutral medium and hydrogen atoms are coupled to protons via charge exchange (see, e.g., Blasi et al.~2012). In either case, the precursor would be expected to produce bulk motions ahead of the shock at velocities of $\sim$5--10\% of the shock velocity (e.g., Blasi et al.~2012; Caprioli \& Spitkovsky 2014). If the blast wave velocity of the shock associated with IC~443 is $\sim$620~km~s$^{-1}$ (as indicated by the very high velocity, highly-ionized component toward HD~43582), then the observed velocity of the $-$63~km~s$^{-1}$ component toward HD~254755 would seem to match this expectation for a shock precursor.\footnote{The extent of the precursor is difficult to predict since it depends on the amount of magnetic field amplification in the upstream direction. For young SNRs, such as SN~1006, the extent of the precursor can be a fraction of a parsec (e.g., Morlino et al.~2010), whereas for older remnants, such as IC~443, the precursor may extend to larger distances (i.e., out to a few parsecs) since $B$-field amplification is no longer effective.}

While the idea of a shock precursor toward HD~254755 is attractive, there may be another explanation for the moderately high velocity material seen in this direction. From an analysis of X-ray data obtained with \emph{ROSAT}, Asaoka \& Aschenbach (1994) proposed the existence of a separate SNR (G189.6+3.3) that partially overlaps with IC~443 (but lies in the foreground) and may be responsible for the faint optical filaments that seem to extend outward from the northeast shell of IC~443 (see also Fesen 1984). While the sight line to HD~254755 lies outside of the northeast boundary of IC~443, it is positioned well inside of the proposed boundary of G189.6+3.3 (see, e.g., Figure~1 of Fesen 1984). Thus, the moderately high velocity gas we observe toward HD~254755 may not be associated with IC~443 but with a separate foreground SNR. Fesen (1984) found that a shock velocity of $\sim$70~km~s$^{-1}$ would be consistent with the relative strengths of the emission lines associated with the faint optical filament that purportedly forms the northern boundary of G189.6+3.3. This velocity is very similar to the observed radial velocity of the component toward HD~254755. (Note that the observed velocity of this component relative to the systemic velocity of the ambient gas would be $|v-v_{\rm sys}|=69\,{\rm km~s}^{-1}$.)

\begin{figure}
\centering
\includegraphics[width=0.8\textwidth]{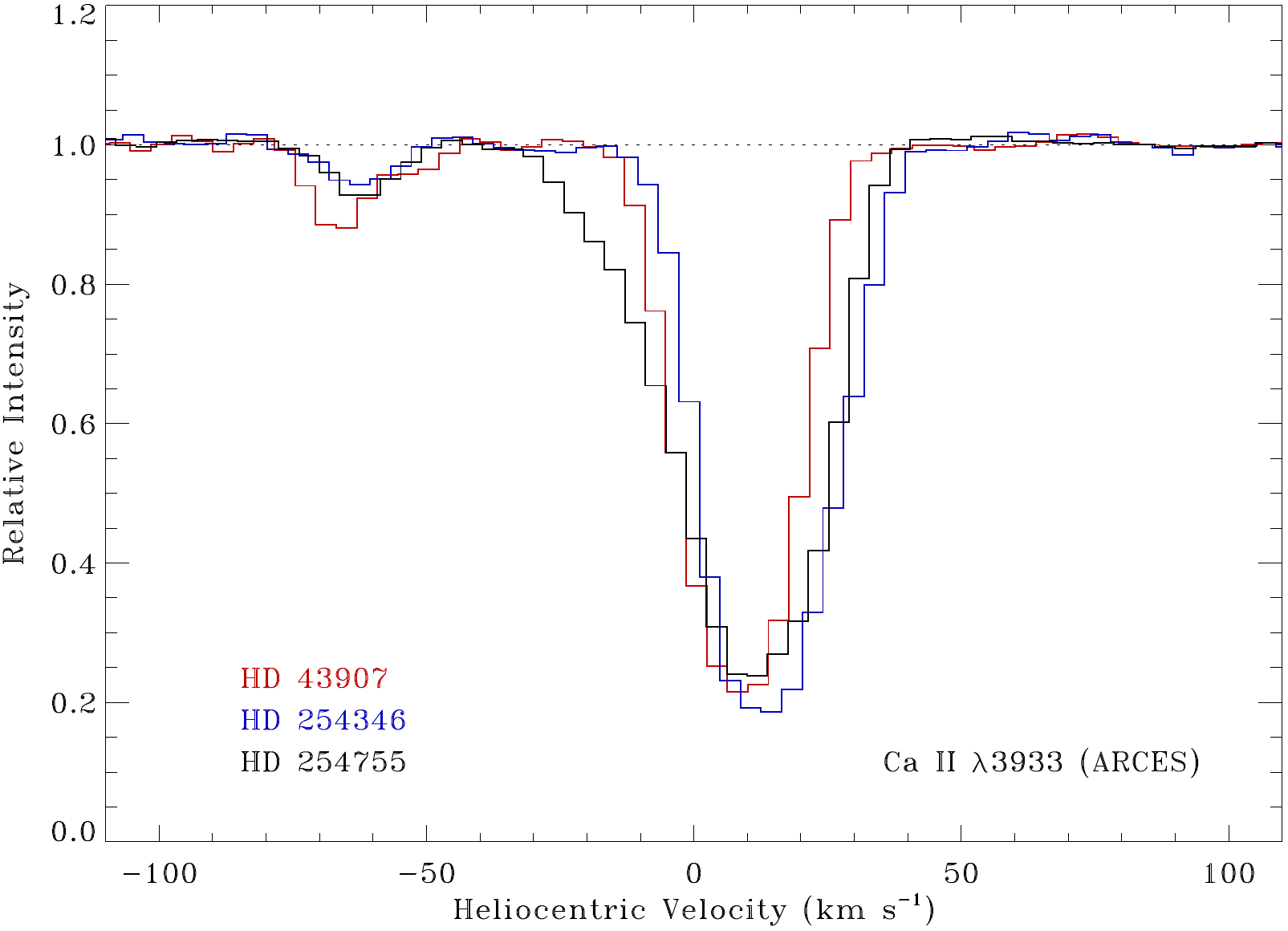}
\caption{Comparison of the Ca~{\sc ii}~$\lambda3933$ profiles toward HD~254755 (black), HD~254346 (blue), and HD~43907 (red) from ARCES data obtained by Hirschauer et al.~(2009). Note the similarity in strength and velocity of the moderately high velocity absorption complexes near $-$63~km~s$^{-1}$ along these three lines of sight.\label{fig:ca2_apo}}
\end{figure}

There is other evidence of an association between the moderately high velocity gas toward HD~254755 and the SNR G189.6+3.3. Hirschauer et al.~(2009) found similar components (at respective heliocentric velocities of $-$63 and $-$67~km~s$^{-1}$) in the Ca~{\sc ii} profiles they obtained toward HD~254346 and HD~43907 (see Figure~\ref{fig:ca2_apo}). While HD~254346 is positioned near the southern boundary of IC~443, it also falls within the proposed border of G189.6+3.3. In contrast, HD~43907 is positioned well outside and to the southeast of IC~443 but is located near the center of the X-ray shell suggested to be associated with G189.6+3.3 (e.g., compare Figure~1 of Asaoka \& Aschenbach 1994 to Figure~1 of Hirschauer et al.~2009). The similarity in both strength and velocity of the moderately high velocity Ca~{\sc ii} components observed toward HD~254755, HD~254346, and HD~43907, and the positions of the sight lines to these stars near the edge or center of G189.6+3.3 suggests that all three stars may be probing material associated with this foreground SNR rather than with IC~443. Moreover, the general agreement between the radial velocities measured in these three directions and the shock velocity obtained by Fesen (1984) from observations of the faint optical filament to the north provides additional support for the interpretation of G189.6+3.3 as a distinct SNR. The positional coincidence between IC~443 and G189.6+3.3 might then suggest that the progenitors of both SNRs belonged to the same association of massive stars (see also Lee et al.~2008).

\section{SUMMARY AND CONCLUSIONS\label{sec:conclusions}}
We obtained high-resolution \emph{HST}/STIS spectra of two stars probing interstellar material in the vicinity of the SNR IC~443. We supplemented the \emph{HST} observations with very high S/N ratio ground-based spectra originally obtained by Taylor et al.~(2012) using the HET/HRS. The \emph{HST} spectra cover numerous interstellar absorption features of neutral and singly-ionized atomic species, molecular species, and highly-ionized species, allowing us to perform a detailed examination of the physical conditions in the various absorption components detected in the two directions. The HET data cover additional atomic and molecular species yielding a more complete picture of the state of the gas seen in absorption along the lines of sight.

The absorption profiles of neutral and singly-ionized atomic species toward HD~43582, a sight line that probes post-shock gas interior to the SNR, comprise numerous distinct absorption components over a wide range in velocity, with much of the absorption occurring at heliocentric velocities between $-$100 and +50~km~s$^{-1}$. Additional high velocity components are found in the absorption profiles of intrinsically strong lines near +96, +122, and +225~km~s$^{-1}$. The sight line to HD~43582 also exhibits absorption from highly-ionized species over nearly the entire velocity range from $-$100~km~s$^{-1}$ to +240~km~s$^{-1}$. In addition, a strong very high velocity component is found in the N~{\sc v}, Si~{\sc iv}, and Si~{\sc iii} profiles near $-$620~km~s$^{-1}$. Corresponding features are seen in C~{\sc ii} and C~{\sc ii}* absorption at $-$534 and $-$425~km~s$^{-1}$. The absorption profiles of low-ionization species toward HD~254755 are mainly confined to heliocentric velocities between 0 and +30~km~s$^{-1}$, although a complex of moderately high velocity gas is seen in some singly-ionized species near $-$63~km~s$^{-1}$.

We studied the physical conditions in the predominantly neutral gas components toward HD~43582 and HD~254755 primarily through an analysis of O~{\sc i} and Si~{\sc ii} fine-structure excitations. Additional information was gleaned from analyses of the fine-structure excitations of C~{\sc i} and the rotational excitations of CO and C$_2$. For the ionized gas components seen at high positive and negative velocity toward HD~43582, we obtained physical conditions from an analysis of C~{\sc ii} fine-structure excitations. The densities and temperatures of the main low velocity absorption components toward HD~43582 and HD~254755 are typical of diffuse atomic and molecular clouds. This low velocity, quiescent gas likely samples the diffuse envelopes of the small molecular clouds identified by Lee et al.~(2012) as the ambient clouds currently interacting with IC~443. The many shocked cloud components observed at moderately high velocity toward HD~43582 exhibit a combination of greatly enhanced thermal pressures and significantly reduced dust-grain depletions. We interpret this material as cooling gas in a recombination zone far downstream from shocks driven into neutral gas clumps characterized by pre-shock densities in the range $\sim$3--53~cm$^{-3}$.

The very high velocity, highly-ionized absorption complex toward HD~43582 (near $-$620~km~s$^{-1}$) may represent the absorption counterpart to the soft X-ray shell discovered by Troja et al.~(2006). The velocity of this material is consistent with the range of shock velocities implied by the temperatures determined from the X-ray emission. The inferred post-shock density is also in good agreement with the results of the X-ray analysis. The shocked clouds at high positive velocity toward HD~43582 (which presumably are located on the far side of the remnant) seem to have been characterized by lower pre-shock densities than the clouds at more moderate velocities (which are predominantly located on the near side). Differences in pressure among the various shocked components toward HD~43582 may be indicative of radiative losses enhanced by evaporation of the clouds embedded in the hot post-shock medium. The moderately high velocity gas seen in the direction of HD~254755 may probe material in a shock precursor. However, a more likely explanation is that this sight line, along with two others observed by Hirschauer et al.~(2009), samples shocked gas from a separate SNR that lies in the foreground along the same line of sight as IC~443.

\acknowledgments
We acknowledge useful conversations about this work with Roger Chevalier, John Raymond, H.-T.~Janka, and other attendees of the Supernova Remnants II conference held in Chania, Crete, Greece in 2019 June. Corbin Taylor performed the original data reduction for the HET/HRS observations presented here. Assistance in reducing the raw STIS spectra was provided by the STIS help desk. We thank Paul Goldsmith for supplying us with additional data needed for the CO excitation analysis. We also acknowledge useful comments made by the anonymous referee. Support for this investigation was provided to the University of Washington, the University of Toledo, and Princeton University through grants HST-GO-13709.001-A, HST-GO-13709.002-A, and HST-GO-13709.003-A, respectively, from the Space Telescope Science Institute, which is operated by the Association of Universities for Research in Astronomy, Inc., under NASA contract NAS5-26555.

\vspace{5mm}
\facilities{\emph{HST} (STIS), HET (HRS).}

\appendix
\section{FINE-STRUCTURE CALCULATIONS FOR O~{\sc i}\label{sec:oi_fine_structure}}
The ground (1s$^2$)2s$^2$2p$^4$ $^3$P term of neutral oxygen is split into three fine-structure levels, $^3$P$_2$ (ground level), $^3$P$_1$ (excitation: 158.27 cm$^{-1}$; $E/k=227.74\,$K), and $^3$P$_0$ (excitation: 226.98 cm$^{-1}$; $E/k=326.61\,$K). These levels can undergo direct excitation and de-excitation by collisions with various particle constituents, and the effects of these collisions are balanced by spontaneous radiative decays. In addition, the levels can be depopulated when neutral oxygen atoms undergo the charge exchange reaction ${\rm O^0+H^+}\rightarrow {\rm O^++H^0}$, or they can be replenished by the reverse reaction ${\rm O^++H^0}\rightarrow {\rm O^0+H^+}$ and the recombination of ${\rm O^+}$ with free electrons. This process can be important at low temperatures because the excitation energy of the (1s$^2$)2s$^2$2p$^3~^4$S$_{3/2}$ ground level of O~{\sc ii} is very close to that of the $^3$P$_1$ level of O~{\sc i}. At high temperatures, electron collisions can excite the singlet $^1{\rm D}_2$ and $^1{\rm S}_0$ metastable levels of O~{\sc i} (with excitation energies of 15867.86 and $33792.58\,{\rm cm}^{-1}$, respectively) and have an indirect effect on the ground-state population ratios, although this effect is influential only at $T\gtrsim 10^4\,$K and $n(e) \gtrsim n({\rm H}^0)$. Excitation by the cosmic microwave background radiation is negligible, except at very low densities and high redshifts (Silva \& Viegas 2002). We describe in this section how we calculated the equilibrium fractions of O~{\sc i} in the three levels for various local conditions. 

We derive the equilibrium relative concentrations of the three O~{\sc i} fine-structure levels, the two metastable levels, and the amount of O~{\sc ii} by solving six simultaneous linear equations. The six unknown quantities reside in a population-fraction column vector {\bf x}$~=[f(^3{\rm P}_2), f(^3{\rm P}_1), f(^3{\rm P}_0), f(^1{\rm D}_2), f(^1{\rm S}_0), f({\rm O}^+)]$, which is multiplied by a $6\times6$ matrix {\bf A} consisting of radiative decay and collision rates from various partners to yield a column vector {\bf  b}~=~[0, 0, 0, 0, 0, 1], i.e.,
\begin{equation}
{\bf A \cdot  x=b}~.
\end{equation}
The first five elements of {\bf b} set the requirement that for each row of {\bf A} the products of the exchange rates and the unknown quantities in {\bf x} balance each other so that a given constituent is in equilibrium with the others. Setting the last element in {\bf b} and all elements in the last row of {\bf A} equal to 1 insures that the sum of all constituent fractions is equal to 1. Coefficients in the matrix {\bf A} are defined by various combinations of the following terms: (1) inelastic collisions for the fine-structure levels, where their rate constants $k_{j,j^\prime}({\rm X},T)$ as a function of temperature $T$ for the changes from $J=j$ to $J=j^\prime$ must be multiplied by the appropriate local particle densities $n({\rm X})$ of various collision partners X, such as H$^0$, ortho- and para-H$_2$, He$^0$, electrons, and protons, (2) spontaneous radiative decays from the more energetic $J$ levels to lower ones with Einstein $A$-coefficients $A_{j,j^\prime}$, (3) charge transfer reaction rates $c_{j,+}(T)n({\rm H}^+)$ that ionize ${\rm O^0}$ and lead to the creation of O$^+$, (4) the reverse charge transfer rates $c_{+,j^\prime}(T)n({\rm H}^0)$ together with recombinations with free electrons $\alpha_{e,j^\prime}(T)n(e)$ that replenish O$^0$ in the $J$-level $j^\prime$, and (5) the metastable level electron collision rate constants $k_{{\rm S},j^\prime}(e,T)$, $k_{{\rm D},j^\prime}(e,T)$,  and $k_{\rm S,D}(e,T)$ multiplied by $n(e)$, together with their radiative decay rates $A_{{\rm S},j^\prime}$, $A_{{\rm D},j^\prime}$, and $A_{\rm S,D}$. For all of the collision rates, we include their reverse counterparts, which are derived from the principle of detailed balance. For completeness we also include in our equations the optical pumping rates $\Gamma_{2,j^\prime}$ from the lowest level caused by an ambient UV radiation field, but in our application the self shielding of the $J=2$ level is so large that we can ignore these terms. The different recombination coefficients $\alpha_{e,j^\prime}(T)$ with free electrons are apportioned from the total rate $\alpha_e(T)$ according to the statistical weights of the respective $j^\prime$ levels. Terms along the diagonal of {\bf A} represent loss rates to other levels, while off-diagonal terms represent gains from those levels. Individual terms in {\bf A} are defined as follows:
\begin{mathletters}
\begin{eqnarray}
A(0,0)&=&-\bigg(  \sum_{\rm X}\Big\{ [k_{2,1}({\rm X},T)+k_{2,0}({\rm X},T)]n({\rm X})\Big\}+c_{2,+}(T)n({\rm H}^+)\nonumber \\
&+&[k_{\rm 2,S}+k_{\rm 2,D}]n(e)+\Gamma_{2,1}+\Gamma_{2,0}\bigg)\\
A(1,0)&=&\sum_{\rm X}\Big[ k_{1,2}({\rm X},T)n({\rm X})\Big] + A_{1,2}\\
A(2,0)&=&\sum_{\rm X}k_{0,2}({\rm X},T)n({\rm X})\\
A(3,0)&=&k_{\rm D,2}(e,T)n(e) + A_{\rm D,2}\\ 
A(4,0)&=&k_{\rm S,2}(e,T)n(e) + A_{\rm S,2}\\
A(5,0)&=&c_{+,2}(T)n({\rm H}^0)+(5/9)\alpha_e(T)n(e)\\
A(0,1)&=&\sum_{\rm X}\Big[ k_{2,1}({\rm X},T)n({\rm X})\Big]+\Gamma_{2,1}\\
A(1,1)&=&-\Bigg( \sum_{\rm X}\Big\{ [k_{1,2}({\rm X},T)+k_{1,0}({\rm X},T)]n({\rm X})\Big\} +A_{1,2}+c_{1,+}(T)n({\rm H}^+)\nonumber \\
&+&[k_{\rm 1,S}+k_{\rm 1,D}]n(e)\Bigg)\\
A(2,1)&=&\sum_{\rm X}\Big[k_{0,1}n({\rm X},T)\Big] + A_{0,1}\\
A(3,1)&=&k_{\rm D,1}(e,T)n(e) + A_{\rm D,1}\\ 
A(4,1)&=&k_{\rm S,1}(e,T)n(e) + A_{\rm S,1}\\
A(5,1)&=&c_{+,1}(T)n({\rm H}^0)+\alpha_e(T)n(e)/3\\
A(0,2)&=&\sum_{\rm X}\Big[ k_{2,0}({\rm X},T)n({\rm X})\Big] +\Gamma_{2,0}\\
A(1,2)&=&\sum_{\rm X}k_{1,0}({\rm X},T)n({\rm X})\\
A(2,2)&=&-\Bigg(\sum_{\rm X}\Big\{ [k_{0,2}({\rm X},T)+k_{0,1}({\rm X},T)]n({\rm X})\Big\} +A_{0,1}+c_{0,+}(T)n({\rm H}^+)\nonumber \\
&+&[k_{\rm 0,S}+k_{\rm 0,D}]n(e)\Bigg)\\ 
A(3,2)&=&k_{\rm D,0}(e,T)n(e) + A_{\rm D,0},~A(4,2)=k_{\rm S,0}(e,T)n(e)\\
A(5,2)&=&c_{+,0}(T)n({\rm H}^0)+\alpha_e(T)n(e)/9\\
A(0,3)&=&k_{\rm 2,D}(T)n(e),~A(1,3)=k_{\rm 1,D}(T)n(e),~A(2,3)=k_{\rm 0,D}(T)n(e)\\
A(3,3)&=&-\Bigg(\sum_{j^\prime}[k_{{\rm D},j^\prime}(T)n(e) +A_{{\rm D},j^\prime}]+k_{\rm D,S}(T)n(e)\Bigg)\\ 
A(4,3)&=&k_{\rm S,D}(T)n(e)+A_{\rm S,D}\\
A(5,3)&=&A(5,4)=0\\
A(0,4)&=&k_{\rm 2,S}(T)n(e),~A(1,4)=k_{\rm 1,S}(T)n(e)\\
A(2,4)&=&k_{\rm 0,S}(T)n(e),~A(3,4)=k_{\rm D,S}n(e)\\
A(4,4)&=&-\Bigg(\sum_{j^\prime}[k_{{\rm S},j^\prime}(T)n(e)] +A_{{\rm S},j^\prime}]+k_{\rm S,D}(T)n(e)+A_{\rm SD}\Bigg)\\
A(0,5)&=&A(1,5)=A(2,5)=A(3,5)=A(4,5)=A(5,5)=1
\end{eqnarray}
\end{mathletters}
We used the collision rate constants listed by Lique et al.~(2018) for the downward transitions $k_{1,2}({\rm H}^0,T)$, $k_{0,2}({\rm H}^0,T)$, and $k_{0,1}({\rm H}^0,T)$ and likewise for collisions with neutral helium and molecular hydrogen in the $J=0$ and 1 rotational levels. For the He$^0$ collision partners, we have adopted $n({\rm He}^0)=0.096[n({\rm H}^0) + 2n({\rm H}_2) + n({\rm H}^+)]$. Values for collisions with free electrons $k_{j,j^\prime}(e,T)$ have been computed by Bell et al.~(1998), and those for the metastable levels are from Zatsarinny \& Tayal (2003). Rates from both of these sources were fitted to analytic expressions as a function of temperature by Draine (2011, Appendix~F)\footnote{For the first through third printings, it is important to note corrections to these formulae given in the errata of Draine's textbook located at \url{http://www.astro.princeton.edu/\~draine/book/index.html}}. Péquignot (1990) computed rate constants for $k_{j,j^\prime}({\rm H}^+,T)$ from the cross sections of Chambaud et al.~(1980) and expressed them in the form of broken power laws with temperature. We have created smooth fits to these power laws to eliminate the discontinuities in the first derivatives of these expressions at the temperature boundaries (mostly for aesthetic reasons).

The spontaneous decay rates $A_{1,2}=8.865\times 10^{-5}\,{\rm s}^{-1}$ and $A_{0,1}=1.772\times 10^{-5}\,{\rm s}^{-1}$ are from Galavís et al.~(1997). Values for $A$ that apply to the metastable levels are from the compilation of van Hoof (2018). Silva \& Viegas (2002) calculated the rates of optical pumping from the ground $J=2$ level to the two upper levels, $\Gamma_{2,1}=3.9\times 10^{-11}\,{\rm s}^{-1}$ and $\Gamma_{2,0}=1.1\times 10^{-11}\,{\rm s}^{-1}$, when oxygen atoms are exposed to the average Galactic UV radiation field and there is no self shielding. (Again, for our particular application, we set the optical pumping rates to zero since the permitted O~{\sc i} transitions out of the $J=2$ level are optically thick.)

\begin{figure}
\centering
\includegraphics[width=0.8\textwidth]{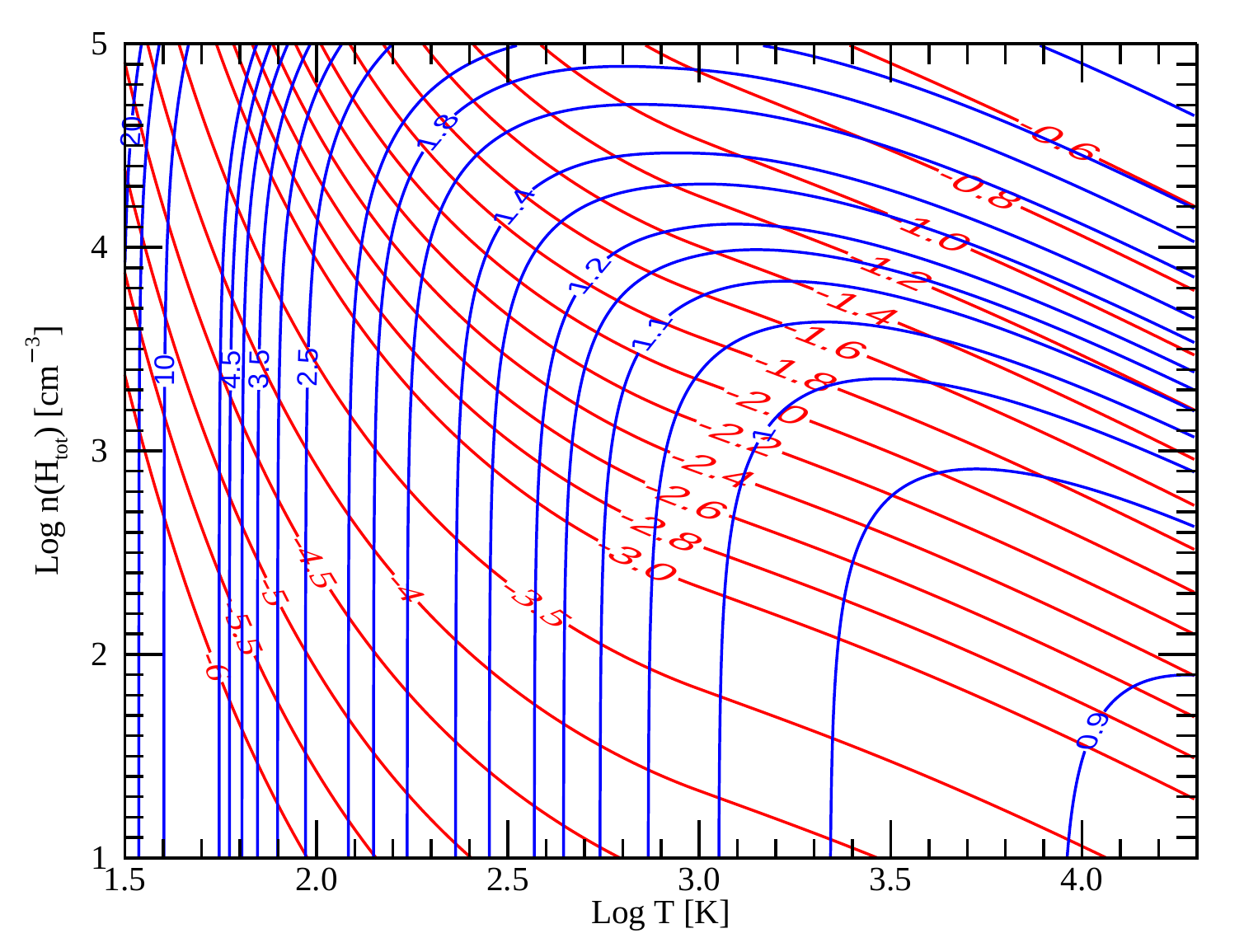}
\caption{Contour diagram illustrating the physical conditions that apply to different levels of $n({\rm O~\textsc{i}\text{*}})/ n({\rm O~\textsc{i}\text{**}})$ (blue contours) and $\log \big[\big(n({\rm O~\textsc{i}\text{*}})+n({\rm O~\textsc{i}\text{**}})\big)\big/\big(n({\rm O~\textsc{i}}) + n({\rm O~\textsc{i}\text{*}}) + n({\rm O~\textsc{i}\text{**}})\big)\big]$ (red contours). In the calculations used to produce this figure, we have adopted $f({\rm H}_2)=0.3$ (with $T_{0,1}=77\,$K) and $x(e)=0.002$. Note the rapid increase in temperature that occurs as $n({\rm O~\textsc{i}\text{*}})/ n({\rm O~\textsc{i}\text{**}})$ approaches a value of 0.9.\label{fig:contours}}
\end{figure}

For the various channels of charge exchange in our equations, we used the coefficients $c_{+,j^\prime} (T)$ given by Stancil et al.~(1999) for populating the $j^\prime$ levels when oxygen ions are neutralized by collisions with H$^0$, and once again used the principle of detailed balance to derive the reverse processes for depopulating the levels of neutral oxygen atoms,
\begin{mathletters}
\begin{eqnarray}
c_{0,+}(T)&=&8c_{+,0}(T)\exp(97\,{\rm K}/T)~,\\
c_{1,+}(T)&=&(8/3)c_{+,1}(T)\exp(-1\,{\rm K}/T)~,\\
c_{2,+}(T)&=&(8/5)c_{+,2}(T)\exp(-229\,{\rm K}/T)~.
\end{eqnarray}
\end{mathletters}
We represent the recombination coefficients $\alpha_e(T)$ in terms of a sum of the radiative recombination coefficients $\alpha_{e,{\rm RR}}(T)$ and the dielectronic recombination coefficients $\alpha_{e,{\rm DR}}(T)$ computed by Badnell (2006). The recombinations with free electrons are much less frequent than the charge exchange reactions with neutral hydrogen, except when the ionization fraction of the gas is unusually large. Also, $\alpha_{e,{\rm DR}}(T) \ll \alpha_{e,{\rm RR}}(T)$ for $T \lesssim {\rm a~few}\,\times 10^3\,$K.

Figure~\ref{fig:contours} demonstrates how the ratios of the two excited O~{\sc i} fine-structure levels with respect to each other, $n({\rm O~\textsc{i}\text{*}})/ n({\rm O~\textsc{i}\text{**}})$, and the logarithm of the sum of the excited level populations with respect to the total amount of O~{\sc i}, $\log \big[\big(n({\rm O~\textsc{i}\text{*}})+n({\rm O~\textsc{i}\text{**}})\big)\big/\big(n({\rm O~\textsc{i}}) + n({\rm O~\textsc{i}\text{*}}) + n({\rm O~\textsc{i}\text{**}})\big)\big]$, change with different combinations of $\log T$ ($x$-axis) and $\log n({\rm H}_{\rm tot})$ ($y$-axis). In producing this figure, we adopted a molecular hydrogen fraction of $f({\rm H}_2)=2n({\rm H}_2)/[n({\rm H}^0)+2n({\rm H}_2)]=0.3$, with equal concentrations of ortho- and para-H$_2$ (i.e., $T_{0,1}=77\,$K), and an electron fraction of $x(e)=n(e)/n({\rm H}_{\rm tot})=0.002$. These values for $f({\rm H}_2)$ and $x(e)$ apply to the main low velocity component toward HD~43582. For the higher velocity, lower column density material toward HD~43582, we assume that $f({\rm H}_2)=0.0$, while, for the line of sight to HD~254755, we adopt a value of $f({\rm H}_2)=0.5$.

\section{TIME VARIABLE NA~{\sc i} ABSORPTION TOWARD HD~43582\label{sec:temp_var}}
\begin{figure}
\centering
\includegraphics[width=1.0\textwidth]{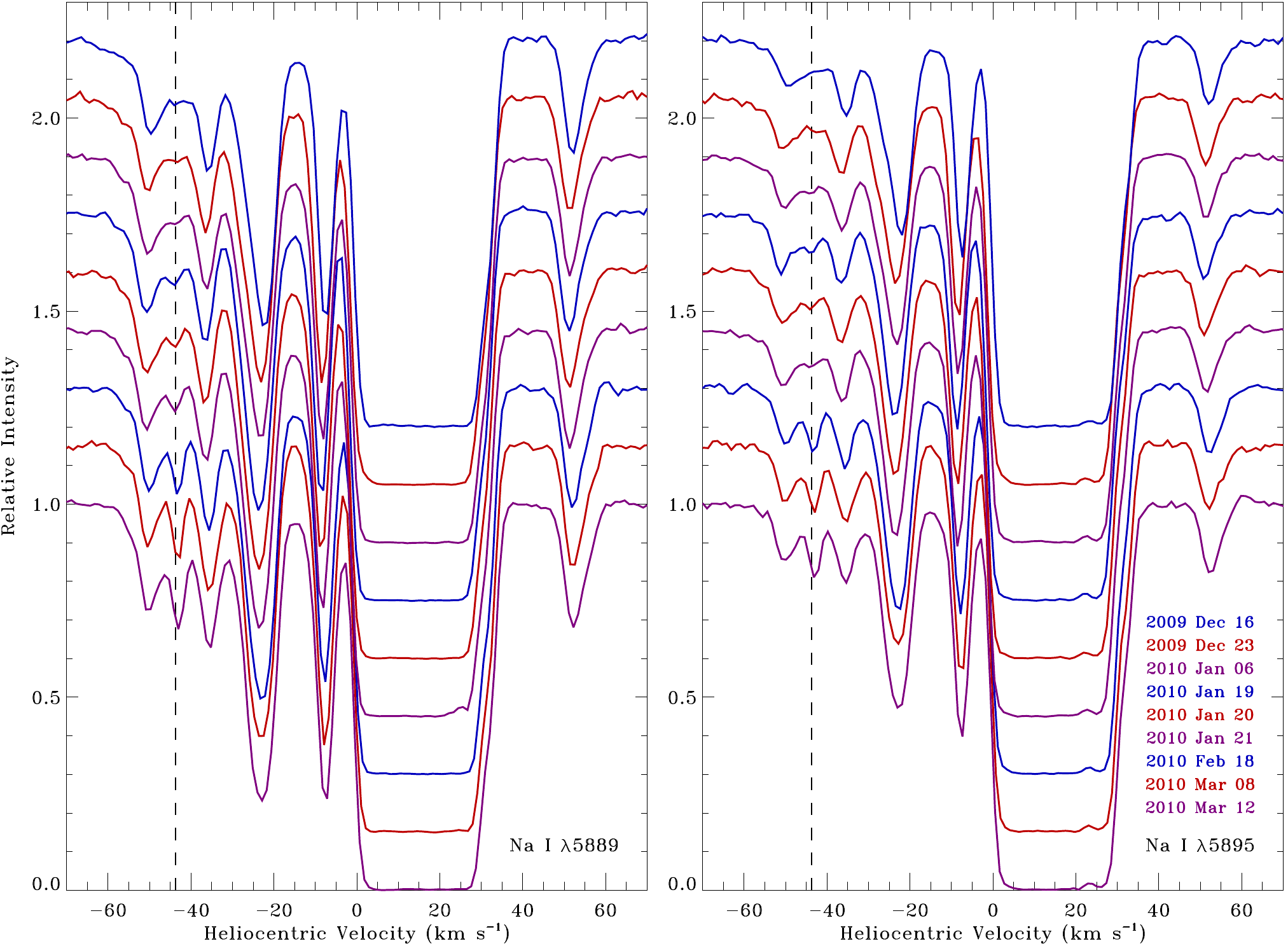}
\caption{Time series of the Na~{\sc i}~$\lambda5889$ and $\lambda5895$ absorption profiles toward HD~43582 (left and right panels, respectively) from HET/HRS data acquired between 2009 December and 2010 March. (The dates shown in the figure correspond to the UT dates of the observations.) A time variable absorption component is identified near $-$44~km~s$^{-1}$ as indicated by the vertical dashed line in each panel.\label{fig:temp_var}}
\end{figure}

In Figure~\ref{fig:temp_var}, we plot the absorption profiles of the Na~{\sc i}~$\lambda\lambda5889,5895$ lines observed toward HD~43582 on nine separate occasions between 2009 December and 2010 March. A time variable absorption component is identified near $-$44~km~s$^{-1}$ (as indicated by the vertical dashed lines in the two panels). The component seems to grow in strength, relative to the two adjacent components that appear at higher and lower velocities, over the course of approximately three months, and the evolution is clearly seen in both lines of the Na~{\sc i} doublet. To quantify the temporal evolution of the variable component, we fit the Na~{\sc i} absorption profiles from each observation separately, adopting the same initial component solution for each fit. (In what follows, we ignore the absorption components that constitute the saturated cores of the profiles.) We find that the column density of the component near $-$44~km~s$^{-1}$ increases by 0.42~dex ($\sim$160\%) over the course of the observations, from a low of $\log N($Na~{\sc i}$)=11.04\pm0.03$ on 2009 December 23 to a high of $\log N($Na~{\sc i}$)=11.46\pm0.03$ on 2010 March 12. The increase in column density may be accompanied by a slight decrease in $b$-value from $\sim$2 to $\sim$1~km~s$^{-1}$, although the velocity of the component is stable at $-$$43.7\pm0.5$~km~s$^{-1}$. None of the other Na~{\sc i} absorption components observed toward HD~43582 exhibit any significant variations.

Time variable Na~{\sc i} (and Ca~{\sc ii}) absorption components have been identified in other SNRs, such as the Vela SNR (Hobbs et al.~1991; Cha \& Sembach 2000; Welty et al.~2008; Rao et al.~2016) and the Monoceros Loop (Dirks \& Meyer 2016). Often, the temporal variations are associated with high-velocity gas, although temporal changes in low-velocity Na~{\sc i} absorption have also been observed (e.g., Rao et al.~2017). The steady increase in the column density of the Na~{\sc i} component at $-$44~km~s$^{-1}$ toward HD~43582 is notable because it occurs on such a short timescale. Most previous observations of temporal variability have tracked changes occurring on timescales of a few years rather than a few months. However, there is evidence that the changes seen in the variable component toward HD~43582 date back to at least 2002. There is a conspicuous lack of absorption near $-$44~km~s$^{-1}$ in the Na~{\sc i} spectra toward HD~43582 obtained by Welsh \& Sallmen (2003) in 2002 December. Those authors fit Na~{\sc i} absorption components at $-$37.2 and $-$48.0~km~s$^{-1}$, similar to the components we find at $-$37.3 and $-$50.5~km~s$^{-1}$, while finding no component near $-$44~km~s$^{-1}$, despite their having higher velocity resolution than we do (i.e., 1.8~km~s$^{-1}$ versus our 3.1~km~s$^{-1}$). On the other hand, the Welsh \& Sallmen (2003) data have much lower S/N than our HET spectra (i.e., $\sim$15 versus our value of $\sim$350 for the summed Na~{\sc i} spectrum toward HD~43582). Still, the lack of a Na~{\sc i} absorption component at $-$44~km~s$^{-1}$ in the Welsh \& Sallmen (2003) spectra is apparent. The steady increase in the Na~{\sc i} column density of the $-$44~km~s$^{-1}$ component toward HD~43582 reinforces our interpretation that this and other similar components trace recombining material far downstream from the SN shock front.

\begin{figure}[!t]
\centering
\includegraphics[width=0.49\textwidth]{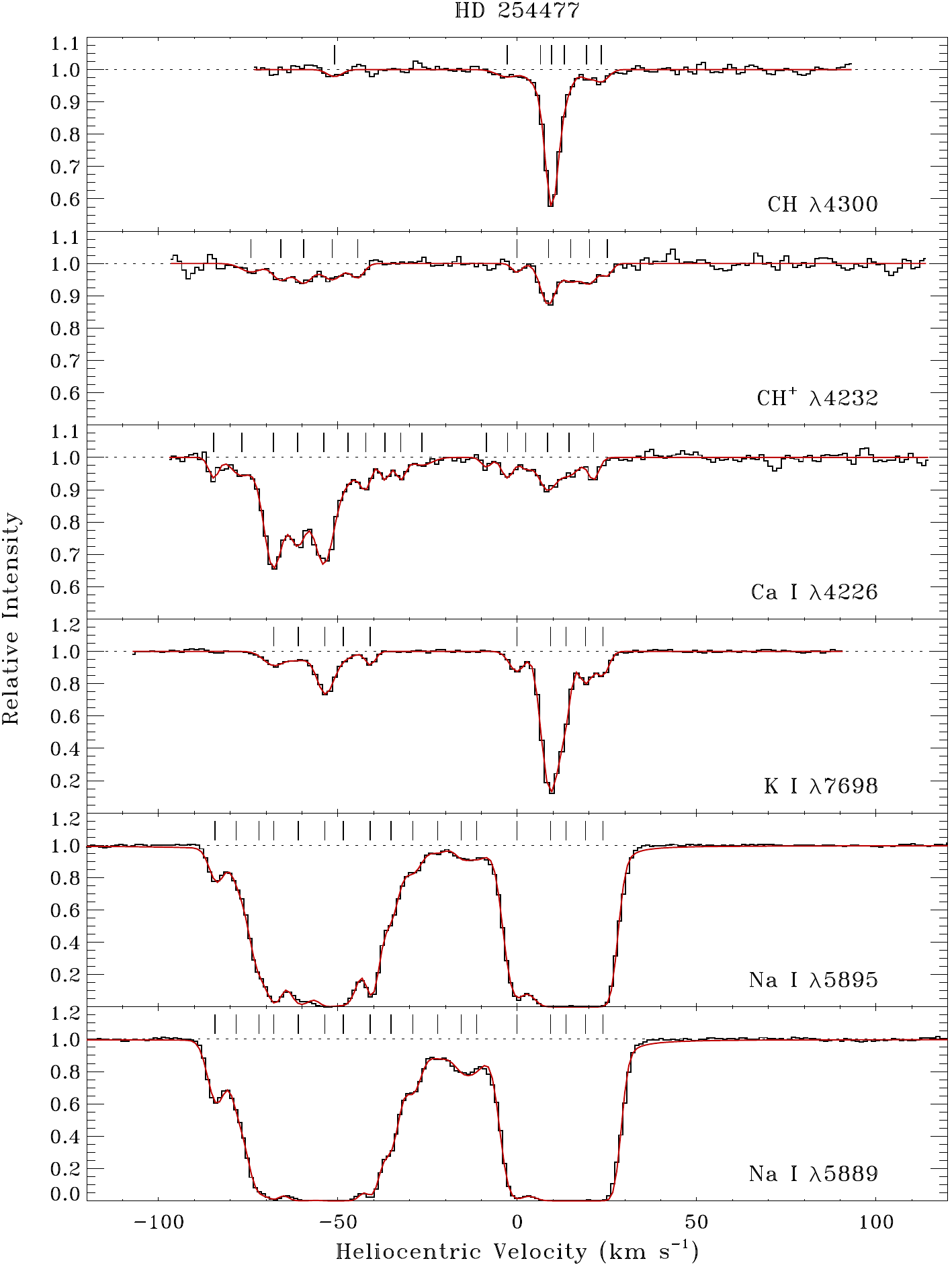}
\includegraphics[width=0.49\textwidth]{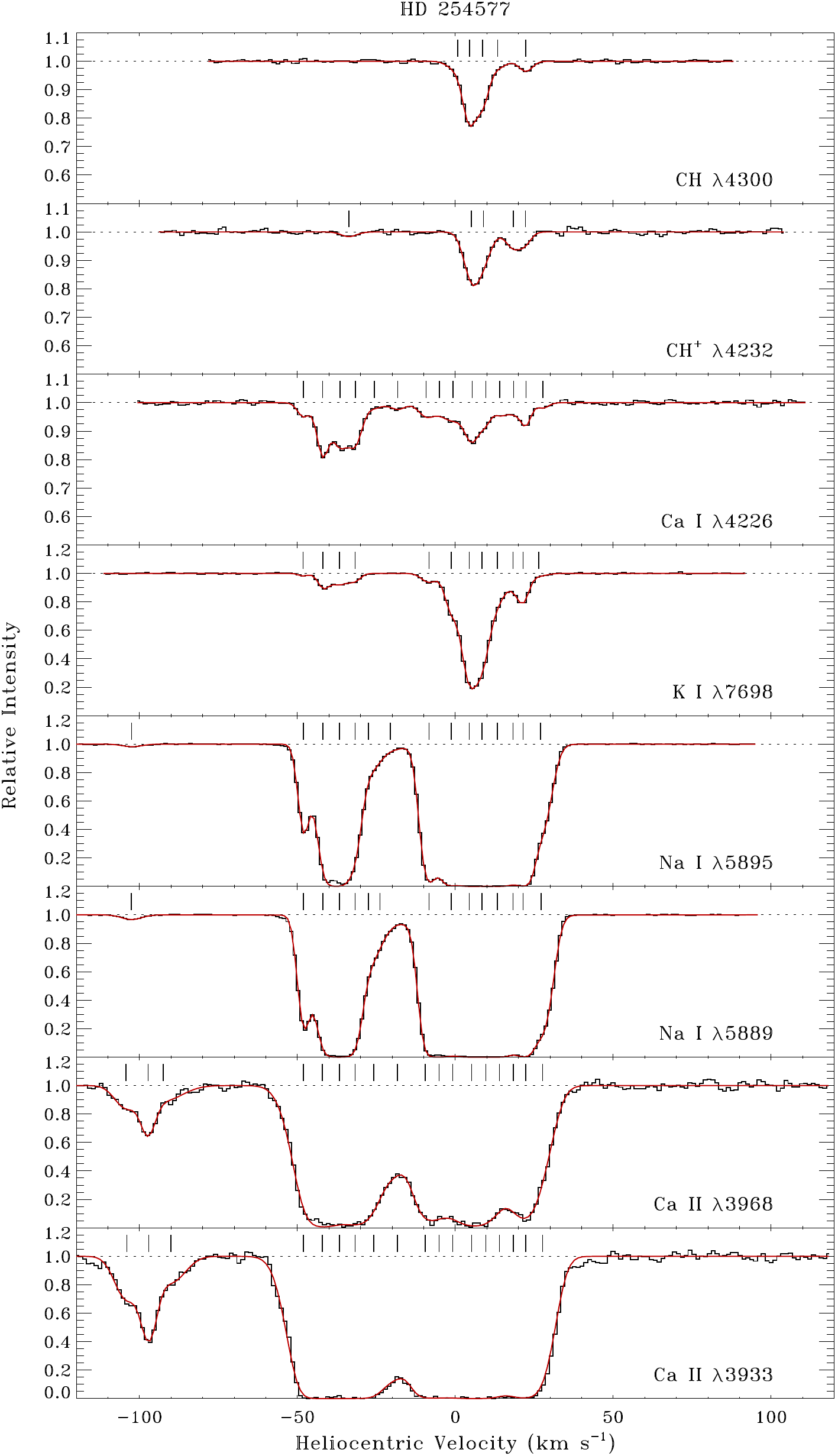}
\caption{Absorption profiles of the atomic and molecular species observed toward HD~254477 (left panels) and HD~254577 (right panels) from ground-based data acquired by Taylor et al.~(2012) using the HET/HRS. The smooth red curves represent multi-component profile synthesis fits to the observed spectra shown as histograms. Tick marks indicate the positions of the velocity components included in the fits. (The HET data for HD~254477 do not include the Ca~{\sc ii}~$\lambda\lambda3933,3968$ lines so those panels have been omitted.)\label{fig:het_profiles}}
\end{figure}

\section{ABSORPTION PROFILES TOWARD HD~254477 AND HD~254577\label{sec:het_profiles}}
The molecular column densities of the main low velocity absorption components toward HD~254477 and HD~254577 were included in the analyses described in Section~\ref{subsec:mol_ex}. However, the HET/HRS observations of these stars (originally obtained by Taylor et al.~2012) contain numerous other atomic and molecular absorption features that may be of interest to researchers investigating the interaction taking place between IC~443 and the interstellar gas in its immediate vicinity. For context, the sight line to HD~254577 passes through a ridge of molecular material located between the shocked molecular clumps labelled C and D by Huang et al.~(1986), while the line of sight to HD~254477 probes the shocked molecular clump labelled B (see also Dickman et al.~1992; van Dishoeck et al.~1993; Hirschauer et al.~2009).

\begin{deluxetable}{lccccccccccccccc}
\tablecolumns{16}
\tablewidth{0pt}
\tabletypesize{\scriptsize}
\tablecaption{Detailed Component Structure from \emph{HST}/STIS Spectra\label{tab:uv_comp_struct}}
\tablehead{ \colhead{Star} & \multicolumn{3}{c}{Si~{\sc ii}*} & \colhead{} & \multicolumn{3}{c}{O~{\sc i}*} & \colhead{} & \multicolumn{3}{c}{O~{\sc i}**} & \colhead{} & \multicolumn{3}{c}{O~{\sc i}} \\
\cline{2-4} \cline{6-8} \cline{10-12} \cline{14-16} \\
\colhead{} & \colhead{$v_{\sun}$} & \colhead{$\log N$} & \colhead{$b$} & \colhead{} &\colhead{$v_{\sun}$} & \colhead{$\log N$} & \colhead{$b$} & \colhead{} & \colhead{$v_{\sun}$} & \colhead{$\log N$} & \colhead{$b$} & \colhead{} & \colhead{$v_{\sun}$} & \colhead{$\log N$} & \colhead{$b$} }
\startdata
HD 43582 & $-$98.3 & $11.53\pm0.19$ & 4.5 && \ldots & \ldots & \ldots && \ldots & \ldots & \ldots && \ldots & \ldots & \ldots \\
 & $-$86.9 & $11.27\pm0.18$ & 2.2 && \ldots & \ldots & \ldots && \ldots & \ldots & \ldots && \ldots & \ldots & \ldots \\
 & $-$80.6 & $12.26\pm0.07$ & 6.0 && \ldots & \ldots & \ldots && \ldots & \ldots & \ldots && \ldots & \ldots & \ldots \\
 & $-$73.0 & $12.27\pm0.07$ & 4.8 && \ldots & \ldots & \ldots && \ldots & \ldots & \ldots && \ldots & \ldots & \ldots \\
 & $-$64.7 & $11.78\pm0.10$ & 2.7 && \ldots & \ldots & \ldots && \ldots & \ldots & \ldots && \ldots & \ldots & \ldots \\
 & $-$52.5 & $12.78\pm0.04$ & 5.8 && $-$54.4 & $12.63\pm0.09$ & 2.6 && $-$56.3 & $12.47\pm0.11$ & 1.6 && \ldots & \ldots & \ldots \\
 & $-$47.5 & $13.10\pm0.03$ & 5.0 && $-$49.7 & $13.52\pm0.04$ & 3.0 && $-$50.2 & $13.47\pm0.04$ & 3.3 && \ldots & \ldots & \ldots \\
 & $-$41.5 & $12.80\pm0.03$ & 2.6 && $-$44.2 & $13.48\pm0.04$ & 2.8 && $-$43.8 & $13.54\pm0.05$ & 3.2 && \ldots & \ldots & \ldots \\
 & $-$37.7 & $12.90\pm0.03$ & 3.8 && $-$37.1 & $13.49\pm0.04$ & 3.7 && $-$38.0 & $13.27\pm0.04$ & 2.2 && \ldots & \ldots & \ldots \\
 & $-$33.8 & $12.51\pm0.05$ & 3.6 && \ldots & \ldots & \ldots && \ldots & \ldots & \ldots && \ldots & \ldots & \ldots \\
 & $-$26.1 & $13.18\pm0.03$ & 3.4 && $-$29.0 & $13.75\pm0.05$ & 2.5 && $-$27.7 & $13.98\pm0.05$ & 5.4 && \ldots & \ldots & \ldots \\
 & $-$22.6 & $13.29\pm0.03$ & 3.3 && $-$22.7 & $14.20\pm0.07$ & 3.2 && $-$23.2 & $14.12\pm0.07$ & 3.3 && \ldots & \ldots & \ldots \\
 & $-$14.3 & $12.79\pm0.04$ & 4.8 && $-$13.0 & $13.60\pm0.04$ & 3.8 && $-$13.8 & $13.60\pm0.04$ & 4.0 && \ldots & \ldots & \ldots \\
 &  \phn$-$7.7 & $13.26\pm0.03$ & 2.2 &&  \phn$-$7.9 & $13.95\pm0.06$ & 2.7 &&  \phn$-$8.6 & $13.87\pm0.06$ & 2.8 && \ldots & \ldots & \ldots \\
 &  \phn$-$2.1 & $12.58\pm0.04$ & 1.9 &&  \phn$-$1.6 & $13.05\pm0.04$ & 1.9 &&  \phn$-$3.4 & $13.32\pm0.05$ & 4.3 && \ldots & \ldots & \ldots \\
 &    \phn+4.2 & $12.36\pm0.05$ & 3.1 &&    \phn+4.5 & $13.63\pm0.05$ & 2.4 &&    \phn+3.8 & $13.37\pm0.04$ & 2.4 &&  \phn+5.7 & $18.00\pm0.03$ & 3.1 \\
 &   +12.6 & $12.45\pm0.07$ & 6.7 &&   +11.5 & $12.86\pm0.07$ & 3.6 &&   +11.0 & $13.07\pm0.08$ & 5.9 &&  \phn+9.2 & $17.64\pm0.05$ & 4.8 \\
 & \ldots & \ldots & \ldots && \ldots & \ldots & \ldots && \ldots & \ldots & \ldots && +14.3 & $17.18\pm0.12$ & 4.8 \\
 & \ldots & \ldots & \ldots && \ldots & \ldots & \ldots && \ldots & \ldots & \ldots && +19.1 & $17.35\pm0.08$ & 4.4 \\
 &   +51.6 & $12.62\pm0.04$ & 2.5 &&   +52.0 & $12.99\pm0.05$ & 2.3 &&   +51.3 & $13.13\pm0.05$ & 3.4 && \ldots & \ldots & \ldots \\
 &   +96.4 & $11.71\pm0.13$ & 4.3 && \ldots & \ldots & \ldots && \ldots & \ldots & \ldots && \ldots & \ldots & \ldots \\
\enddata
\tablecomments{Units for $v_{\sun}$ and $b$ are km~s$^{-1}$.}
\tablecomments{(This table is available in its entirety in machine-readable form.)}
\end{deluxetable}

In Figure~\ref{fig:het_profiles}, we present fits to the absorption profiles of the atomic and molecular transitions covered by the HET observations of HD~254477 and HD~254577. (The HET data for HD~254477 do not include the Ca~{\sc ii}~$\lambda\lambda3933,3968$ lines so those panels have been omitted from the figure.) Both sight lines exhibit extensive absorption at high to moderately high negative velocity (although neither direction shows absorption at high positive velocity). Of particular note is the high-velocity complex seen in Na~{\sc i} and Ca~{\sc ii} absorption near $-$100~km~s$^{-1}$ toward HD~254577. The low $N$(Na~{\sc i})/$N$(Ca~{\sc ii}) ratio for this absorption complex suggests efficient dust destruction by shocks (similar to what we find for the high pressure components toward HD~43582). Also noteworthy are the clusters of relatively strong Ca~{\sc i} components at moderately high negative velocity toward both HD~254477 and HD~254577. The high $N$(Ca~{\sc i})/$N$(K~{\sc i}) ratios for these components, relative to the components at low velocity, are another indication of the destruction of dust grains by SN shocks. Hirschauer et al.~(2009) first reported the detection of moderately high velocity CH$^+$ absorption toward HD~254477. We confirm this detection and report tentative detections of a corresponding component in CH toward HD~254477 and a moderately high negative velocity component in CH$^+$ toward HD~254577. (For the details concerning these, and other, profile synthesis fits, see Appendix~\ref{sec:comp_structures}.)

\begin{deluxetable}{lccccccccccccccc}
\tablecolumns{16}
\tablewidth{0pt}
\tabletypesize{\scriptsize}
\tablecaption{Detailed Component Structure from HET/HRS Spectra\label{tab:vis_comp_struct}}
\tablehead{ \colhead{Star} & \multicolumn{3}{c}{Ca~{\sc ii}} & \colhead{} & \multicolumn{3}{c}{Ca~{\sc i}} & \colhead{} & \multicolumn{3}{c}{Na~{\sc i}} & \colhead{} & \multicolumn{3}{c}{K~{\sc i}} \\
\cline{2-4} \cline{6-8} \cline{10-12} \cline{14-16} \\
\colhead{} & \colhead{$v_{\sun}$} & \colhead{$\log N$} & \colhead{$b$} & \colhead{} &\colhead{$v_{\sun}$} & \colhead{$\log N$} & \colhead{$b$} & \colhead{} & \colhead{$v_{\sun}$} & \colhead{$\log N$} & \colhead{$b$} & \colhead{} & \colhead{$v_{\sun}$} & \colhead{$\log N$} & \colhead{$b$} }
\startdata
HD~43582 & $-$98.6 & $11.13\pm0.06$ & 4.6 && \ldots & \ldots & \ldots && \ldots & \ldots & \ldots && \ldots & \ldots & \ldots \\
 & $-$84.8 & $11.46\pm0.04$ & 4.9 && \ldots & \ldots & \ldots && \ldots & \ldots & \ldots && \ldots & \ldots & \ldots \\
 & $-$77.9 & $10.59\pm0.08$ & 1.2 && \ldots & \ldots & \ldots && \ldots & \ldots & \ldots && \ldots & \ldots & \ldots \\
 & $-$71.5 & $11.88\pm0.03$ & 4.6 && \ldots & \ldots & \ldots && \ldots & \ldots & \ldots && \ldots & \ldots & \ldots \\
 & $-$61.6 & $11.12\pm0.07$ & 4.6 && \ldots & \ldots & \ldots && \ldots & \ldots & \ldots && \ldots & \ldots & \ldots \\
 & $-$54.2 & $12.22\pm0.03$ & 6.6 && $-$55.6 &  \phn$8.84\pm0.27$ & 0.5 && $-$57.3 & $10.25\pm0.04$ & 0.6 && \ldots & \ldots & \ldots \\
 & $-$48.8 & $12.55\pm0.04$ & 5.3 && $-$49.0 & $10.08\pm0.05$ & 3.8 && $-$50.5 & $11.49\pm0.02$ & 3.2 && \ldots & \ldots & \ldots \\
 & $-$45.0 & $12.51\pm0.04$ & 5.5 && $-$42.8 &  \phn$9.79\pm0.05$ & 1.6 && $-$43.7 & $11.24\pm0.03$ & 2.0 && \ldots & \ldots & \ldots \\
 & $-$38.2 & $12.68\pm0.05$ & 4.2 && $-$37.4 &  \phn$9.81\pm0.06$ & 2.1 && $-$37.0 & $11.44\pm0.03$ & 2.3 && \ldots & \ldots & \ldots \\
 & $-$34.3 & $11.75\pm0.04$ & 2.8 && $-$33.8 &  \phn$9.41\pm0.10$ & 1.2 && $-$35.3 & $11.05\pm0.03$ & 3.2 && \ldots & \ldots & \ldots \\
 & $-$26.8 & $12.58\pm0.05$ & 3.8 && $-$27.6 &  \phn$9.38\pm0.10$ & 0.5 && $-$29.6 & $10.99\pm0.03$ & 1.7 && \ldots & \ldots & \ldots \\
 & $-$23.1 & $12.55\pm0.06$ & 3.5 && $-$24.1 & $10.31\pm0.02$ & 2.9 && $-$24.4 & $12.10\pm0.04$ & 2.9 && $-$24.2 &  \phn$9.94\pm0.08$ & 0.5 \\
 & $-$21.4 & $12.18\pm0.03$ & 6.5 && $-$21.9 & $10.01\pm0.03$ & 1.4 && $-$21.6 & $11.51\pm0.03$ & 1.8 && $-$20.7 &  \phn$9.58\pm0.16$ & 0.5 \\
 & $-$17.7 & $12.41\pm0.04$ & 7.6 && $-$17.7 &  \phn$9.61\pm0.07$ & 0.8 && $-$11.7 & $10.81\pm0.03$ & 5.0 && \ldots & \ldots & \ldots \\
 &  \phn$-$7.6 & $12.58\pm0.05$ & 3.9 &&  \phn$-$7.2 & $10.10\pm0.03$ & 2.4 &&  \phn$-$8.2 & $11.86\pm0.04$ & 1.3 && \ldots & \ldots & \ldots \\
 &    \phn+1.1 & $11.69\pm0.04$ & 2.5 &&    \phn+1.4 &  \phn$9.75\pm0.05$ & 1.3 && \ldots & \ldots\tablenotemark{a} & \ldots &&    \phn+1.0 & $10.96\pm0.02$ & 0.5 \\
 &    \phn+6.4 & $12.45\pm0.05$ & 3.3 &&    \phn+7.1 & $10.12\pm0.03$ & 2.3 && \ldots & \ldots\tablenotemark{a} & \ldots &&    \phn+6.4 & $11.97\pm0.05$ & 1.4 \\
 &    \phn+9.9 & $12.28\pm0.05$ & 2.6 &&   +10.3 &  \phn$9.83\pm0.06$ & 2.4 && \ldots & \ldots\tablenotemark{a} & \ldots &&    \phn+9.9 & $11.57\pm0.05$ & 0.6 \\
 &   +15.1 & $12.33\pm0.04$ & 4.8 && \ldots & \ldots & \ldots && \ldots & \ldots\tablenotemark{a} & \ldots &&   +15.0 & $11.08\pm0.01$ & 1.6 \\
 &   +19.9 & $12.19\pm0.03$ & 4.6 &&   +19.0 & $10.02\pm0.06$ & 4.3 && \ldots & \ldots\tablenotemark{a} & \ldots &&   +19.8 & $10.86\pm0.02$ & 2.0 \\
 &   +26.1 & $12.26\pm0.04$ & 4.5 && \ldots & \ldots & \ldots && \ldots & \ldots\tablenotemark{a} & \ldots &&   +25.2 & $10.78\pm0.02$ & 0.9 \\
 &   +29.3 & $11.17\pm0.05$ & 4.4 &&   +28.4 &  \phn$9.49\pm0.08$ & 0.9 && \ldots & \ldots\tablenotemark{a} & \ldots &&   +28.7 & $10.41\pm0.03$ & 0.5 \\
 &   +52.0 & $12.02\pm0.04$ & 2.2 &&   +52.2 &  \phn$9.33\pm0.11$ & 0.5 &&   +51.4 & $11.23\pm0.03$ & 1.8 && \ldots & \ldots & \ldots \\
 &   +56.0 & $11.50\pm0.03$ & 2.1 &&   +54.6 &  \phn$9.10\pm0.18$ & 0.7 &&   +55.2 & $10.54\pm0.03$ & 1.1 && \ldots & \ldots & \ldots \\
 &   +94.3 & $10.83\pm0.06$ & 2.1 && \ldots & \ldots & \ldots && \ldots & \ldots & \ldots && \ldots & \ldots & \ldots \\
 &   +99.4 & $10.98\pm0.07$ & 4.0 && \ldots & \ldots & \ldots && \ldots & \ldots & \ldots && \ldots & \ldots & \ldots \\
\enddata
\tablecomments{Units for $v_{\sun}$ and $b$ are km~s$^{-1}$.}
\tablenotetext{a}{Strongly saturated component.}
\tablecomments{(This table is available in its entirety in machine-readable form.)}
\end{deluxetable}

\section{DETAILED COMPONENT STRUCTURES\label{sec:comp_structures}}
In Table~\ref{tab:uv_comp_struct}, we present detailed profile fitting results for many of the neutral and singly-ionized atomic species seen in our \emph{HST}/STIS spectra of HD~43582 and HD~254755. In Table~\ref{tab:vis_comp_struct}, we provide the results of fitting the absorption profiles of the atomic and molecular species covered by the HET/HRS observations of HD~43582, HD~254477, HD~254577, and HD~254755. For species with more than one available transition (see Table~\ref{tab:column_densities}), the velocities and $b$-values listed in Tables~\ref{tab:uv_comp_struct} and \ref{tab:vis_comp_struct} are average values, and the column densities are weighted mean values, from (in most cases) independent fits to the different transitions. (For C~{\sc i}, C~{\sc i}*, and C~{\sc i}**, a global profile synthesis fit to all available multiplets yielded a single solution for the component structure in each of the three fine-structure levels.) Results for strongly saturated absorption components (i.e., for the main low velocity components in S~{\sc ii} and Na~{\sc i}) are omitted from Tables~\ref{tab:uv_comp_struct} and \ref{tab:vis_comp_struct}. A detailed description of the techniques used in fitting the various absorption profiles is provided in Section~\ref{subsubsec:components}. The fits themselves are presented in Figures~\ref{fig:dominant1}--\ref{fig:ci2} and Figure~\ref{fig:het_profiles}.

\end{document}